\documentclass[singlespace]{myeasychithesis2}

\newcommand{\da}{d_A}

\newcommand{\tot}{{\rm t}}
\newcommand{\hal}{{\rm h}}

\newcommand{\lin}{{\rm lin}}
\newcommand{\lens}{{\rm lens}}
\newcommand{\ang}{{\rm def}}
\newcommand{\shell}{{\rm s}}
\newcommand{\dirac}{{\rm D}}
\newcommand{\halo}{{\rm h}}
\newcommand{\ngau}{{\rm NG}}
\def\C{{\bf C}}
\def\Ch{\widehat{\C}}
\newcommand{\nm}{\frac{d\bar{n}}{dM}}
\newcommand{\nma}{\frac{d\bar{n}}{dM_1}}
\newcommand{\nmb}{\frac{d\bar{n}}{dM_2}}
\newcommand{\nmc}{\frac{d\bar{n}}{dM_3}}
\newcommand{\nmd}{\frac{d\bar{n}}{dM_4}}

\newcommand{\bp}{{\cal C}} 

\newcommand{\bfx}{{\mathbf{x}}}
\newcommand{\bfl}{{\mathbf{l}}}

\newcommand{\bfk}{{\mathbf{k}}}

\newcommand{\veck}{{\bf k}}
\newcommand{\vecl}{{\bf l}}
\newcommand{\vecr}{{\bf r}}
\newcommand{\vecx}{{\bf x}}

\newcommand{\rms}{{\it rms}}
\newcommand{\cmb}{\Theta}

\newcommand{\Cov}{{\rm Cov}}

\newcommand{\fsky}{f_{\rm sky}}

\newlength{\tskip}
\newlength{\colwidth}

\newcommand{\beq}{\begin{equation}}
\newcommand{\eeq}{\end{equation}}
\newcommand{\beqa}{\begin{eqnarray}}
\newcommand{\eeqa}{\end{eqnarray}}
\def\simgt{\gtrsim}
\newcommand{\fore}{{\rm f}}  
 \newcommand{\wj}{\left(
                          \begin{array}{ccc}
                          l_1  &  l_2  & l_3 \\
                            0  &  0    &  0
                          \end{array}
                          \right)}
\newcommand{\wjmp}[3]{\left(
                       \begin{array}{ccc}
       l_#1 & l_#2  & l_#3  \\
         m_#1 & m_#2  & m_#3
                         \end{array}
                   \right)}
\newcommand{\wjm}{\left(
                          \begin{array}{ccc}
                          l_1  &  l_2  & l_3 \\
                           m_1  &  m_2   &  m_3
                          \end{array}
                          \right)}
\newcommand{\wjma}[6]{\left(
                           \begin{array}{ccc}
         #1 & #2  & #3  \\
         #4 & #5  & #6
                           \end{array}
                   \right)}

\newcommand{\bi}{B_{l_1 l_2 l_3}}

\newcommand{\deld}{\delta^{\rm D}}
\newcommand{\bn}{\hat{\bf n}}
\newcommand{\bm}{\hat{\bf m}}
\newcommand{\bl}{\hat{\bf l}}
\newcommand{\bk}{\hat{\bf k}}
\newcommand{\rad}{r}    % comoving radial distance

\newcommand{\sky}{{\rm sky}}

\newcommand{\se}{{\rm S}}
\newcommand{\ri}{{\rm ri}}

\newcommand{\Ylm}[1]{Y_{l_#1}^{m_#1}}
\newcommand{\Ylmn}{Y_{l}^{m}}
\newcommand{\alm}[1]{a_{l_#1 m_#1}}
\newcommand{\almn}{a_{l m}}

\newcommand{\dsz}{{\rm kSZ}}
\newcommand{\sz}{{\rm SZ}}
\newcommand{\sn}{\frac{{\rm S}}{{\rm N}}}
\newcommand{\noise}{{\rm noise}}
\newcommand{\g}{{\rm G}}
\newcommand{\gal}{{\rm gal}}

\begin{document}

\title{Applications of Halo Approach to Non-Linear Large Scale
Structure Clustering}
\author{Asantha Roshan Cooray}
\date{June 2001}
\department{Astronomy \& Astrophysics}
\division{Physical Sciences}
\degree{Doctor of Philosophy}
\maketitle

\dedication
\vspace{5in}
\begin{center}
Copyright $\copyright$ 2001 by Asantha Roshan Cooray\\
All rights reserved.
\end{center}

\topmatter{Abstract}
We present astrophysical applications of the recently popular halo
model to describe large scale structure clustering. We formulate the
power spectrum, bispectrum and trispectrum of dark matter density
field in terms of correlations within and between dark matter
halos. The halo approach uses results from  numerical simulations and involves
a profile for dark matter, a mass function for halos, 
and a description of halo biasing with respect to the
linear density field. This technique can easily be extended to
describe clustering of any property of the large scale structure, such
as galaxies, baryons and pressure, provided that one formulate the
relationship between such properties and dark matter. 
We discuss applications of the halo model for 
several observational probes of the local universe involving  weak
gravitational lensing, thermal Sunyaev-Zel'dovich (SZ) effect and the
kinetic SZ effect.

With respect to weak gravitational lensing, we study the generation of
non-Gaussian signals which are potentially observable in galaxy shear
data. We study the three and four-point statistics, 
specifically the bispectrum and trispectrum, of the convergence
using the dark matter halo approach. Our approach 
allows us to study the effect of the mass distribution in
observed fields, in particular the bias induced by the lack of rare massive
halos (clusters). At low redshifts, the non-linear gravitational
evolution of large scale structure also produces a non-Gaussian covariance
in the shear power spectrum measurements that affects their
translation into cosmological  parameters.  Using the dark matter halo
approach, we study the covariance of binned band power spectrum estimates.
We compare this semi-analytic estimate to results from N-body
numerical simulations and find a good agreement. We find that for a
survey out to z $\sim$ 1, the power spectrum
covariance increases the errors on cosmological parameters determined under the
Gaussian assumption by about 15\%. Through a description of galaxies
in halos, we comment  on the recent measurement of weak
lensing tangential shear-galaxy correlation function.

Extending applications of the halo model to cosmic microwave
background temperature fluctuations, we discuss non-Gaussian effects
associated with the thermal SZ effect.
The non-Gaussianities here arise from the existence
of a four-point correlation function in large scale pressure fluctuations.
Using the  pressure trispectrum calculated under the halo model,
we discuss the full covariance of the SZ thermal power spectrum,
beyond the Gaussian sample variance. We
use this full covariance matrix to study the astrophysical uses of
the SZ effect and discuss the extent to which gas properties can be
derived from the SZ power spectrum. With the SZ thermal effect
separated in CMB temperature fluctuations using
its frequency information, a map with a thermal spectrum is 
expected to be dominated at small angular
scales by the kinetic SZ  effect. The kinetic SZ effect arises from
the density modulation of the Doppler effect due to the motion of
scatterers in the rest frame of CMB photons. 
The presence of the SZ kinetic effect
can be determined through a cross-correlation between
frequency-separated SZ and CMB
maps; since the SZ kinetic effect is second order, contributions to
such a cross-correlation arise, to the lowest order, in the form of a 
bispectrum. Here, we suggest an additional statistic
involving the power spectrum of the squared temperatures, instead of
the usual temperature itself. Through a signal-to-noise
calculation, we show that future small angular scale multi-frequency
CMB experiments, sensitive to multipoles of a few thousand,
 will be able to measure the cross-correlation of pressure traced by
SZ thermal and baryons traced by SZ kinetic effect through 
a power spectrum of the squared temperatures.

In addition to measures involving
statistical properties of the individual effects,
we also consider the astrophysical uses of the dark matter halo
spatial distribution in
wide-field survey images and propose the measurement of the angular
power spectrum involved with halo clustering.
 Using the shape of the linear power spectrum as a standard ruler,
we find that a survey on 4000 deg.$^2$ scales
provide enough information for a useful determination of the angular
diameter distance as a function of redshift, independent of any
unknowns that may be associated with the halo mass function or halo
bias.
Under a cosmological model and reasonable prior knowledge on halo
bias, we show that adequate ($\sim$ 20\%) information can be
obtained on the equation of state of
 an additional energy density component.

\topmatter{Acknowledgments}

I am grateful to my advisor, Wayne Hu, for
suggesting many of the problems  and calculations presented in this thesis.
The large number of hours I have spent with him, initially over e-mail
and phone while he  was at the Institute and later at the chalk board in his LASR office, has certainly
been helpful over the last two years. I am also extremely grateful to
him for his role in my understanding of theoretical issues related to cosmic
microwave background and large scale structure. 

I thank my other thesis committee members, John Carlstrom, Scott
Dodelson and Don York for their guidance and helpful suggestions.
I thank John for introducing me to the Sunyaev-Zel'dovich
effect during his SZ experiments and Don for
introducing me to ultraviolet spectroscopy and absorption lines during
the work related to FUSE observations of low redshift AGNs. I also
thank Don for helping me out during a dark period of my graduate
student life, in between my failed attempts at experimental work
and the recovery to do some theory based studies. During the last two
summers, Don played a major role in my outreach activities with minority high school students from
Chicago Public Schools.  This was probably the best teaching experience I ever had over these
years; I am grateful to him and Duel Richardson for giving me the opportunity.
I found another Don (Lamb) to be helpful on various number of
issues time to time.  As usual, Sandy (Heinz) provided full support
with all administrative matters. Daily computer related questions went
to John (Valdes) who was always around, especially at 9 pm when he
was most needed.

I thank Andr\'e Fletcher, formerly a graduate student at MIT, 
for introducing me to basics of astrophysical
research during my undergraduate years. Most of my early work  related to planetary sciences was
conducted under Jim Elliot at MIT and he has made sure since then that I
continue to work in some field of astronomy. I am grateful for his occasional, but
sometimes much needed, advice and help. I thank Jean Quashnock and Coleman
Miller for early work on gravitational lensing statistics and Dan
Reichart, my former officemate, for long discussions ranging from galaxy clusters to
gamma-ray bursts. My other officemate, Shanquin Zhan, takes credit
for introducing me to the Nasdaq 100, well before the subsequent
burst. I also thank Daniel Eisenstein,
Lloyd Knox and Zoltan Haiman for collaborative work related to the
far-infrared background.

With regards to topics discussed in this thesis, I acknowledge useful
discussions  and/or collaborative work with  Jordi 
Miralda-Escud\'e, Gil Holder, Dragan Huterer, Joe Mohr, Ryan Scranton,
Roman Scoccimarro, Uros Seljak, Ravi Sheth,  Max Tegmark and Matias Zaldarriaga.
Max Tegmark and Bhuvnesh Jain refereed two of the papers presented in this 
thesis and suggested some additional work which we have since then considered.
Max is also acknowledged for his help during the writing of our paper
involving the separation of the SZ effect in CMB data using its
frequency dependence. Our series of publications using the halo model began with a collaborative project 
involving Jordi, and I grateful for him to suggesting the halo approach to 
large scale structure statistics.  I am also grateful 
to Ravi Sheth for lengthy discussions and his indirect contributions
to our papers.  I thank Marc Kamionkowski for inviting us to 
submit a review article on the halo model to be published in 
Physics Reports; clearly, he has given me a good reason to write 
this thesis.

Finally, Djuna made my daily life at Chicago perfect.
She has always forgiven me for extra long hours I spent with my laptop when
writing this thesis and many papers on it. 
Daily walks with Aubila (our dog) gave me enough
opportunities to reflect on research. Just as Chicago was an
amazing adventure for all of us, we are now looking forward to
camping trips in San Gabriels. 

During the four years at Chicago, I was supported by grants to John
Carlstrom, Don York, a McCormick Fellowship, many teaching
assistantships, and a Grant-In-Aid of
Research from Sigma Xi, the national science honor society.

\tableofcontents

%
% List of figures
%

\listoffigures

%
% List of tables
%

\listoftables

%
% Begin Body
%
\mainmatter

\chapter{General Overview}

\section{Introduction}

This thesis presents astrophysical applications of a novel approach to study the non-linear clustering of
dark matter and other physical properties 
of the low redshift large scale structure. We use the spatial distribution
of halos to write correlation functions of various properties, such as
the dark matter, through clustering within and between halos. Underlying this so-called 
halo approach is the assertion that dark matter halos are 
locally biased tracers of density perturbations in the linear
regime. Necessary ingredients for this technique comes from numerical
simulations and involve halo profiles (e.g., \cite{Navetal96} 1996), 
mass functions (e.g., \cite{PreSch74} 1974; \cite{SheTor99} 1999) and a
description of bias (e.g., \cite{Moetal97} 1997)
for these halos with respect to the linear density field.

This so-called halo model dates back to early 1950s with the publication of a
paper by  Neyman \& Scott (1952) where they described the
clustering of galaxies as a realization of a random distribution. 
The method has been developed over the years by Peebles (1974),
McClelland \& Silk (1978) and Scargle (1981), though
most of the early work was limited with respect to their predictive
power given the limited knowledge on the distribution of dark matter
and galaxies in individual halos 
(see, \cite{Pee01} 2001 for a historical overview on the
developments related to clustering studies of large scale structure).
The modern version of the halo-model was first written down by
\cite{SchBer91} (1991) and included the fact that halos themselves are clustered following the linear density field, though  a complete
description of halo biasing did not exist till the late 90s (e.g.,
\cite{Moetal97} 1997). Further work related to the halo approach
includes in a series of papers by Sheth including \cite{SheJai97}
(1997) and \cite{SheLem99} (1999).
The advent of high resolution and larg e volume numerical
simulations, especially over the last few years, has now provided
necessary ingredients for detailed halo-based calculations. These high
resolution simulations have now provided adequate knowledge on the
halo dark matter profiles while large volume simulations have tested
halo mass functions over many decades in mass. Thus, it should not be a
surprise that the halo approach has resurfaced to become a popular
semianalytical tool for detailed studies on the clustering properties, and related
statistics, of the large scale structure. The recent
activities  with respect to the halo model began with publication of a
series of papers by \cite{Sel00} (2000), \cite{MaFry00b} (2000b), 
\cite{Cooetal00b} (2000b), and \cite{Scoetal00} (2000), among others.

During the last year, we (\cite{Cooetal00b} 2000b; \cite{Coo00} 2000; \cite{CooHu01a}
2001a; \cite{CooHu01b} 2001b) have extended the applications of the
halo model to consider clustering of dark matter and, thereby, make
observable predictions associated with weak gravitational lensing
observations. We have also applied this halo model for cosmic
microwave background (CMB) studies involving the 
thermal Sunyaev-Zel'dovich (SZ; \cite{SunZel80} 1980) effect associated with local large
scale structure pressure and potentially 
observable in CMB experiments sensitive to arcminute scale temperature
fluctuations.  Additionally, we have now
extended this model to consider the non-Gaussian effects associated
with both the thermal and the kinetic SZ effects. We will
present a detailed account of these applications in the present study.

This thesis is organized as following: In Chapter 1, we introduce the halo 
approach to clustering and discuss the dark matter density field power 
spectrum, bispectrum and trispectrum. We compare predictions related to 
power spectrum covariance with results from numerical simulations by 
\cite{MeiWhi99} (1999). In Chapters 2 and 3, we extend the discussion on 
dark matter clustering to discuss statistics of weak gravitational lensing 
and its covariance. Implications for cosmology are discussed in 
Chapter~3. In Chapters 4 to 6, we discuss applications of the halo model
to secondary effects in cosmic microwave background. In particular, we
discuss the thermal Sunyaev-Zel'dovich effect (Chapter 4), The kinetic
Sunyaev-Zel'dovich effect (Chapter 5) and the correlations between
thermal and kinetic Sunyaev-Zel'dovich effects (Chapter 6).
In Chpater~7, we briefly introduce a new cosmological test involving
the clustering properties of halos through the angular power spectrum.

The relevant work related to Chapters 1 to 3 could be found
in following papers:\\
Weak lensing power spectrum: Cooray, A., Hu, W., \& Miralda-Escud\'e, J.
2000, ApJ, 536, L9.\\
Weak lensing bispectrum: Cooray, A. \& Hu, W. 2001, ApJ, 548, 7.\\
Weak lensing trispectrum and covariance: Cooray, A. \& Hu, W. 2001,
ApJ in press (astro-ph/0012087). 

Related to Chapters 4 to 6, we refer the reader to following papers:\\
For an initial application of the halo model to thermal Sunyaev-Zel'dovich
effect: Cooray, A. 2000, Phys. Rev. D., 62, 103506.\\
For issues related to frequency separation of the SZ effect, in
multifrequency CMB experiments: Cooray, A., Hu, W., Tegmark, M. 2001,
ApJ, 540, 1.\\
For a detailed discussion of bispectra formed through non-linear mode correlations associated
with certain secondary effects, such as gravitational lensing of CMB
photons and the Ostriker-Vishniac effect: Cooray, A. \& Hu, W. 2000, ApJ, 534, 533.

The recent work related to non-Gaussianities in the thermal SZ effect and
the cross-correlations between thermal SZ and kinetic SZ effects, in
Chapters 4 to 6, will be published in a separate paper. The work
related to Chapter 7 is submitted for publication by Cooray, Hu, Huterer
and Joffre.

\section{General Properties}

We first review the properties of adiabatic CDM models relevant to
the present calculations. We then discuss the general
properties of the halo model as applied to the calculation of the
non-linear dark matter, baryon and pressure density field power
spectra of the local large scale structure.

\subsection{Adiabatic CDM Model}
The expansion rate for adiabatic CDM cosmological models with a
cosmological constant is
\begin{equation}
H^2 = H_0^2 \left[ \Omega_m(1+z)^3 + \Omega_K (1+z)^2
              +\Omega_\Lambda \right]\,,
\end{equation}
where $H_0$ can be written as the inverse
Hubble distance today $H_0^{-1} = 2997.9h^{-1} $Mpc.
We follow the conventions that
in units of the critical density $3H_0^2/8\pi G$,
the contribution of each component is denoted $\Omega_i$,
$i=c$ for the CDM, $g$ for the baryons, $\Lambda$ for the cosmological
constant. We also define the
auxiliary quantities $\Omega_m=\Omega_c+\Omega_g$ and
$\Omega_K=1-\sum_i \Omega_i$, which represent the matter density and
the contribution of spatial curvature to the expansion rate
respectively.

Convenient measures of distance and time include the conformal
distance (or lookback time) from the observer
at redshift $z=0$
\begin{equation}
\rad(z) = \int_0^z {dz' \over H(z')} \,,
\end{equation}
and the analogous angular diameter distance
\begin{equation}
\da = H_0^{-1} \Omega_K^{-1/2} \sinh (H_0 \Omega_K^{1/2} \rad)\,.
\end{equation}
Note that as $\Omega_K \rightarrow 0$, $\da \rightarrow \rad$
and we define $\rad(z=\infty)=\rad_0$.

The adiabatic CDM model possesses a two, three and four-point
correlations of the dark matter density field as defined in the usual
way
\begin{eqnarray}
\langle \delta(\veck_1) \delta(\veck_2) \rangle& = &(2\pi)^3
\delta_\dirac (\veck_{12}) P(k_1) \, , \\
\langle \delta(\veck_1) \delta(\veck_2)\delta(\veck_3)\rangle &=&
(2\pi)^3 \delta_\dirac (\veck_{123})
B(\veck_1,\veck_2,\veck_3) \, , \\
\langle \delta(\veck_1) \ldots \delta(\veck_4)\rangle_c &=&
(2\pi)^3 \delta_\dirac (\veck_{1234})
T(\veck_1,\veck_2,\veck_3,\veck_4) \, ,
\end{eqnarray}
where $\veck_{i\ldots j} = \veck_i + \ldots + \veck_j$ and
$\delta_\dirac$ is
the delta function not to be confused with the density perturbation.
Note that
the subscript $c$ denotes the connected piece, i.e. the
trispectrum is defined to be identically zero for a Gaussian field.
Here and throughout, we occasionally suppress the redshift dependence
where no confusion will arise.

In linear perturbation theory\footnote{It should be understood that
``$\lin$'' denotes here the
lowest non-vanishing order of perturbation theory for the object in
question.
For the power spectrum, this is linear perturbation theory; for the
bispectrum, this is second order perturbation theory, etc.},
\begin{equation}
\frac{k^3P^\lin(k)}{2\pi^2} = \delta_H^2 \left({k \over
H_0} \right)^{n+3}T^2(k) \, .
\end{equation}
We use the fitting formulae of Eisenstein \& Hu (1999) in evaluating
the
transfer function $T(k)$ for CDM models.
Here, $\delta_H$ is the amplitude of present-day density fluctuations
at the Hubble scale; we adopt the COBE normalization for
$\delta_H$ (Bunn \& White 1997).

The bispectrum in perturbation theory is given by
\footnote{The
kernels $F_n^{\rm s}$ are derived in \cite{Goretal86} (1986) (see,
equations A2 and A3 of \cite{Goretal86} 1986; note that their
$P_n\equiv F_n$), and we have written such that the symmetric form of
$F_n$'s are used. The use of the symmetric form accounts for the
factor of 2 in Eqs.~\ref{eqn:bpt} and factors of 4 and 6 in
(\ref{eqn:tript}).}
\begin{eqnarray}
B^\lin(\veck_p,\veck_q,\veck_r) &=& 2 F_2^{\rm
s}(\veck_p,\veck_q)P(k_p)P(k_q)
+ 2\; {\rm Perm.} \, , \nonumber \\
\label{eqn:bpt}
\end{eqnarray}
with $F_2^{\rm s}$ term given by second order gravitational
perturbation calculations.

Similarly, the perturbation theory trispectrum is
(\cite{Fry84} 1984)
\begin{eqnarray}
&& T^\lin =4\left[F_2^{\rm s}(\veck_{12},-\veck_1) F_2^{\rm
s}(\veck_{12},\veck_3\
)P(k_1)P(k_{12})P(k_3)
+ {\rm Perm.}\right] \nonumber \\
&&\quad + 6 \left[F_3^{\rm
s}(\veck_1,\veck_2,\veck_3)P(k_1)P(k_2)P(k_3) + {\rm Perm.}\right]
\, .
\label{eqn:tript}
\end{eqnarray}
The permutations involve a total
of 12 terms in the first set and 4 terms in the second set.
For the Sunyaev-Zel'dovich effect discussed here, we are more
interested in the clustering properties of pressure, rather than the
dark matter density field. We do not have a reliable way to calculate
the pressure power spectrum and higher order correlations
analytically. We will introduce the semi-analytic halo model for this
purpose following \cite{Coo00} (2000). The same is also true for the
baryon power spectrum, which is relevant for the kinetic SZ effect.

In linear theory, the density field may be scaled backwards to higher
redshift
by the use of the growth function $G(z)$, where
$\delta(k,r)=G(r)\delta(k,0)$ (Peebles 1980)
\begin{equation}
G(r) \propto {H(r) \over H_0} \int_{z(r)}^\infty dz' (1+z') \left(
{H_0
\over H(z')} \right)^3\,.
\end{equation}
Note that in the matter dominated epoch $G \propto a=(1+z)^{-1}$.

For fluctuation spectra and growth rates of interest here,
reionization of the universe
is expected to occur rather late $z_\ri \la 50$ such that the
reionized media is optically thin to Thomson scattering of CMB photons
$\tau \la 1$.
The probability of last scattering within $d \rad$ of $\rad$ (the
visibility function) is
\begin{equation}
g =  \dot \tau e^{-\tau} = X H_0 \tau_H (1+z)^2 e^{-\tau}\,.
\end{equation}
Here
$\tau(r) = \int_0^{\rad} d\rad \dot\tau$ is the optical depth out to
$r$,
$X$ is the ionization fraction,
\begin{equation}
       \tau_H = 0.0691 (1-Y_p)\Omega_g h\,,
\end{equation}
is the optical depth to Thomson
scattering to the Hubble distance today, assuming full
hydrogen ionization with
primordial helium fraction of $Y_p$.
Note that the ionization
fraction can exceed unity:
$X=(1-3Y_p/4)/(1-Y_p)$  for singly ionized helium,
$X=(1-Y_p/2)/(1-Y_p)$ for fully ionized helium.

Although we maintain generality in all derivations, we
illustrate our results with the currently favored $\Lambda$CDM
cosmological model. The parameters for this model
are $\Omega_c=0.30$, $\Omega_g=0.05$, $\Omega_\Lambda=0.65$, $h=0.65$,
$Y_p = 0.24$, $n=1$, $X=1$, with a normalization such that
mass fluctuations on the $8 h$ Mpc$^{-1}$
scale is  $\sigma_8=0.9$, consistent with observations on the
abundance of galaxy clusters (\cite{ViaLid99} 1999).  A reasonable
value is important since  higher order correlations is nonlinearly dependent
on the amplitude of the density field. We also use this $\Lambda$CDM
cosmology as the inputs for some of our calculations come from
numerical simulations for this or similar cosmology.

\section{Angular Spectra}
\label{sec:bispectrum}

In this thesis, we will discuss higher order correlations associated
with effects such as the weak gravitational lensing and the SZ effect.
The  bispectrum $\bi$ is the spherical harmonic transform of the three-point
correlation function just as the angular power spectrum $C_\ell$
is the transform of the two-point function.
In terms of the multipole moments of the
temperature fluctuation field $T(\hat{\bf n})$,
\begin{equation}
a_{lm} = \int d\bn T(\bn) \Ylmn {}^*(\bn)\,,
\end{equation}
the two point correlation function is given by
\begin{eqnarray}
C(\bn,\bm) &\equiv& \langle T(\bn) T(\bm) \rangle  \nonumber\\
           &=& \sum_{l_1 m_1 l_2 m_2} \langle \alm{1}^* \alm{2}
\rangle
               \Ylmn{}^*(\bn) \Ylmn(\bm)\,.
\label{eqn:twopoint}
\end{eqnarray}
Under the assumption that the temperature field is statistically
isotropic, the correlation is independent of $m$
\begin{eqnarray}
\langle \alm{1}^* \alm{2}\rangle = \deld_{l_1 l_2} \deld_{m_1 m_2}
        C_{l_1}\,,
\end{eqnarray}
and called the angular power spectrum.
Likewise the three point correlation function is given by
\begin{eqnarray}
B(\bn,\bm,\bl) &\equiv& \langle T(\bn)T(\bm)T(\bl) \rangle \\
               &\equiv&
                \sum % \wjm \bi
                \langle \alm{1} \alm{2} \alm{3} \rangle
                \Ylm{1}(\bn) \Ylm{2}(\bm)  \Ylm{3}(\bl)\,,\nonumber
\end{eqnarray}
where the sum is over $(l_1,m_1),(l_2,m_2),(l_3,m_3)$.
Statistical isotropy again allows us
to express the correlation in terms an $m$-independent function,
\begin{eqnarray}
\langle \alm{1} \alm{2} \alm{3} \rangle  = \wjm \bi\,.
\end{eqnarray}
Here the quantity in parentheses is the Wigner-3$j$ symbol.
Its orthonormality relation
\begin{eqnarray}
\sum_{m_1 m_2} \wjm \wjmp{1}{2}{4} = {
\deld_{\ell_3 \ell_4} \deld_{m_3    m_4    } \over 2\ell_3+1}\,,
\nonumber \\
\label{eqn:ortho}
\end{eqnarray}
implies
\begin{eqnarray}
\bi = \sum_{m_1 m_2 m_3}  \wjm
                \langle \alm{1} \alm{2} \alm{3} \rangle \,.
\label{eqn:bispectrum}
\end{eqnarray}

The angular bispectrum, $\bi$, contains all the information available
in the three-point correlation function.  For example, the skewness,
the collapsed three-point function of \cite{Hinetal95} (1995) and the
equilateral configuration statistic of \cite{Feretal98} (1998) can all
be expressed as linear combinations of the bispectrum
terms (see \cite{Ganetal94} 1994 for explicit expressions).

It is also useful to note its relation to the bispectrum defined on a
small flat section of the sky. In the flat sky approximation, the
spherical polar coordinates $(\theta,\phi)$ are replaced with
radial coordinates on a plane $(r=2\sin\theta/2 \approx \theta,\phi)$.
The Fourier variable conjugate to these coordinates is a 2D vector
${\bf l}$ of length $l$ and azimuthal angle $\phi_l$.  The expansion
coefficients of the Fourier transform of a given ${\bf l}$ is
a weighted sum over $m$ of the spherical harmonic moments of the
same $l$ (\cite{Whietal99} 1999)
\begin{equation}
a({\bf l}) = \sqrt{ 4\pi \over 2l+1} \sum_m i^{-m} a_{l m}
e^{im\phi_l}\,,
\end{equation}
so that
\begin{eqnarray}
\langle a^*({\bf l}_1) a({\bf l}_2) \rangle
 & = &  {2\pi \over l_1}\deld_{l_1,l_2} C_{l_1}
                \sum_m e^{im(\phi_{l_1}-\phi_{l_2})}
        \nonumber\\
 &\approx& (2\pi)^2 \deld({\bf l}_1 + {\bf l}_2) C_{l_1}\,.
\end{eqnarray}
Likewise  the 2D bispectrum is defined as
\begin{eqnarray}
\langle a({\bf l}_1) a({\bf l}_2) a({\bf l}_3) \rangle
&\equiv& (2\pi)^2 \deld({\bf l}_1+{\bf l}_2 + {\bf l}_3) B({\bf l}_1,
{\bf l}_2\
, {\bf l}_3)
        \\
&\approx&
        {(2\pi)^{3/2} \over (l_1 l_2 l_3)^{1/2}} \bi
\sum_{m_1,m_2} e^{i m_1 (\phi_{l_1} -\phi_{l_3})}
\nonumber\\
&&\times
                       e^{i m_2 (\phi_{l_2} -\phi_{l_3})}
        \wjma{l_1}{l_2}{l_3}{m_1}{m_2}{-m_1-m_2} \,. \nonumber
\end{eqnarray}
The triangle inequality of the Wigner-3$j$ symbol
becomes a triangle equality relating the 2D vectors.  The implication
is
that the triplet ($l_1$,$l_2$,$l_3$) can be considered to contribute
to
the triangle configuration ${\bf l}_1$,${\bf l}_2$,${\bf l}_3=-{\bf
l}_1+
{\bf l}_2$ where the multipole number is taken as the length of the
vector. The correspondance between the all-sky angular bispectrum
given by $\bi$ and the flat-sky vectorial representation of the
bispectrum by $B({\bf l}_1, {\bf l}_2, {\bf l}_3)$ is
\begin{equation}
\bi = \sqrt{\frac{\prod_{i=1}^3 (2l_i+1)}{4\pi}} \wj B(l_1,l_2,l_3) \, ,
\end{equation}
and follows the discussion in \cite{Hu00b} (2000b).

Similarly, we con formulate the trispectrum, or the Fourier analog of
the four-point correlation  function. In this thesis, we will only
encounter specific configurations of the trispectrum that contribute
to the covariance of the power spectrum and to the power spectrum of
squared quantities. The issues related to the general trispectrum will
be discussed in a separate paper.

\section{How to Describe Large Scale Structure Properties Using Halos?}
\label{sec:halomodel}

Throughout this thesis, we will be interested in observational probes
of large scale structure properties involving dark matter, pressure
and baryons. To make detailed predictions on observational statistics,
we make use of the halo model which is now fully described in
\cite{CooHu01a} (2001a; see also, \cite{Cooetal00b} 2000b;
\cite{MaFry00b} 2000b; \cite{Scoetal00} 2000). 
In the context of standard cold dark matter (CDM) models for
structure formation, the dark matter halos that are responsible for lensing
have properties that have been intensely studied by numerical
simulations.  In particular, analytic scalings and fits now exist
for the abundance, profile, and correlations of halos of
a given mass. We show how the dark matter power spectrum predicted in these
simulations can be constructed from these halo properties.
The critical ingredients are: the Press-Schechter formalism (PS;
\cite{PreSch74} 1974) or a variant for the mass function; the NFW
profile of \cite{Navetal96} (1996) or a variant to describe the dark matter 
halo distribution, and the halo bias
model of \cite{MoWhi96} (1996).

Underlying the halo approach is the assertion that dark matter halos
of virial mass $M$ are
locally biased tracers of density perturbations in the linear regime.
In this case, functional relationship between the over-density of
halos and
mass can be expanded
in a Taylor series
\begin{equation}
\delta_\hal({\bf x},M;z) = b_0 +
b_1(M;z)\delta_\lin({\bf x};z)   + {1 \over 2} b_2(M;z)\delta^2_\lin({\bf x};z)
+ \ldots\
\label{eqn:bias}
\end{equation}

The over-density of halos can be related to more familiar mass
function
and the halo density profile by assuming that we can model the fully
non-linear density field as a set of correlated discrete objects or
halos with profiles
\begin{equation}
\rho(\bfx;z) = \sum_{i} \rho_\halo (\bfx -  \bfx_i;M_i;z) \, ,
\end{equation}
where the sum is over all positions.
The density fluctuation in Fourier space, as a function of redshift,  is
\begin{eqnarray}
\delta(\veck;z)& = & \sum_i e^{i \veck \cdot \bfx_i}
\delta_{\halo}(\veck;M_i;z) \,.
\end{eqnarray}

Following \cite{Pee80} (1980), we divide space into
sufficiently
small volumes $\delta V$ that they contain only one or zero halos of a
given mass and convert the sum over halos to a sum over the
volume elements and masses
\begin{eqnarray}
\delta(\veck;z)
             & = & \sum_{V_1,M_1}
                        n_1 e^{i \veck \cdot \bfx_1}
\delta_\halo(\veck,M_1;z)\,.
\label{eqn:fourierdelta}
\end{eqnarray}
By virtue of the small volume element
$n_1=n_1^2=n_1^\mu = $ $1$ or $0$ following \cite{Pee80} (1980).

\begin{figure}[t]
\begin{center}
\includegraphics[width=4in,angle=-90]{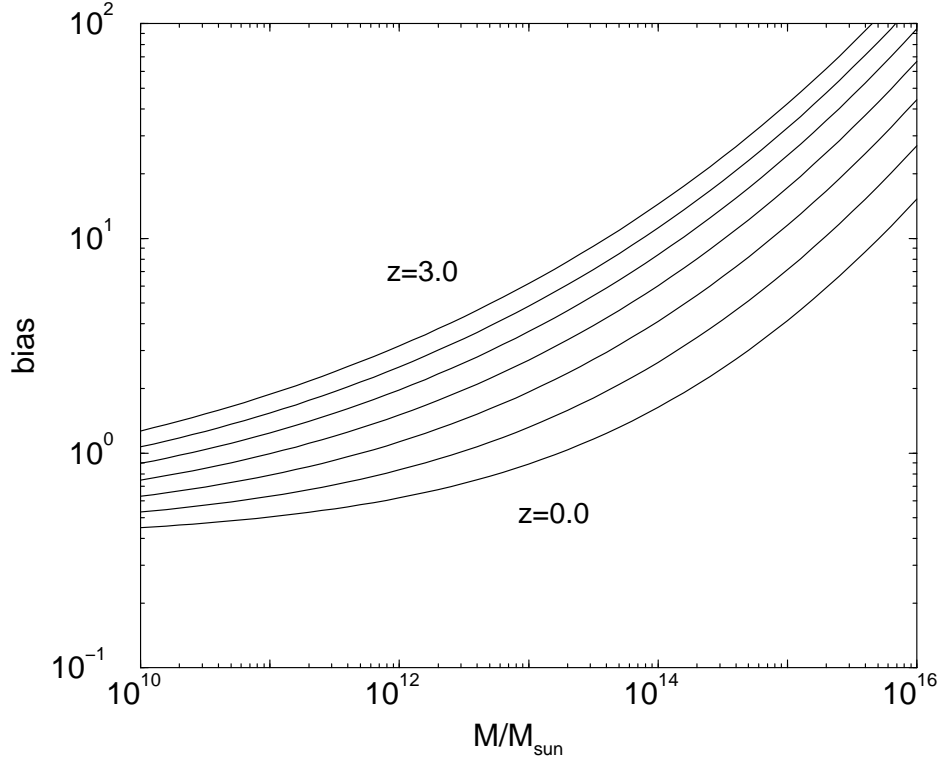}
\end{center}
\caption[Dark matter halo bias]{The dark matter halo bias as a
function of the halo mass and redshift. The curves show the redshift
evolution of bias from $z=0$ to $z=3$ at steps of 0.5.}
\label{fig:biasz}
\end{figure}

As written above, we take the halos to be biased tracers of the linear
density field such that their number density fluctuates as
\begin{equation}
{d^2 n \over d M d c}(\bfx;z) =
{d^2 \bar n \over d M d c}
        [b_0 + b_1(M;z) \delta_\lin(\bfx;z) + {1 \over 2} b_2(M;z)
\delta_\lin^2(\bf x;z)
        \ldots] \, .
\label{eqn:fluctuate}
\end{equation}
Thus, \begin{eqnarray}
\left< n_1 \right> &=& { d^2 \bar n \over dM dc} \delta M_1 \delta c_1
\, ,\\
\left< n_1 n_2 \right> &=&
        \left<n_1\right> \delta_{12} +
        \left<n_1\right> \left<n_2 \right>
        [b_0^2 + b_1(M_1;z)b_1(M_2;z) \, ,
        \nonumber\\
        &&
                \times \left< \delta_\lin(\bfx_1;z)\delta_\lin
        (\bfx_2;z) \right>] \, . \nonumber\\
\left< n_1 n_2 n_3 \right> &=& \ldots \, .
\label{eqn:expectation}
\end{eqnarray}
In Eq.~\ref{eqn:fluctuate}, $b_0 \equiv 1$, $\delta_{12}$ is the
Dirac delta function, and we have only considered the lowest order
contributions.  
The halo bias parameters given in \cite{Moetal97} (1997):
\begin{eqnarray}
b_1(M;z) &=& 1 + \frac{\nu^2(M;z) - 1}{\delta_c} \nonumber \\
b_2(M;z) &=& \frac{8}{21}[b_1(M;z)-1] + { \nu^2(M;z) -3 \over
\sigma^2(M;z)}\, .
\label{eqn:biasparams}
\end{eqnarray}
Here, $\nu(M;z) = \delta_c/\sigma(M;z)$ and 
$\sigma(M;z)$ is the rms fluctuation within a top-hat filter at the
virial radius corresponding to mass $M$,
and $\delta_c$ is the threshold over-density of spherical
collapse (see \cite{Hen00}  2000) for useful fitting functions). 
In Fig.~\ref{fig:biasz}, we show the mass dependence and the redshift
evolution of bias, $b_1(M;z)$.

The derivation of the higher point functions in Fourier space is
now a straightforward but tedious exercise in algebra.  The Fourier
transforms inherent in Eq.~\ref{eqn:fourierdelta} convert the correlation
functions in Eq.~\ref{eqn:expectation} into the power spectrum,
bispectrum, trispectrum, etc., of perturbation theory. We outline this description
in the Appendix.

Following \cite{CooHu01a} (2001a) and \cite{Coo00} (2000), it is now
convenient to define a
general integral over the halo mass function and profile distribution
$d^2\bar{n}/dMdc$. Though we presented the description of halo
clustering for dark matter, we can generalize this discussion to
consider any physical property associated with halo; one simply
relates the over-density of halos through the density profile
corresponding to the property of interest in Eq.~\ref{eqn:bias}.
Since we will encounter dark matter, pressure and baryon density
fields through out this thesis, we write a general integral that
applies to all these three properties as
\begin{eqnarray}
&& I_{\mu,i_1\ldots i_\mu}^{\beta,\eta}(k_1,\ldots,k_\mu;z)
\equiv
\int dM  \frac{d^2\bar{n}}{dMdc}(M,z) b_\beta(M;z)
\nonumber\\
&& \times
T_e(M;z)^\eta y_{i_1}(k_1,M;z)\ldots y_{i_\mu}(k_\mu,M;z)\,.
\label{eqn:I}
\end{eqnarray}
Here, in addition to the dark matter, to account for clustering
properties of
pressure associated with baryons in large scale structure, we have
introduced the electron temperature, $T_e(M;z)$. 

In Eq.~\ref{eqn:I}, the three-dimensional Fourier transform of the density
fluctuation through the halo profile of the density distribution,
$\rho_i(r,M;z)$, of any physical property is
\begin{equation}
y_i(k,M;z) = \frac{1}{\rho_{bi}} \int_0^{r_v} dr\, 4 \pi r^2 \rho_i(r,M;z)
\frac{\sin(kr)}{kr} \, ,
\end{equation}
with the background mean density of the same quantity given by $\rho_{bi}$.
Since in this thesis we discuss the dark matter, pressure and baryons,
the index $i$ will be used to represent either the density, $\delta$
(with $y \equiv y_\delta$), the baryons, $g$ (with $y \equiv y_g$),
or pressure, $\Pi$ (with $\equiv y_g$). Note that for both dark matter density and baryon
clustering, $\eta=0$, as there is no temperature contribution, but for
clustering of pressure, $\eta=\mu$ when $i_1\ldots i_\mu$ describes
pressure. One additional note here is that the profile used for
baryons will be the same as the profile that we will use for
pressure. The only difference between baryon clustering and pressure
clustering is that we weigh the latter with the electron temperature,
leading to a selective contribution from electrons with the highest temperature,
while the former includes all baryons.

\subsection{Correlation Functions in Fourier Space}

For the calculations presented in this thesis, we will
encounter the power spectrum, bispectrum and trispectrum involving
these properties. We now write down these Fourier space correlations
under the halo approach. To generalize the discussion, we will use the
index $i$ to represent the property of interest.

\subsubsection{Power Spectrum}

In general, the power spectrum of these three quantities
under the halo model now becomes (\cite{Sel00} 2000)
\begin{eqnarray}
P_i(k) &=& P^{1h}_i(k) +  P^{2h}_i(k) \,, \\
P^{1h}_i(k) & = & I_{2,ii}^0(k,k) \,, \\
P^{2h}_i(k) & = &\left[  I_{1,i}^1(k) \right]^2 P^\lin(k)\,,
\end{eqnarray}
where the two terms represent contributions from two points in
a single halo (1h) and points in different halos (2h)
respectively. 

Similar to above, we can also define the cross power spectra between
two fields as
\begin{eqnarray}
P_{ij}(k) &=& P^{1h}_{ij}(k) +  P^{2h}_{ij}(k) \,, \\
P^{1h}_{ij}(k) & = & I_{2,ij}^0(k,k) \,, \\
P^{2h}_{ij}(k) & = &I_{1,i}^1(k)I_{1,j}^1(k) P^\lin(k)\,.
\end{eqnarray}
It is also useful to define the bias of one field relative to the dark
matter density field
as \begin{equation}
b_{i}(k) = \sqrt{\frac{P_i(k)}{P_\delta(k)}} \, .
\end{equation}
We can also define a dimensionless
correlation coefficient between the two fields as
\begin{equation}
r_{ij}(k) = \frac{P_{ij}(k)}{\sqrt{P_i(k)P_j(k)}} \, .
\end{equation}
During the course of this paper, we will encounter, and use,
cross-power spectra as the one involving baryon and pressure,
$P_{g\Pi}$, and dark matter and pressure, $P_{\delta\Pi}$.

Following \cite{TegPee98} (1998), one can define a covariance matrix
in Fourier space containing the full information on scale dependence
of bias and correlations such that
\begin{equation}
\Ch(k )\equiv \left(\begin{array}{cc}
P_{ii}(k) & P_{ij}(k) \\
P_{ij}(k) & P_{jj}(k)
\end{array}\right) = P_{\delta\delta}(k)\left(\begin{array}{cc}
b_i^2 & r_{ij} b_i b_j \\
r_{ij} b_i b_j & b_j^2
\end{array}\right) \, .
\end{equation}
For example, the observation measurement of pressure bias, $b_\Pi$,
and pressure-dark matter correlation $r_{\delta\Pi}$, can be considered
by an inversion of the SZ-SZ, lensing-lensing and SZ-lensing
power spectra as a function of redshift bins in which lensing-lensing
or SZ-lensing power spectra are constructed.

\subsubsection{Bispectrum}

Similarly, we decompose the bispectrum into
terms involving one, two and three halos 
(see \cite{SchBer91} 1991; \cite{MaFry00b} 2000b):
\begin{eqnarray}
B_i &=& B_i^{1h}  + B_i^{2h}+  B_i^{3h} \, ,
\end{eqnarray}
where  here and below the argument of the bispectrum is understood to
be $(\veck_1,\veck_2,\veck_3)$. The term involving the  single halo
contribution is
\begin{eqnarray}
B_i^{1h} = I_3^0(k_1,k_2,k_3)\, .
\label{eqn:b1h}
\end{eqnarray}
Similarly, the term involving two halos trace the linear density field
power spectrum
\begin{eqnarray} 
B_i^{2h} = I_2^1(k_1,k_2) I_1^0(k_3) P^\lin (k_3) + {\rm Perm.} \, ,
\label{eqn:b2h}
\end{eqnarray}
while the term involving three halos trace the linear density field bispectrum
\begin{eqnarray}
B_i^{3h} = I_1^1(k_1) I_1^1(k_2) \left[ B^\lin(\veck_1,\veck_2,\veck_3) I_1^1(k_3) + I_1^2(k_3)
P^\lin(k_1)P^\lin(k_2) \right] + {\rm Perm.} 
\label{eqn:b3h}
\end{eqnarray}
for triple halo contributions. Here the 2 permutations are $k_3 \leftrightarrow k_1$, $k_2$.

\subsubsection{Trispectrum}

In the appendix, as an example on how these Fourier spaced correlation
functions are obtained, we detail the derivation of the trispectrum under the
halo model. As described there (see, also, \cite{CooHu01b} 2001b),
the contributions to the trispectrum
may be separated into those involving one to four halos
\begin{equation}
T_i = T^{1h}_i +  T^{2h}_i + T^{3h}_i + T^{4h}_i\,,
\end{equation}
where here and below the argument of the trispectrum is understood
to be $(\veck_1,\veck_2,\veck_3,\veck_4)$.
The term involving a single halo probes correlations of the physical
property $i$ within that halo
\begin{equation}
T^{1h}_i =
I_{4,iiii}^0(k_1,k_2,k_3,k_4) \, ,
\end{equation}
and is independent of configuration due to the assumed
spherical symmetry for our halos.

The term involving two halos can be further broken up into two parts
\begin{equation}
T^{2h}_i = T^{2h}_{31,iiii} + T^{2h}_{22,iiii}\,,
\end{equation}
which represent taking three or two points in the first halo
\begin{eqnarray}
T^{2h}_{31,iiii} = P^\lin(k_1)I_{3,iii}^1(k_2,k_3,k_4)I_{1,i}^1(k_1) +
3\; {\rm
Perm.,} \\
T^{2h}_{22,iiii} =
P^\lin(k_{12})I_{2,ii}^1(k_1,k_2)I_{2,i}^1(k_3,k_4)+ 2\; {\rm Perm.}
\end{eqnarray}
The permutations involve the 3 other choices of $k_i$ for the
$I_{1,i}^1$
term in
the first equation and the two other pairings of the $k_i$'s for the
$I_{2,ii}^1$ terms in the second.
Here, we have defined $\veck_{12} =
\veck_1+\veck_2$; note that $k_{12}$ is the length of one of the
diagonals
in the configuration.

The term containing three halos can only arise with two points in one
halo and one in each of the others
\begin{eqnarray}
T^{3h}_i &=&
B^\lin(\veck_1,\veck_2,\veck_{34})I_{2,ii}^1(k_3,k_4)I_{1,i}^1(k_1)I_{1,i}^1(k_2)
\nonumber \\
&&+
P^\lin(k_1)P^\lin(k_2)I_{2,ii}^2(k_3,k_4)I_{1,i}^1(k_1)I_{1,i}^1(k_2)
+ 5\;
{\rm Perm.}\, , \nonumber \\
\nonumber
\end{eqnarray}
where the permutations represent the unique pairings of the $k_i$'s in
the $I_{2,ii}$ factors.  This term also depends on the configuration.

Finally for four halos, the contribution is
\begin{eqnarray}
T^{4h}_i &=&  I_{1,i}^1(k_1)I_{1,i}^1(k_2)I_{1,i}^1(k_3)I_{1,i}^1(k_4)
\Big\{ T\
^\lin
        + \Big[ {I_{2,ii}^2(k_4) \over I_{1,i}^1(k_4) }
\nonumber\\ &&\quad \times
P^\lin(k_1)P^\lin(k_2) P^\lin(k_3)+ 3\; {\rm Perm.}\Big] \Big\},
\end{eqnarray}
where the permutations represent the choice of $k_i$ in the
$I_{1,i}^1$'s in
the brackets. We now discuss the results from this modeling for a
specific choice of
halo input parameters and cosmology.

Because of the closure condition expressed by the delta function,
the trispectrum may be viewed as a four-sided figure with sides
$\veck_i$.
It can alternately be described by the length of the four sides $k_i$
plus the diagonals.  We occasionally refer to elements of the
trispectrum
that differ by the length of the diagonals as different configurations
of the trispectrum. In the rest of this thesis, we will encounter the
dark matter density field trispectrum $T_\delta \equiv T_{\delta
\delta \delta \delta}$, pressure trispectrum $T_\Pi \equiv T_{\Pi\Pi\Pi\Pi}$ and the
pressure-baryon cross trispectrum $T_{g\Pi g\Pi}$.

\begin{figure}[t]
\begin{center}
\includegraphics[width=4in,angle=-90]{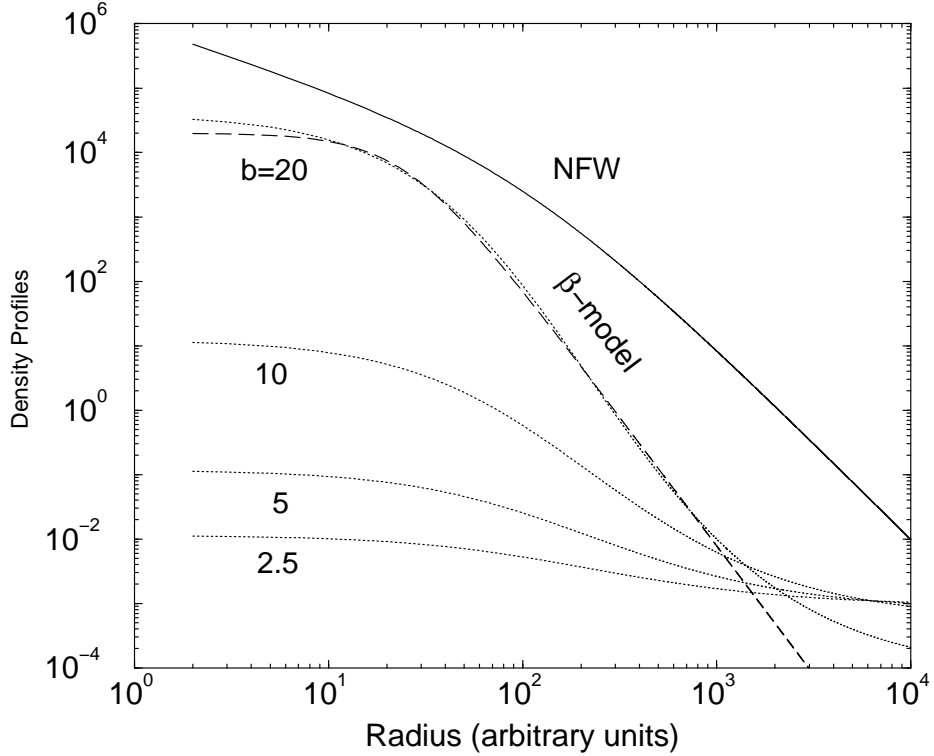}
\end{center}
\caption[Dark matter and gas profiles of halos]{The dark matter (NFW) profile and the ones predicted
by the hydrostatic equilibrium for gas, as a function of the $b$
parameter (see, Eq.~\ref{eqn:b}) with $r_s=100$. The relative
normalization between individual parameters is set using a  gas
fraction value of 0.1, though the NFW profile is arbitrarily
normalized with $\rho_s=1$; the gas profiles scale with the same
factor. For comparison, we
also show a typical example of the so-called $\beta$ model
$(1+r^2/r_c^2)^{-3\beta/2}$
which is generally used as a fitting function for X-ray and SZ
observations of clusters. We refer the reader to \cite{Maketal98}
(1998) and \cite{Sutetal98} (1998) for a detailed comparison of
$\beta$ models and the NFW-gas profiles.}
\label{fig:profiles}
\end{figure}

\subsection{Halo Parameters}

To calculate the power spectrum and higher order Fourier-space
correlation function of the dark matter density field and other
properties of the large scale structure we need several inputs as
outlined in the introduction. We detail these ingredients, which we
take obtain following results from numerical simulations.

\subsubsection{Dark Matter Profile}

The dark matter profile of collapsed halos are taken to be the NFW
\cite{Navetal96} with a density distribution
\begin{equation}
\rho_\delta(r) = \frac{\rho_s}{(r/r_s)(1+r/r_s)^{2}} \, .
\end{equation}
The density profile can be integrated and related to the total dark
matter mass of the halo within $r_v$
\begin{equation}
M_\delta =  4 \pi \rho_s r_s^3 \left[ \log(1+c) - \frac{c}{1+c}\right]
\label{eqn:deltamass}
\end{equation}
where the concentration, $c$, is $r_v/r_s$.
Choosing $r_v$ as the virial radius of the halo, spherical
collapse tells us that
$M = 4 \pi r_v^3 \Delta(z) \rho_b/3$, where $\Delta(z)$ is
the over-density of collapse and $\rho_b$ is the background matter
density today. We use comoving coordinates throughout.
By equating these two expressions, one can
eliminate $\rho_s$ and describe the halo by its mass $M$ and
concentration $c$. Following the results from $\Lambda$CDM simulations by
\cite{Buletal00} (2000), we take a concentration-mass relationship
such that
\begin{eqnarray}
{d \bar n \over dM dc} &=& \left({d n \over dM}\right)_{\rm PS} p(c)\,
, \\
p(c) dc &=& \frac{1}{\sqrt{2 \pi \sigma_c^2}} \exp\left[-\frac{(\ln c
-
\ln \bar{c})^2}{2\sigma_{\ln c}^2}\right] d\ln c \, ,\nonumber
\end{eqnarray}
where PS denotes the Press-Schechter mass function (\cite{PreSch74}
1974), which we use to describe the mass function of halos (see, below).

From the simulations of \cite{Buletal00} (2000), the
mean and width of the concentration distribution is
taken to be
\begin{eqnarray}
\bar{c}(M,z)   & = &  9 (1+z)^{-1} \left[
\frac{M}{M_*(z)}\right]^{-0.13}\,,\\
\sigma_{\ln c} & = &  0.2\,,
\label{eqn:concentration}
\end{eqnarray}
where $M_*(z)$ is the non-linear mass scale at which the peak-height
threshold, $\nu(M,z)=1$.

In describing pressure, due to computational limitations, we will
ignore the distribution of concentrations and only use the mean value:
\begin{equation}
c(M;z)= 9(1+z)^{-1}\left[\frac{M}{M_\star(z)}\right]^{-0.13} \, .
\label{eqn:conc}
\end{equation}
Additionally, in \cite{Cooetal00b} (2000b), we suggested a 
concentration-mass relation for the
$\Lambda$CDM model such that it will reproduce approximately the Peacock \&
Dodds (PD; \cite{PeaDod96} 1996) fitting function for the non-linear
power spectrum. We can write this relation as
\begin{equation}
c(M,z) = a(z) \left[\frac{M}{M_{\star}(z)}\right]^{-b(z)} \, ,
\end{equation}
 such that $a(z)=10.3(1+z)^{-0.3}$ and $b(z)=0.24(1+z)^{-0.3}$.
The dark matter power spectrum is well reproduced with these
parameters when using a NFW profile in a $\Lambda$CDM model, to
within 20\% for $0.0001 < k < 500$ Mpc$^{-1}$, out to a redshift of 1.
These values also agree with the ones given by Seljak (2000) for the
NFW profile at $z=0$. The two power spectra differ increasingly with scale at $k > 500$
Mpc$^{-1}$, but the Peacock and Dodds (1996) power spectrum is not reliable there
due to the resolution limit of the simulations from which the non-linear
power spectrum was derived.

\subsubsection{Gas Density Profile}

The gas density profile, $\rho_g(r)$, is calculated assuming
the hydrostatic equilibrium between the gas distribution and the dark
matter density field with in a halo. This is a valid assumption given
that current observations of halos, mainly galaxy clusters, suggest
the existence of regularity relations, such as size-temperature (e.g.,
\cite{MohEvr97} 1997), between physical properties of dark matter and
baryon distributions.

The hydrostatic equilibrium implies,
\begin{equation}
\frac{kT_e}{\mu m_p} \frac{d\log \rho_g}{dr} = -
\frac{GM_\delta(r)}{r^2} \, ,
\end{equation}
with $\mu=0.59$, corresponding to a hydrogen mass fraction of 76\%.
Here, $M_\delta (r)$ is the mass only out to a radius of $r$.
Note that we have assumed here an isothermal temperature for the gas
distribution.  Solving for the the equations above, we can
 analytically calculate the baryon density profile $\rho_g(r)$
\begin{equation}
\rho_g(r) = \rho_{g0} e^{-b} \left(1+\frac{r}{r_s}\right)^{br_s/r} \,
,
\label{eqn:gasprofile}
\end{equation}
where $b$ is a constant, for a given mass,
\begin{equation}
b = \frac{4 \pi G \mu m_p \rho_s r_s^2}{k_B T_e} \, ,
\label{eqn:b}
\end{equation}
with the Boltzmann constant, $k_B$ (\cite{Maketal98} 1998;
\cite{Sutetal98} 1998). This is derived
only under the assumption of hydrostatic equilibrium for the gas
distribution in a dark matter profile given by the NFW equation.
In above, the normalization $\rho_{go}$ is determined under the
assumption of
a constant gas mass fraction for halos comparable with the universal
baryon to dark matter ratio: $f_g \equiv M_g/M_\delta
=\Omega_g/\Omega_m$. When investigating astrophysical uses of the SZ
effect, we will vary this parameter and consider variations of gas fraction as
a function of mass and redshift.

The electron temperature can be calculated based on the virial
theorem or similar arguments as discussed in \cite{Coo00} (2000).
Using the virial theorem, we can write
\begin{equation}
k_B T_e = \frac{\gamma G \mu m_p M_\delta}{3 r_v} \, ,
\end{equation}
with $\gamma=3/2$. Since $r_v \propto M_\delta^{1/3}(1+z)^{-1}$ in
physical coordinates, $T_e \propto
M^{2/3}(1+z)$. The average density weighted temperature is
\begin{equation}
\left<T_e\right>_\delta  = \int dM\, \frac{M}{\rho_b} \frac{dn}{dM}(M,z) T_e(M,z) \,
.
\label{eqn:etemp}
\end{equation}

The total gas mass present in a dark matter halo within $r_v$ is
\begin{equation}
M_g(r_v) = 4 \pi \rho_{g0} e^{-b} r_s^3 \int_0^{c} dx \, x^2
(1+x)^{b/x} \, .
\label{eqn:gasmass}
\end{equation}

In Fig.~\ref{fig:profiles}, we show the NFW profile for the dark
matter and arbitrarily normalized gas profiles predicted by the
hydrostatic equilibrium and virial theorem for several values of $b$.
As $b$ is decreased, such that the temperature is increased, the turn
over radius of the gas distribution shifts to higher radii. As an
example, we also show the so-called $\beta$ model that is commonly
used to describe X-ray and SZ observations of galaxy clusters and for
the derivation purpose of the Hubble constant by combined SZ/X-ray
data. The $\beta$ model describes the underlying gas distribution
predicted by the gas profile used here in equilibrium with the NFW
profile, though, we find differences especially at the outer most
radii of halos. This difference can be used as a way to establish the
hydrostatic equilibrium of clusters, though, any difference of gas
distribution at the outer radii should be accounted in the context of
possible substructure and mergers.

A discussion on the comparison
between the gas profile used here and the $\beta$ model is available
in \cite{Maketal98} (1998) and \cite{Sutetal98} (1998).
In addition, we refer the reader to \cite{Coo00} (2000)
for full detailed discussion on issues
related to modeling of pressure power spectrum using halo and
associated systematic errors. Comparisons of the halo model
predictions with numerical simulations are available in
\cite{Seletal00} (2000) and \cite{RefTey01} (2001).
Similarly, issues related to modeling of
the dark matter clustering using halos is discussed in \cite{CooHu01a}
(2001a) for the bispectrum and \cite{CooHu01b} (2001b) for the trispectrum

\begin{figure}[!h]
\begin{center}
\includegraphics[width=5.9in]{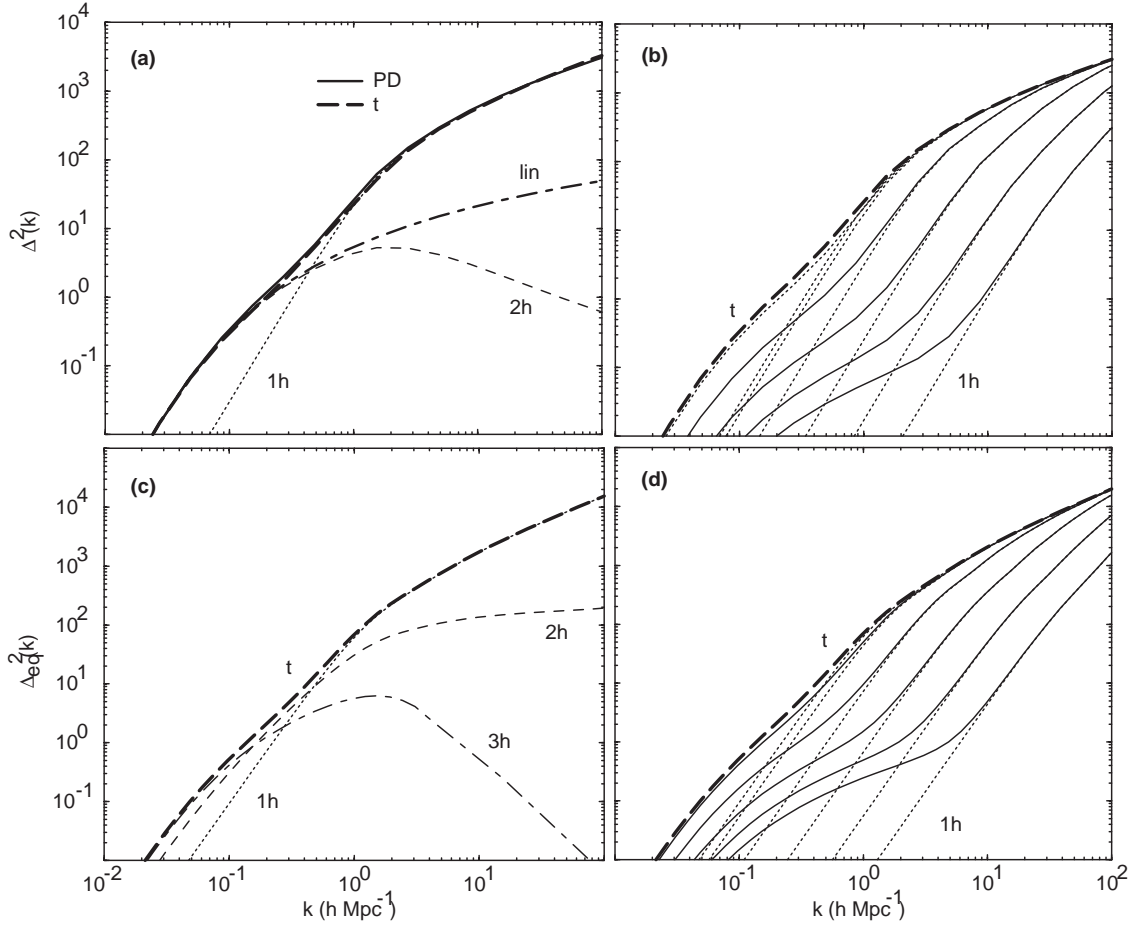}
\end{center}
\caption[Dark matter power spectrum and bispectrum under the halo model]{Present day dark matter density (a) power spectrum and
(c) equilateral bispectrum under the halo prescription.
The power spectrum shown in (a)
is compared with the PD fitting function and the
linear $P(k)$. We have decomposed the power spectrum and
bispectrum to individual contributing terms under the halo approach.
The mass cut off effects on the present day dark matter density
power spectrum (b) and bispectrum (d) under the  halo approach.
From bottom to top, the maximum mass used in the calculation is
$10^{11}$, $10^{12}$, $10^{13}$, $10^{14}$, $10^{15}$ and $10^{16}$ M$_{\sun}$.}
\label{fig:dmpower}
\end{figure}

\subsubsection{Mass Function}

In order to describe the dark matter halo mass distribution, in general,
we can consider two analytical forms commonly found in the literature.
These are the Press-Schechter (PS; \cite{PreSch74} 1974)
and Sheth-Tormen (ST; \cite{SheTor99} 1999) mass functions and
are both parameterized by
\begin{eqnarray}
\frac{dn}{dM} dM = \frac{\rho_b}{M} f(\nu) d\nu
\end{eqnarray}
with $f(\nu)$ taking the general form of
\begin{eqnarray}
\nu f(\nu) = A\sqrt{\frac{2}{\pi}}\left[1+(a\nu^2)^{-p}\right] (a\nu)
\exp \left(-a\nu^2/2\right)
\, .
\label{eqn:massfunction}
\end{eqnarray} Here, $\nu = \delta_c/\sigma(M,z)$, where
$\sigma(M,z)$ is the rms fluctuation within a top-hat filter at the
virial radius corresponding to mass $M$,
and $\delta_c$ is the threshold overdensity of spherical
collapse.

The normalization $A$  in Eq.~\ref{eqn:massfunction}
is set by requiring the mass conservation, such that
the average mass density from the mass function is same as the average
mass density of the universe:
\begin{eqnarray}
\int \frac{dn}{dM} \frac{M}{\rho_b} dM = \int f(\nu) d\nu = 1 \, ,
\end{eqnarray}
and takes values of 0.5 and 0.383 when the PS $(p=0,a=1)$ or ST
$(p=0.3,a=0.707)$ mass functions are used respectively. The two mass
functions behave such that when $\nu$ is small,
 $\nu f(\nu) \propto \nu^{1.0}$ and $\propto \nu^{0.4}$ for PS and ST
mass functions, respectively. Note that the difference in mass
functions can be compensated by a difference in the concentration-mass
relation (see, e.g., \cite{Sel00} 2000; \cite{CooHu01a} 2001a).
Thus, we will simply use the PS mass function throughout here.
We take the minimum mass to be $10^3$ M$_{\sun}$ while the maximum mass is varied to
study the effect of massive halos on lensing convergence statistics.
In general, masses above $10^{16}$ M$_{\sun}$ do not contribute to low
order statistics due to the exponential decrease in the number 
density of such massive halos.

\section{Dark Matter Power Spectrum and  Bispectrum}

In Fig.~\ref{fig:dmpower}(a-b), we show the density field power
spectrum today ($z=0$), written  such that $\Delta^2(k)=k^3 P(k)/2\pi^2$ 
is the power per logarithmic interval in
wavenumber. In Fig~\ref{fig:dmpower}(a), we show individual contributions
from the single and double halo terms  and a comparison to the non-linear power
spectrum as predicted by the PD fitting function.
In Fig.~\ref{fig:dmpower}(b), we show the dependence of density field
power as a function of maximum mass used in the calculation. Here, we show 
the power spectrum and bispectrum such that the concentration-mass formula 
is modified to match the PD fitting function with parameters as listed in
under Eq.~\ref{eqn:conc}.

In general, the behavior of dark matter power spectrum due to halos
can be understood in the following way.
The linear portion of the dark matter power spectrum, $k < 0.1$
h Mpc$^{-1}$, results from the correlation
between individual dark matter halos and reflects the
bias prescription.  The fitting formulae of \cite{MoWhi96} (1996)
adequately describes this regime for all redshifts.
The mid portion of the power spectrum, around $k \sim 0.1-1$
h Mpc$^{-1}$ corresponds to the non-linear scale
$M \sim M_{\star}(z)$, where the Poisson and correlated
term contribute comparably. At higher $k$'s, the power arises mainly from
the contributions of individual halos. Similarly, at the same high scales 
when $k \gtrsim$ few tens h Mpc$^{-1}$, the PD fitting function is not 
reliable due to resolution limit of the simulations from which the 
fitting function for the non-linear power spectrum was derived. In addition 
to the NFW profile, one can consider variants, however, with the freedom 
to change the concentration-mass relation, such variations do not produce 
recognizable differences in the power spectrum and the bispectrum (see, 
\cite{Sel00} 2000 and \cite{CooHu01a} 2001a for a discussion).
We also refer the reader to \cite{Sel00} 2000 for a
discussion of the detailed properties of galaxy power spectra due to 
halos; we briefly discuss the subject of galaxy power spectra in
\S~\ref{sec:galaxy} using the PCSZ redshift-space galaxy power
spectrum from \cite{HamTeg00} (2000).

\begin{figure}[!h]
\begin{center}
\includegraphics[width=4.2in]{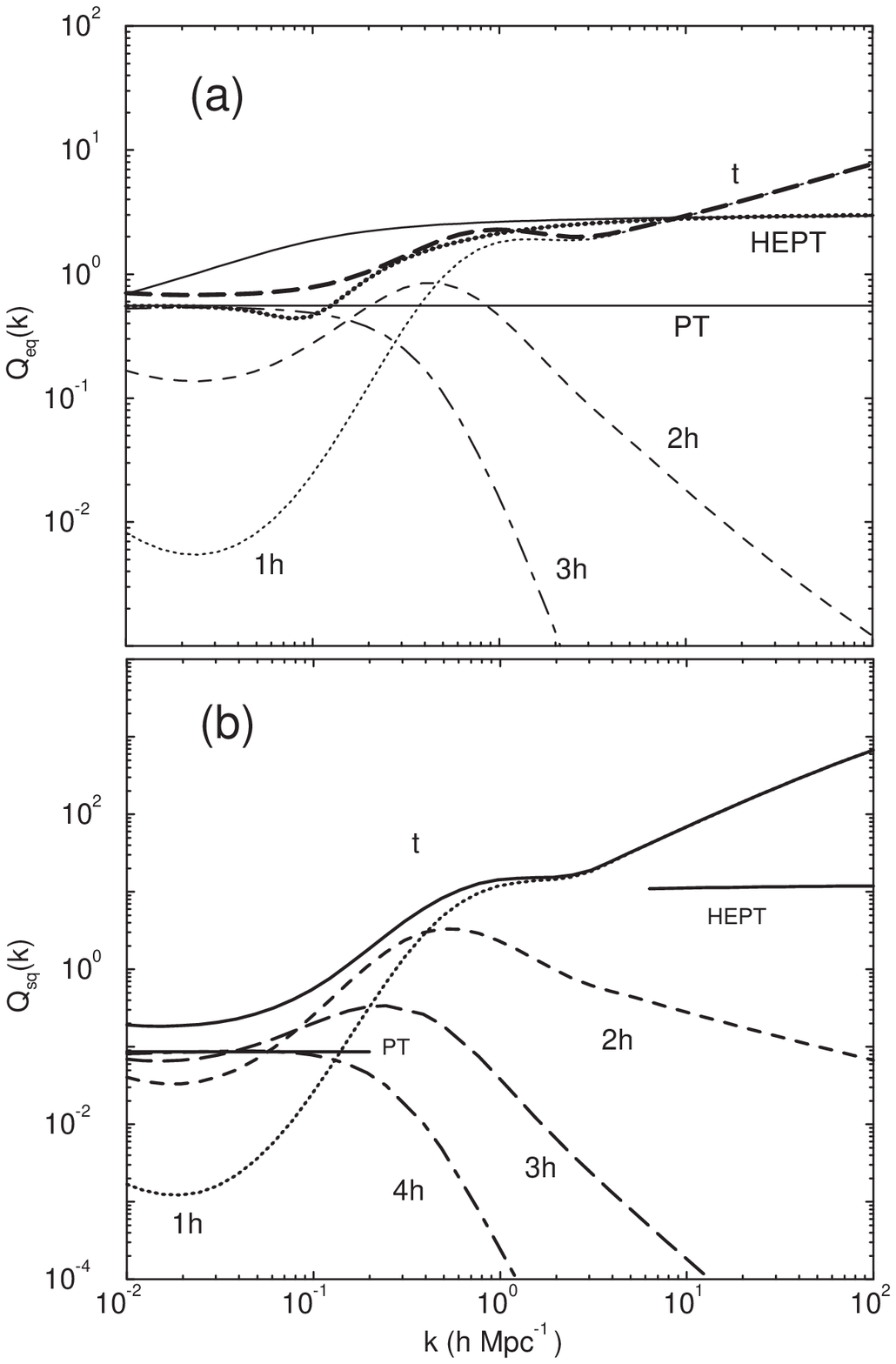}
\end{center}
\caption[Q$_{\rm eq}$(k) and Q$_{\rm sq}$ for the dark matter
bispectrum and trispectrum, respectively.]{(a) $Q_{\rm eq}(k)$ 
and (b) $Q_{\rm sq}$  at present broken into individual
contributions under the halo description and compared with second order
perturbation theory (PT) and hyper-extended perturbation theory
(HEPT). In (a), the thick dotted line shows the $Q_{\rm eq}$ based on the
fitting function of \cite{ScoCou00} (2000) that combines HEPT at small
scales and PT at large scales. In (b), in the linear regime,
the perturbation theory (PT) prediction
is reproduced by the $4$ halo term which is only $\sim 1/2$ of the
total.   See text for a discussion of discrepancies.}
\label{fig:dmq}
\end{figure}

Since the bispectrum generally scales as the square of the power
spectrum, it is useful to define 
\begin{equation}
\Delta_{\rm eq}^2(k) \equiv \frac{k^3}{2\pi^2} \sqrt{B(k,k,k)} \,,
\end{equation}
which represents equilateral triangle configurations,
and its ratio to the power spectrum 
\begin{equation}
Q_{\rm eq}(k) \equiv {1 \over 3}
\left[ {\Delta_{\rm eq}^2(k) \over \Delta^2(k)} \right]^2\,.
\end{equation}
In second order perturbation theory,
\begin{equation}
Q_{\rm eq}^{\rm PT} = 1 - \frac{3}{7}\Omega_m^{-2/63}
\end{equation}
and under hyper-extended perturbation theory (HEPT; \cite{ScoFri99}
1999), 
\begin{equation}
Q_{\rm eq}^{\rm HEPT}(k) = \frac{4 - 2^{n(k)}}{1+2^{n(k)+1}} \, ,
\label{Q3}
\end{equation}
which is claimed to be valid in the deeply nonlinear regime.
Here, $n(k)$ is the {\it linear} power spectral index at $k$. 

In Fig.~\ref{fig:dmpower}(c-d), we show $\Delta_{\rm eq}^2(k)$ 
separated into its various contributions (c) and as a function of
maximum mass (d).  Since the power spectra and equilateral bispectra
share similar features, it is more instructive to examine
$Q_{\rm eq}(k)$ (see Fig.~\ref{fig:dmq}a).
Here we also compare it with the second order perturbation theory (PT)
and the HEPT prediction.
In the halo prescription, $Q_{\rm eq}$ 
at $k \simgt 10 k_{\rm nonlin} \sim 10 h$Mpc$^{-1}$ 
arises mainly from the single halo term. 
We also show $Q_{\rm eq}(k)$ predicted by the fitting function of
\cite{ScoCou00} (2000) based on simulations in the
range of $0.1 \lesssim k \lesssim 3$ h Mpc$^{-1}$. This function is
designed such that it converges to HEPT value at small scales and PT
value at large scales. The HEPT prediction, however, falls short on smaller
scales; further work with numerical simulations, especially at scales with $k
\gtrsim 10$ h Mpc$^{-1}$, where the predictions based on HEPT and halo
models differ, will be useful to distinguish between various
clustering hypotheses (see, e.g., \cite{MaFry00c} 2000c). The scales
where the two predictions significantly differ is unlikely to be
probed by weak lensing observations as such scales only contribute
at angular scales of few arcseconds ($l \sim 10^{4}$).

\begin{figure}[!h]
\begin{center}
\includegraphics[width=4in]{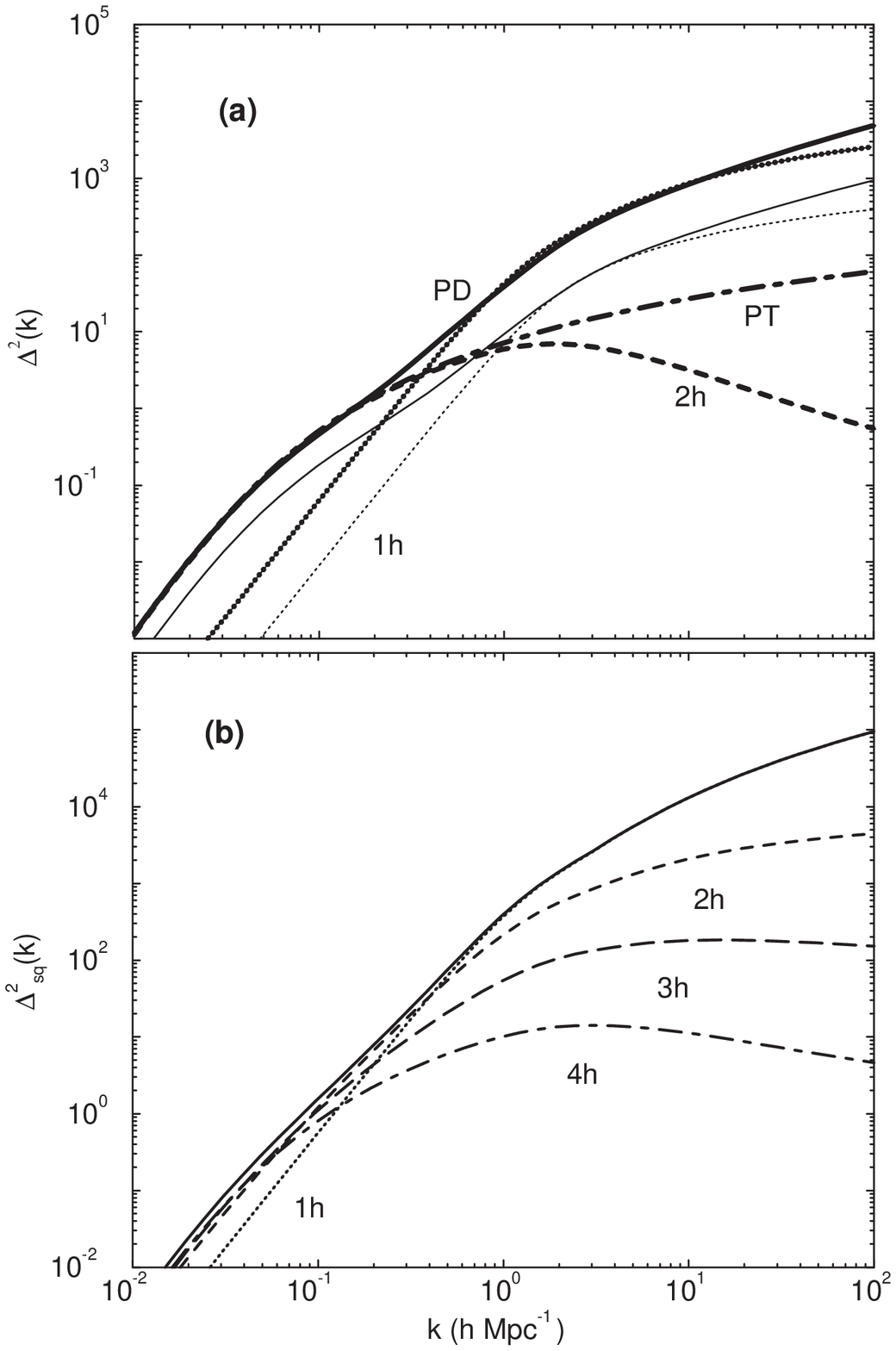}
\end{center}
\caption[Dark matter power spectrum and trispectrum under the halo model]{The 
dark matter power spectrum (a) and square-configuration
trispectrum (b) broken into individual contributions under the halo description.
The lines labeled 'PD' shows the dark matter power spectrum under the
\cite{PeaDod96} (1996) non-linear fitting function while the curve
labeled 'PT' is the linear dark matter power spectrum (at redshift of
0). In (a), we show the power spectrum at redshifts of 0 and 1. In (b), we
show the square configuration trispectrum (see text).
In both (a) and (b), at small scales
the single halo term dominates while at large scales halo
correlations contribute. }
\label{fig:dmtripower}
\end{figure}

\section{Dark Matter Power Spectrum Covariance}
 
Following \cite{Scoetal99} (1999), we can relate the trispectrum
to the variance of the estimator of the binned power spectrum
\begin{equation}
\hat P_i = {1 \over V } \int_{\shell i} {d^3 k \over V_{\shell i}}
\delta^*(-\veck) \delta(\veck)  \, ,
\end{equation}
where the integral is over a shell in $k$-space centered around $k_i$,
$V_{\shell i} \approx 4\pi k_i^2 \delta k$ is the volume of the shell
and $V$
is the volume of the survey.  Recalling that $\delta({\bf 0})\rightarrow V/(2\pi)^3$
for a finite volume,
\begin{eqnarray}
C_{ij} &\equiv& \left< \hat P_i \hat P_j \right> - 
      \left< \hat P_i \right> 
      \left< \hat P_j \right>  \nonumber\\
       &=& {1 \over V} \left[ {(2\pi)^3 \over V_{\shell i} } 2 P_i^2
\delta_{ij}
+
      T_{ij} \right]  \, ,
\end{eqnarray}
where
\begin{eqnarray}
T_{ij} &\equiv& \int_{\shell i} {d^3 k_i \over V_{\shell i}}
      \int_{\shell j} {d^3 k_j \over V_{\shell j}}
      T(\veck_i,-\veck_i,\veck_j,-\veck_j) \,.
\label{eqn:covarianceij}
\end{eqnarray}
Notice that though both terms
scale in the same way with the volume of the survey, only the Gaussian
piece
necessarily decreases with the volume of the shell.  For the Gaussian
piece,
the sampling error reduces to a simple root-N mode counting of
independent modes
in a shell.  The trispectrum quantifies the non-independence of the
modes both within a shell
and between shells.  Calculating the covariance matrix of the power
spectrum
estimates reduces to averaging the elements of the trispectrum across
configurations
in the shell.  It is to the description of the trispectrum that
we now turn.

\subsection{Trispectrum}

In Fig.~\ref{fig:dmtripower}(a), we show the logarithmic 
power spectrum $\Delta^2(k)=k^3 P(k)/2\pi^2$ with
contributions broken down to the $1h$ and $2h$ terms today and the 
$1h$ term at redshift of 1. Here, we use the concentration-mass relation as 
found by \cite{Buletal00} (2000) in their numerical simulations in the
$\Lambda$CDM cosmology. We have taken the width of concentration-mass
distribution to be $\sigma_{\rm ln c}=0.2$. Our prediction for the 
non-linear power spectrum is compared with the PD fitting function.
The same prediction here with the concentration-mass relation from 
simulation and the one obtained by fitting for the PD function can be 
compared through Fig.~\ref{fig:dmpower}(a).
When compared to PD fitting function, and using 
results from numerical simulations for 
concentration, we find that there is an
slight overprediction of power at scales  corresponding to
$1 \lesssim k \lesssim 10$ h Mpc$^{-1}$ at redshifts of 0 and 1, 
and a more substantial underprediction at small scales
with $k \gtrsim 10$ h Mpc$^{-1}$. Since the non-linear
power spectrum has only been properly studied out to
overdensities $\Delta^2 \sim 10^3$ with numerical simulations
it is unclear whether the small-scale disagreement is significant. 
Fortunately, it is on sufficiently small scales so as not to
affect weak gravitational lensing observables.

For the trispectrum, and especially the contribution of trispectrum to 
the covariance, we are mainly interested in terms
involving $T(\veck_1,-\veck_1,\veck_2,-\veck_2)$, i.e. parallelograms 
which are defined by either the length $k_{12}$ or the angle
between $\veck_1$ and $\veck_2$.  For illustration purposes
we will take $k_1=k_2$ and the angle to be $90^\circ$
($\veck_2=\veck_\perp$)
such that
the parallelogram is a square.
It is then convenient to define
\begin{equation}
\Delta^2_{\rm sq}(k) \equiv \frac{k^3}{2\pi^2}
T^{1/3}(\veck,-\veck,\veck_\perp,-\veck_\perp) \, ,
\end{equation}
such that this quantity scales roughly as the logarithmic
power spectrum itself $\Delta^2(k)$.  This spectrum is
shown in Fig.~\ref{fig:dmtripower}(b) with the individual
contributions from the 1h, 2h, 3h, 4h terms shown.

\begin{figure}[!h]
\begin{center}
\includegraphics[width=4in]{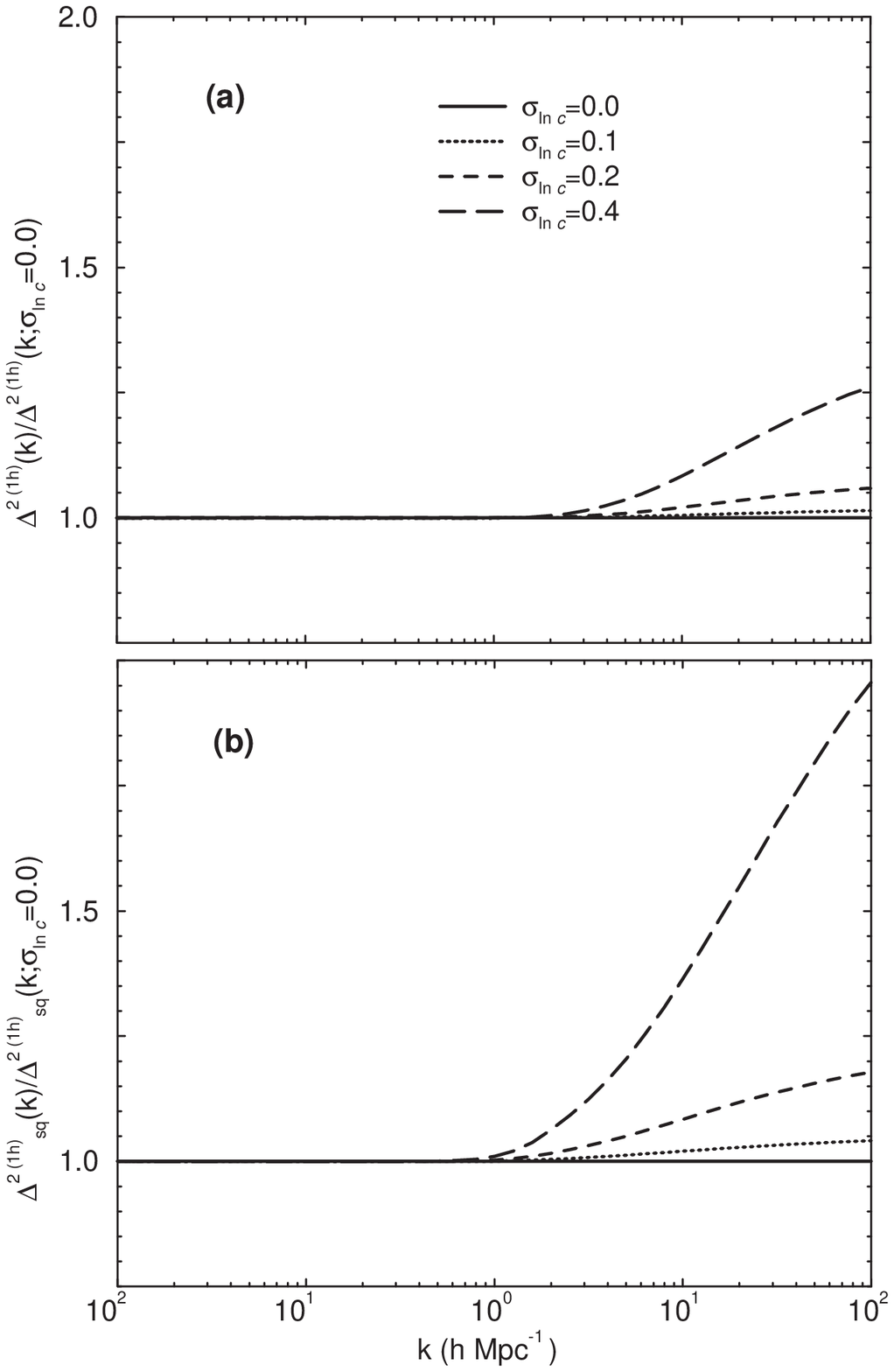}
\end{center}
\caption[Effect of concentration distribution on the single halo terms]{The ratio of the single halo term contribution to that for 
a concentration width $\sigma_{\ln c}\rightarrow 0$ for the
(a) power spectrum and (b) trispectrum. 
The small scale behavior is increasingly sensitive to the high
concentration
tails for the higher order statistics.}
\label{fig:conc}
\end{figure}

We test the sensitivity of our calculations to the width of the distribution in
Fig.~\ref{fig:conc}, where we
show the ratio between single halo contribution, 
as a function of the
concentration  distribution width, to the halo term with a delta
function distribution $\sigma_{\rm ln c}=0$.  The fiducial value of the width 
suggested by simulations is $\sigma_{\rm ln c}=0.2$.
As in the power spectrum
the effect of increasing the width is to increase the amplitude at
small scales due to the high concentration tail of the distribution.
Notice that the width effect
is stronger in the trispectrum than the power spectrum since the
tails of the distribution are weighted more heavily in higher point
statistics.

To compare the specific scaling predicted by perturbation theory
in the linear regime 
and the hierarchical ansatz in the deeply non-linear regime,
it is useful to define the quantity
\begin{equation}
Q_{\rm sq}(k) \equiv
\frac{T(\veck,-\veck,\veck_\perp,-\veck_\perp)}{[8P^2(k)P(\sqrt{2}k)][4P^3(k)]}\, .
\end{equation}
In the halo prescription, $Q_{\rm sq}$
at $k \simgt 10 k_{\rm nonlin} \sim 10 h$Mpc$^{-1}$
arises mainly from the single halo term. 
In perturbation theory $Q_{\rm sq} \approx 0.085 $. 
The $Q_{\rm sq}$ does not approach the perturbation theory
prediction as $k \rightarrow 0$ since that contribution appears
only as one term in the 4 halo piece. 
Our model therefore does not recover
the true trispectrum of the density field in the linear regime.
The problem is that in modeling the density field with discrete objects,
here halos, there is an error associated with shot noise.  A more familiar
example of the same effect comes from the use of galaxies as tracers
of the dark matter density field.
While this error appears large
in the $Q_{\rm sq}$ statistic, it does not affect the calculations of the
power spectrum covariance since in this regime, it is the Gaussian piece
errors that dominate.

The hierarchical ansatz
predicts that $Q_{\rm sq}=$ const. in the deeply non-linear regime.
Its value is unspecified by the ansatz but is given 
as 
\begin{equation}
Q_{\rm sq}^{\rm sat} = \frac{1}{2}\left[\frac{54 - 27\cdot 2^n +
2\cdot 3^n + 6^
n}{1+6\cdot2^n + 3\cdot 3^n + 6\cdot 6^n}\right]
\end{equation}
under hyperextended perturbation theory (HEPT; \cite{ScoFri99}).
Here $n=n(k)$ is the linear  power spectral index at $k$. As
shown in Fig.~\ref{fig:dmq}(b), the halo model predicts $Q_{\rm sq}$ 
increases at high $k$.  
This behavior, also present at the three point level for the dark
matter density field bispectrum, 
suggests disagreement between the halo approach and hierarchical
clustering ansatz
(see, \cite{MaFry00b} 2000b), though numerical simulations do not yet
have enough resolution  to test this disagreement.  Fortunately
the discrepancy is also outside of the regime important for
lensing.

\subsection{Further Tests of the Dark Matter Covariance}

To further test the accuracy of our halo trispectrum, we compare
dark matter correlations predicted by our method to those from
numerical simulations by \cite{MeiWhi99} (1999).  For this purpose, we
calculate the covariance matrix $C_{ij}$ from Eq.~\ref{eqn:covarianceij}
with the bins centered at $k_i$ and volume $V_{\shell i} =
4\pi k_i^2 \delta {k_i}$ corresponding to their scheme. 
We also employ the parameters of their $\Lambda$CDM cosmology
and assume that the parameters that defined the halo
concentration properties from our fiducial $\Lambda$CDM model holds
for this cosmological model also. The physical differences between the
two cosmological model are minor, though normalization differences can 
lead to large changes in the correlation coefficients.

\begin{table}[!h]
\begin{flushleft}
\begin{tabular}{lrrrrrrrrr} \hline
$k$   &  0.06 & 0.07 & 0.09 & 0.11 & 0.14 & 0.17 & 0.21 &
 0.25 & 0.31\\
\hline
0.06 & 1.00 & 0.06 & 0.12 & 0.18 & 0.25 & 0.30 & 0.33 & 0.34 & 0.33 \\
0.07 & (0.04) & 1.00 & 0.10 & 0.19 & 0.30 & 0.37& 0.41 & 0.42 & 0.41 \\
0.09 & (0.03) & (0.08) & 1.00 & 0.16 & 0.29 &0.40 &  0.47 & 0.48 & 0.48\\
0.11 & (0.09) & (0.09) & (0.03) & 1.00 & 0.28 & 0.43 & 0.54 & 0.58 & 0.57 \\
0.14  & (0.15) & (0.20) & (0.08) & (0.20) & 1.00 & 0.43 & 0.58 & 0.69 & 0.70 \\
0.17 & (0.14) & (0.23) & (0.18) & (0.25) & (0.28) & 1.00 & 0.59 & 0.74 & 0.78 \\
0.21 & (0.18) & (0.32) & (0.19) & (0.31) & (0.40) & (0.48) & 1.00 & 0.75 & 0.84 \\
0.25 & (0.21) & (0.34) & (0.26) & (0.35) &(0.49) & (0.61) & (0.65) & 1.00 & 0.86 \\ 
0.31 & (0.20) & (0.37) & (0.26) & (0.40) & (0.51) & (0.62) & (0.72) & (0.82) & 1.00 \\
\hline
$\sqrt{\frac{C_{ii}}{C_{ii}^{G}}}$ & 
1.02& 1.03& 1.04& 1.07& 1.14& 1.23& 1.38& 1.61& 1.90\\
\end{tabular}
\caption[Dark matter covariance matrix]{
Diagonal normalized covariance matrix of the binned dark matter
density field power spectrum with $k$ values in units of h Mpc$^{-1}$.
Upper triangle displays the covariance found under the halo model.
Lower triangle (parenthetical numbers) displays the covariance found
in numerical simulations by \cite{MeiWhi99} (1999).  Final line shows
the fractional increase in the errors (root diagonal covariance) due to
non-Gaussianity as calculated under the halo model.}
\label{tab:dmcorr}
\end{flushleft}
\end{table}

In Table \ref{tab:dmcorr}, we compare the predictions 
for the correlation coefficients
\begin{equation}
\hat C_{ij} = {C_{ij} \over \sqrt{C_{ii} C_{jj}}}
\end{equation}
with the simulations.  Agreement in the off diagonal elements
is typically better than $\pm 0.1$, even in the region where
non-Gaussian effects dominate, and the qualitative
features such as the increase in correlations across the
non-linear scale are preserved.

A further test on the accuracy of the halo approach is to consider 
higher order real-space moments such as skewness and kurtosis. In
\cite{CooHu00} (2000), we discussed the weak lensing convergence skewness 
under the halo model and found it to be in agreement with numerical predictions
from \cite{WhiHu99} (1999). The fourth
moment of the density field, under certain approximations, was
calculated by \cite{Scoetal99} (1999) using dark matter halos and
was found to be in good agreement with N-body simulations.
Given that density field moments have already been studied by
\cite{Scoetal99} (1999), we no longer consider them here other than to
suggest that the halo model has provided, at least qualitatively,
a consistent description better than any of the perturbation theory
arguments.

\section{From Dark Matter to Galaxies}
\label{sec:galaxy}

\begin{figure}[!h]
\begin{center}
\includegraphics[width=4.2in,angle=-90]{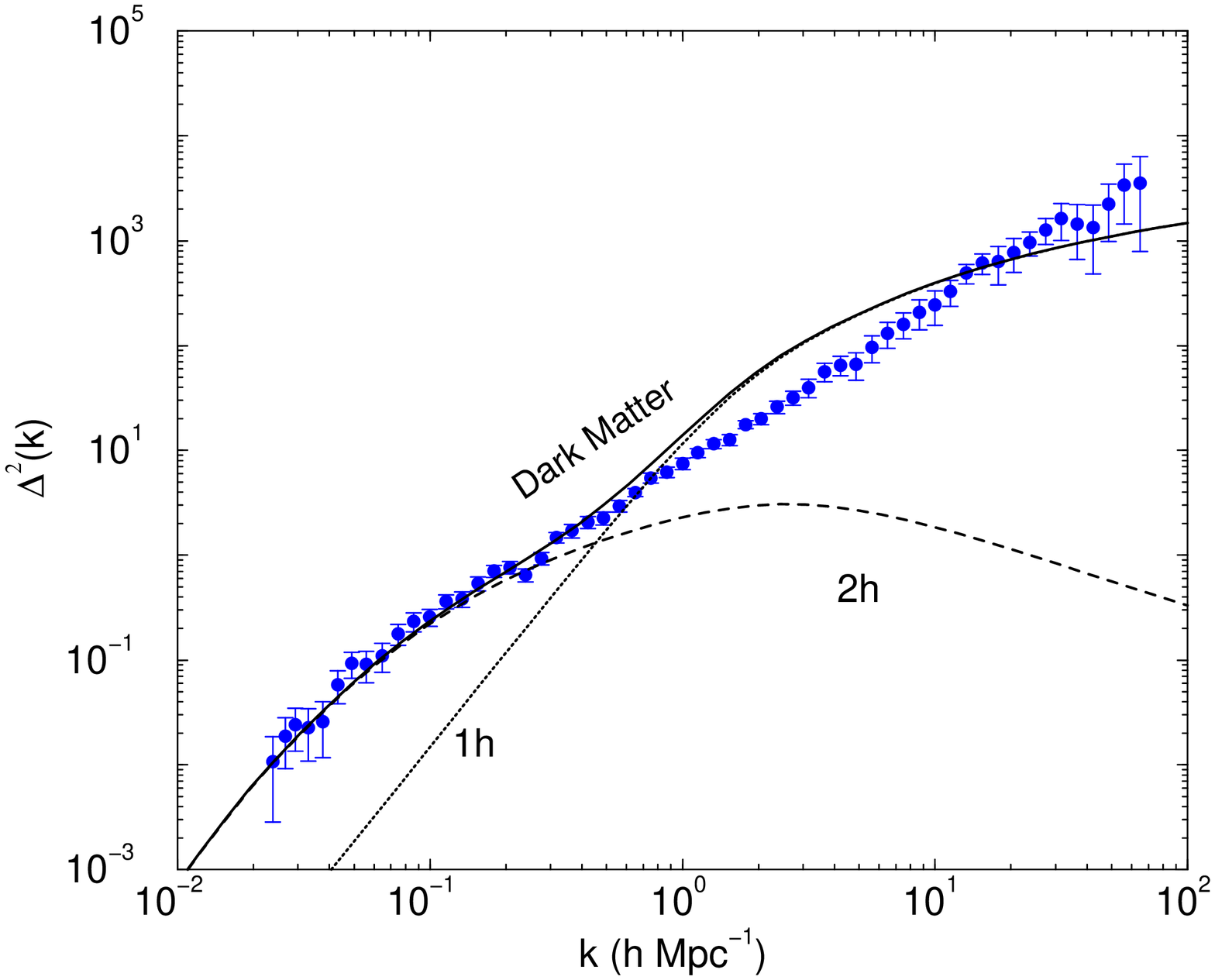}
\end{center}
\caption[PCSZ galaxy power spectrum vs. dark matter]{The PCSZ galaxy
power spectrum compared to the dark matter power spectrum. The galaxy
power spectrum comes from \cite{HamTeg00} (2000), and we have scaled
the dark matter power spectrum with a linear scale-independent bias
factor. At non-linear scales, the dark matter clustering cannot
reproduce the galaxy power.}
\label{fig:pcszpower}
\end{figure}

In Fig.~\ref{fig:pcszpower}, we show the redshift-space galaxy power
spectrum from the PCSZ survey as derived by \cite{HamTeg00}
(2000). For comparison, we show the non-linear dark matter power
spectrum with the galaxy power spectrum scaled with a constant bias in
the linear regime following the analysis given in \cite{HamTeg00}
(2000). In the mildly  to deeply non-linear regime, the galaxy power spectrum cannot be
simply reproduced through an overall scaling of the non-linear dark
matter power spectrum. This disagreement provides a strong argument
against a scale independent bias for galaxy at all scales. 

To understand the behavior of the galaxy power spectrum under
the halo model, we follow discussions in \cite{Sel00}
(2000) and \cite{Scoetal00} (2000). The basic idea here is that the galaxies can
be considered as a tracer of the dark matter. Thus, its clustering
properties inside a halo will simply follow the distribution of dark
matter in that halo. Since the clustering measurements only involve
the galaxies, one can relate the galaxy population in halos to the
dark matter, as a function of the halo mass, through a relation that
involves the mean number of galaxies per halo. This is essentially
similar to the idea we presented to describe pressure, which  involves
a similar mean relation through the temperature-mass description for
electrons in halos.

Following \cite{Sel00} (2000), we describe the average number of galaxies per
halo, $\left< N_\gal \right>$ in Eq.~\ref{eqn:I},  such that
\begin{equation}
\left< N_\gal \right> = \left(\frac{M}{M_{\rm min}}\right)^{0.8}\\
\end{equation}
where $M_{\rm min}$, the minimum dark matter halo mass in which a
galaxy is found, is taken to be $5.3 \times 10^{11} h^{-1}$ M$_{\sun}$
for our fiducial $\Lambda$CDM cosmological model following
\cite{Benetal99} (1999). The above relation is consistent with
semi-analytical models, however, we ignore scatter in the observed
distribution on the mean number of galaxies per halo. In addition to
semi-analytic work, the above mean number of galaxies is consistent
with the relation found by \cite{Scoetal00} (2000) under the halo
approach when compared to clustering of galaxies in the APM survey. 
The reason why the number of galaxies scales as $M^{0.8}$, instead of
simply mass, can be understood by noting that the galaxy formation is
suppressed in large mass halos due to the significantly higher cooling
time when compared to the cooling times for gas in low mass
halos. Thus, low mass halos, such as galaxy groups, have a higher 
efficiency for galaxy formation than high mass halos, such as
massive clusters of galaxies. Such a mass dependent efficiency for
galaxy formation can be easily used to explain the excess of entropy
in galaxy clusters relative to smaller groups (see, e.g., \cite{Bry00}
2000).

To calculate the 1-halo term of the galaxy power spectrum, in addition to the mean number of
galaxies, one also require information on the second moment of the
galaxy distribution. Using semi-analytic models, \cite{Scoetal00} (2000), advocate
\begin{equation}
\left< N_\gal (N_\gal -1)\right> = \alpha(M)^2 \left< N_\gal \right>^2
\end{equation}
where $\alpha(M)$ is used to quantify the deviations from Poisson
statistics. 

In \cite{Scoetal00} (2000), $\alpha(M) \sim log(M/10^{11}
h^{-1} M_{\sun})^{0.5}$ out to a mass of $10^{13}$ $h^{-1}$ M$_{\sun}$
while $\alpha(M)=1$ thereafter. Other variants to this approach are
considered in \cite{Sel00} (2000).

\begin{figure}[!h]
\begin{center}
\includegraphics[width=4.2in,angle=-90]{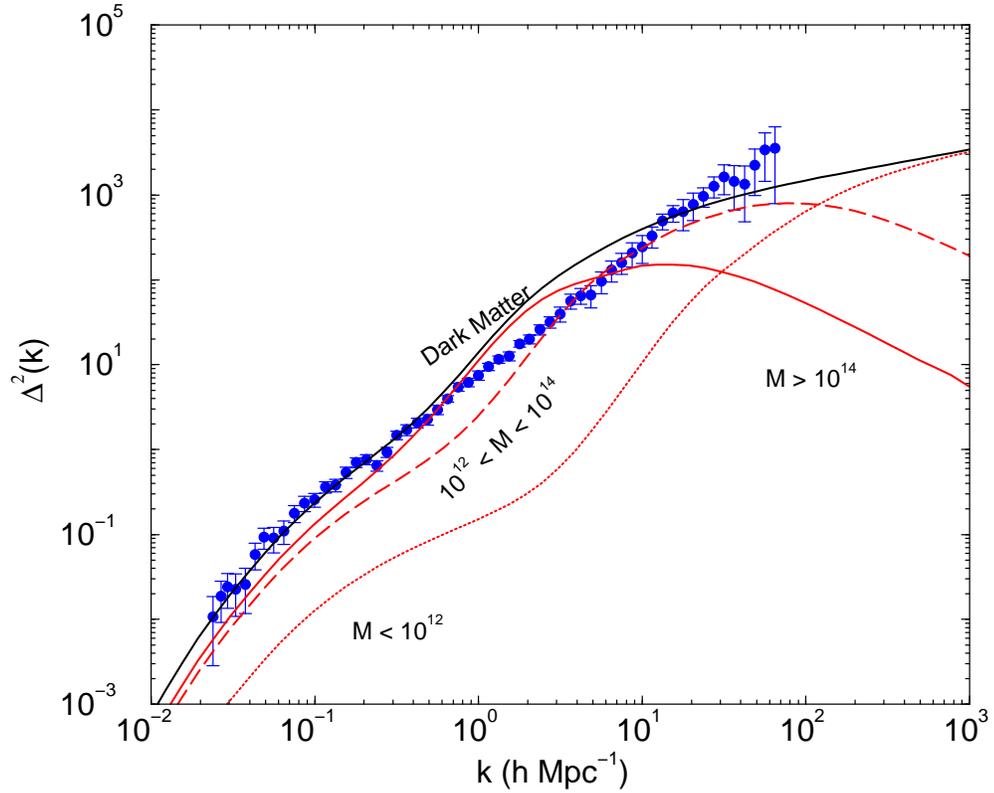}
\end{center}
\caption[Dark matter power spectrum as a function of mass]{The
non-linear dark matter power spectrum under the halo approach broken
to contributions as a function of mass. The large scale power is
produced by massive halos while the small scales power is produced by
the
small mass halos. The galaxy power spectrum is produced through
appropriate scaling of the contributions to the dark matter power
spectrum as a function of mass through the average number of galaxies
vs. mass relation.}
\label{fig:galaxymass}
\end{figure}

Using the information on the galaxy distribution within halos, 
the 2-halo and 1-halo terms for the galaxy power spectrum is
\begin{equation}
P^{2h}_{\gal}(k) = P^\lin(k) \left[ \int dM \frac{d\bar{n}}{dM} b_1(M)
\frac{\left< N_\gal\right>}{\bar{n}_\gal} y_\delta(k,M)\right]^2 \, ,
\end{equation}
and
\begin{equation}
P^{1h}_{\gal}(k) = \left[ \int dM \frac{d\bar{n}}{dM} 
\frac{\left< N_\gal(N_\gal-1)\right>}{\bar{n}_\gal^2} y_\delta(k,M)\right]^2 \, ,
\end{equation}
respectively. With the the mean number of galaxies per halo, as a function of
mass, the mean number density of galaxies can be written as an
integral over
the PS mass function
\begin{equation}
\bar{n}_\gal = \int dM \, \left< N_\gal \right>\frac{dn}{dM}(M,z)
\, .
\end{equation}

At large scales, since the galaxy power spectrum can be written as a
simply scaling of the linear power spectrum
\begin{equation}
P_{\gal}(k) = b_\gal^2 P^\lin(k),
\end{equation}
we can write the galaxy bias at such linear scales as a mass weighted halo bias
\begin{equation}
b_\gal = \int dM \frac{d\bar{n}}{dM} b_1(M) \frac{\left< N_\gal \right>}{\bar{n}_\gal} \, .
\end{equation}
With sufficient statistics, a measurement of the galaxy power spectrum
at linear scales, as a function of galaxy type or environment, allows
one to relate the observed bias to a mean mass of halos in which
galaxies under study reside. It is likely that such studies can easily
be carried out with wide-field surveys such as the Sloan Digital Sky
Survey (SDSS).

\begin{figure}[!h]
\begin{center}
\includegraphics[width=4.2in,angle=-90]{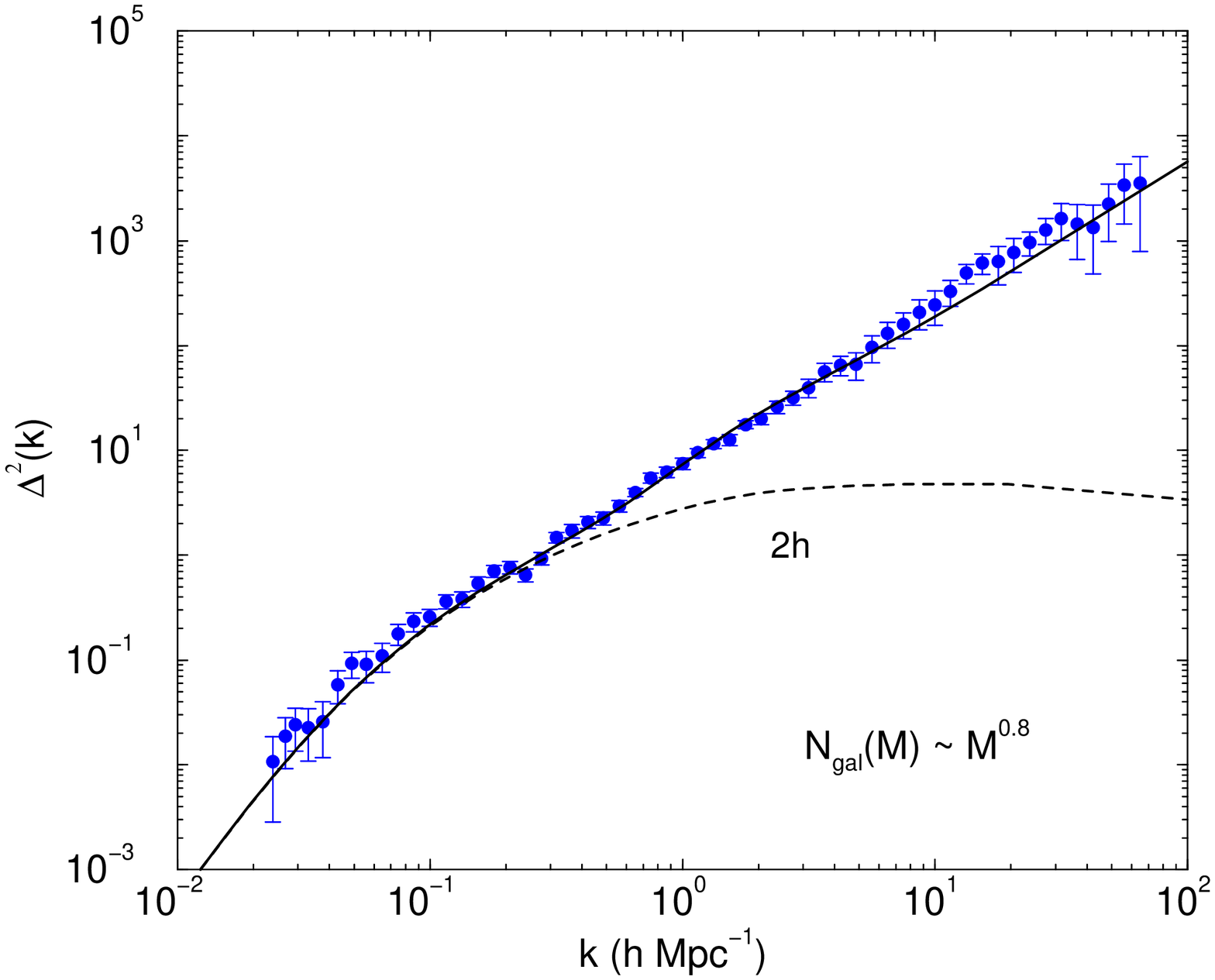}
\end{center}
\caption[Galaxy power spectrum under the halo model]{The PCSZ galaxy
power spectrum produced through the mean number of galaxies as a
function of mass relation. With appropriate scaling, the halo model
produces the almost power law galaxy power spectrum measured in the
PCSZ survey.}
\label{fig:galaxypower}
\end{figure}

To construct the galaxy power spectrum, through the relation involving  the mean number of galaxies as a
function of mass, one essentially rescales the contribution to the
dark matter power spectrum. The scaling through $N_\gal \sim M^{0.8}$  is such that one weighs
the high mass end of dark matter halos relatively higher than the low mass end.
In Fig.~\ref{fig:galaxymass}, we show the dark
matter power spectrum such that contributions are  separated as a
function of mass. In Fig.~\ref{fig:galaxypower}, we show the prediction for the galaxy
power spectrum, with parameters for the galaxy distribution as defined
above. Note that the we have not tried to vary the parameters for the
galaxy prescription so as to fit the PCSZ redshift-space galaxy power
spectrum. Given that there are still discrepancies between this
prediction and the measured galaxy power spectrum, it is likely that
some variants of the parameters can lead to a better model. We leave
these detailed issues to future studies, since we are primarily
interested in here for a simple description of galaxy power spectrum
under the halo approach.

\section{Discussion}

Even though the dark matter halo formalism provides a physically
motivated means for calculating the statistics of the dark matter density field,
there are several limitations of the approach that should be borne in
mind when interpreting the results.  

The approach assumes all halos to be spherical with a single
profile shape. Any variations in the profile through halo mergers and
resulting substructure can affect the power spectrum and higher order
correlations. Also, real halos are not perfectly
spherical which affects the configuration dependence of the
bispectrum.  Furthermore, there are parameter degeneracies in the formalism that
prevent a straightforward interpretation of observations in terms of halo
properties. For example, one might think that the power spectrum and bispectrum 
can be used to measure any mean deviation from the assumed NFW profile
form. However as pointed out by
\cite{Sel00} (2000), changes in the slope of the inner profile 
can be compensated by changing the concentration as a function of
mass; this degeneracy is also preserved in the bispectrum. 

In the case of the trispectrum and power specrum covariance, we have attempted to 
include variations in the halo profiles with the addition of a
distribution function for concentration parameter based on results from  
numerical simulations. Also, for the calculation involving dark matter trispectrum 
and covariance, we have not modified the concentration-mass relation to fit the PD 
non-linear power spectrum, but rather have taken results directly 
from simulations as inputs. Though we have partly accounted for halo
profile variations,  the assumption that halos are spherical is 
likely to affect detailed results on the configuration dependence 
of the bispectrum and trispectrum. 

We do not expect these issues
 to affect our  qualitative results.
If this technique is to be used for precision studies of cosmological
parameters, however, more work will be required in testing it quantitatively against simulations.
Studies by \cite{MaFry00a} (2000a) show that
the bispectrum predictions of the halo formalism are in good agreement
with simulations, at least when averaged over configurations.
\cite{Scoetal00} (2000) find that there are discrepancies at the $\sim 20-30\%$
level in the mildly non-linear regime that show up most markedly in the
configuration dependence; uncertainties in the mass function, with
respect to the mass functions produced in simulations, also produce variations at
this level.  The replacement of individual halos found in numerical simulations
with synthetic smooth halos with NFW profiles by \cite{MaFry00b}
(2000b) show that the smooth profiles can regenerate the measured
power spectrum and bispectrum in simulations. This agreement, at least
at scales less than 10$k_{\rm nonlin}$,
suggests that mergers and substructures may not be important at such
scales.

The agreement between the power spectrum and bispectrum for a given halo
prescription is also significant in that, as we shall see, the two statistics
weight high mass halos very differently.   The agreement serves as a test
that the halo prescription correctly captures the halo mass dependence of
the statistics.  We conclude that the halo model is useful in that it provides
a means to study the halo mass dependence of two, three and four point
statistics and an approximate means to bridge the gap
between the linear regime where PT is valid
and the non-linear regime where extensions such as HEPT can be used.

In the deeply non-linear regime (here $k \simgt 10 h$Mpc$^{-1}$)
there are qualitative differences between the halo predictions and HEPT.  
Unfortunately, current state-of-the-art simulations do not have the resolution to
address the differences \cite{Scoetal00} (2000).  For weak lensing purposes,
the differences are less relevant since in the deeply non-linear regime
shot-noise from the intrinsic ellipticities of the galaxies will
likely dominate. We will now discuss applications of the halo model to
weak gravitational lensing.

\chapter{Weak Gravitational Lensing}

\section{Introduction}

Weak gravitational lensing of faint galaxies probes the
distribution of matter along the line of sight.  Lensing by
large-scale structure (LSS) induces
correlation in the galaxy ellipticities at the percent level
(e.g., \cite{Blaetal91} 1991; \cite{Mir91} 1991; 
\cite{Kai92} 1992).  Though challenging to measure, these
correlations provide important cosmological information that is
complementary to that supplied by
the cosmic microwave background and potentially as precise
(e.g., \cite{JaiSel97} 1997;
\cite{Beretal97} 1997; \cite{Kai98} 1998; \cite{Schetal98}
1998; \cite{HuTeg99} 1999; \cite{Coo99} 1999; \cite{Vanetal99} 1999;
see \cite{BarSch00} 2000 for a recent review).
Indeed several recent studies have provided the first clear evidence
for weak lensing in so-called blank fields (e.g., \cite{Vanetal00}
2000; \cite{Bacetal00} 2000; \cite{Witetal00} 2000; \cite{Kaietal00}
2000), though more work is
clearly needed to understand even the statistical errors 
(e.g. \cite{Cooetal00b} 2000b).

Weak lensing surveys are currently limited to small fields which may
not be representative of the universe as a whole, owing to sample
variance.  In particular, rare massive objects can contribute
strongly to the mean power in the shear or convergence but not
be present in the observed fields.  The problem is compounded
if one chooses blank fields subject to the condition that they do not
contain known clusters of galaxies. The objective with halo approach is to
(1) quantify these effects and to understand what fraction of the
total convergence power spectrum and higher order correlations arise
from lensing by individual massive clusters as a function of scale
and (2) understand how the sample variance effects affect the cosmological
interpretation of weak lensing convergence observations through galaxy 
shear data. In this chapter, we address the first issue while the second 
issue is discussed in the next chapter.

Given that weak gravitational lensing results from the projected mass
distribution, the statistical properties of weak lensing convergence
reflect those of the dark matter. 
Non-linearities in the mass distribution induce non-Gaussianity
in the convergence distribution. These non-Gaussianities contribute to
the covariance of power spectrum measurements, especially in the case
when observations are limited to a finite field of view and the
measurements are binned in multipole space.
Here, we present an analytical estimate on the covariance of binned
power spectrum, based on the non-Gaussian contribution.
The calculation of the full convergence covariance requires detailed
knowledge of the dark matter density bispectrum, which can be
obtain analytically through perturbation theory (e.g.,
\cite{Beretal97} 1997) or numerically through simulations (e.g.,
\cite{JaiSelWhi00} 2000; \cite{WhiHu99}
1999). Perturbation theory, however, is not applicable at all scales
of
interest, while numerical simulations are limited by computational
expense
to a handful of realizations of cosmological models with modest
dynamical
range.   Here, we use a recent popular approach to obtain the density
field bispectrum analytically by describing the underlying three point
correlations as due to contributions from (and correlations between)
individual dark matter halos. 

Techniques for studying the dark matter density field through
halo contributions have recently been developed 
(\cite{Sel00} 2000; \cite{MaFry00b} 2000b; \cite{Scoetal00} 2000) 
and applied to two-point and three-point lensing statistics
(\cite{Cooetal00b} 2000b; \cite{CooHu00} 2000).
The critical ingredients are: a mass function for the halo
distribution, such as the Press-Schechter (PS; \cite{PreSch74} 1974) 
or Sheth-Tormen (ST; \cite{SheTor99} 1999) mass function; a profile
for the dark matter halo, e.g., the profile of \cite{Navetal96} (1996;
NFW),  and a description of halo biasing (\cite{Moetal97} 1997;
extensions in \cite{SheLem99} 1999 and \cite{SheTor99} 1999). 
The dark matter
halo approach provides a physically motivated method to calculate the
correlation functions.   
Since lensing probes scales
ranging from linear to deeply non-linear, this is an important
advantage
over perturbation-theory calculations.

\begin{figure}[!h]
\begin{center}
\includegraphics[width=5.9in]{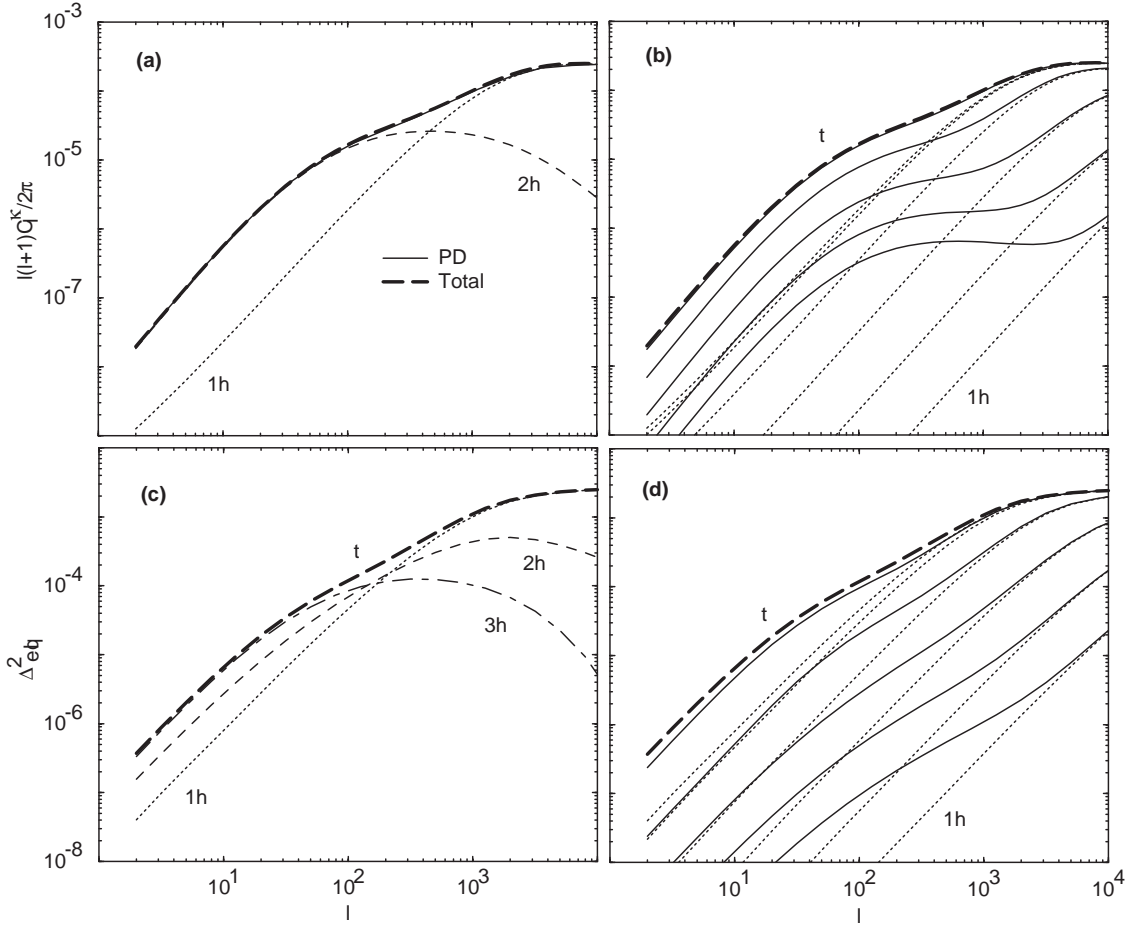}
\end{center}
\caption[Weak lensing power spectrum and bispectrum]{Weak lensing convergence (a) power spectrum and (c)
bispectrum under the halo description.
Also shown in (a) is the prediction from the PD nonlinear power
spectrum fitting function. We have separated individual contributions
under the halo approach to weak lensing angular power spectrum and bispectrum.
The mass cut off effects on the weak lensing convergence power
spectrum (d) and bispectrum (d).  The maximum mass used is same as
in Fig.~\ref{fig:dmpower}(b \& d). We have assumed that all sources are at $z_s\
=1$.}
\label{fig:weakpower}
\end{figure}

\section{Convergence Power Spectrum}
\label{sec:convergence}

The angular power spectrum of the convergence is defined in
terms of the multipole moments $\kappa_{l m}$ as
\begin{equation}
\left< \kappa_{l m}^* \kappa_{l' m'}\right> = C_l^\kappa \delta_{l l'}
\delta_{m
 m'}\,.
\end{equation}
$C_l$ is numerically equal to the flat-sky power spectrum
in the flat sky limit.
It is related to the dark matter power spectrum by (\cite{Kai92} 1992;
1998)
\begin{equation}
C^\kappa_l = \int d\rad \frac{W^\lens(\rad)^2}{d_A^2} 
P^\tot\left(\frac{l}{d_A};\rad\right) \, ,
\label{eqn:lenspower}
\end{equation}
where $\rad$ is the comoving distance and  $d_A$ is the angular
diameter
distance.  When all background sources are at a distance of 
$\rad_s$, the weight function becomes
\begin{equation}
W^\lens(\rad) = \frac{3}{2} \Omega_m \frac{H_0^2}{c^2 a} \frac{
d_A(\rad) d_A(\rad_s -\rad)}{d_A(\rad_s)} \, ; 
\label{eqn:weight}
\end{equation}
for simplicity, we will assume $\rad_s = r(z_s=1)$ throughout.
In deriving Eq.~\ref{eqn:lenspower}, we have used the
Limber approximation (\cite{Lim54} 1954) by setting $k=l/d_A$ and
the flat-sky approximation. A potential problem in
using the Limber approximation is that 
we implicitly integrate over the unperturbed photon paths
(Born approximation).   The Born approximation has been tested
in numerical simulations by \cite{JaiSelWhi00} (2000; see
their Fig.~7) and found to be an excellent approximation for
the two point statistics. The same approximation can
also be tested through lens-lens coupling involving lenses
at two different redshifts. For higher order correlations, analytical
calculations in the mildly non-linear regime by 
\cite{vanetal00} (2000b; also, \cite{Beretal97}
1997; \cite{Schetal98} 1998) indicate that corrections are again 
less than a few percent. Thus, our use of the Limber approximation
by ignoring the lens-lens coupling is not expected to 
change the final results significantly.

In Fig.~\ref{fig:weakpower}(a), we show the convergence power
spectrum of the dark matter halos compared with that predicted by
the \cite{PeaDod96} (1996) power spectrum.
The lensing power spectrum due to halos has the same behavior as the
dark matter power spectrum. At large angles ($l \lesssim 100$),
the correlations between halos dominate.
The transition from linear to non-linear is at $l \sim
500$ where halos of mass similar to $M_{\star}(z)$ contribute.
The single halo contributions start dominating at $l > 1000$.
When $l \gtrsim$ few thousand, at small scales corresponding to deeply
non-linear regime, the intrinsic correlations
between individual background galaxy shapes can complicate the
accurate recovery of lensing signal (\cite{CroMet00} 2000;
\cite{Heaetal00} 2000; \cite{Catetal00} 2000). Therefore, it is
unlikely that the lensing observations can be used to test various
clustering models that are relevant to such non-linear regimes.

As shown in Fig.~\ref{fig:weakpower}(b), and discussed in \cite{Cooetal00b}
(2000b), if there is a lack of massive halos in the observed fields
convergence measurements will be biased low compared with the cosmic 
mean.  The lack of massive halos affect the
single halo contribution more than the halo-halo correlation term,
thereby changing the shape of the total power spectrum in addition to
decreasing the overall amplitude. Since the lensing power spectrum is simply a 
projected measure of the dark
matter power spectrum, the variations in the weak lensing angular power 
spectra are
consistent with the behavior observed in the dark matter power
spectrum.

\begin{figure}[!h]
\begin{center}
\includegraphics[width=5.9in]{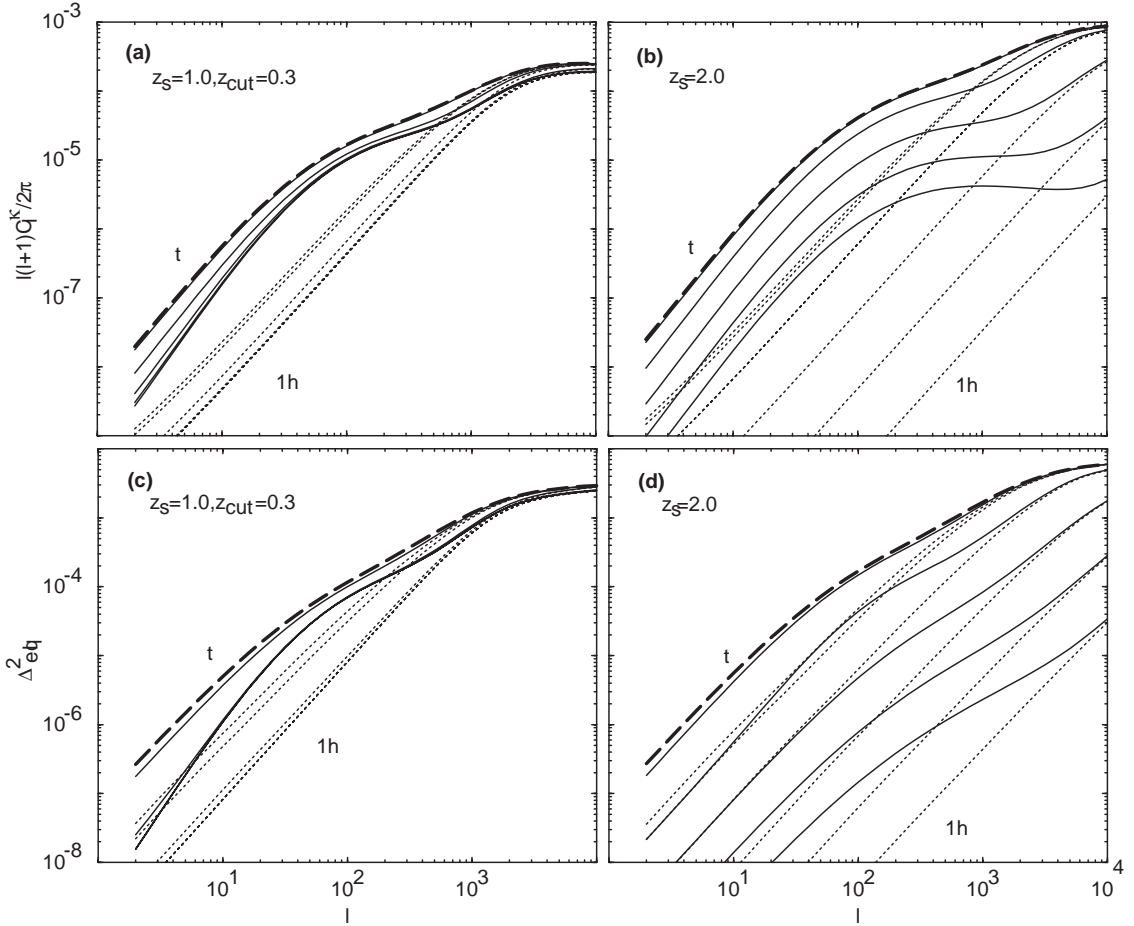}
\end{center}
\caption[Weak lensing convergence power spectrum and bispectrum as a 
function of source redshift]{Weak lensing convergence spectra under the halo description
for sources at $z_s=1$ with a mass cut off only applied to halos at
$z_c = z < 0.3$ and for source $z_s=2$ with mass cut off to the same
redshift: (a) \& (b) angular power spectrum and (c) \& (d) Equilateral
bispectrum. 
The mass cuts are the same as in Fig.~\protect\ref{fig:dmpower}(b \& d).
A significant fraction of the
effect comes from rare massive halos at high redshift.}
\label{fig:zcut}
\end{figure}

It is interesting to study the origin of this result in terms of the
physical parameters to see how they depend on assumptions.
The lensing convergence weight function (Eq.~\ref{eqn:weight}) peaks
at half the angular diameter distance to background sources,\footnote{The physical scale
in the halos roughly corresponds to the angular scale times half the angular diameter
distance to the source.  For example at one arcmin, the scale corresponding
to sources at $z_s=1$ is $\sim$ 400 kpc.}
which for our fiducial $\Lambda$CDM model with sources at $z_s = 1$ corresponds
to $z \approx 0.4$ with the growth of structures shifting this peak
redshift to a slightly lower value.
In Fig.~\ref{fig:zcut}(a \& c), we show the
result of the mass cuts where only those halos for which $z<0.3$ and
$M<M_{\rm cut}$ are excluded.   
Note that the sensitivity to the mass threshold is reduced
indicating that a substantial fraction of the effect comes from rare
massive halos at high redshift.  As shown in
Fig.~\ref{fig:zcut}(b \& d) when $z_s=2$,
changing the source redshift therefore does not affect the
results qualitatively.

\begin{figure}[!h]
\begin{center}
\includegraphics[width=4.5in]{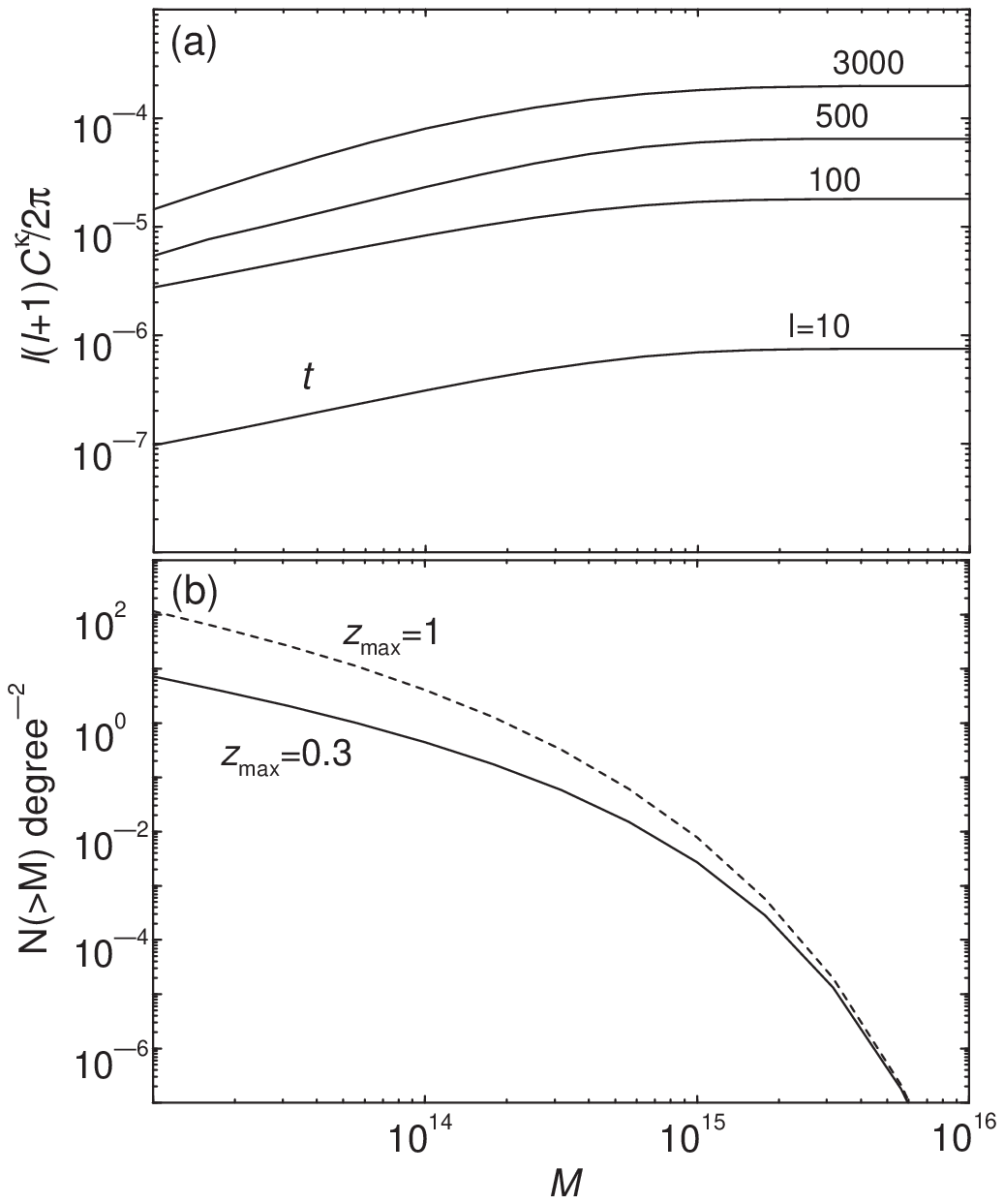}
\end{center}
\caption[Convergence power as a function of mass]{(a)
Total lensing convergence $C_\kappa^{\rm tot}$ as a function of
maximum mass for several $l$-values and sources at $z_{s}=1$.
As shown, contributions from halos with masses $>10^{15}$ $M_{\sun}$
are negligible.
(b) Surface density of halo masses as a function of minimum
mass using PS formalism out to $z_{\rm max}=0.3$ and $z_{\rm max}=1$. This determines
the survey area needed to ensure a fair sample of halos greater than
a given mass.}
\label{fig:press}
\end{figure}

In Fig.~\ref{fig:press}(a), we show the dependence of $C_{\kappa}$, 
for several $l$ values.  If halos
$<10^{15}$ $M_{\sun}$ are well represented in a survey, then
the power spectrum will track the LSS convergence power spectrum
for all $l$ values of
interest.  The surface number density of halos determines
how large a survey should be to possess a fair sample of halos
of a given mass.  We show this in Fig.~\ref{fig:press}(b)
as predicted by PS formalism for our fiducial
cosmological model for halos out to
($z=0.3$ and $z=1.0$). Since the surface number density of
$>10^{15} M_{\sun}$ halos out to a redshift of 0.3 and 1.0 is
$\sim$ 0.03 and 0.08 degree$^{-2}$
respectively, a survey of  order
$\sim$ 30 degree$^2$ should be sufficient to contain a fair
sample of the universe for recovery of the full LSS convergence
power spectrum.

One caveat is that mass cuts may affect the higher moments
of the convergence differently so that a fair sample for
a quantity such as skewness will require a different survey
strategy.  From numerical simulations
(\cite{WhiHu99} 1999), we know that $S_{3}\equiv \left< \kappa^{3}
\right> /\left< \kappa^{2}\right>^{2}$ shows substantial sample
variance, implying that it may be dominated by
rare massive halos. As we find later, when calculated using density field bispectrum
constructed using dark matter halos,
the skewness decreased by  a factor of $\sim$ 10 with a mass cut off at
$\sim$ $10^{13}$ M$_{\sun}$ at an angular
scale of $10'$ from the maximum value with masses out to $\sim$
10$^{16}$ M$_{\sun}$. We will discuss issues related to
non-Gaussianities in the next section.

While upcoming wide-field weak lensing surveys, such as the MEGACAM
experiment at Canada-France-Hawaii Telescope (\cite{Bouetal98} 1998), and the
proposed wide field survey by Tyson et al. (2000, private
communication) will cover areas up to $\sim$ 30 degree$^2$ or more,
the surveys that have been so far published, e.g.,
\cite{Witetal00} (2000), only cover at most 4 degree$^2$ in areas
without known clusters. The observed convergence
in these fields should be biased low compared with the mean
and vary widely from field to field due to sample variance from
the Poisson contribution of the largest mass halos in the fields,
which are mainly responsible for the sample variance below $10'$ (see
\cite{WhiHu99} 1999).

Our results can also be used proactively.
If properties of the mass distribution such as the maximum mass
halo in the observed lensing fields are known, say through prior optical,
X-ray, SZ or even internally in the lensing observations (see
\cite{KruSch99} 1999), one can
make a fair comparison of the observations to theoretical model
predictions
with a mass cut off in our formalism.  Even for larger surveys,
the identification and extraction of large halo contributions can
be beneficial: most of the sample variance in the fields will be
due to rare massive halos. The dependence of massive halos in
producing a large non-Gaussian signal can also be used to identify their
presence and perhaps correct the possible non-fair sampling  of
observing fields and variance of convergence measurement.
A reduction in the sample variance increases the precision with which the power spectrum
can be measured and hence the cosmological parameters upon which
it depends. In the next Chapter, we will address the effect on
cosmological parameters  due to non-Gaussianities and the associated
sample variance.

\begin{figure}[!h]
\begin{center}
\includegraphics[width=5.9in]{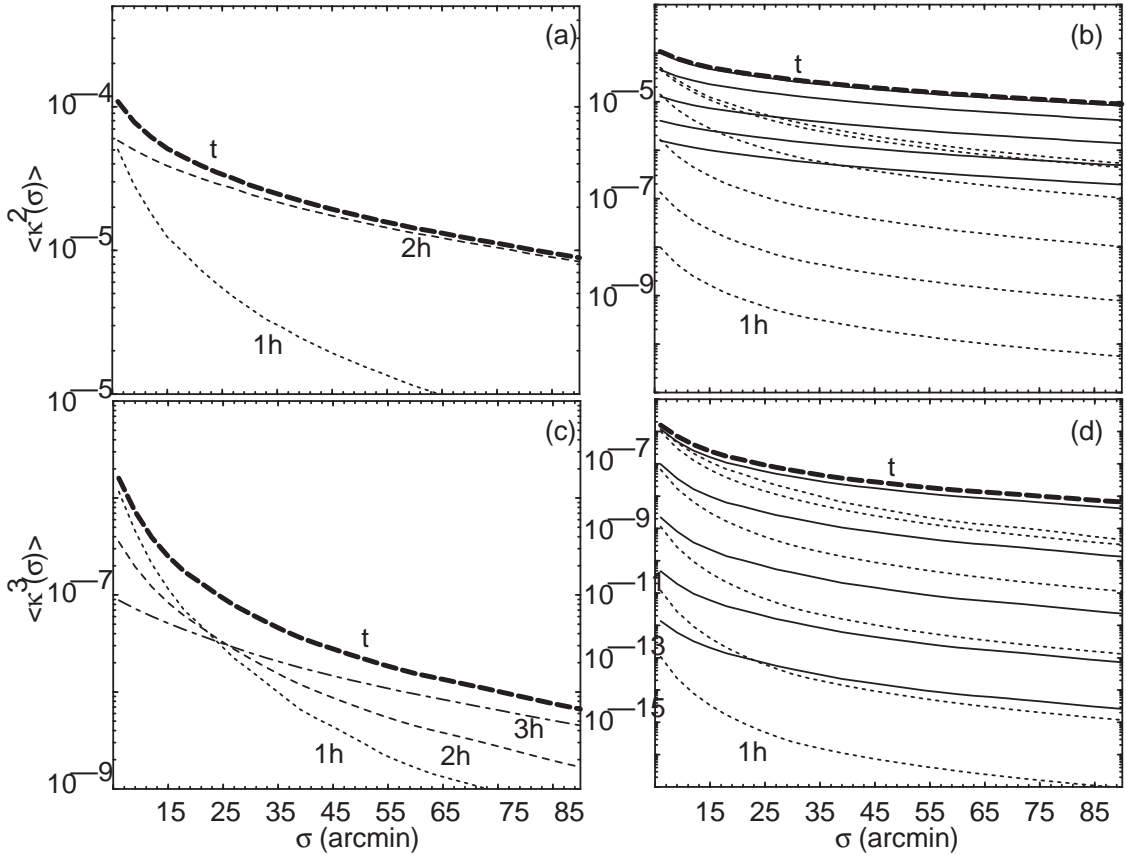}
\end{center}
\caption[Weak lensing convergence moments in real space]{Moments of the convergence field as a function of
top-hat smoothing scale $\sigma$.
(a) Second moment broken into individual contributions.
(b) Mass cut off effects on the second moments.
(a) Third moment broken into individual contributions.
(b) Mass cut off effects on the third moments.
The mass cuts are the same as in Fig.~\protect\ref{fig:dmpower}.
}
\label{fig:moments}
\end{figure}
  
In the case of the two point function, one can also consider the
second moment, or variance, in addition to the power spectrum.
The variance of a map smoothed with a window is
related to the power spectrum by
\begin{equation}
\left< \kappa^2(\sigma) \right> =
{1 \over 4\pi} \sum_l (2l+1) C_l^\kappa W_l^2(\sigma)\,.
\label{eqn:secondmom}
\end{equation}
where $W_l$ are the multipole moments (or Fourier transform in a
flat-sky approximation) of the window.   For simplicity, we will
choose 
a window which is a two-dimensional top hat in real space with a 
window function in
multipole space of $W_l(\sigma) = 2J_1(x)/x$ with $x = l\sigma$.

In Fig.~\ref{fig:moments}(a-b), we show the second moment as a
function of
smoothing scale $\sigma$. Here, we have considered angular scales
ranging from 5$'$ to 90$'$, which are likely to be probed by ongoing
and upcoming weak lensing experiments.
As shown, most of the contribution to the second
moment comes from the double halo correlation term and is mildly
affected by a 
mass cut off. 
  
\begin{figure}[!h]
\begin{center}
\includegraphics[width=5.2in]{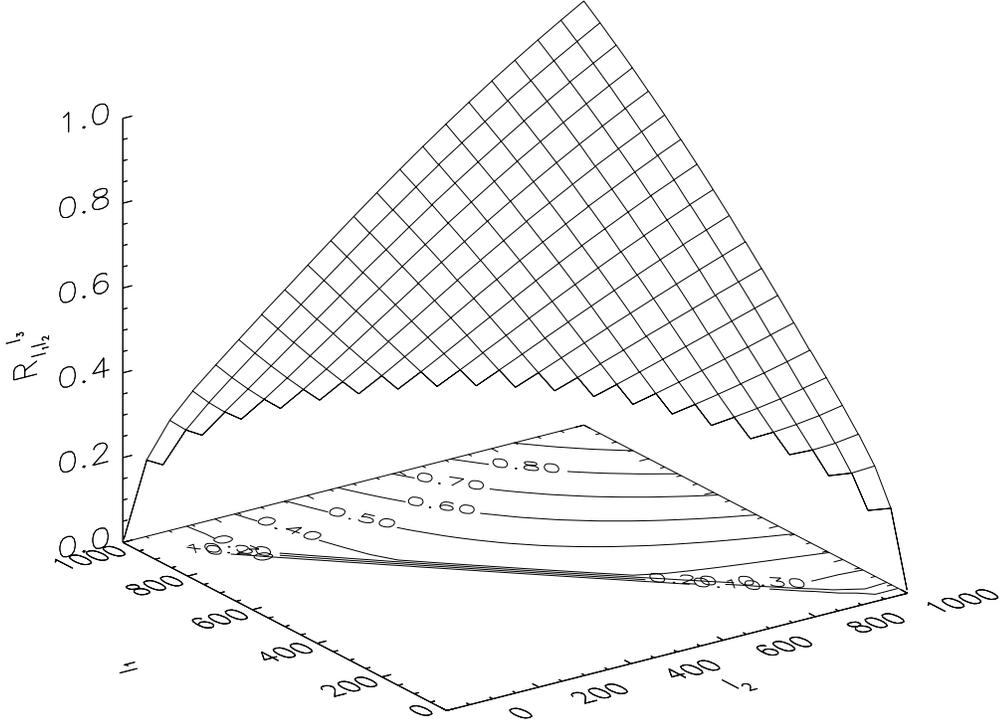}
\end{center}
\caption[Weak lensing convergence bispectrum surface]{The bispectrum configuration dependence $R_{l_1l_2}^{l_3}$
 as a function of $l_1$ and $l_2$ with
$l_3=1000$.  Due to triangular conditions associated with $l$'s, only the
upper triangular region in $l_1$-$l_2$ space contribute to the
bispectrum.}
\label{fig:bispecsurface}
\end{figure}

\section{Convergence Bispectrum}
The angular bispectrum of the convergence is defined as
\begin{equation}
\left< \kappa_{l_1 m_1} \kappa_{l_2 m_2} \kappa_{l_3 m_3} \right> 
= \wjm B_{l_1 l_2 l_3}^\kappa \,, 
\end{equation}
where 
\begin{equation}
\kappa(\bn) = \sum \kappa_{l m} \Ylmn(\bn) \, ,
\end{equation}
with spherical moments of the convergence field defined such that
\begin{eqnarray}
\kappa_{lm} &=& i^l \int \frac{d^3\veck}{2 \pi^2}
\delta_\delta(\veck)  I_l^\lens(k) \Ylmn(\hat{\veck}) \, , \nonumber\\
I_l^\lens(k) &=& \int d\rad  W^\lens(k,\rad)j_{l}(k\rad) \, ,
\label{eqn:secondaryform}
\end{eqnarray}
where $W(k,\rad)$ is the source function associated with weak lensing
(see, Eq.~\ref{eqn:weight}). 
Here, we have simplified using the Rayleigh expansion of a plane wave
\begin{equation}
e^{i{\bf k}\cdot \hat{\bf n}\rad}=
4\pi\sum_{lm}i^lj_l(k\rad)Y_l^{m \ast}(\bk) \Ylmn(\bn)\, .
\label{eqn:Rayleigh}
\end{equation}

The bispectrum can be written through
\begin{eqnarray}
&& \left< \kappa_{l_1 m_1}\kappa_{l_2 m_2}\kappa_{l_3 m_3}  \right>
=  i^{l_1+l_2+l_3} \nonumber \\
&\times& \int \frac{d^3\veck_1}{2 \pi^2}
\int \frac{d^3\veck_2}{2 \pi^2} \int \frac{d^3\veck_3}{2 \pi^2}
\left<\delta_\delta(\veck_1)  \delta_\delta(\veck_2)  \delta_\delta(\veck_3)  \right>
\nonumber \\ &\times& I_l^\lens(k_1) I_l^\lens(k_2) I_l^\lens(k_3)
\Ylm{1}(\hat{\veck_1})  \Ylm{2}(\hat{\veck_2})  \Ylm{3}(\hat{\veck_3})
\, ,
\end{eqnarray}
and can be simplified further by using the bispectrum of density fluctuations
\begin{eqnarray}
&& \left< \kappa_{l_1 m_1}\kappa_{l_2 m_2}\kappa_{l_3 m_3}  \right>
=  i^{l_1+l_2+l_3} \int \frac{d^3\veck_1}{2 \pi^2}
\int \frac{d^3\veck_2}{2 \pi^2} \int \frac{d^3\veck_3}{2 \pi^2}
\nonumber \\ &\times&
(2 \pi)^3 B_\delta(k_1,k_2,k_3) \delta_D(\veck_{123})
\nonumber \\
&\times& I_l^\lens(k_1) I_l^\lens(k_2) I_l^\lens(k_3)
\Ylm{1}(\hat{\veck_1})  \Ylm{2}(\hat{\veck_2})  \Ylm{3}(\hat{\veck_3})
\, ,
\end{eqnarray}
the expansion of a delta function
\begin{equation}
\delta_D(\veck_{123})= \int
\frac{d^3\vecx}{(2\pi)^3} e^{i \vecx \cdot (\veck_1+\veck_2+\veck_3)}\, ,
\end{equation}
and the Rayleigh expansion (Eq.~\ref{eqn:Rayleigh}), to write
\begin{eqnarray}
&&\left< \kappa_{l_1 m_1}\kappa_{l_2 m_2}\kappa_{l_3 m_3}  \right>
=  \frac{2^3}{\pi^3} \int k_1^2 dk_1 \int k_2^2 dk_2 \int k_3^2 dk_3 \nonumber \\
&\times& B_\delta(k_1,k_2,k_3)
 I_l^\lens(k_1) I_l^\lens(k_2) I_l^\lens(k_3) \nonumber \\
&\times& \int x^2 dx  j_{l_1}(k_1x) j_{l_2}(k_2x) j_{l_3}(k_3x) 
\int d{\bf \hat{x}}\Ylm{1}({\bf \hat{x}})
\Ylm{2}({\bf\hat{x}})  \Ylm{3}({\bf \hat{x}}) \, .
\end{eqnarray}

Here, the density bispectrum should be understood as arising from the
full unequal time correlator
\begin{equation} 
\left< \delta_\delta(\bfk_1;r_1) \delta_\delta(\bfk_2;r_2) \delta_\delta(\bfk_3;r_3) \right> \,,
\end{equation}
where the temporal coordinate is introduced to the source functions through individual $I_l^\lens$'s.

Using the Gaunt integral
\begin{eqnarray}
\int d\bn
        \Ylm{1}
        \Ylm{2}
        \Ylm{3}
=
\sqrt{
        \prod_{i=1}^3(2 l_i+1)\over 4\pi }
        \wj \wjm \,,  \nonumber \\
\label{eqn:harmonicsproduct}
\end{eqnarray}
we can write the convergence bispectrum  as
\begin{eqnarray}
B^\kappa_{l_1 l_2 l_3)} &=& \sum_{m_1 m_2 m_3} \wjm
\left< \kappa_{l_1 m_1}\kappa_{l_2 m_2}\kappa_{l_3 m_3}  \right>  \nonumber \\
&=&  \sqrt{\frac{\prod_{i=1}^3(2l_i +1)}{4 \pi}}
\left(
\begin{array}{ccc}
l_1 & l_2 & l_3 \\
0 & 0  &  0
\end{array}
\right) b_{l_1,l_2,l_3} \, ,
\label{eqn:bigeneral}
\end{eqnarray}
with
\begin{eqnarray}
&& b_{l_1,l_2,l_3} = \frac{2^3}{\pi^3}\int k_1^2 dk_1 \int k_2^2 dk_2
\int k_3^2 dk_3 B_\delta(k_1,k_2,k_3) \nonumber  \\
&\times& I^\lens_{l_1}(k_1) I^\lens_{l_2}(k_2) I^\lens_{l_3}(k_3)
\int x^2 dx  j_{l_1}(k_1x) j_{l_2}(k_2x) j_{l_3}(k_3x) \, . \nonumber
\\
\end{eqnarray}

In general, the calculation of $b_{l_1,l_2,l_3}$ involves seven
integrals involving the mode coupling integral and three integrals
involving distances and Fourier modes, respectively.
For efficient calculational purposes, we can simplify further by using the Limber approximation.
Here, we employ a version based on
the completeness relation of spherical Bessel functions (see,
\cite{CooHu00} 2000 for details)
\begin{equation}
\int dk k^2 F(k) j_l(kr) j_l(kr')  \approx {\pi \over 2} \da^{-2}
\deld(r-r')
                                                F(k)\big|_{k={l\over
d_A}}\,,
\label{eqn:ovlimber}
\end{equation}
where the assumption is that $F(k)$ is a slowly-varying function. This
is in fact  the well known Limber approximation under the weak coupling
approximation (see, \cite{HuWhi96} 1996). Under this assumption,
the contributions to the bispectrum come only from correlations at
equal time surfaces. 

Applying this to the integrals involving $k_1$, $k_2$ and $k_3$
allows us to write the angular bispectrum of lensing convergence as
\begin{eqnarray}
\bi^\kappa &=& \sqrt{\prod_{i=1}^3(2l_i+1) \over 4\pi} \wj \nonumber
\\
&\times&
      \left[ \int dr {[W^\lens(r)]^3 \over \da^4}  B_\delta\left({l_1 \over
        \da},{l_2 \over \da},{l_3\over \da};r\right)\right] \, .
\label{eqn:szbispectrum}
\end{eqnarray}
The more familiar flat-sky bispectrum is simply the expression in
brackets (\cite{Hu00b} 2000b). 
The basic properties of Wigner-3$j$ symbol
can be found in \cite{CooHu00} (2000).

Similar to the density field bispectrum,
we define
\begin{equation}
\Delta^2_{{\rm eq}l} = \frac{l^2}{2 \pi}
\sqrt{B^\kappa_{l l l}} \, ,
\end{equation}
involving equilateral triangles in $l$-space.

In Fig.~\ref{fig:weakpower}(b), we show $\Delta^2_{{\rm eq}l}$.
The general behavior of the lensing bispectrum can be
understood through the individual contributions to the
density field bispectrum: at small multipoles, the triple halo
correlation term  dominates, while at high multipoles,
the single halo term dominates. The double halo term
contributes at intermediate $l$'s corresponding to angular scales of a
few tens of arcminutes.  The variations in the weak lensing bispectrum
as a function of maximum mass  is shown in
Fig.~\ref{fig:weakpower}(d). Here, again, the variations and
consistent with the behavior seen in dark matter bispectrum and
produce qualitatively consistent results regardless of the exact halo
profile or mass function.

\subsection{Skewness}

As discussed in the case of the second moment, it is likely that the
first measurements of higher order correlations in lensing would be
through real space statistics. Thus, in addition to the bispectrum, we
also consider skewness which is
associated with the third moment  of the smoothed map (c.f. 
Eq.~[\ref{eqn:secondmom}])
\begin{eqnarray}
\left< \kappa^3(\sigma) \right> &=&
                {1 \over 4\pi} \sum_{l_1 l_2 l_3}
                \sqrt{\prod_{i=1}^3(2l_i+1) \over 4\pi} \wj \nonumber\\
                &&\times  \bi^\kappa
                W_{l_1}(\sigma)W_{l_2}(\sigma)W_{l_3}(\sigma)
                \,. \nonumber \\
\end{eqnarray}
We then construct the skewness as
\begin{equation}
S_3(\sigma) =
\frac{\left<\kappa^3(\sigma)\right>}{\left<\kappa^2(\sigma)\right>^2}
\, .
\end{equation}

\begin{figure}[!h]
\begin{center}
\includegraphics[width=5.2in]{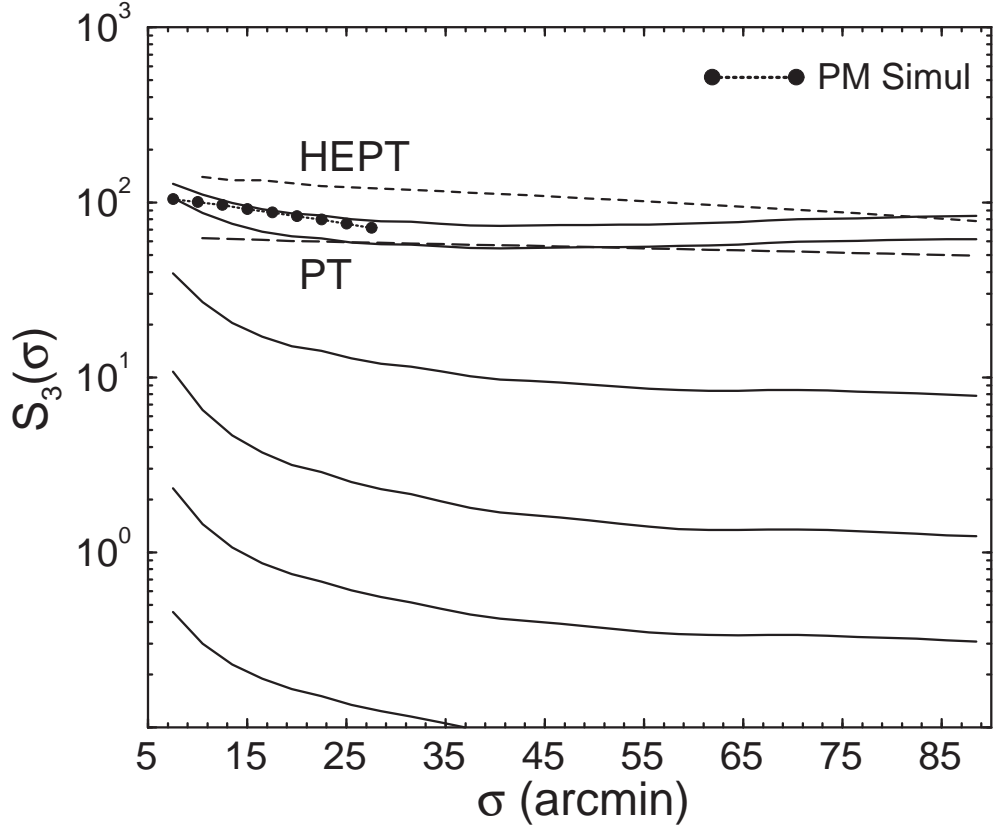}
\end{center}
\caption[Weak lensing convergence skewness]{The skewness, $S_3(\sigma)$, as a function of angular scale.
Shown here is the skewness values with varying maximum mass
as in Fig.~\protect\ref{fig:dmpower}(c-d).
For comparison, we also show
skewness values as measured in particle-mesh (PM) simulations of
\cite{WhiHu99} (1999), as predicted by hyper-extended perturbation
theory (HEPT; dashed line) and
second-order perturbation theory (PT; long-dashed line).}
\label{fig:skewness}
\end{figure}

The effect of the mass cut off is dramatic in the third moment.
As shown  in Fig~\ref{fig:moments}(c-d), most of the contributions to the
third moment come from the single halo term, with those
involving halo  correlations contributing
significantly only at angular scales greater than $\sim$ 25$'$.
With a mass cut off,
the total third moment decreases rapidly and is suppressed
by more than three orders of magnitude when the maximum mass drops to
$10^{13}$ M$_{\sun}$. The skewness only saturates
when the maximum mass is raised to a few times $10^{15}$
M$_{\sun}$. Even though a small change in the maximum mass does not
greatly change the convergence power spectrum (Fig.~3 of
\cite{Cooetal00b} 2000b), the third moment, or the bispectrum, is
strongly sensitive to the rarest or most massive dark matter halos.

In Fig.~\ref{fig:skewness} we plot the
skewness as a function of maximum mass, ranging from
$10^{11}$ to $10^{16}$ M$_{\sun}$.
Our total maximum skewness agrees with what is predicted by numerical
particle mesh simulations (\cite{WhiHu99} 1999) and yields
a value of $\sim$ 116 at 10$'$.
However, it is lower than predicted by HEPT arguments and
simulations of \cite{JaiSelWhi00}
(2000), which suggest a skewness of $\sim$ 140 at angular
scales of 10$'$. The skewness based on second-order PT is factor of
$\sim$ 2 lower than the maximum skewness predicted by halo
calculation. As shown, the PT skewness decreases slightly from angular
scales of few arcminutes to 90$'$ and increases thereafter.
Our halo based calculation of skewness differs from both \cite{Hui99} (1999)
and \cite{Beretal97} (1997) as these authors used HEPT and PT
respectively to calculate lensing skewness.

\begin{figure}[!h]
\begin{center}
\includegraphics[width=4.5in]{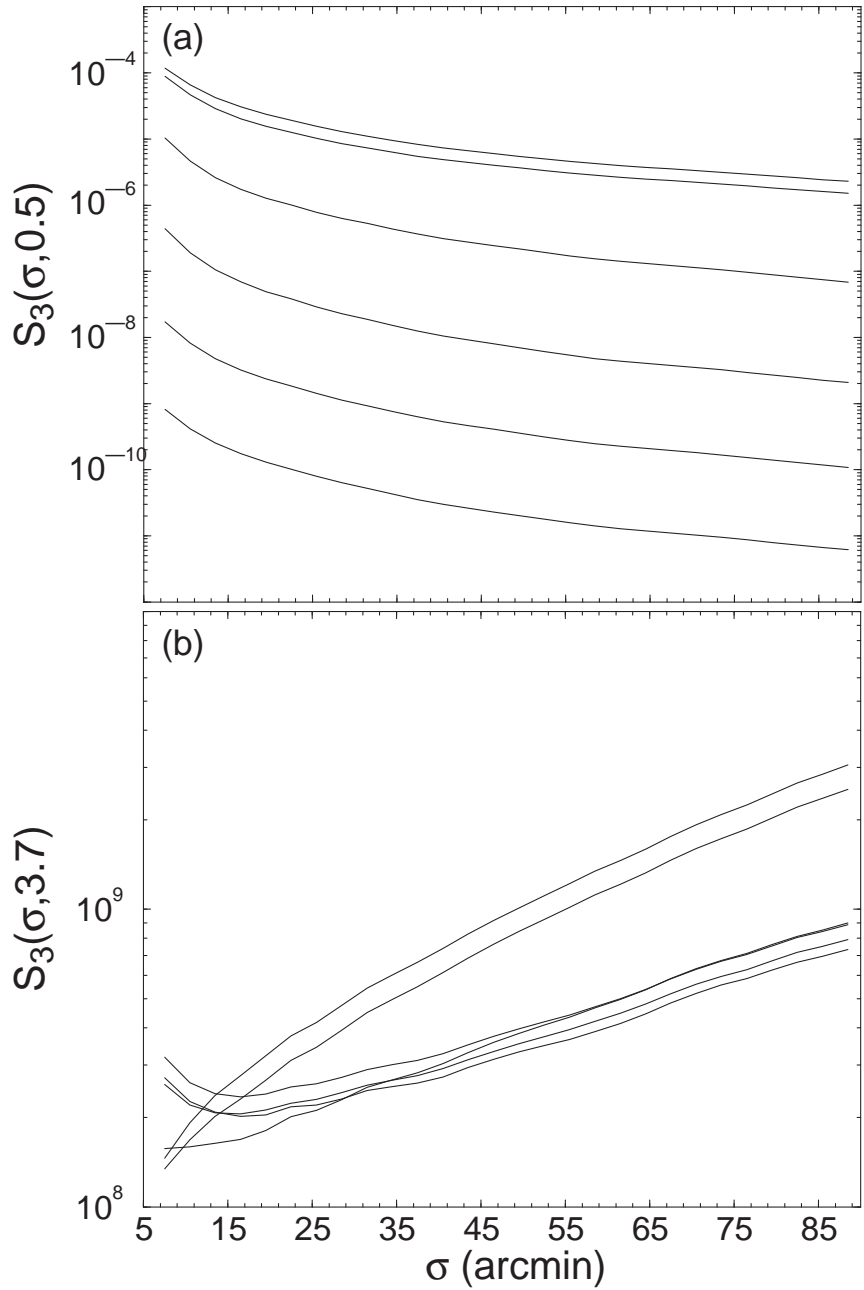}
\end{center}
\caption[Generalized skewness]{Generalized skewness statistic $S_3(\sigma,m)$. (a) $m=1/2$
following \cite{JaiSelWhi00} (2000). (b)
$m=3.7$ chosen to minimize the mass cut off dependence.
}
\label{fig:skewmod}
\end{figure}

The effect of maximum mass on the skewness is interesting. 
When the maximum mass is decreased to
$10^{15}$ M$_{\sun}$ from the maximum mass value where skewness
saturates ($\sim 3\times10^{15}$ M$_{\sun}$), 
the skewness decreases from $\sim$ 116 to 98 at an angular scale of
10$'$, though the convergence power spectrum only changes by less than
few percent when the same change on the maximum mass used 
is made. 
When the maximum mass used in the calculation is
$10^{13}$ M$_{\sun}$, the skewness at 10$'$ is $\sim 8$, which is
roughly a factor of 15 decrease in the skewness from the total.

The variation in skewness as a function of angular scale is due to
the individual contribution to the second and third moments. The
increase
in the skewness at angular scales less than $\sim$ 30$'$ is
due to the single halo contributions for the third moment.
 The triple halo correlation terms dominate angular scales greater
than 50$'$, 
leading to a slight increase toward large angles, e.g.  from
$\sim$ 74 at 40$'$ to $\sim$ 85 at 90$'$. However, this increase is
not present when the 
maximum mass used in the calculation is less than $\sim 10^{14}$
M$_{\sun}$.
Even though mass cut off affects the single halo contributions more
than
the halo contribution, at such masses, the change in halo contribution
with mass cut off prevents an increase in skewness at large angular
scales.

\begin{figure}[!h]
\begin{center}
\includegraphics[width=4.8in]{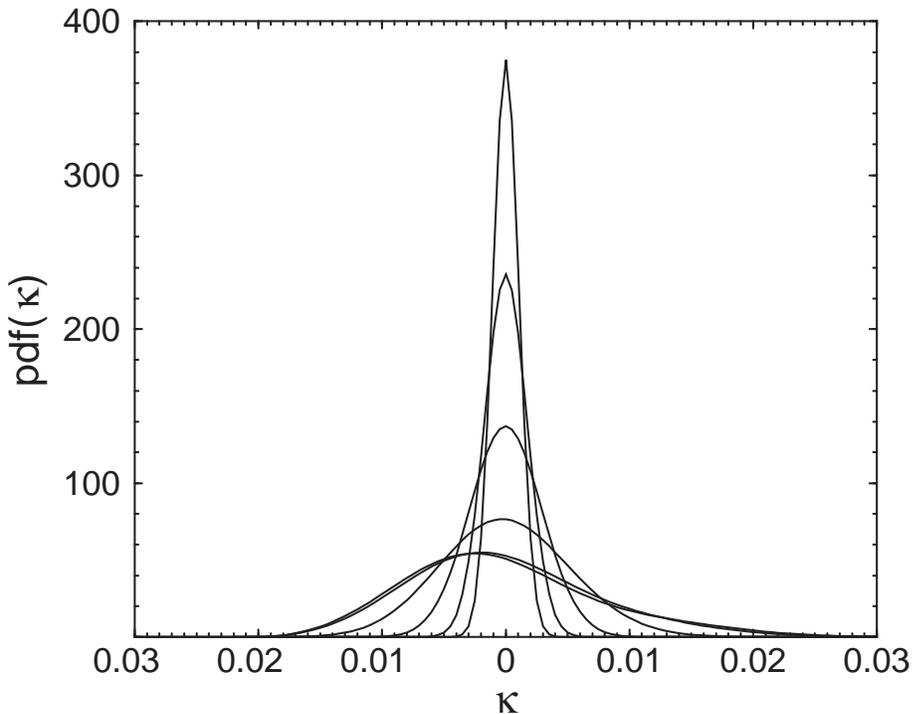}
\end{center}
\caption[Convergence probability distribution function]{The probability distribution function of the weak lensing
convergence as a function of maximum mass used in the calculation at
an angular scale of $12'$. From top to bottom, the curves range from
$10^{11}$ to $10^{16}$ M$_{\sun}$.}
\label{fig:pdf}
\end{figure}

The absence of rare and massive halos in observed fields will
certainly bias 
the skewness measurement from the cosmological mean.  One therefore
needs
to exercise caution in using the skewness to constrain cosmological
models \cite{Hui99} 1999). 
In \cite{Cooetal00b} (2000b), we suggested that lensing observations
in a field of $\sim$ 30 deg$^2$ may be adequate for an unbiased
measurement of the convergence power spectrum. For the skewness,
observations within a similar area may be biased by as much as
$\sim$ 25\%.  This is consistent with the sampling errors found in
numerical simulations: 1$\sigma$ errors of 24\% at $10'$ with a 36
deg$^{2}$ field (\cite{WhiHu99} 1999).
To obtain the skewness within few percent of the total,
one requires a fair sample of halos out to
$\sim 3 \times 10^{15}$ M$_{\sun}$,  requiring observations
of $\sim$ 1000 deg$^2$, which is within the reach of upcoming lensing
surveys involving wide-field cameras,
such as the MEGACAM at Canada-France-Hawaii-Telescope
(\cite{Bouetal98} 1998),
 and proposed dedicated telescopes (e.g., Dark Matter Telescope;
Tyson, private communication).

Still, this does not mean that non-Gaussianity measured in smaller
fields
will be useless.   With this halo approach one can calculate the
expected
skewness if one knows that the most massive halos are not present in 
the observed fields. 
This knowledge may come from external information such as X-ray data
and
Sunyaev-Zel'dovich measurements or internally from the lensing data.

\subsection{Related Statistics}

The halo description in general allows one to test the effect of rare
massive halos 
on any statistic related to the two and three point functions.
In particular, it can be used to design more robust statistics.

Generalized three point statistics have been considered previously by
\cite{JaiSelWhi00} (2000) following \cite{NusDek93} (1993) and 
\cite{Jusetal95} (1995). One such statistic is the $\left< \kappa
|\kappa| \right>_{\kappa>0}$, which is expected to reduce the
sampling variance from rare and massive halos (see, \cite{JaiSelWhi00}
2000 for details). This statistic is proportional to $\left<
\kappa^3\right>/\left< \kappa^2\right>^{1/2}$.
In Fig.~\ref{fig:skewmod}(a), we show this statistic as a function of
maximum mass used in the calculation.  We still find strong 
variations with changes to the maximum mass. Similar variations were
also present in other statistics considered by \cite{JaiSelWhi00}
(2000).

Consider instead the generalized statistic 
\begin{equation}
S_3(\sigma,m) = \left<\kappa^3\right>/\left< \kappa^2 \right>^{m}\,
\end{equation}
where $m$ is an arbitrary index. We varied $m$ 
such that the effect of mass cuts
are minimized on skewness. In Fig.~\ref{fig:skewmod}(b), we show such
an example with $m=3.7$. Here, the values are separated to two groups
involving
with most massive and rarest halos and another with halos of masses
$10^{14}$ M$_{\sun}$ or less. Though the values from the two groups
agree with each other on small angular scales, they depart
significantly
above $25'$ reaching a difference of 2.5 at 80$'$.  
Statistics involving such a high index $m$,
weigh the single halo contributions highly when the most massive
halos are
present, whereas they weight the halo correlation terms more strongly
for $M<10^{14}$ M$_{\sun}$.  
To some extent this may be useful to identify the presence of rare
halos in the observations. 

However the consequence of using these generalized statistics
is that one progressively loses their independence on the details
of the cosmological model, e.g. the shape and amplitude of the underlying
density power spectrum, as one departs from $m=2$, thereby contaminating
the probe of dark matter and dark energy.
The correction for noise bias in the generalized skewness statistic
also depends on $m$.  The distribution also changes but in a way that it is
predictable from the distributions of second and third moments.
Further work is necessary find the optimal trade off between
robustness, cosmological independence and noise properties
 of these and other generalized statistics.

Another observable statistic is the probability distribution function 
(pdf) of the convergence maps smoothed on the scale $\sigma$.
This possibility has been recently studied by
\cite{Jaivan99} (1999), where the reconstruction of pdf using peak
statistics were considered. Using the Edgeworth
expansion to capture small deviations from Gaussianity, 
one can write the pdf of convergence to second order as
\begin{eqnarray}
p(\kappa) &=& \frac{1}{\sqrt{2 \pi \left<\kappa^2(\sigma)\right>}} \;
e^{-\kappa(\sigma)^2/2\left<\kappa^2(\sigma)\right>} \\
&& \times \left[1+\frac{1}
{6}S_3(\sigma)\sqrt{\left<
\kappa^2(\sigma)\right>}H_3\left(\frac{\kappa(\sigma)}{\sqrt{\left<\kappa^2(\sigma)\right>}}\right)\right]
\, ,\nonumber
 \end{eqnarray}
where $H_3(x)=x^3-3x$ is the third order Hermite polynomial (see,
\cite{Jusetal95} 1995 for details).  

In Fig.~\ref{fig:pdf}, we show the pdf of convergence at 12$'$ as a
function of maximum mass used in the calculation. As shown, the
greatest departures from Gaussianity begin to occur when the maximum
mass included is greater than $10^{14}$ M$_{\sun}$. Given that we have
only constructed the pdf using terms out to skewness, the presented
pdfs should only be considered as approximate. With increasing
non-Gaussian behavior, the approximated pdfs are likely to depart from
this form especially in the tails.
As studied in \cite{Jaivan99} (1999), the measurement of 
the full pdf can potentially be used a a probe of cosmology.  
Its low order properties describe deviations from Gaussianity near
the peak as opposed to the skewness which is more weighted to the
tails.

\section{The Galaxy-Mass Cross-Correlation}

\begin{figure}[!h]
\begin{center}
\includegraphics[width=4.2in,angle=-90]{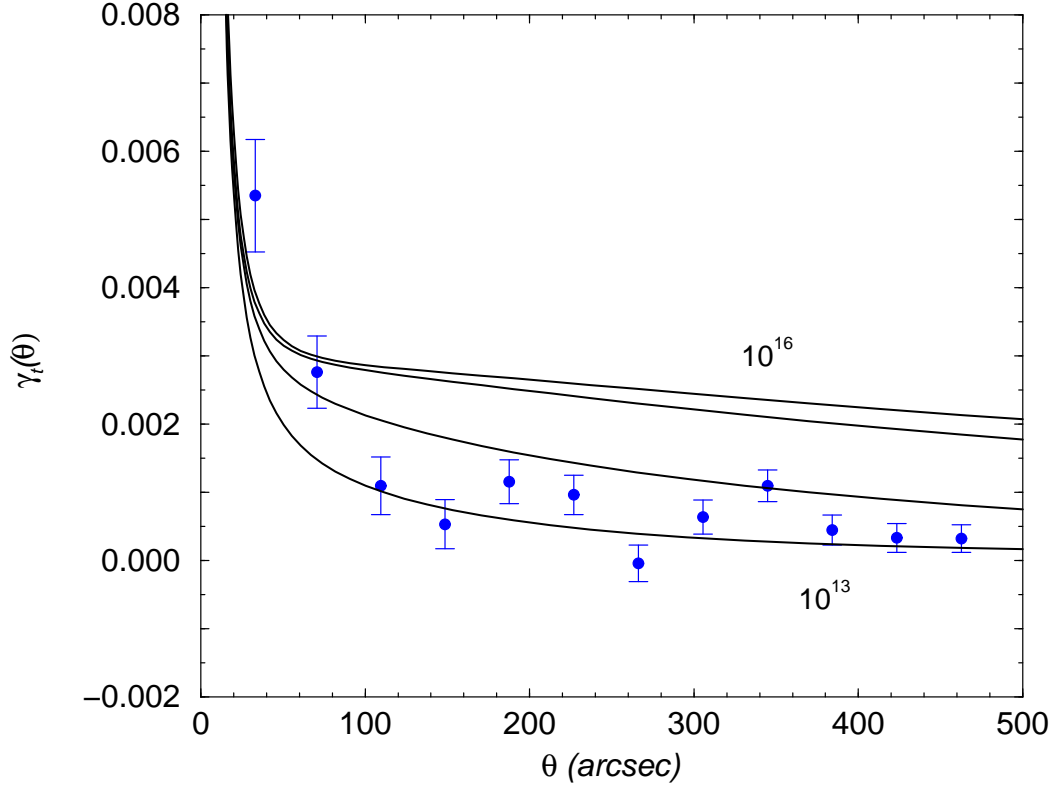}
\end{center}
\caption[Galaxy-mass cross-correlation]{The SDSS galaxy-mass
cross-correlation using galaxy-shear correlation function. The halo
model predictions are shown here as a function of maximum mass for the
dark matter halo used in the calculation: from top to bottom curves
are for $10^{16}, 10^{15}, 10^{14}$ and 10$^{13}$ M$_{\sun}$. The data are from
\cite{Fisetal00} (2000).}
\label{fig:sloanshear}
\end{figure}

Our description for the galaxy power spectrum, see \S~\ref{sec:galaxy},
allows us to extend the discussion to also consider cross-correlation
between galaxies and mass. Such a cross-power spectrum can be probed through
the weak lensing shear-galaxy correlation function. Here, observations
involve the mean tangential shear due to graviational lensing given by
\begin{equation}
\left< \gamma_t(\theta) \right> = -\frac{1}{2} \frac{d
\bar{\kappa}(\theta)}{d {\rm ln} \theta}  \, ,
\end{equation}
where $\bar{\kappa}(\theta)$ is the mean convergence within a circular
radius of $\theta$ (\cite{SquKai96} 1996). Since the shear is
correlated with foreground galaxy positions, one essentially probe the
galaxy-mass correlation such that
\begin{equation}
\bar{\kappa}(\theta) = \int d\rad W^\lens(\rad)W^\gal(\rad)
\int dk k P_{\gal\delta}(k) \frac{2 J_1(k d_A \theta) }{k d_A \theta}
\end{equation}
and
\begin{equation}
\left<\gamma(\theta)\right> = \int d\rad W^\lens(\rad)W^\gal(\rad)
\int dk k P_{\gal\delta}(k) J_2(k d_A \theta) \, .
\end{equation}
Here, $W^\lens$ is the lensing windown function introduced in 
Eq.~\ref{eqn:lenspower}, while $W^\gal$ is the normalized redshift
distribution of foreground galaxies. Note that, in general, $W^\lens$
involves the redshift distribution of background sources beyond the
simple single source redshift assumption. 

The tangential shear-foreground galaxy correlation has been measured
in the Sloan survey by
\cite{Fisetal00} (2000) and we compare these measurements with
predictions in Fig.~\ref{fig:sloanshear}. The observed
measurements from the Sloan survey come only for field galaxies, and
thus, it is likely that the correlation function does not include any
contributions from massive halos and rather from
medium to small mass halos that contain one to few galaxies.  Our
predictions, where we see a lack of significant correlation at large
angular distances with the inclusion of massive halos, are consistent
with this observation. A more thorough study of the weak lensing
shear-galaxy cross-correlation, under the halo model, is available in
\cite{GuzSel00} (2000) and we refer the reader to this paper for
further details.

\section{Summary}
\label{sec:conclusions}

We have presented an efficient method to calculate the non-Gaussian
statistics of lensing convergence at the three point level based on
a description of the underlying density field in terms of dark matter
halos. The bispectrum contains all of the three point information, 
including the skewness.
The prior attempts at calculating lensing bispectrum and
skewness were limited by the accuracy of perturbative approximations
and the
dynamic range and sample variance of simulations.  

Though the present technique 
provides a clear and an efficient method
to calculate the statistics of the convergence field,
it has its own shortcomings.  Halos
are not all spherical, which can to some extent affect the
configuration dependence in moments higher than the two point level. 
Substructures due to mergers of halos can
also introduce scatter. Though such effects unlikely to
dominate our calculations, further work using numerical simulations
will be necessary to determine to what extent present method can be
used as a precise tool to study the higher order statistics associated
with
weak gravitational lensing.

The dark matter halo approach also allows one to
study possible selection effects that may be present in weak lensing
observations due to the presence or absence of rare massive halos in
the small fields that are observed. 
We have shown that the weak lensing skewness is mostly
due to the most massive and rarest dark matter halos in the
universe. The effect of such halos is stronger at the three point
level than the two point level. The absence of massive halos, with
masses
greater than $10^{14}$ M$_{\sun}$, leads
to a strong decrease in skewness, suggesting that a straightforward
use of measured skewness values as a test of cosmological models may
not be appropriate unless prior observations are available on the
distribution of masses in observed lensing fields.

One can correct for such biases using the halo approach, however. 
 To implement such a correction in practice, further work will be needed to
calibrate the technique precisely against simulations across a wide range of
cosmologies. Efficient techniques to correct for mass biases both in
the lensing power spectrum and bispectrum will be needed.
Alternatively, this technique can be used to search for
generalized three point statistics that are more robust to sampling
issues. Given the great potential to study the dark matter distribution
through weak lensing, this issues merit further study.

\chapter{Weak Gravitational lensing Covariance}
\label{sec:covariance}

\section{General Definitions}

As discussed in the previous section, weak lensing probes the statistical properties of the
shear field on the sky which is a weighted projection of
the matter distribution along the line of sight to the
source galaxies.  As such, observables can be
reexpressed as a scalar quantity, the convergence $\kappa$, on
the sky.  

The power spectrum and trispectrum of convergence are 
defined in the flat sky approximation in the usual way
\begin{eqnarray}
\left< \kappa(\bfl_1)\kappa(\bfl_2)\right> &=& 
      (2\pi)^2 \delta_\dirac(\bfl_{12}) C_l^\kappa\,,\nonumber\\
\left< \kappa(\bfl_1) \ldots
       \kappa(\bfl_4)\right>_c &=& (2\pi)^2 \delta_\dirac(\bfl_{1234})
      T^\kappa(\bfl_1,\bfl_2,\bfl_3,\bfl_4)\,.
\end{eqnarray}

These are related to the density power spectrum and trispectrum
by the projections (\cite{Kai92} 1992; \cite{Scoetal99} 1999)
\begin{eqnarray}
C^\kappa_l &=& \int d\rad \frac{W(\rad)^2}{d_A^2} P\left
(\frac{l}{d_A};\rad\right) \, , \\
T^\kappa   &=& \int d\rad \frac{W(\rad)^4}{d_A^6} T\left( 
\frac{\bfl_1}{d_A},
\frac{\bfl_2}{d_A},
\frac{\bfl_3}{d_A},
\frac{\bfl_4}{d_A},
;\rad\right) \, ,
\label{eqn:lenstripower}
\end{eqnarray}                
where $\rad$ is the comoving distance and  $d_A$ is the angular
diameter distance with the weight function defined in  
Eq.~\ref{eqn:weight}. For simplicity, we will assume $\rad_s = r(z_s=1)$. 

For the purpose of this calculation, we assume that upcoming weak
lensing convergence power spectrum will measure binned logarithmic
band  powers at several $l_i$'s in multipole space with bins of
thickness $\delta l_i$.
\begin{equation}
\bp_i = 
\int_{\shell i} 
{d^2 l \over{A_{\shell i}}} 
\frac{l^2}{2\pi} \kappa(\bf l) \kappa(-\bf l) \, ,
\end{equation}
where $A_\shell(l_i) = \int d^2 l$ is the area of the two-dimensional shell in 
multipole and can be written as $A_\shell(l_i) = 2 \pi l_i \delta l_i 
+ \pi (\delta l_i)^2$.

We can now write the signal covariance matrix
as
\begin{eqnarray}
C_{ij} &=& {1 \over A} \left[ {(2\pi)^2 \over A_{\shell i}} 2 \bp_i^2
+ T^\kappa_{ij}\right]\,,\\
T^\kappa_{ij}&=&
\int {d^2 l_i \over A_{\shell i}} 
\int {d^2 l_j \over A_{\shell j}} {l_i^2 l_j^2 \over (2\pi)^2}
T^\kappa(\bfl_i,-\bfl_i,\bfl_j,-\bfl_j)\,,
\label{eqn:variance}
\end{eqnarray}
where 
$A$ is the area of the survey in steradians.  Again the first
term is the Gaussian contribution to the sample variance and the
second
the non-Gaussian contribution.
A realistic survey will also have shot noise variance due to
the finite number of source galaxies in the survey.  
We will return to this point in the \S \ref{sec:parameters}.

\begin{figure}[!h]
\begin{center}
\includegraphics[width=4.2in]{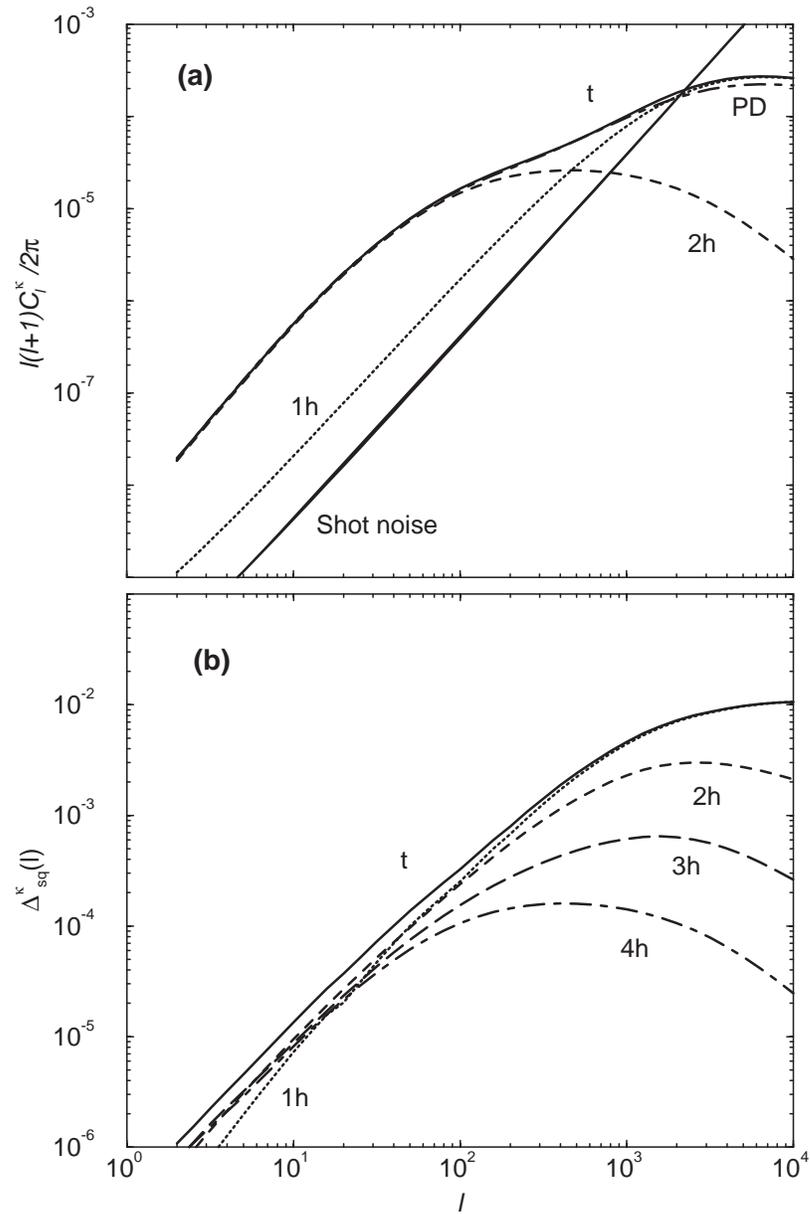}
\end{center}
\caption[Weak lensing convergence power spectrum and trispectrum]{Weak lensing 
convergence (a) power spectrum and (b)
trispectrum under the halo description. Also shown in (a) is the
prediction from the PD nonlinear power spectrum fitting function. We
have separated individual contributions under the halo approach and
have assumed that all sources are at $z_s=1$. We have also shown the
shot noise contribution to the power spectrum assuming a survey down
to a limiting magnitude of R $\sim$ 25 with an intrinsic rms shear of
0.4 in
each component.}
\label{fig:weaktripower}
\end{figure}

\section{Comparisons}
\label{sec:discussion}

Using the halo model, we can now calculate contributions to lensing
convergence power spectrum and trispectrum. 
The power spectrum, shown in Fig.~\ref{fig:weaktripower}(a),
shows the same behavior as the density field when compared with
the PD results: a slight overprediction of power when $l \gtrsim
10^3$. This results through the distribution of the concentration-mass 
relation from simulations by \cite{Buletal00} (2000); In comparison to the 
previous chapter, we now include the full information on concentration from 
simulations to be complete instead of the fitting function for concentration 
which results in the recovery of PD non-linear power spectrum for dark 
matter. As shown in Fig.~\ref{fig:weaktripower}(a), the differences 
arising from variations in the concentration-mass relations are
 not likely to be observable given
the shot noise from the finite number of galaxies at small scales.

In Fig~\ref{fig:weaktripower}(b), we show the scaled trispectrum 
\begin{equation}
\Delta^\kappa_{\rm sq}(l) = \frac{l^2}{2\pi}
T^\kappa(\vecl,-\vecl,\vecl_\perp,-\vecl_\perp)^{1/3} \, .
\end{equation}
where $l_\perp=l$ and $\vecl \cdot \vecl_\perp=0$.
The projected lensing trispectrum again shows the same behavior as the
density
field trispectrum with similar conditions on $\veck_i$'s. 

\begin{figure}[!h]
\begin{center}
\includegraphics[width=4.2in]{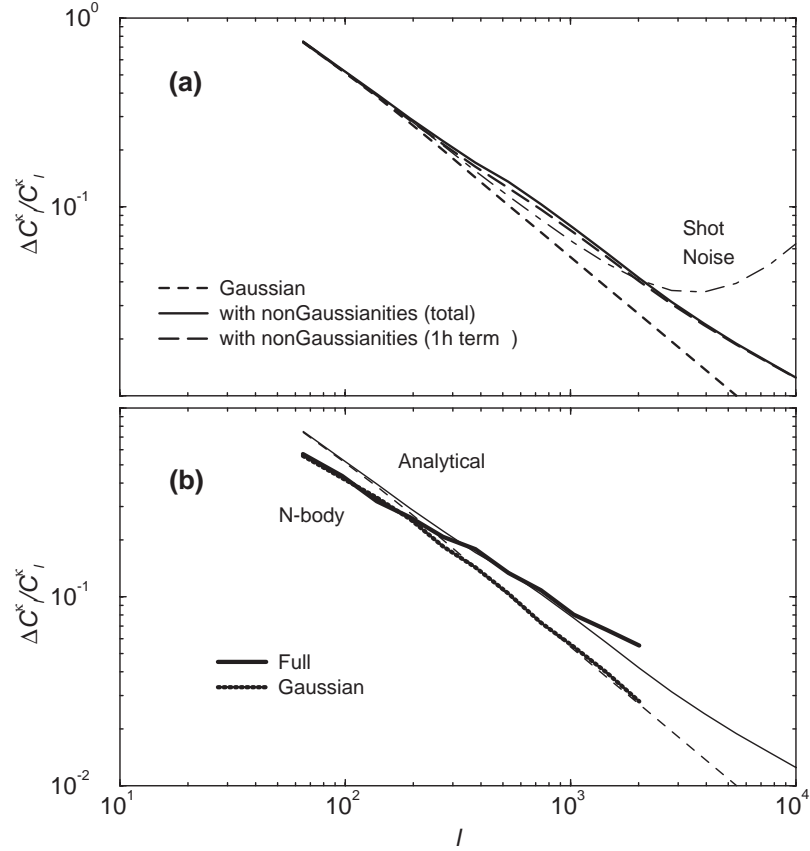}
\end{center}
\caption[Weak lensing convergence power spectrum errors]{The 
fractional errors in the measurements of the
convergence band powers.
In (a), we show the fractional errors under the Gaussian
approximation,
the full halo description, the Gaussian plus single halo term, and the
Gaussian plus shot noise term (see \S \ref{sec:parameters}). 
As shown, the additional variance can be modeled with the single halo
piece
while shot noise generally becomes dominant before non-Gaussian
effects
become large.
In (b), we compare the halo model with 
simulations from \cite{WhiHu99} (1999). The decrease in the variance
at small $l$ in the simulations is due to the conversion of variance
to covariance by the finite box size of the simulations.  } 
\label{fig:variance}
\end{figure}

We can now use this trispectrum to study the 
contributions to the covariance, which is what we are primarily
concerned here. In Fig.~\ref{fig:variance}a, we show the
fractional error, 
\begin{equation}
{\Delta \bp_i  \over \bp_i} \equiv {\sqrt{C_{ii}}  \over \bp_i} \, ,
\end{equation}
for bands $l_i$ given in Table~\ref{tab:cov} following the
binning scheme used by \cite{WhiHu99} (1999) on $6^\circ \times
6^\circ$ fields.

The dashed line compares that with the Gaussian errors, 
involving the
first term in the covariance (Eq.~\ref{eqn:variance}).
At multipoles of a few hundred and
greater, the non-Gaussian term begins to dominate the
contributions.  For this reason, the errors are well approximated by
simply taking the Gaussian and single halo contributions.

In Fig.~\ref{fig:variance}(b), we compare these results
with those of the \cite{WhiHu99} (1999) simulations.  The
decrease in errors from the simulations at small $l$ reflects
finite box effects that convert variance to covariance
as the fundamental mode in the box becomes comparable to 
the bandwidth. 

\begin{figure}[!h]
\begin{center}
\includegraphics[width=4.2in]{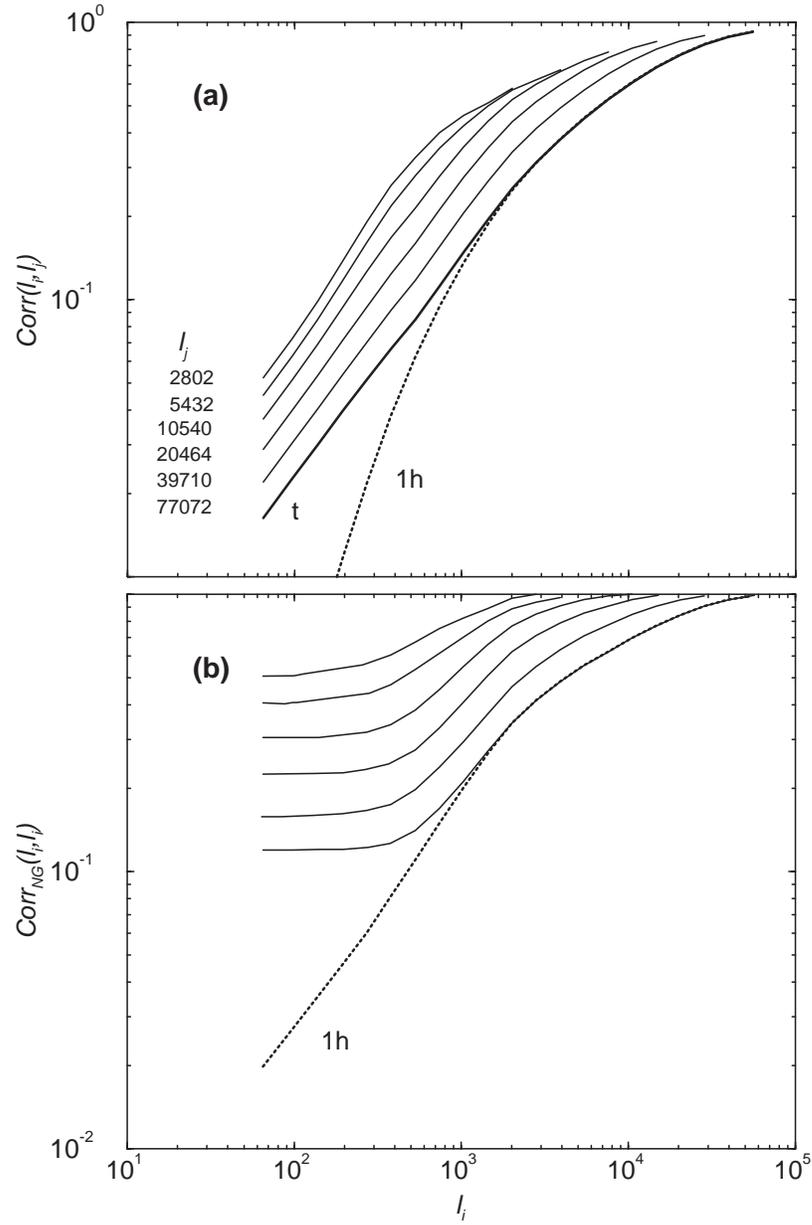}
\end{center}
\caption[Correlation coefficients]{(a) The correlation coefficient, $\hat C_{ij}$ as a function
of the multipole $l_i$ with $l_j$ as shown in the figure.  We show the
correlations
calculated with the full halo model and also with only the single halo
term for
$l_j=77072$.
In (b), we show
the non-Gaussian correlation coefficient $\hat C_{ij}^{\rm NG}$, 
which only involves the trispectrum (see,
Eq.~\ref{eqn:ng}). The transition to full correlation is due to the 
domination of the single halo contribution. }
\label{fig:corr}
\end{figure}

The correlation between the bands is given by
\begin{equation}
\hat C_{ij} \equiv \frac{C_{ij}}{\sqrt{C_{ii} C_{jj}}} \, .
\end{equation}
In Table \ref{tab:cov} we compare the halo predictions to 
the simulations by \cite{WhiHu99} (1999). 
The upper triangle here is the
correlations under the halo approach, while the lower triangle shows
the correlations found in numerical simulations.
The correlations along individual columns increase (as one goes to
large $l$'s or small angular scales) consistent with simulations.
In Fig.~\ref{fig:corr}, we show the correlation coefficients with (a)
and without (b) the Gaussian contribution to the diagonal. 

\begin{table}[!h]
\begin{flushleft}
\begin{tabular}{cccccccccc}
\hline 
$\ell_{\rm bin}$
       & 138     & 194     & 271     & 378     & 529     &
739 & 1031 
   & 1440   & 2012 \\
\hline
   138 & 1.00   & 0.08    & 0.10   & 0.11    & 0.12    &0.12
& 0.12   
 & 0.11 & 0.11\\
   194 & (0.31) & 1.00   & 0.14    & 0.17    & 0.18    &0.18
& 0.17    
& 0.16 & 0.15\\
   271  & (0.21)  & (0.26) & 1.00  & 0.24   & 0.25     &0.25
& 0.24   &
 0.22   & 0.21\\
   378 & (0.09)  & (0.24) & (0.38) & 1.00    & 0.33   &0.33
& 0.32    
& 0.30   & 0.28\\
   529 & (0.14)  & (0.28) & (0.33) & (0.45) & 1.00    &0.42
& 0.40    
& 0.37  & 0.35\\
   739 & (0.16)  & (0.17)  & (0.34) & (0.38) & (0.50) & 1.00
& 0.48   
 & 0.45   & 0.42\\
  1031  & (0.18)  & (0.15) & (0.27) & (0.33) & (0.48) & (0.54)
& 1.00   
 & 0.52  & 0.48\\
  1440 & (0.15) & (0.19) & (0.19) &(0.32) & (0.36) & (0.53) &
(0.57) & 
1.00  & 0.54\\
  2012 & (0.22) & (0.16) & (0.32) & (0.27) & (0.46) & (0.50)
& (0.61) & (0.65) & 1.00\\
\hline
\end{tabular}
\caption[Weak lensing convergence power spectrum covariance]{
Covariance of the binned power spectrum when sources are at a redshift
of 1.
Upper triangle displays the covariance found under the halo model.
Lower triangle (parenthetical numbers) displays the covariance found
in numerical simulations by \cite{WhiHu99} (1999). To be consistent
with these simulations, we use the same binning scheme as the one used
there.}
\label{tab:cov}
\end{flushleft}
\end{table}

We show in Fig.~\ref{fig:corr}(a) the behavior of the correlation
coefficient between a fixed $l_j$ as a function of $l_i$.  When
$l_i=l_j$
the coefficient is 1 by definition.  Due to the presence of
the dominant Gaussian contribution at $l_i=l_j$, the coefficient has
an apparent
 discontinuity between $l_i=l_j$ and $l_i = l_{j-1}$ that decreases
as $l_j$ increases and non-Gaussian effects dominate.

To better understand this behavior it is useful to isolate
the  purely non-Gaussian correlation
coefficient 
\begin{equation}
\hat C^{\rm NG}_{ij} =
\frac{T_{ij}}{\sqrt{T_{ii} T_{ij}}} \,.
\label{eqn:ng}
\end{equation}
As shown in Fig.~\ref{fig:corr}(b), 
the coefficient remains constant for $l_i \ll l_j$ and smoothly
increases
to unity across a transition scale that is related to where the
single halo terms starts to contribute. 
A
comparison of Fig.~\ref{fig:corr}(b) and \ref{fig:weakpower}(b), shows
that this transition happens around $l$ of few hundred to 1000.
Once the power spectrum is dominated by correlations in single halos,
the fixed profile of the halos will correlate the power in all the
modes.
The multiple halo terms on the other hand correlate linear and
non-linear
scales but at a level that is generally negligible compared with the
Gaussian variance. 

The behavior seen in the halo based covariance, however, is not
present when the covariance is
calculated with hierarchical arguments for the trispectrum (see,
\cite{Scoetal99} 1999). With hierarchical arguments, which are by
construction only valid in the deeply nonlinear regime, one predicts
correlations which are, in general, constant across all scales and
shows no decrease in correlations between very small and very large
scales.
Such hierarchical models also violate the  Schwarz inequality with
correlations greater than 1 between large and small scales (e.g.,
\cite{Scoetal99} 1999; \cite{Ham00} 2000).
The halo model, however, shows a decrease in correlations similar
to numerical simulations suggesting that the
halo model, at least qualitatively, provides a better
approach to studying non-Gaussian correlations in the translinear
regime.

\section{Effect on Parameter Estimation}
\label{sec:parameters}

Modeling or measuring the 
covariance matrix of the power spectrum estimates will be
essential for interpreting observational results.
In the absence of many fields where the covariance can be
estimated directly from the data, the halo model provides
a useful, albeit model dependent, quantification of the
covariance.  As a practical approach one could imagine
taking the variances estimated from the survey under
a Gaussian approximation,  but which accounts for uneven 
sampling and edge effects (\cite{HuWhi00} 2000), 
and scaling it up by the non-Gaussian
to Gaussian variance ratio of the halo model along with
inclusion of the band power correlations. Additionally, it is in
principle
possible to use the expected correlations 
from the halo model to decorrelate individual band power measurements,
similar to studies involving CMB temperature anisotropy and galaxy
power spectra (e.g., \cite{Ham97} 1997; \cite{HamTeg00} 2000).

We can estimate the resulting effects on cosmological parameter
estimation with an analogous procedure on the Fisher matrix.
In \cite{HuTeg99} (1999), the potential of wide-field lensing
surveys to measure cosmological parameters was investigated
using the Gaussian approximation of a diagonal covariance
and Fisher matrix techniques.
The Fisher matrix is simply a projection of the covariance
matrix onto the basis of cosmological parameters $p_i$
\begin{equation}
{\bf F}_{\alpha\beta} = \sum_{ij} 
      {\partial \bp_i \over \partial p_\alpha} (C_{\rm tot}^{-1})_{ij}
{\partial \bp_j \over \partial p_\beta} \, ,
\label{eqn:fisher}
\end{equation}
where the total covariance includes both the signal
and noise covariance.  Under the approximation of Gaussian shot
noise, this reduces to replacing $C^\kappa_l \rightarrow
C^\kappa_l + C^{\rm SN}_l$ in the expressions leading up
to the covariance Eq.~\ref{eqn:variance}.  
The shot noise power spectrum is given by 
\begin{equation}
C^{\rm SN}_l = \frac{\langle \gamma_{\rm int}^2\rangle}{\bar{n}} \, ,
\end{equation}
where $\langle \gamma_{\rm int} \rangle^{1/2} \sim 0.4$ is the
rms noise per component introduced by intrinsic ellipticities and
measurement errors and $\bar{n} \sim 6.6
\times 10^{8}$ sr$^{-1}$ is the surface number density of background
source galaxies. The numerical values here are appropriate
for surveys that reach a  
limiting magnitude in $R\sim 25$ (e.g., \cite{Smaetal95} 1995).

Under the approximation that there are a sufficient number
of modes in the band powers that the distribution of power
spectrum estimates is approximately Gaussian, the Fisher matrix
quantifies
the best possible errors on cosmological parameters that can
be achieved by a given survey.  In particular $F^{-1}$ is
the optimal covariance matrix of the parameters and
$(F^{-1})_{ii}^{1/2}$
is the optimal error on the $i$th parameter.
Implicit in this approximation of the Fisher matrix is the neglect of
information
from the cosmological parameter dependence of the covariance matrix
of the band powers themselves.  Since the covariance is much less
than the mean power, we expect this information content to be
small. 

In order to estimate the effect of non-Gaussianities on the
cosmological parameters, we calculate the Fisher matrix elements using
our fiducial $\Lambda$CDM cosmological model and define the dark
matter
density field, today, as
\begin{equation}
\Delta^2(k) = A^2 \left( \frac{k}{H_0} \right)^{n_s+3} T^2(k) \, .
\end{equation}
Here, $A$ is the amplitude of the present day density fluctuations 
and $n_s$ is the tilt at the Hubble scale. The density power spectrum
is evolved to higher redshifts using the growth function $G(z)$
(\cite{Pee80} 1980) and the transfer function $T(k)$ is calculated
using  the fitting functions from \cite{EisHu99} (1999). Since we are
only interested in the relative effect of non-Gaussianities, 
we restrict ourselves to a small subset of the cosmological parameters
considered by \cite{HuTeg99} (1999) and assume a full sky survey with
$f_\sky=1$.

\begin{table}[!h]
\begin{center}
\begin{tabular}{lrrrrr}
\hline 
$p_{i}$
      & $\Omega_\Lambda$     & $\ln$ A     & $\Omega_K$     & $n_s$
& $\Omega_mh^2$\\
\hline
$\Omega_\Lambda$ & 1.57  & -5.96  & -1.39    & 4.41 & -1.76\\
$ \ln$ A &  & 25.89   & 5.83    & -17.34 & 6.74\\
$\Omega_K$ & & & 1.41   & -3.81 & 1.43\\
$n_s$ & & & & 14.01 & -6.03\\
$\Omega_mh^2$ & & & & & 2.67\\
\hline
\end{tabular}
\begin{tabular}{lrrrrr}
\hline 
$p_{i}$
      & $\Omega_\Lambda$     & $\ln$ A     & $\Omega_K$     & $n_s$
& $\Omega_mh^2$ \\
\hline
$\Omega_\Lambda$ & 2.03  & -7.84  & -1.82    & 5.76 & -2.30 \\
$ \ln$ A &  & 33.92   & 7.65    & -22.79 & 8.91\\
$\Omega_K$ & & & 1.78   & -5.01 & 1.95 \\
$n_s$ & & & & 18.43 & -7.85\\
$\Omega_mh^2$ & & & & & 3.44 \\
\hline
\end{tabular}
\caption[Inverse Fisher matrix for weak lensing]{
Inverse Fisher matrix under the Gaussian assumption (top) and the
halo model (bottom). The error on an individual parameter is the
square root of the diagonal element of the Fisher matrix for the
parameter while off-diagonal entries of the inverse Fisher matrix
shows correlations, and, thus, degeneracies, between parameters. We
have assumed a full sky survey ($f_\sky=1$) with parameters as
described in \S~\ref{sec:parameters}.}
\label{tab:fisher}
\end{center}
\end{table}

In Table~\ref{tab:fisher}, we show the inverse Fisher matrices determined under the
Gaussian and non-Gaussian covariances, respectively. 
For the purpose of this calculation, we adopt the binning scheme as 
shown in Table~\ref{tab:cov}, following \cite{WhiHu99} (1999).
The Gaussian errors are computed using the same scheme by setting
$T^\kappa =0$.  As shown in Table~\ref{tab:fisher}, the
inclusion of non-Gaussianities lead to an increase in the inverse
Fisher matrix elements.
We compare the errors on individual parameters, mainly
$(F^{-1})_{ii}^{1/2}$, between the Gaussian and
non-Gaussian assumptions in Table~{\ref{tab:errors}}.  
The errors increase typically by $\sim 15$\%.  
Note also that band power correlations do not necessarily increase
cosmological parameter errors.  Correlations induced by non-linear 
gravity introduce larger errors in the overall amplitude of the power 
spectrum measurements but have a much smaller effect on those 
parameters controlling the shape of the spectrum.

For a survey of this assumed depth, the shot noise power becomes
the dominant error before the non-Gaussian signal effects dominate
over the Gaussian ones.   For a deeper survey with better imaging,
such as the one planned with Large-aperture Synoptic Survey Telescope
(LSST; \cite{TysAng00} 2000)\footnote{http://www.dmtelescope.org}, the
effect of shot noise
decreases and non-Gaussianity is potentially more
important. However, the non-Gaussianity itself also decreases with
survey depth, and as we now discuss, in terms of the effect of
non-Gaussianities, deeper surveys should be preferred over the shallow
ones.

\begin{figure}[!h]
\begin{center}
\includegraphics[width=4.2in]{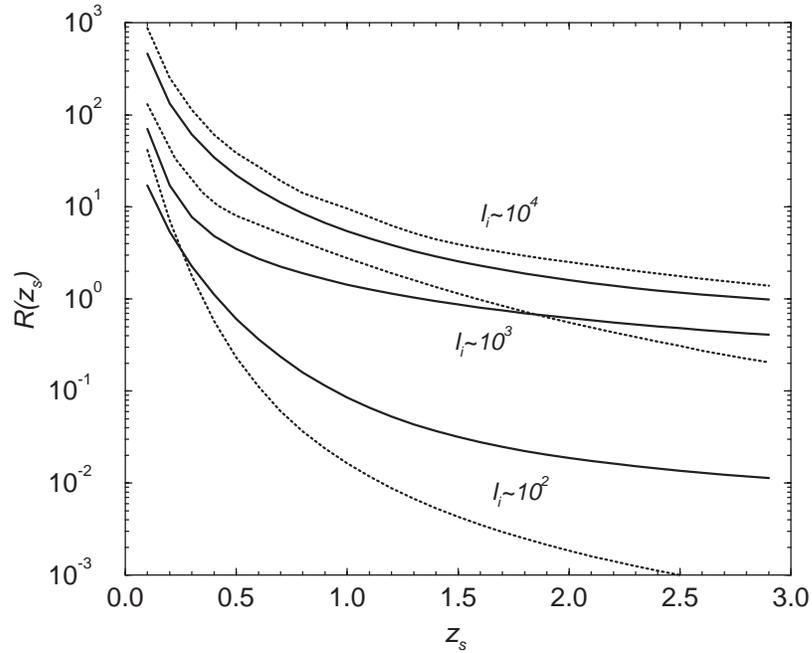}
\end{center}
\caption[Ratio of non-Gaussian to Gaussian contribution]{The ratio of non-Gaussian to Gaussian contributions,
$R$, as a  function of source redshift ($z_s$). The solid lines
are through the exact calculation (Eq.~\ref{eqn:rexact}) while the
dotted lines are using the approximation given in
Eq.~\ref{eqn:rapprox}. Here, we show the ratio $R$ for three
multipoles corresponding to large, medium and small angular
scales. The multipole binning is kept constant such that $\delta l \sim
l$. Decreasing this bin size will linearly decrease the value of $R$.}
\label{fig:r}
\end{figure}

\section{Scaling Relations}

To better understand how the non-Gaussian contribution scale with our
assumptions,  we consider the ratio of
non-Gaussian variance to the Gaussian variance 
\begin{equation}
\frac{C_{ii}}{C_{ii}^{\rm G}} = 1 + R \, ,
\end{equation}
with
\begin{equation}
R \equiv \frac{A_{si} T_{ii}^\kappa}{(2
\pi)^2 2 C_i^2} \, .
\label{eqn:rexact}
\end{equation}
Under the assumption that contributions to lensing convergence can be
written through an effective distance $r_\star$, at half the angular
diameter distance to background sources, and a width $\Delta r$
for the lensing window function,
the ratio of lensing convergence trispectrum and power
spectrum contribution to the variance  can be further simplified to
\begin{equation}
R \sim \frac{A_{si}}{(2 \pi)^2V_{\rm eff}}\frac{
\bar{T}(r_\star)}{2\bar{P}^2(r_\star)} \, .
\label{eqn:rapprox}
\end{equation}
Since the lensing window function peaks at $r_\star$, we have
replaced the integral over the window function of the
density field trispectrum and power spectrum by its value at the peak.
This ratio shows how the relative contribution from non-Gaussianities
scale with survey parameters: (a) increasing the bin size, through
$A_{si}$ ($\propto \delta l$), leads to an increase in the 
non-Gaussian contribution linearly,
(b) increasing the source redshift, through the effective volume of
lenses in the survey 
($V_{\rm eff} \sim r_\star^2 \Delta r$), decreases the non-Gaussian
contribution, while (c)
the growth of the density field trispectrum and power spectrum,
through the ratio $\bar{T}/\bar{P}^2$,
decreases the contribution as one moves to a higher redshift. The
volume factor quantifies the number of foreground halos in the survey
that effectively act as gravitational lenses 
for background sources; as the number of such halos is increased, 
the non-Gaussianities are reduced by the central limit theorem.

In Fig.~\ref{fig:r}, we summarize our results as a function of
source redshift with $l_i \sim 10^2,10^3$ and 10$^4$ and setting
the bin width such that $A_s(l_i) \sim l_i^2$, or $\delta l \sim l$.
As shown, increasing the source redshift leads to a decrease in the
non-Gaussian contribution to the variance.
The prediction based on the simplifications in Eq.~\ref{eqn:rapprox}
tend to
overestimate the non-Gaussianity at lower
redshifts while underestimates it at higher redshifts, though the
exact transition depends on the angular scale of interest; this
behavior can be understood due to the fact that we do not
consider the full lensing window function but  only the
contributions at an effective redshift, midway between the 
observer and sources.

In order to determine whether its the increase in volume or the
decrease in the growth of
structures that lead to a decrease in the 
relative importance of non-Gaussianities
as one moves to a higher source redshift, we calculated
the non-Gaussian to Gaussian variance ratio under the halo model 
for several source redshifts and
survey volumes. Up to source redshifts $\sim$ 1.5, the increase in
volume decreases the non-Gaussian contribution significantly. When
surveys are sensitive to sources at redshifts beyond 1.5, 
the increase in volume becomes less significant
and the decrease in the growth of structures
begin to be important in
decreasing the non-Gaussian contribution. Since, in the deeply
non-linear
regime, $\bar{T}/\bar{P}^2$ scales with 
redshift as the cube of the growth factor,
this behavior is consistent with the overall redshift scaling of the
volume and growth. 

The importance of the non-Gaussianity to the variance also scales
linearly with bin width.  As one increases the bin width the
covariance
induced by the non-Gaussianity manifests itself as increased variance
relative
to the Gaussian case.  The normalization of $R$ is therefore somewhat
arbitrary in that it depends on the binning scheme, i.e. $R \ll 1$ 
does not necessarily mean non-Gaussianity can be entirely neglected 
when summing over all the bins.  The scaling with redshift and the
overall
scaling of the variance with the survey area $A$ is not.
One way to get around the increased non-Gaussianity associated with
shallow 
surveys, is to have it sample a wide patch of sky since 
$C_{ii} \propto (1+R)/A$.
This relation tells us the trade off between designing an survey to
go wide instead of deep.   One should bear in mind though that not
only will
shallow surveys have decreasing number densities of source galaxies
and hence
increasing shot noise, they will also suffer more from the decreasing
amplitude of the signal itself and the increasing import
ance of systematic effects, 
including the intrinsic correlations of galaxy shapes (e.g.,
\cite{Catetal00}
2000; \cite{CroMet00} 2000; \cite{Heaetal00} 2000). These problems 
tilt the balance more towards deep but narrow surveys than the naive
statistical scaling would suggest.

\begin{table}[!h]
\begin{center}
\begin{tabular}{lrrrrr}
\hline 
      & $\Omega_\Lambda$     & $\ln$ A     & $\Omega_k$     & $n_s$
& $\Omega_m h^2$ \\
\hline
Gaussian  & 0.039  & 0.160    & 0.037    & 0.118 & 0.051\\
Full & 0.045  & 0.184   & 0.042    & 0.135 & 0.058\\
Increase (\%) & 15.3 & 15 & 13.5   & 14.4 & 13.7\\
\hline
\end{tabular}
\caption[Parameter Errors]{Parameter errors, $(F^{-1})_{ii}^{1/2}$, under the Gaussian assumption
(top) and the
halo model (bottom) and following the inverse-Fisher matrices in
Table~3. We have assumed a full sky survey ($f_\sky=1$) with
parameters as
described in \S~\ref{sec:parameters}.}
\label{tab:errors}
\end{center}
\end{table}

\section{Conclusions}
\label{sec:conclusion}  

Weak gravitational lensing due to large scale structure
provides important information on the evolution
of clustering and angular diameter distances and therefore,
cosmological parameters.  
This
information complements what can be learned from cosmic microwave
background anisotropy observations. The tremendous progress on the
observational front warrants detailed studies of the statistical
properties
of the lensing observables and their use in constraining cosmological
models.

The non-linear growth of large-scale structure induces
high order correlations in the derived shear and convergence fields.
In this work, we have studied the four point correlations in the
fields.  Four point statistics are special in that they quantify 
the errors in the determination of the two point statistics.
To interpret future lensing measurements on the power spectrum, it
will
be essential to have an accurate assessment of the correlation 
between the measurements. 

Using the halo model for clustering, we have provided a
semi-analytical
method to calculate the four point function of the
lensing convergence as well as the dark matter density field. 
We have tested this model
against numerical $N$-body simulations of the power spectrum
covariance
in both the density and
convergence fields and obtained good agreement.  As such, this
method provides a practical means of estimating the error matrix
from future surveys in the absence of sufficiently large fields where
it may be estimated directly from the data or large suites of $N$-body

simulations where it can be quantified in a given model context.
Eventually a test of whether the covariance matrix estimated from
the data and the theory agree may even provide further cosmological
constraints. This method may also be used to study other aspects of the four point
function in lensing and any field whose relation to the dark matter
density field can be modeled.  Given the approximate nature of these
approximations, each potential use must be tested against simulations.
Nonetheless, the halo model provides the most intuitive and
extensible means to study non-Gaussianity in the cosmological context
currently known.

\chapter{Thermal Sunyaev-Zel'dovich Effect}

\section{Introduction}

In recent years, increasing attention has been given to the physical properties
of the intergalactic warm and hot plasma gas distribution
associated with large scale structure and the possibility of its
detection (e.g., \cite{CenOst99} 1999).
It is now widely believed that at least $\sim$ 50\% of the
present day baryons, when compared to the total baryon density through
big bang nucleosynthesis, are present in this warm gas distribution
and have remained undetected given its nature (e.g.,
\cite{Fuketal98} 1998). Currently 
proposed methods for the detection of this gas with
include observations of the
thermal diffuse X-ray emission (e.g., \cite{Pieetal00} 2000), associated
X-ray and UV absorption
and emission lines (e.g., \cite{Trietal00} 2000) and resulting
Sunyaev-Zel'dovich (SZ;
\cite{SunZel80} 1980) effect (e.g., \cite{Cooetal00a} 2000a).

The SZ effect arises from the  inverse-Compton scattering of CMB
photons by hot electrons
along the line of sight. This effect has now been directly imaged
towards massive galaxy clusters (e.g., \cite{Caretal96} 1996;
\cite{Jonetal93} 1993), where the temperature of the scattering medium
can reach as high as
10 keV, producing temperature changes in the CMB of order 1 mK at
Rayleigh-Jeans wavelengths. Previous analytical predictions of the
resulting SZ effect due to large scale structure have been based on
either through a Press-Schechter (PS; \cite{PreSch74} 1974) description of the
contributing galaxy clusters  (e.g., \cite{ColKai88} 1988;
\cite{KomKit99} 1999) or using a biased description of the pressure
power spectrum with respect to the dark matter
density field (e.g., \cite{Cooetal00a} 2000a). Numerical simulations
(e.g., \cite{daS99} 1999; \cite{Refetal99} 1999; \cite{Seletal00}
2000; \cite{Spretal00} 2000)
are beginning to improve some of these analytical predictions,
but are still limited to handful of simulations with limited dynamical
range and resolution. Therefore, it is important that one consider
improving analytical models of the large scale structure SZ effect,
and provide predictions which can be easily tested through simulations.

Our present study on the large scale baryon pressure and the resulting
SZ effect is timely for two main reasons. First, the improvements in
hydrodynamical simulations now allow detailed predictions on the
statistics of pressure power spectrum and resulting SZ effect (e.g.,
\cite{Spretal00} 2000; \cite{RefTey01} 2001). The numerical studies
are easily extendable to higher order correlations through models such
as the halo based one advocated here.  Here,  we extend previous
analytical and numerical studies by considering the
full power spectrum, bispectrum, and trispectrum 
of pressure fluctuations. The pressure correlations
contain all necessary information on the large scale
distribution of temperature weighted baryons, whereas, the thermal SZ
angular power spectrum is only a redshift projected measurement of the pressure power
spectrum. This can be compared to weak gravitational lensing,
where lensing is a  direct probe of the projected dark matter density
distribution. The bispectrum of pressure fluctuations, and SZ
bispectrum, contains all the information present at the three-point
level, whereas conventional statistics, such as skewness, do not.
An useful advantage of using three-dimensional statistics, such as the pressure
power spectrum, is that they can directly compared to numerical
simulations, while only two-dimensional statistics, such as the projected pressure
power spectrum along the line of sight, basically the SZ power
spectrum, can only be observed. Our approach here is to consider both
such that our calculations can eventually be compared to both
simulations and observations.

The calculation of pressure power spectrum and higher order
correlations requires detailed knowledge on the baryon distribution, which can eventually be
obtained numerically through hydrodynamical simulations. Here, we provide an
analytical technique to obtain the pressure power spectrum,
bispectrum and trispectrum by extending the  dark matter halo
approach.  The baryons are assumed to be in
hydrostatic equilibrium with respect to dark matter distribution,
which is a valid assumption, at least for the high mass halos that have been
observed with X-ray instruments, given the existence of regularity
relations between cluster baryon and dark matter physical properties
(e.g., \cite{MohEvr97} 1997).
We take a description of the temperature structure of electrons involving
 the virial temperature. When estimating astrophysical parameters from
the SZ effect, we will consider an  an additional source
of non-gravitational energy, 
independent of mass and redshift. The latter consideration allows the
possibility for a secondary source of energy for baryons, such as
due to preheating through stellar formation and feedback processes.
Numerical simulations (e.g., \cite{CenOst99} 1999; \cite{Pen99} 1999), as well
observations (e.g., \cite{Davetal95} 1995; \cite{Ren97} 1997), suggest the
existence of such an energy source. 

The second reason why this study is timely is that the progress in the
experimental front strongly suggests possibilities for detailed
observational studies of the SZ power spectrum and higher order correlations.
Given that the SZ effect also bears a
spectral  signature that differs from other temperature fluctuations,
SZ contribution can be separated in upcoming
multifrequency CMB data. As discussed in detail in \cite{Cooetal00a} (2000a),
a multi-frequency approach can easily be applied to current Boomerang 
(\cite{deBetal00} 2000), and upcoming
MAP\footnote{http://map.nasa.gsfc.gov} and Planck
surveyor\footnote{http://astro.estec.esa.nl/Planck/; also, ESA
D/SCI(6)3.} missions.  At small angular scales, though a wide-field SZ image,
is yet to be produced, several experiments are now
working towards this goal. These experimental attempts include the
interferometric survey by \cite{Caretal96} (1996) at the combined
BIMA/OVRO array (CARMA), the MINT interferometer (Lyman Page, private
communication), and the BOLOCAM array on the Caltech Submillimeter Observatory (Andrew
Lange, private communication).

In the present Chapter, we  discuss the SZ effect and address
what astrophysical properties can be deduced with a measurement of the
SZ power spectrum. For this, we require detailed knowledge on the
covariance of the SZ power spectrum beyond the simple Gaussian
assumptions for the variance. Given that the SZ effect probes the
projected pressure distribution in the local universe, its statistical
properties reflect those of pressure. As discussed in detail in
\cite{Coo00} (2000), the statistics of large scale structure pressure
is highly non-Gaussian due to the associated non-linearities. We have
previously shown that the SZ effect has a significant skewness
associated with it, primarily given that the large scale pressure
power spectrum is associated with massive halos, which are rare and
discrete. The same non-Gaussianities also induce a four-point
correlation function in pressure, which in return, can be used to
study the
correlations in the pressure power spectrum due to resulting
non-Gaussian covariance.
 This is is analogous to the dark matter 
covariance correlations discussed in \cite{MeiWhi99} (1999) and
\cite{Scoetal99} (1999), using numerical simulations, and in
\cite{CooHu01b} (2001b)  using the halo model.
The pressure trispectrum is also of interest since it
determines the covariance of the thermal SZ power spectrum
measurements. Again, this is analogous to the covariance we recently
discussed for weak gravitational lensing (\cite{CooHu01b} 2001b)
resulting from the trispectrum of dark matter due to 
non-linear clustering at low redshifts and the covariance discussed in
\cite{EisZal01} (2001) for the APM galaxy power spectrum.

In order to calculate the covariance associated with SZ power spectrum
measurements, we extend the semi-analytical model presented in
\cite{Coo00} (2000) and calculate the pressure trispectrum. Here, we
show that the SZ effect is highly non-Gaussian at all scales of
interest and that these non-Gaussianities correlate the SZ power
spectrum
measurements significantly.  The full covariance matrix now allows us
to quantify the astrophysical abilities of SZ measurements as a probe
of gas and its temperature properties. Previous to this study, we were
unable to perform a detailed calculation on how well SZ effect 
probe gas and temperature properties due to the unknown covariance
associated with the effect. We also briefly discuss some aspects of the clustering of
SZ halos and suggest a useful way to obtain an average value of halo
mass, as a function of redshift, through the correlation function and bias.

\begin{figure}[!h]
\begin{center}
\includegraphics[width=4.2in,angle=-90]{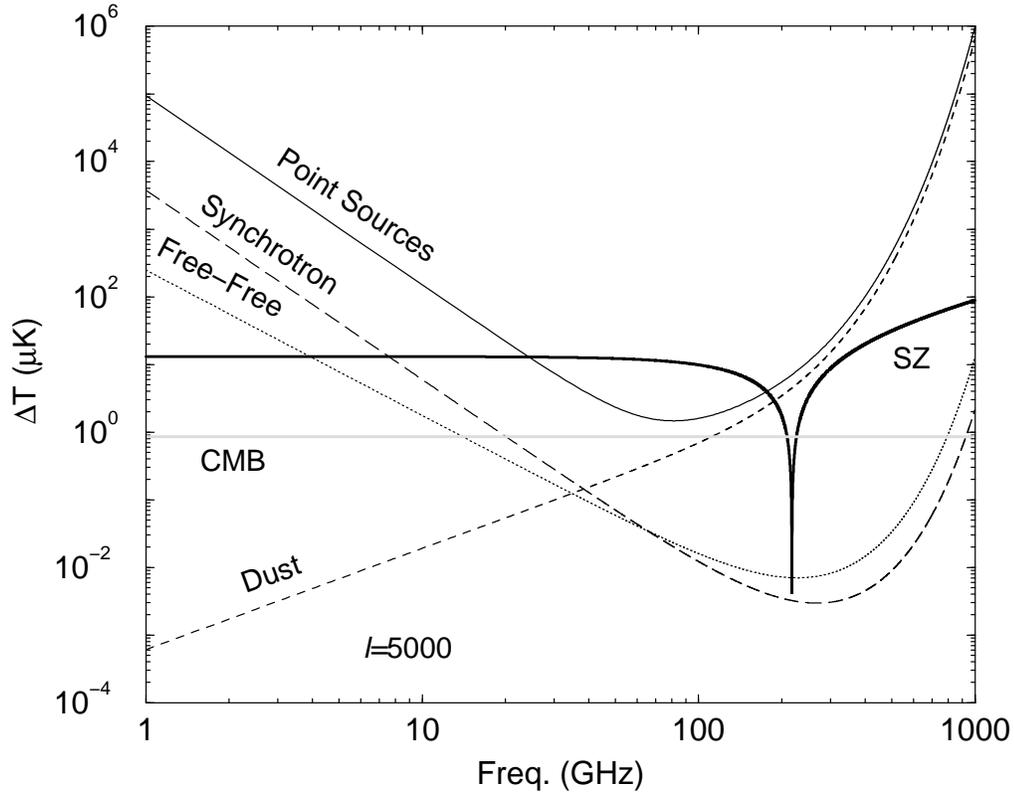}
\end{center}
\caption[Frequency dependence of the SZ effect]{Frequency dependence
of the SZ effect at a multipole of $l \sim 5000$. Here, we show the
absolute value of temperature relative to the thermal CMB spectrum. 
For comparison, we also
show the temperature fluctuations due to point sources (both radio at
low frequencies and fra-infrared sources at high frequencies; solid
line), galactic synchrotron (long dashed line), galactic free-free
(dotted line) and galactic dust (short dashed line). At small angular
scales, frequencies around 50 to 100 GHz is ideal for a SZ experiment.}
\label{fig:szfreq}
\end{figure}

\section{Frequency Separation}
\label{sec:cleaning}

\begin{figure}[!h]
\begin{center}
\includegraphics[width=5.8in]{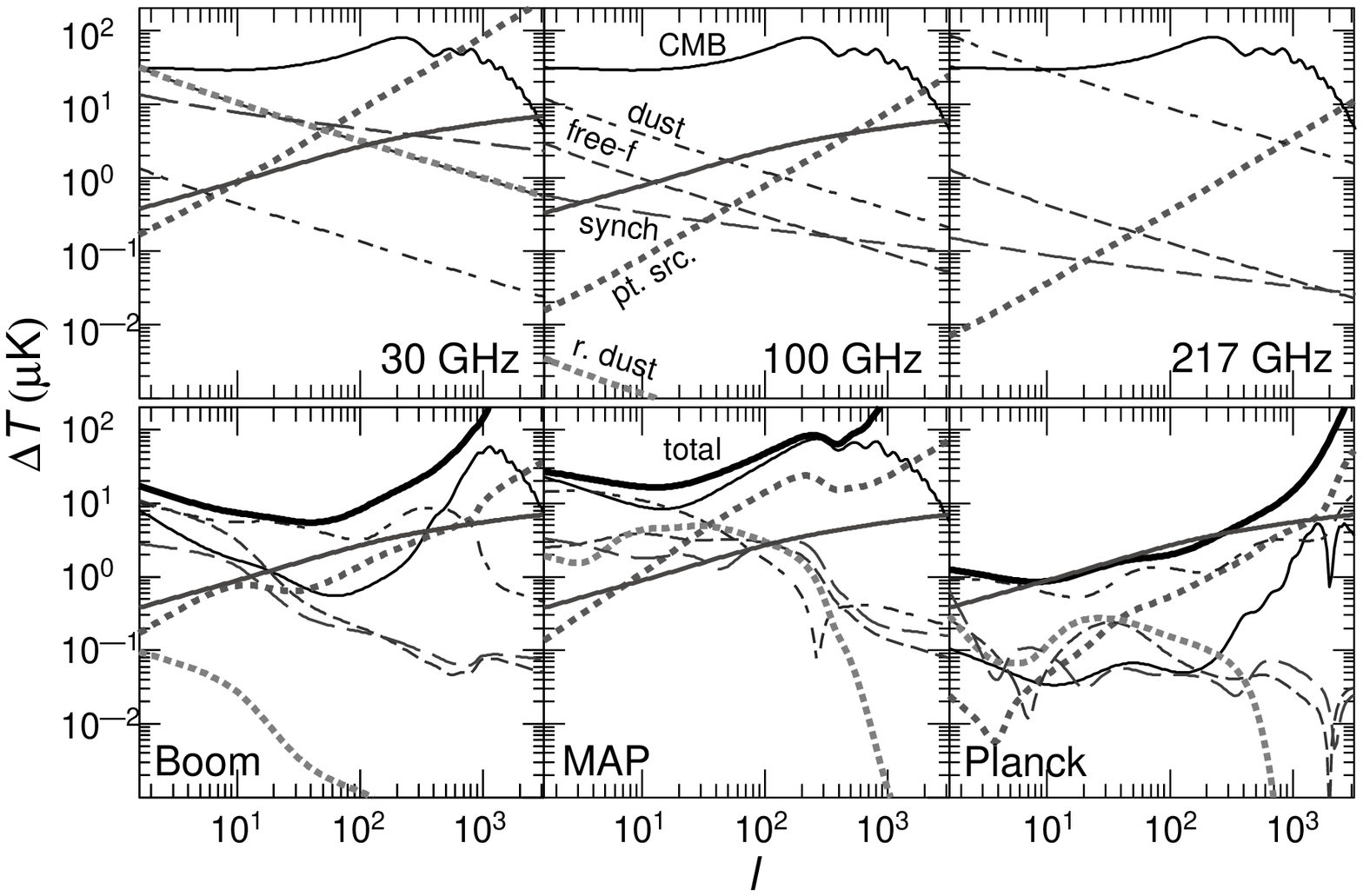}
\end{center}
\caption[Contributions to temperature anisotropy]{Top: foreground contributions to temperature anisotropies
$(\Delta T/T)^2 = l(l+1)C_l/2\pi$ from the various foregrounds
(dust, free-free, synchrotron, radio and infrared point sources,
and rotating dust) at three fiducial frequencies as labeled.
The SZ signal (solid, unlabeled) is estimated with a simplified biased
tracer model (see, \cite{Cooetal00a} 2000a).
Bottom: residual foregrounds after multifrequency subtraction for
Boomerang, MAP and Planck. The total includes detector noise and
residual CMB.
}
\label{fig:clean}
\end{figure}

The main obstacle for the detection of the SZ effect from large-scale
structure for angular scales above a few arcminutes is the CMB itself.
Here the primary anisotropies dominate the SZ effect for frequencies
near and below the peak in the CMB spectrum (see Fig.~\ref{fig:clean}).
Fortunately, the known frequency dependence and statistical properties of primary
anisotropies allows for extremely effective subtraction of their contribution (e.g.,
\cite{Hobetal98} 1998; \cite{BouGis99} 1999).
In particular, primary anisotropies obey Gaussian statistics
and follow the blackbody spectrum precisely.

Perhaps more worrying are the galactic and extragalactic foregrounds,
some of which are expected to to be at least comparable to the SZ
signal in
each frequency band.  These foregrounds typically have spatial and/or
temporal
variations in their frequency dependence leading to imperfect
correlations
between
their contributions in different frequency bands.   We attempt here to
provide
as realistic an estimate as possible of the prospects for CMB and
foreground removal, given our incomplete understanding of
the foregrounds. In Fig.~\ref{fig:szfreq}, we summarize our knowledge
on the frequency dependence of the SZ, CMB and other foreground
contaminants. Here, we consider a small angular scale experiments and
all effects are scaled relative to the thermal CMB spectrum. Other
than CMB, the only well known spectral dependence in this plot is the
SZ effect. To avoid complications in plotting, we only show the
absolute value of temperature here, but, it should be understood that
the SZ effect produces a decrement below the null frequency ($\sim$
217 GHz) and an increment thereafter.

\begin{table}[!h]
\begin{center}
\begin{tabular}{rcccc}
\hline
Experiment & $\nu$ & FWHM & $10^6 \Delta T/T$ &  \\
\hline
Boomerang
& 90 & 20 & 7.4 \\
& 150 & 12 & 5.7 \\
& 240 & 12 & 10 \\
& 400 & 12 & 80 \\
\hline
MAP
& 22 & 56 & 4.1  \\
& 30 & 41 & 5.7  \\
& 40 & 28 & 8.2  \\
& 60 & 21 & 11.0 \\
& 90 & 13 & 18.3 \\
\hline
Planck
& 30  & 33 & 1.6 \\
& 44  & 23 & 2.4 \\
& 70  & 14 & 3.6 \\
& 100 & 10 & 4.3 \\
& 100 & 10.7 & 1.7 \\
& 143 & 8.0 & 2.0  \\
& 217 & 5.5 & 4.3  \\
& 353 & 5.0 & 14.4 \\
& 545 & 5.0 & 147  \\
& 857 & 5.0 & 6670 \\
\hline
\end{tabular}
\caption[Experimental data]{
Specifications used for
Boomerang, MAP and Planck.
Full width at half maxima (FWHM) of the beams are in arcminutes and
should be converted to radians for the noise formula.
Boomerang covers a fraction $\sim$ 2.6\% of the sky, while we assume
a usable fraction of 65\% for MAP and Planck. }
\end{center}
\end{table}

\begin{figure}[!h]
\begin{center}
\includegraphics[width=4.2in]{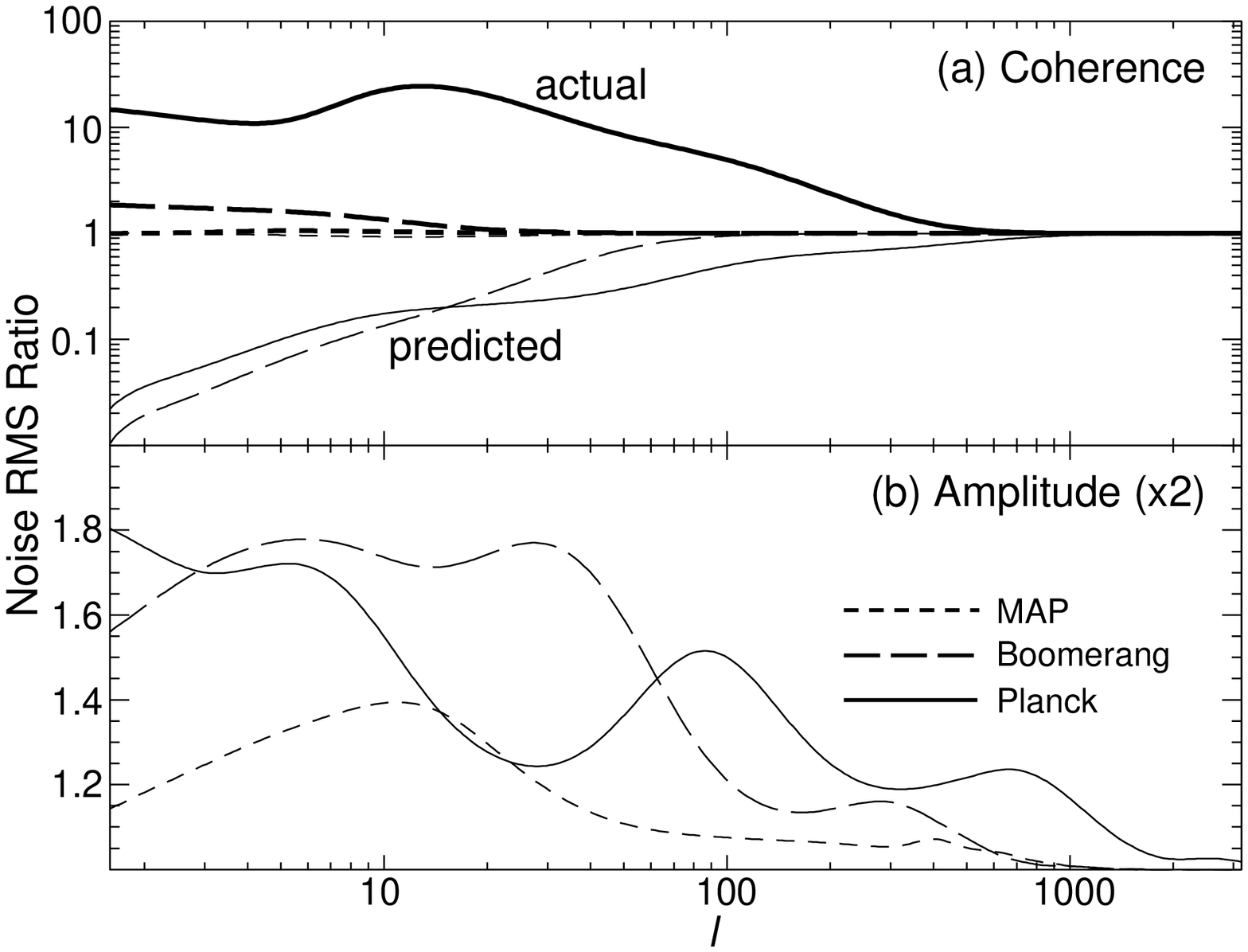}
\end{center}
\caption[Dependence of residual noise on assumptions]{Dependence of the residual noise rms on foreground
assumptions expressed as a ratio to the fiducial model of
Fig.~\ref{fig:clean}.  
(a) Falsely assuming the foregrounds have perfect
frequency coherence not only underpredicts the residual noise by a
substantial
factor but also leads to substantially more actual residual noise.
(b) Multiplying the foreground amplitudes by 2 (power by
4) produces less than a factor of 2 increase in the residual noise.}
\label{fig:simul}
\end{figure}

\subsection{Foreground Model and Removal}
\label{sec:foregmodel}

We use the ``MID'' foreground model of
\cite{Tegetal99} (1999) and adapt the subtraction techniques found
there for the purpose of extracting the SZ signal.
The assumed level of the foreground signal in the power spectrum
for three fiducial frequencies is shown in Fig.~\ref{fig:clean}.

The foreground model is defined in terms of the covariance between
the multipole moments at different frequency
bands\footnote{A potential caveat for this type of modeling is that it
assumes the foregrounds are statistically
isotropic whereas we know that the presence of the Galaxy violates
this assumption at least for the low order multipoles.   We assume
that
$1-f_\sky \sim 0.35$ of the sky is lost to this assumption even with
an all-sky experiment. }
\begin{equation}
\left< a_{l' m'}^{\fore *}(\nu') a_{l m}^{\fore} (\nu) \right> =
C_l^{\fore}(\nu',\nu)
               \delta_{l l'} \delta_{m m'}\,,
\end{equation}
in thermodynamic temperature units as set by the CMB blackbody.
In this section, we will speak of the primary
anisotropies and detector noise simply as other foregrounds with very
special
properties:
\begin{eqnarray}
C_l^{\rm CMB}(\nu',\nu)&=& C_l\,, \nonumber\\
C_l^{\rm noise}(\nu',\nu)&=& 8\ln 2 \theta(\nu)^2 e^{\theta^2(\nu)
l(l+1)}
        \left({\Delta T\over T}\right)^2\Big|_{\rm noise}
\delta_{\nu,\nu'}\,. \nonumber \\
\label{eqn:clnoise}
\end{eqnarray}
The FWHM$=\sqrt{8\ln 2} \theta$ and noise specifications
of the Boomerang, MAP and Planck frequency channels
are given in Tab.~1.  True foregrounds generally fall in
between these extremes of perfect and no frequency correlation.

The difference between extracting the SZ signal and the primary signal
is
simply
that one performs the subtraction referenced to the
SZ frequency dependence
\begin{equation}
s(\nu) = 2 - {x \over 2}\coth {x \over 2}\,,
\end{equation}
where $x = h\nu/kT_{\rm cmb} \approx \nu/56.8$GHz.   Note that
in the RJ limit
$s(\nu) \rightarrow 1$ such that
\begin{equation}
C_l^{\rm SZ}(\nu,\nu') = s(\nu)s(\nu') C_l^{\rm SZ}
\end{equation}
where $C_l^{\rm SZ}$ is the SZ power spectrum in the RJ
limit.

Consider an arbitrary linear combination of the channels,
\begin{equation}
b = \sum_{\nu_{i}} {1 \over s(\nu_i)} w(\nu_i) a(\nu_i)\,,
\end{equation}
where we will normalize the sum of the weights to unity
$\sum w(\nu_i)=1$ to obtain an unbiased estimator of the
RJ multipoles.
Since the subtraction is done multipole by multipole, we have
temporarily
suppressed the multipole index.
The covariance of this quantity is
\begin{equation}
\left< b^2 \right> = C^{\sz}[\sum_{\nu_i} w(\nu_i)]^2
        + \sum_{\nu_i,\nu_j} w(\nu_i) w(\nu_j)
        \tilde C(\nu_i,\nu_j)\,,
\end{equation}
%        \sum_{\fore} {C^\fore(\nu_i,\nu_j) \over
%s(\nu_i)s(\nu_j)}
%\,.
%\end{equation}
where
the scaled foreground covariance matrix is
\begin{eqnarray}
\tilde C(\nu_i,\nu_j) \equiv
\sum_{\fore} \tilde C^\fore(\nu_i,\nu_j)
 &=&
\sum_{\fore} {C^\fore(\nu_i,\nu_j) \over s(\nu_i)s(\nu_j)}
\,.
\end{eqnarray}
Minimizing the variance contributed by the foregrounds subject to the
constraint that the SZ estimation be unbiased, we obtain
\begin{equation}
\sum_{\nu_i} w(\nu_j) {\tilde C(\nu_i,\nu_j)} =
{\rm const.}\,
\end{equation}
whose solution is ${\bf w} \propto \tilde{\bf C}^{-1} {\bf e}$,
where $e(\nu_i)=1$.
The constant of proportionality is fixed by the condition
$\sum w(\nu_i)=1$, i.e.
\begin{equation}
w(\nu_i) = {\sum_{\nu_j} {\tilde C}^{-1}(\nu_i,\nu_j)
        \over
            \sum_{\nu_k,\nu_j} {\tilde C}^{-1}(\nu_k,\nu_j)} \,.
\label{eqn:weights}
\end{equation}

\begin{figure}[!h]
\begin{center}
\includegraphics[width=4.2in]{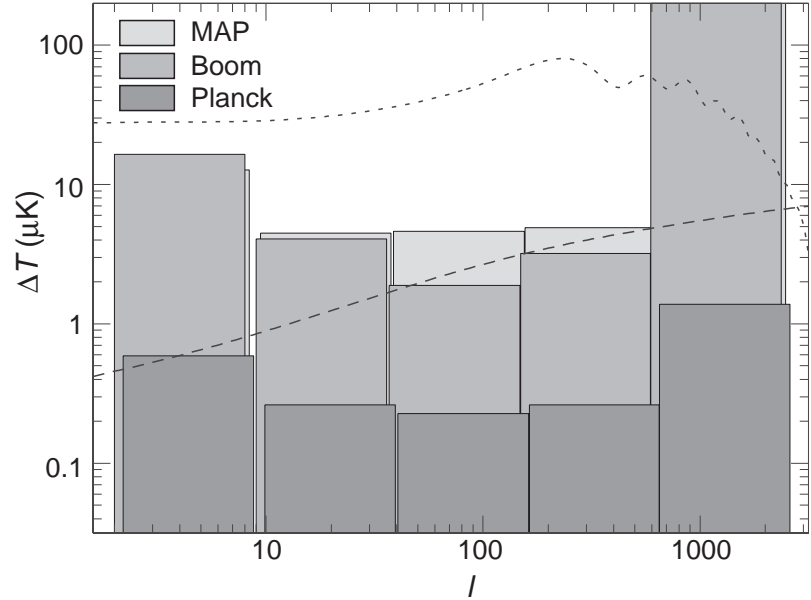}
\end{center}
\caption[Detection thresholds for the SZ effect]{Detection thresholds for the SZ effect.   Error boxes
represent the
1-$\sigma$ rms residual noise in multipole bands and can be
interpreted
as the detection threshold.  Also shown (dotted) is the level of the
primary
anisotropies that have been subtracted with the technique and the
signal (dashed) expected in the simplified model of \cite{Cooetal00a} (2000a).}
\label{fig:error}
\end{figure}

\subsection{Detection Threshold}
The residual noise variance from each foreground component
is then
\begin{equation}
N_l^\fore % = {\bf w}_l^t {\tilde {\bf C}_l^\fore} {\bf w}_l
= \sum_{\nu_i,\nu_j} w_l(\nu_i) w_l(\nu_j) \tilde
C_l^\fore(\nu_i,\nu_j)\,.
\label{eqn:residualcomponent}
\end{equation}
with the total
\begin{equation}
N_l = \sum_{\fore} N_l^\fore\,,
\label{eqn:nl}
\end{equation}
where we have restored the multipole index.

Note that the residual noise in the map is independent of assumptions
about the
SZ signal including whether it is Gaussian or not.   However if the
foregrounds
themselves are non-Gaussian, then this technique only minimizes the
variance
and may not be optimal for recovery of non-Gaussian features in the SZ
map
itself. \cite{Bouetal95} (1995) have shown that similar techniques are
quite
effective
even
when confronted with non-Gaussian foregrounds.
This is a potential caveat especially for cases in which
the residual noise is not dominated by the primary anisotropies
or detector noise.  

The residual noise sets the detection threshold for the
SZ effect for a given experiment.
In Fig.~\ref{fig:clean},
we show the rms of the residual noise after
foreground subtraction for the Boomerang, MAP and Planck
experiments assuming the ``MID'' foreground model from
\cite{Tegetal99} (1999).  With the Boomerang and Planck channels,
elimination of the primary anisotropies is excellent up to the beam
scale where detector noise dominates.  As expected, the MAP channels,
which are all on the RJ side of the spectrum, do not
allow good elimination of the primary anisotropies.

It is important not to assume that the foregrounds are
perfectly correlated in frequency, which is the usual
assumption in the literature (e.g., \cite{Hobetal98} 1998).  There are two types of errors
incurred by doing so.  The first is that one underpredicts
the amount of residual noise in the SZ map (see Fig.~\ref{fig:simul}).
The second is that if one calculates the optimal weights in
Eq.~\ref{eqn:weights} based on this assumption the actual
residual noise increases.  For Planck it can actually increase the
noise
beyond the level in which it appears in the $100$GHz maps with no
foreground subtraction at all.  The reason is that the cleaning
algorithm then erroneously uses the contaminated high and low
frequency channels to subtract out the small foreground contamination
in the central channels.  In Planck, the difference between the
predicted
and actual rms noise from falsely assuming perfect frequency coherence
can be more than two orders of magnitude.

For Boomerang and Planck, the largest residual noise component,
aside from detector noise, is dust emission and is sufficiently
large that one might worry that current uncertainties in our knowledge
of the foreground model may affect
the implications for the detection of the SZ effect.  It is therefore
important to explore variations on our fiducial foreground
model.

Multiplying
the foreground rms amplitudes uniformly by a factor of 2 (and hence
the power by a factor of 4), produces less than a factor of 2 increase
in the residual noise rms as shown in Fig.~\ref{fig:simul}.
Likewise, as discussed in \cite{Tegetal99} (1999), minor
variations in the frequency coherence do not effect the residual noise
much
in spite of the fact that it is crucial not to assume perfect
correlation.
We conclude that uncertainties in the properties of currently known
foregrounds are unlikely to change our conclusions qualitatively.
There is however always the possibility that some foreground that does
not
appear in the currently-measured frequency bands will affect our
results.

The fact that the residual dust contributions are comparable to those
of the detector noise for Boomerang and Planck is problematic for
another reason.  Since the algorithm minimizes
to total residual variance, it attempts to keep these two main
contributors roughly comparable.  However the dust will clearly
be non-Gaussian to some extent and one may prefer instead to trade
more residual detector noise for dust contamination.  One can
modify the subtraction algorithm to account for this by artificially
increasing the rms amplitude of the dust when calculating the weights
in Eq.~\ref{eqn:weights} while using the real amplitude
in calculating the residual noise in
Eq.~\ref{eqn:residualcomponent}. For example we have explored increasing the amplitude by a factor of
4 (power by 16) for the weights.  The result is an almost negligible
increase
in total residual noise rms but an improvement in dust rejection by
a factor of 3-4 in rms.   For Planck this brings the ratio of
dust to total rms to $\sim 10\%$ and recall that the noise adds
in quadrature so that the total dust contribution is really
$\sim 1\%$ of the total.
This more conservative approach is thus advisable but since
it leaves the total residual noise rms essentially unchanged, we
will adopt the minimum variance noise to estimate the
detection threshold.

Fig.~\ref{fig:clean} directly tells us the detection threshold per
$(l,m)$ multipole moment.  Since the SZ signal is likely to have a
smooth
power spectrum in $l$,
one can average over bands in $l$ to beat down the residual noise.
Assuming Gaussian-statistics, the residual noise variance $ 2N_l^2$
for
the power spectrum estimate is then given by
\begin{equation}
N_l^{-2}\Big|_{\rm band} =
{f_\sky}  \sum_{l_{\rm band}} (2l+1) N_l^{-2}\,,
\end{equation}
where $f_\sky$ accounts for the reduction of the number of independent
modes due to the fraction of sky covered.
The result for the three experiments is shown in Fig.~\ref{fig:error}.
In the absence of a detection, they can be interpreted as the optimal
1 $\sigma$ upper limits on SZ bandpowers achievable by the experiment.
Boomerang and MAP can place upper limits on the SZ signal in the
interesting
$\mu$K regime whereas Planck can detect signals well below a $\mu$K.

This noise averaging procedure in principle implicitly assumes that
the
statistical properties of the residual noise, and by implication the
full covariance matrix of the other foregrounds, is precisely
known.  In reality, they too must be estimated from the multifrequency
data itself through
either through the subtraction techniques discussed here or
by direct modeling of the foregrounds in the maps.  \cite{Tegetal99}
(1999) found that direct modeling of the foregrounds with hundreds of fitted
parameters did not appreciably degrade our ability to extract the
properties of the primary anisotropies.  The main source of variance
there was the cosmic variance of the primary anisotropies themselves whose
properties are precisely known.
Similarly here the main source of residual variance is either the
primary anisotropies (for MAP) or detector noise (for Boomerang and
Planck)
and their statistical properties may safely be considered known.

\subsection{Discussion}

We have studied the prospects for extracting the statistical
properties of the Sunyaev-Zel'dovich (SZ) effect associated with hot gas in
large-scale structure using upcoming multifrequency CMB experiments.
This gas currently remains undetected but may comprise a substantial
fraction of the present day baryons.
The SZ effect has a distinct spectral dependence with a null at a
frequency of $\sim$ 217 GHz compared with true temperature anisotropies.
This frequency dependence is what allows for effective separation
of the SZ contribution with multifrequency
CMB measurements.

\begin{figure}[!h]
\begin{minipage}[t]{5.92in}
\includegraphics[width=2.96in]{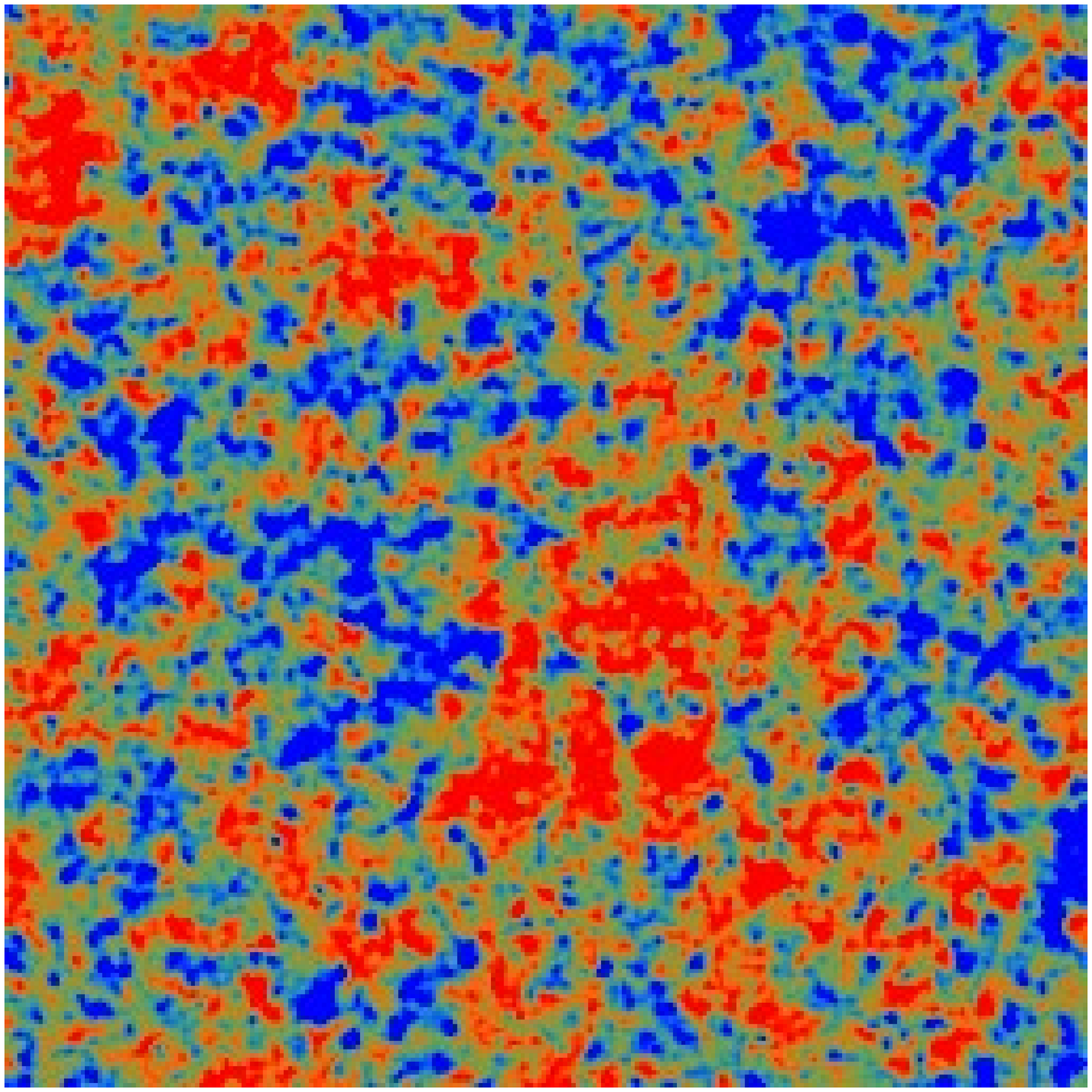}
\includegraphics[width=2.96in]{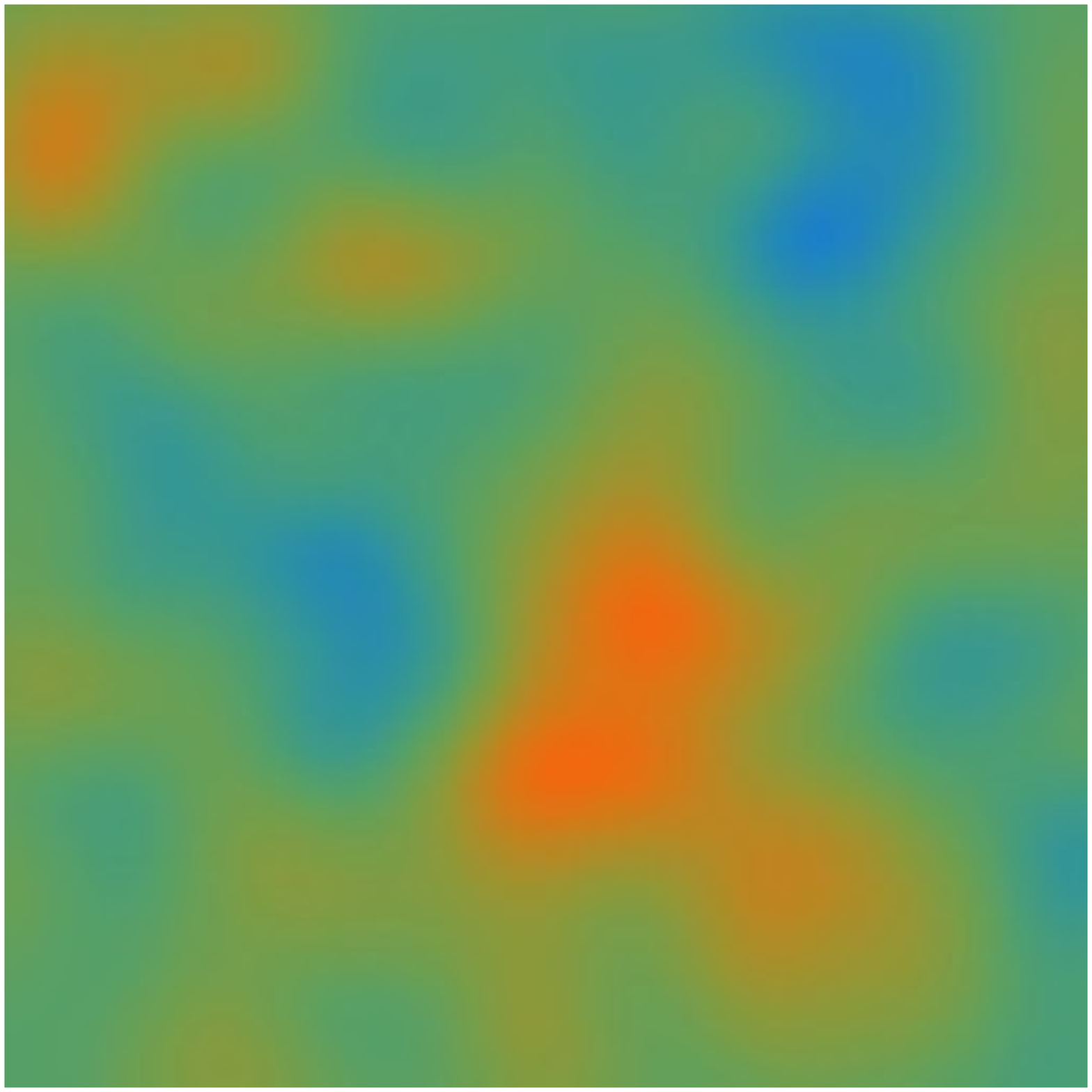}
\begin{minipage}[t]{5.92in}
\includegraphics[width=2.96in]{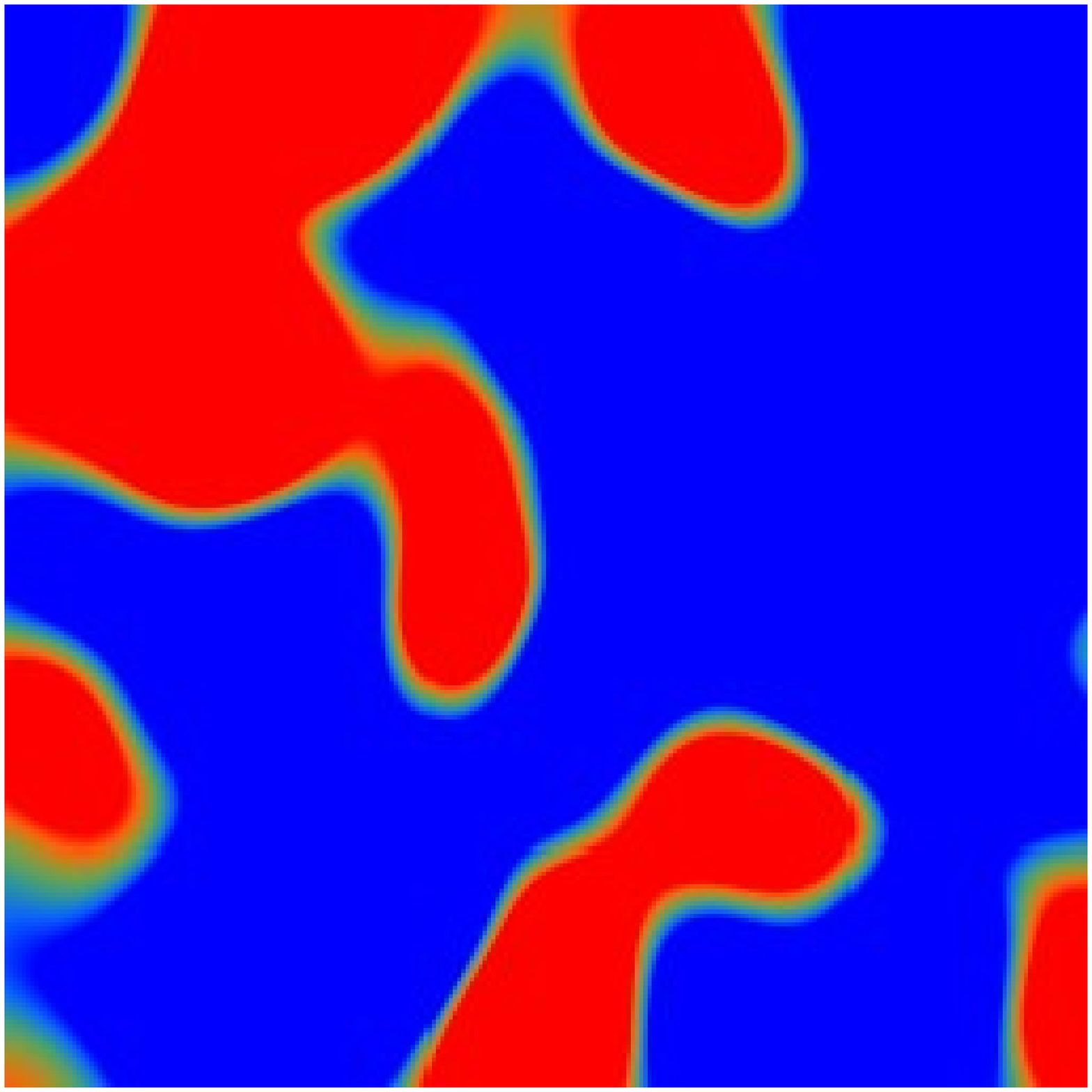}
\includegraphics[width=2.96in]{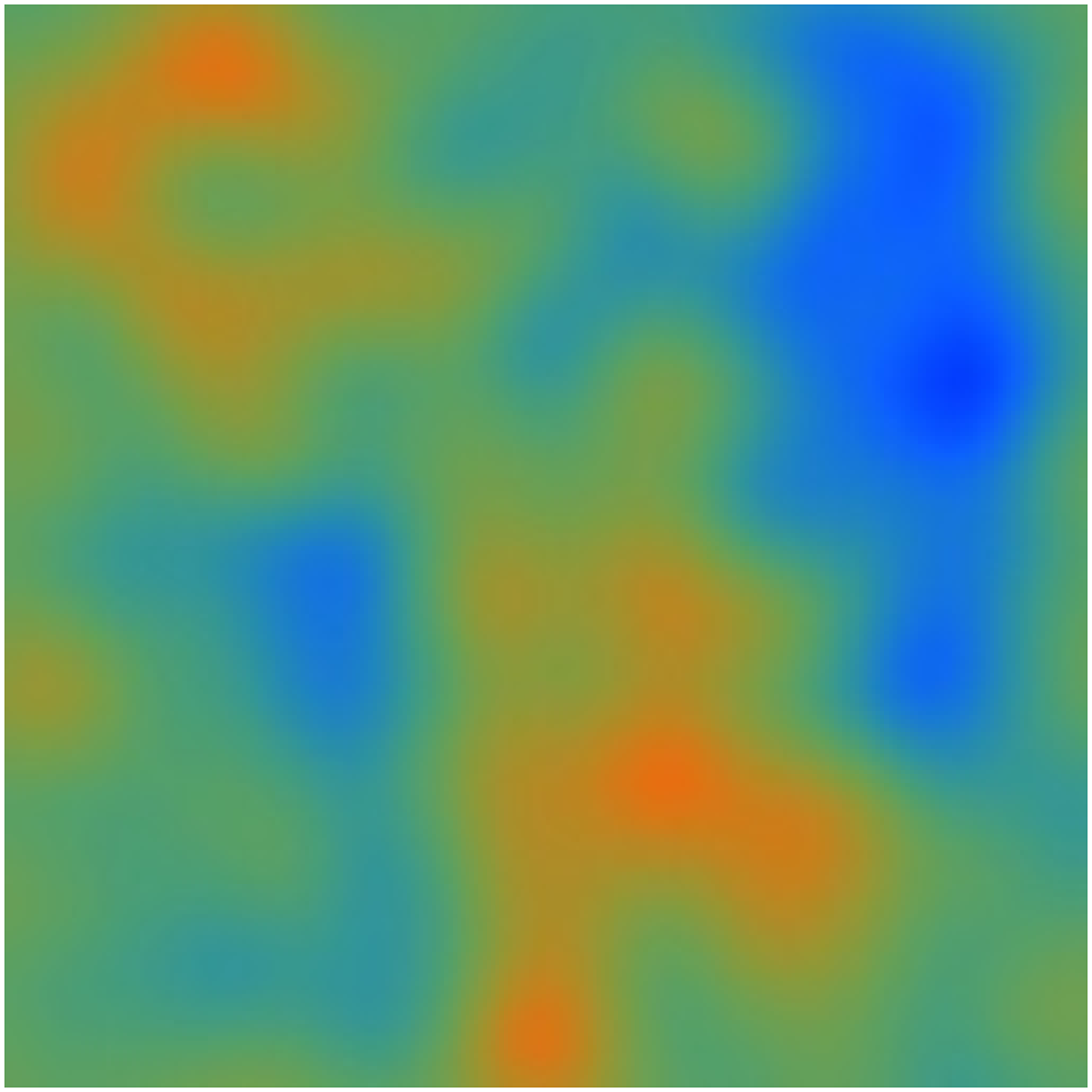}
\end{minipage}
\caption[Recovery of the SZ signal in Planck]{Recovery of the SZ
signal with Planck. Top-Left: SZ  effect in the
$\Lambda$CDM assuming pressure traces dark matter density field with a
scale independent bias. The field is $6^\circ\times 6^\circ$  and
the range of the map is $-100\mu K , 25\mu K$ with an
rms of $9\mu K$ and has an approximate angular resolution of $2'$.
Note the lack of obvious filamentary structures. Form top-left to
 to bottom-left, model
SZ signal with first map smoothed with a top-hat of radius $20'$, 
signal $+$ noise from primary anisotropies and foregrounds,
and final recovered map from Planck.}
\label{fig:recovery}
\end{minipage}
\end{figure}

As examples, we have employed the frequency and noise specifications
of the Boomerang, MAP, Planck experiments.
The MAP satellite only covers frequencies at
RJ part of the frequency spectrum.  Consequently, only
Boomerang and Planck can take full advantage of multifrequency
separation of the SZ and primary anisotropies.  We have evaluated the
detection threshold for SZ power
spectrum measurements (see  Fig.~\ref{fig:error}). In
Fig.~\ref{fig:recovery}, we demonstrate the ability of Planck mission
to produce a map of the SZ effect using the frequency information.
Boomerang and MAP should provide limits on the degree
scale fluctuations at the several
$\mu$K level in rms; Planck should be able to detect sub $\mu$K
signals. This statement is independent of our assumptions about the SZ
effect, including the non-Gaussianity. The exact detection threshold,
however, depends on the assumptions associated with foreground
distribution, including whether the foregrounds can be modeled as a
Gaussian distribution.

\begin{figure}[!h]
\begin{center}
\includegraphics[width=4.2in,angle=-90]{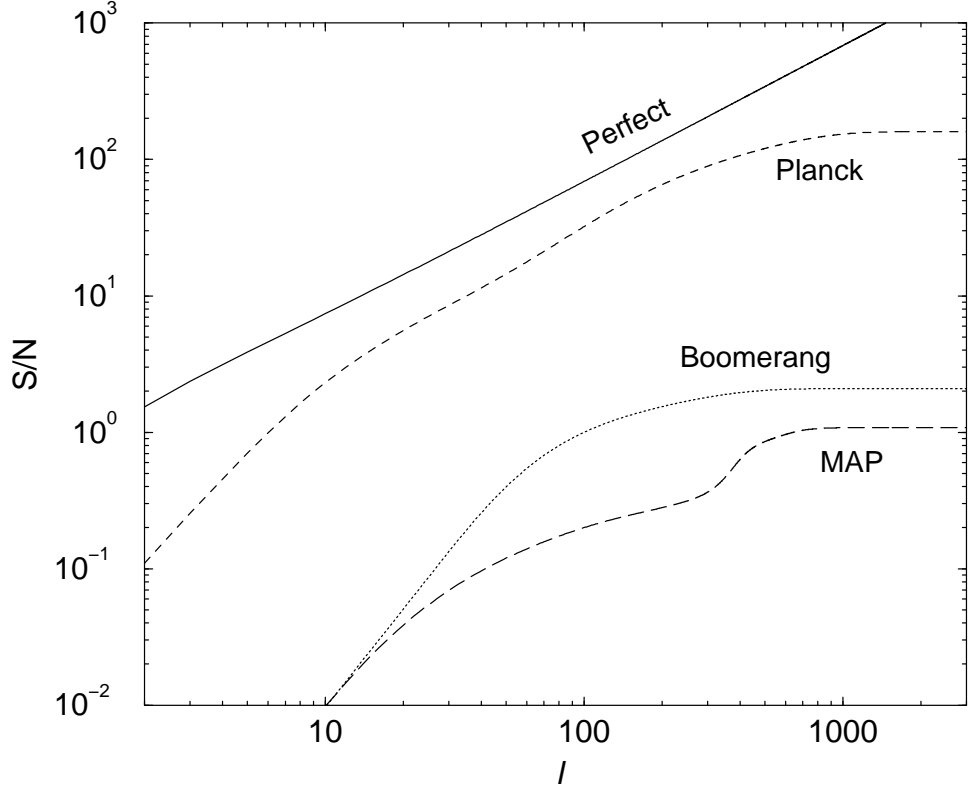}
\end{center}
\caption[Cumulative signal-to-noise for the SZ power spectrum
detection]{Cumulative 
signal-to-noise in the measurement of the  SZ
power spectrum with Boomerang, MAP and Planck as a  function of
maximum
$l$. The solid line
is the maximum signal-to-noise achievable in a perfect experiment with
no instrumental noise and full-sky coverage.}
\label{fig:clsn}
\end{figure}

In the next few sections, we will turn to the question on the exact SZ
contribution. Since at large scales, the SZ effect can be well modeled
with a model in which gas traces the density field with a bias, we can
use such a simplified approach to obtain an order of  magnitude
estimate on the signal-to-noise for the detection of the SZ power
spectrum (see, \cite{Cooetal00a} 2000a). We summarize our results in Fig.~\ref{fig:clsn}. 
As shown, Planck mission as the greatest potential to detect the SZ
effect at large scales while MAP and Boomerang allow useful upper
limits on few sigma detections. 

In the next section, we use the basic fact that SZ effect traces
pressure fluctuations in the universe and study clustering properties
of large scale pressure under the halo model. We use these predictions
for pressure power spectrum and higher order correlations to study the
angular power spectrum, bispectrum and trispectrum of the SZ effect.

\section{Clustering Properties of Pressure}
\label{sec:pressure}

Following \cite{Scoetal99} (1999), we can relate the trispectrum
to the variance of the estimator of the binned power spectrum
\begin{equation}
\hat P_i = {1 \over V } \int_{\shell i} {d^3 k \over V_{\shell i}}
\Pi^*(-\veck) \Pi(\veck)  \, ,
\end{equation}
where the integral is over a shell in $k$-space centered around $k_i$,
$V_{\shell i} \approx 4\pi k_i^2 \delta k$ is the volume of the shell
and $V$
is the volume of the survey.  Recalling that $\delta({\bf 0})
\rightarrow V/(2\pi)^3$
for a finite volume,
\begin{eqnarray}
C_{ij} &\equiv& \left< \hat P_i \hat P_j \right> -
        \left< \hat P_i \right>
        \left< \hat P_j \right>  \nonumber\\
       &=& {1 \over V} \left[ {(2\pi)^3 \over V_{\shell i} } 2 P_i^2
        \delta_{ij}+
        T_{ij}^\Pi \right]  \, ,
\end{eqnarray}
where
\begin{eqnarray}
T_{ij}^\Pi &\equiv& \int_{\shell i} {d^3 k_i \over V_{\shell i}}
        \int_{\shell j} {d^3 k_j \over V_{\shell j}}
        T_\Pi(\veck_i,-\veck_i,\veck_j,-\veck_j) \,.
\label{eqn:covariancepiij}
\end{eqnarray}

\begin{figure}[!h]
\begin{center}
\includegraphics[width=4.2in]{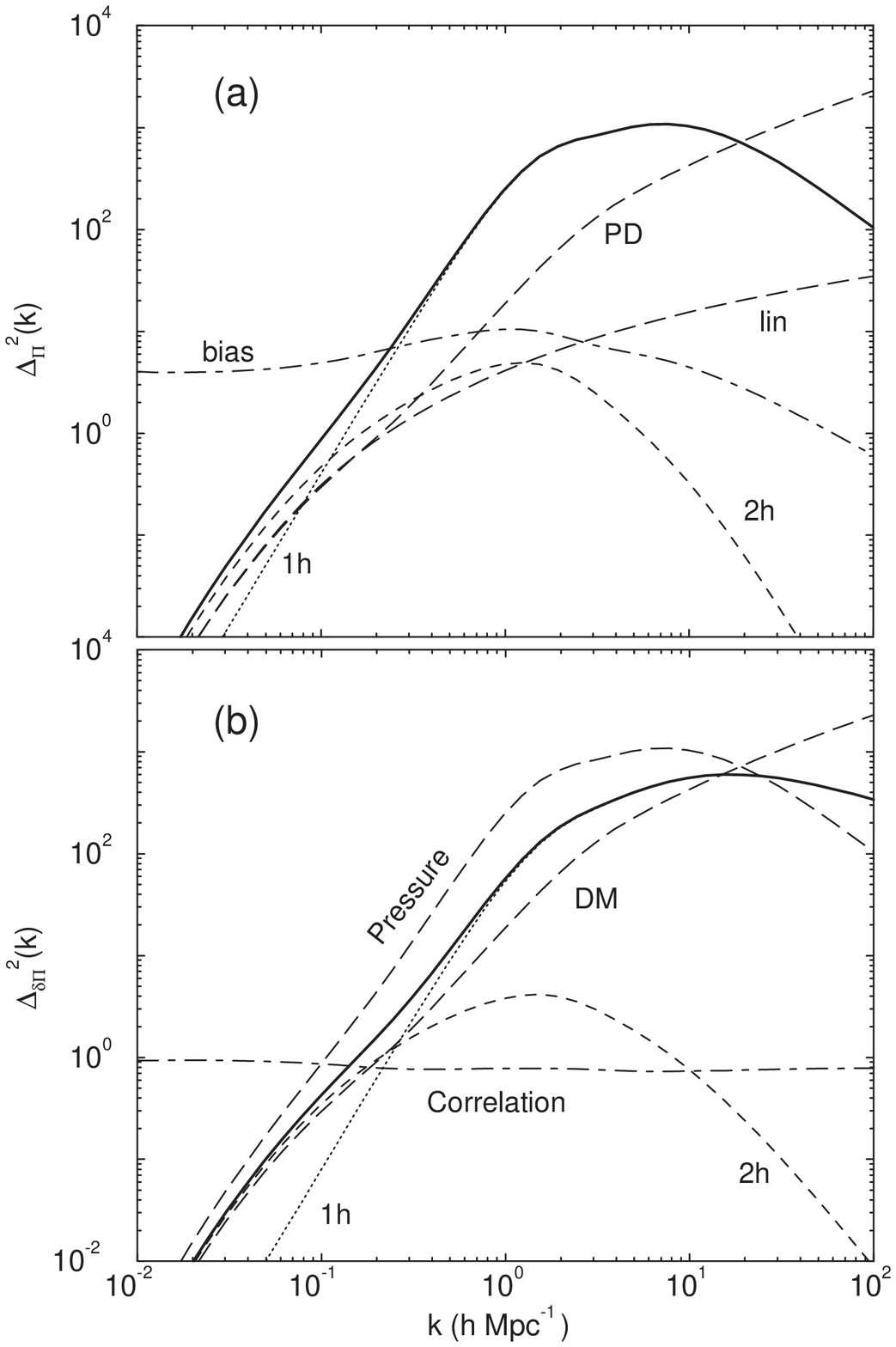}
\end{center}
\caption[The pressure power spectrum and the pressure-dark matter
cross power spectrum]{The pressure power spectrum (a) and the
cross-power spectrum between pressure and dark matter (b) today ($z=0$)
broken into individual contributions under the halo description.
The lines labeled 'bias' and ``correlations'' shows the pressure bias
and correlation relative to the
dark matter power spectrum under the halo model. The pressure behaves
such that it correlates well with dark matter at large scales while
there is a breakdown in this correlation at small scales.}
\label{fig:pressurepower}
\end{figure}

Notice that though both terms
scale in the same way with the volume of the survey, only the Gaussian
piece
necessarily decreases with the volume of the shell.  For the Gaussian
piece,
the sampling error reduces to a simple root-N mode counting of
independent modes
in a shell.  The trispectrum quantifies the non-independence of the
modes both within a shell
and between shells.  Calculating the covariance matrix of the power
spectrum estimates reduces to averaging the elements of the
trispectrum across
configurations
in the shell.

\begin{figure}[!h]
\begin{center}
\includegraphics[width=4.2in]{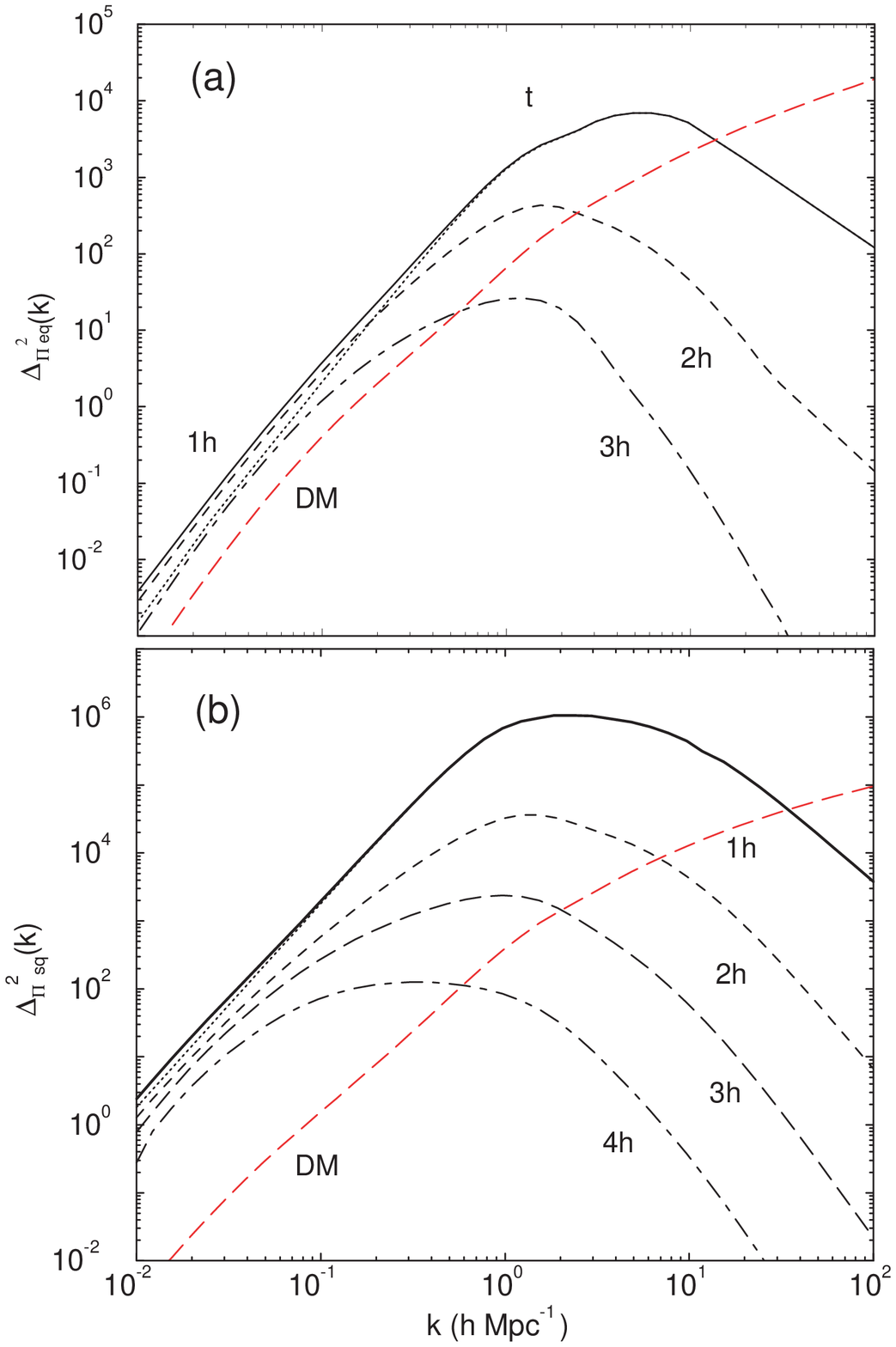}
\end{center}
\caption[The pressure bispectrum and trispectrum]{The pressure
bispectrum  (a) and the
trispectrum  today ($z=0$)
broken into individual contributions under the halo description. 
Here, we show the equilateral configuration for the bispectrum and
square configuration for the trispectrum. For
comparison, we also show the dark matter bispectrum and trispectrum,
under the halo model, for the same configurations.}
\label{fig:pressuretri}
\end{figure}

\begin{table}[!h]
\begin{flushleft}
\begin{tabular}{lrrrrrrrrr}
\hline
$k$   & 0.06 & 0.07 & 0.09 & 0.11 & 0.14 & 0.17
& 0.21 & 0.25 & 0.31\\
\hline
0.05 & 1.00 & 0.65 & 0.73 & 0.74 & 0.74 &
0.71 & 0.70 & 0.68 & 0.68 \\
0.07 & (0.06) & 1.00 & 0.81 & 0.87 & 0.87
&0.85 & 0.83 & 0.82& 0.81  \\
0.09 & (0.12) & (0.10) & 1.00 & 0.90 & 0.93
& 0.92 & 0.90 & 0.89 & 0.88\\
0.11 & (0.18) & (0.19) & (0.16) & 1.00 &
0.95 & 0.96 & 0.95 & 0.93 & 0.93 \\
0.14 & (0.25) & (0.30) & (0.29)& (0.28) &
1.00 & 0.98 & 0.97 & 0.96 & 0.95 \\
0.17 & (0.30) & (0.37) & (0.37) & (0.43)&
(0.43) & 1.00 & 0.98 & 0.97 & 0.96 \\
0.21 & (0.33) & (0.33) & (0.41) & (0.54) &
(0.58) & (0.59) & 1.00 & 0.98 & 0.97 \\
0.25 & (0.34) & (0.34) & (0.42) & (0.58) &
(0.69) & (0.74) & (0.75) & 1.00 & 0.99 \\
0.31 & (0.33) & (0.33) & (0.41) & (0.57) &
(0.70)& (0.78)& (0.84) & (0.86) & 1.00 \\
\hline
$\sqrt{\frac{C_{ii}}{C_{ii}^{G}}}_{\rm \Pi}$ & 1.29&
1.64& 1.95& 2.24& 3.58& 6.09& 9.27& 16.4& 21.2\\
\hline
$\sqrt{\frac{C_{ii}}{C_{ii}^{G}}}_{\rm \delta}$ & 
 1.02& 1.03& 1.04& 1.07& 1.14& 1.23& 1.38& 1.61& 1.90 \\
\end{tabular}
\caption[Pressure power spectrum covariance]{
Diagonal normalized covariance matrix of the binned pressure (upper
triangle) and dark matter density field (lower triangle)
power spectrum with $k$ values in units of h Mpc$^{-1}$.
Lower triangle (parenthetical numbers) displays the covariance from
halo model in \cite{CooHu01b} (2001b). Final line shows
the fractional increase in the errors (root diagonal covariance) due
to
non-Gaussianity as calculated under the halo model for pressure and
dark matter density field.}
\label{tab:pressurecorr}
\end{flushleft}
\end{table}

\subsection{Discussion}

In Fig.~\ref{fig:pressurepower}(a), we show the logarithmic
power spectrum of pressure and dark matter such that
$\Delta^2(k)=k^3 P(k)/2\pi^2$ with
contributions broken down to the $1h$ and $2h$ terms today. 
As shown, the pressure power spectrum depicts an increase in power
relative to the dark matter at scales out to few h
Mpc$^{-1}$, and a decrease thereafter. The decrease in power at small
scales can be understood through the relative contribution to pressure
as a function of the halo mass.

\begin{figure}[!h]
\begin{center}
\includegraphics[width=4.2in]{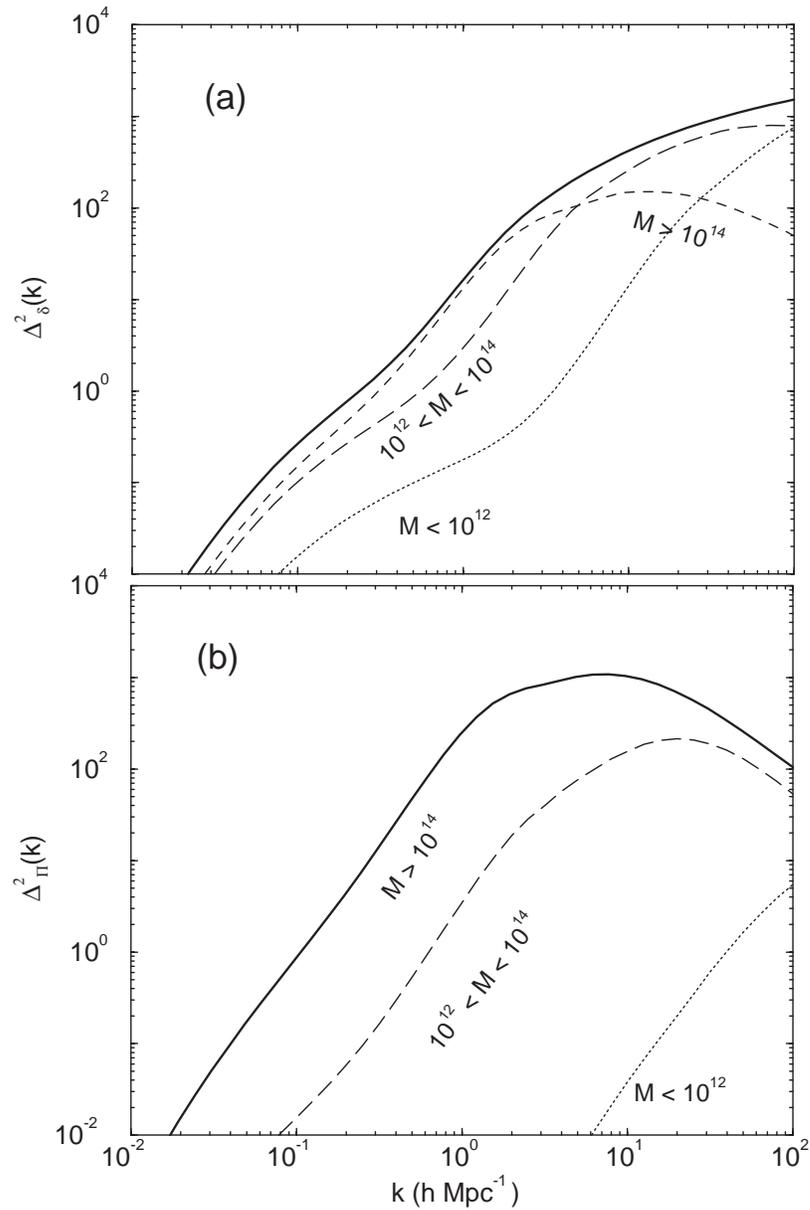}
\end{center}
\caption[Mass dependence on the dark matter and pressure
power spectra]{
The mass dependence on the dark matter power spectrum (a) and
pressure power spectrum (b). Here, we show the total contribution
broken in mass limits as written on the figure.
As shown in (a), the large scale contribution to the dark matter power
comes from massive halos while small mass halos contribute at small
scales. For the pressure, in (b), only massive halos above a mass of
$10^{14}$ M$_{\rm sun}$ contribute to the power.}
\label{fig:powermass}
\end{figure}

In Fig.~\ref{fig:powermass}, we break the total dark matter power
spectrum
(a) and the total pressure power spectrum (b), to a function of
mass. As
shown, contributions to both dark matter and pressure comes from
massive halos at large scales and by small mass halos at small scales.
The pressure power spectrum is such that through temperature weighing,
with $T_e \propto M^{2/3}$ dependence, the contribution from
low mass halos to pressure is suppressed relative to that from the
high
mass end. Thus, the pressure power spectrum, at all scales of
interest, can be easily described with halos of mass greater than
$10^{14}$ $M_{\sun}$. A comparison of the dark matter and pressure
power spectra, as a
function of mass, in Fig.~\ref{fig:powermass} reveals that the turn
over in the pressure power spectrum results at an effective scale
radius for halos with mass greater than $10^{14}$ M$_{\sun}$.
We refer the reader to \cite{Coo00} (2000) for  further
details on the pressure power spectrum and its properties.

Similar to the electron temperature dependence on mass in pressure, 
the description of the galaxy power spectrum under the
halo model, as discussed in \S~\ref{sec:galaxy}, 
is such that the number of galaxies as a function of mass
has a functional description given by $N_\gal \propto M^{\alpha}$ with
$\alpha \sim 0.7$ to 0.8 (see, \cite{Scoetal00} 2000). However, there
is a significant contribution from low mass halos at small angular
scales since the description  allows the formation of at least one
galaxy in halo down to some low mass limit. Thus, contrary to the behavior in pressure, one finds that
there is significant low halo mass contribution to the galaxy power
spectrum. We can increase the contribution from low mass halos if we
allow for a minimum temperature in electrons, such as due to
non-gravitational or so-called preheating. Though we do not include
such a minimum temperature here, we will address how one can
observationally determine such a minimum temperature for electrons
later.

For the covariance of pressure, and also for the covariance of the SZ
power spectrum,
we are mainly interested in terms of the pressure trispectrum
involving configurations that result in
$T_\Pi(\veck_1,-\veck_1,\veck_2,-\veck_2)$, i.e. parallelograms
which are defined by either the length $k_{12}$ or the angle
between $\veck_1$ and $\veck_2$.  For illustration purposes
we will take $k_1=k_2$ and the angle to be $90^\circ$
($\veck_2=\veck_\perp$) such that
the parallelogram is a square.
It is then convenient to define
\begin{equation}
\Delta^2_{\Pi \rm sq}(k) \equiv \frac{k^3}{2\pi^2}
T_\Pi^{1/3}(\veck,-\veck,\veck_\perp,-\veck_\perp) \, ,
\end{equation}
such that this quantity scales roughly as the logarithmic
power spectrum itself $\Delta^2(k)$.  This spectrum is
shown in Fig.~\ref{fig:pressuretri}(b) with the individual
contributions from the 1h, 2h, 3h, 4h terms shown.
As shown, almost all contributions to the pressure trispectrum come
from the single halo term.

Using the pressure trispectrum, we can now predict the pressure
covariance and, more appropriately, correlations in the binned
measurements of the pressure. The predictions made here with the halo
model to describe pressure can easily be tested in numerical
simulations and the accuracy
of the halo model can be further studied. For this purpose, we
calculate the covariance matrix $C_{ij}$ from
Eq.~\ref{eqn:covariancepiij}
with the bins centered at $k_i$ and volume $V_{\shell i} =
4\pi k_i^2 \delta {k_i}$. The binning scheme used here is the one 
we utilized in \cite{CooHu01b} (2001b) to calculate the binned dark
matter power spectrum correlations. 

In Table~\ref{tab:pressurecorr}, we tabulate the pressure (upper
triangle), and for comparison dark matter (lower triangle),
correlation coefficients
\begin{equation}
\hat C_{ij} = {C_{ij} \over \sqrt{C_{ii} C_{jj}}} \, .
\end{equation}
The dark matter correlations are from the halo based predictions by
 \cite{CooHu01b} (2001b). There, for the dark matter,
 we suggested that the
halo model predicted correlations agree with numerical simulations of
\cite{MeiWhi99} (1999) typically 
better than $\pm 0.1$, even in the region where
non-Gaussian effects dominate, and that the the qualitative
features such as the increase in correlations across the
non-linear scale are preserved. As we do not have measurements of the
pressure correlations from simulations, we cannot perform a detailed
comparison on the accuracy of the halo model predictions for pressure
here. 

A further test on the accuracy of the halo approach is to consider
higher order real-space moments such as the skewness and kurtosis. In
\cite{Coo00} (2000), we discussed the SZ skewness under the
halo model. As discussed in detail by \cite{RefTey01} (2001), halo
model predictions agree remarkably well with numerical simulations,
especially for the pressure and SZ power spectra, though, detailed
comparisons still remain to be made  with respect to bispectrum and
trispectrum.

Even though the dark matter halo formalism provides a physically
motivated means of calculating the statistics of the dark matter
density field and associated properties such as pressure, 
there are several limitations of the approach that should be borne in
mind when interpreting results. The approach assumes all halos to
share a parameterized
spherically-symmetric profile and this assumption is
likely to affect detailed results on the configuration dependence of
the bispectrum and trispectrum.
Since we are considering a weighted average of configurations, our
predictions presented here may be insufficient to establish the
validity of the trispectrum modeling in general. Further numerical
work is required to
quantify to what extent the present approach reproduces simulation
results for the full trispectrum. We do not consider such comparisons
here, other than to  suggest that the halo model has provided, at
least qualitatively,
a consistent description better than any of the arguments involving a
biased description of gas tracing the dark matter etc.

\begin{figure}[!h]
\begin{center}
\includegraphics[width=6.0in]{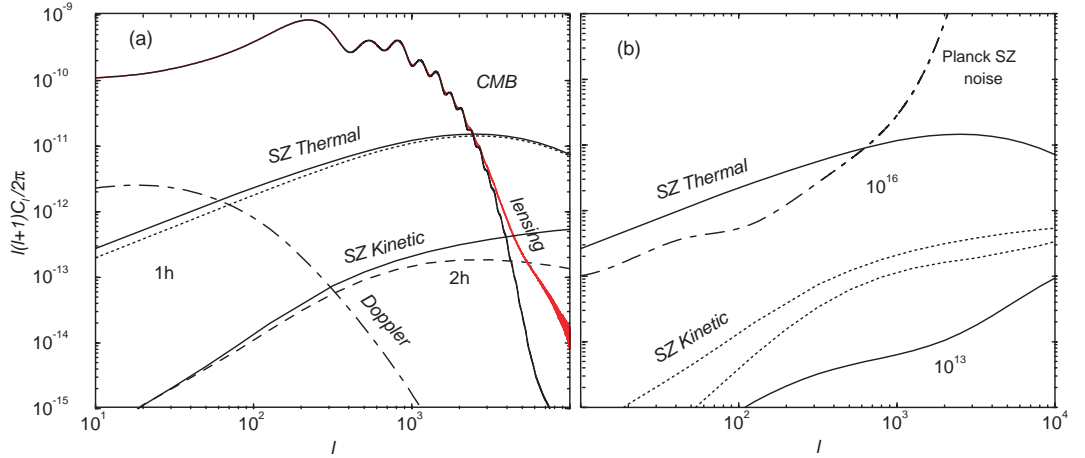}
\end{center}
\caption[SZ thermal and SZ kinetic power spectra]{
The angular power spectra of SZ thermal and kinetic
effects. As shown in (a), the thermal
SZ effect is dominated by individual halos, and thus, by the single
halo term, while the kinetic effect is dominated by the large scale
structure correlations depicted by the 2-halo term. In (b), we show
the mass dependence of the SZ thermal and kinetic effects with a
maximum mass of $10^{16}$ and $10^{13}$ M$_{\sun}$. The SZ thermal
effect is strongly dependent on the maximum mass, while due to large
scale correlations, kinetic effect is not.}
\label{fig:szpower}
\end{figure}

\section{SZ Power Spectrum, Bispectrum and Trispectrum}
\label{sec:szpower}

The Sunyaev-Zel'dovich (SZ; \cite{SunZel80} 1980) effect arises
from the  inverse-Compton scattering of CMB photons by hot electrons
along the line of sight.
The temperature decrement along the line of sight  due to SZ effect
can be written as the integral of pressure along the same line of
sight
\begin{equation}
y\equiv\frac{\Delta T}{T_{\rm CMB}} = g(x) \int  d\rad  a(\rad)
\frac{k_B
\sigma_T}{m_e c^2} n_e(\rad) T_e(\rad) \,
\end{equation}
where $\sigma_T$ is the Thomson cross-section, $n_e$ is the electron
number density, $\rad$ is the comoving distance, and $g(x)=x{\rm
coth}(x/2) -4$ with $x=h \nu/k_B
T_{\rm CMB}$ is the spectral shape of SZ effect. At Rayleigh-Jeans
(RJ) part of the CMB, $g(x)=-2$.
For the rest of this paper, we assume observations in the
Rayleigh-Jeans
regime of the spectrum; an experiment such as Planck with sensitivity
beyond the peak of the spectrum can separate out these contributions
based on the spectral signature, $g(x)$ (\cite{Cooetal00a} 2000a).

The SZ power spectrum, bispectrum and trispectrum are
defined in the flat sky approximation
in the usual way
\begin{eqnarray}
\left< y(\bfl_1)y(\bfl_2)\right> &=&
        (2\pi)^2 \delta_\dirac(\bfl_{12}) C_l^\sz\,,\nonumber\\
\left< y(\bfl_1) y(\bfl_2) 
       y(\bfl_3)\right>_c &=& (2\pi)^2 \delta_\dirac(\bfl_{123})
        B^\sz(\bfl_1,\bfl_2,\bfl_3)\,, \nonumber \\
\left< y(\bfl_1) \ldots
       y(\bfl_4)\right>_c &=& (2\pi)^2 \delta_\dirac(\bfl_{1234})
        T^\sz(\bfl_1,\bfl_2,\bfl_3,\bfl_4)\,.
\end{eqnarray}
These can be written as a redshift projection of the
pressure power spectrum, bispectrum and trispectrum, respectively:
\begin{eqnarray}
C_l^\sz &=& \int d\rad \frac{W^\sz(\rad)^2}{d_A^2}
P_{\Pi}^\tot\left(\frac{l}{d_A},\rad\right) \, , \\
B^\sz  &=& \int d\rad \frac{W^\sz(\rad)^3}{d_A^4} B_\Pi\left(
\frac{\bfl_1}{d_A},
\frac{\bfl_2}{d_A},
\frac{\bfl_3}{d_A},
;\rad\right) \, , \nonumber \\
T^\sz  &=& \int d\rad \frac{W^\sz(\rad)^4}{d_A^6} T_\Pi\left(
\frac{\bfl_1}{d_A},
\frac{\bfl_2}{d_A},
\frac{\bfl_3}{d_A},
\frac{\bfl_4}{d_A},
;\rad\right) \, .
\label{eqn:szpower}
\end{eqnarray}
Here, $d_A$ is the angular diameter distance. At RJ part of the
frequency spectrum,  the SZ weight function is
\begin{equation}
W^\sz(\rad) = -2 \frac{k_B \sigma_T \bar{n}_e}{a(\rad)^2 m_e c^2}
\end{equation}
where $\bar{n}_e$ is the mean electron density today. In deriving
Eq.~\ref{eqn:szpower},
we have used the Limber approximation \cite{Lim54} by setting
$k = l/d_A$ and flat-sky approximation.

\subsection{Discussion}

\begin{figure}[!h]
\begin{center}
\includegraphics[width=4.2in,angle=-90]{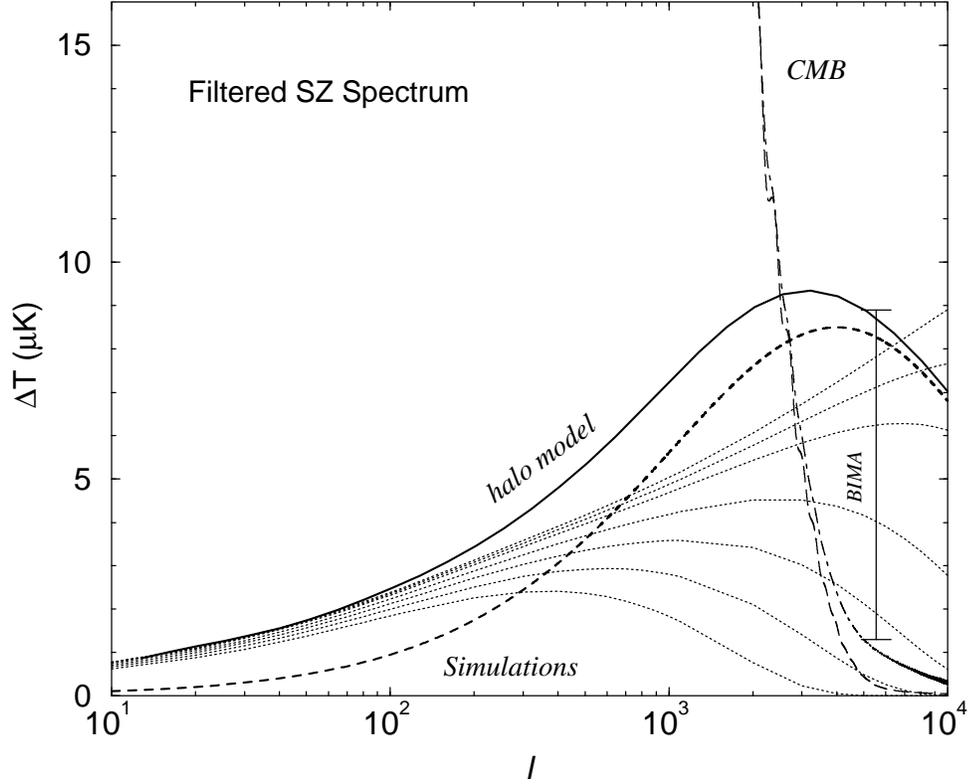}
\end{center}
\caption[Filtered approach to the SZ effect]{Flat band power in the
fiducial $\Lambda$CDM model with the pressure  power spectrum related
to a filtered version of the non-linear dark matter power spectrum
(dotted lines). From top to bottom $k_F=\infty,10,5,2,1,0.5,0.2$.
With a dashed  line, we have also shown the SZ power spectrum 
derived in typical numerical
simulations (e.g., \cite{Seletal00} 2000). The solid line shows the
prediction based on the halo approach. We also plot the 68\%
confidence on a preliminary detection of anisotropies at small angular
scales using the BIMA interferometer by  \cite{Dawetal00} (2000).}
\label{fig:szfilpower}
\end{figure}

In Fig.~\ref{fig:szpower}(a), we show the  SZ power spectrum due to
baryons present in virialized halos.
As shown, most of the contributions to SZ power
spectrum comes from individual massive halos, while the halo-halo
correlations only contribute at a level of 10\% at large angular
scales. This is contrary to, say, the lensing convergence power
spectrum discussed in \cite{Cooetal00b} (2000b), where most of the
power at large angular scales is due to the halo-halo correlations. The
difference can be understood by noting that the SZ effect is strongly
sensitive to the most massive halos due to $T \propto M^{2/3}$
dependence in temperature and to a lesser, but somewhat related,
extent that its weight function increases towards
low redshifts. The lensing weight function selectively probes
the large scale dark matter density power spectrum at comoving
distances half to that of background sources ($z \sim 0.2$ to 0.5 when
sources are at a redshift of 1), but has no extra dependence on mass.
The fact that the SZ power spectrum results mainly from the single
halo term also results in a sharp reduction of power when the maximum
mass used in the calculation is varied. For example, as discussed in
\cite{Coo00} (2000) and illustrated in Fig~\ref{fig:szpower}(b), with
the maximum mass decreased from $10^{16}$ to $10^{13}$ M$_{\sun}$, the
SZ power spectrum reduced by a factor nearly two orders of magnitude
in large scales and an order of magnitude at $l \sim 10^{4}$.

\begin{figure}[!h]
\begin{center}
\includegraphics[width=4.2in]{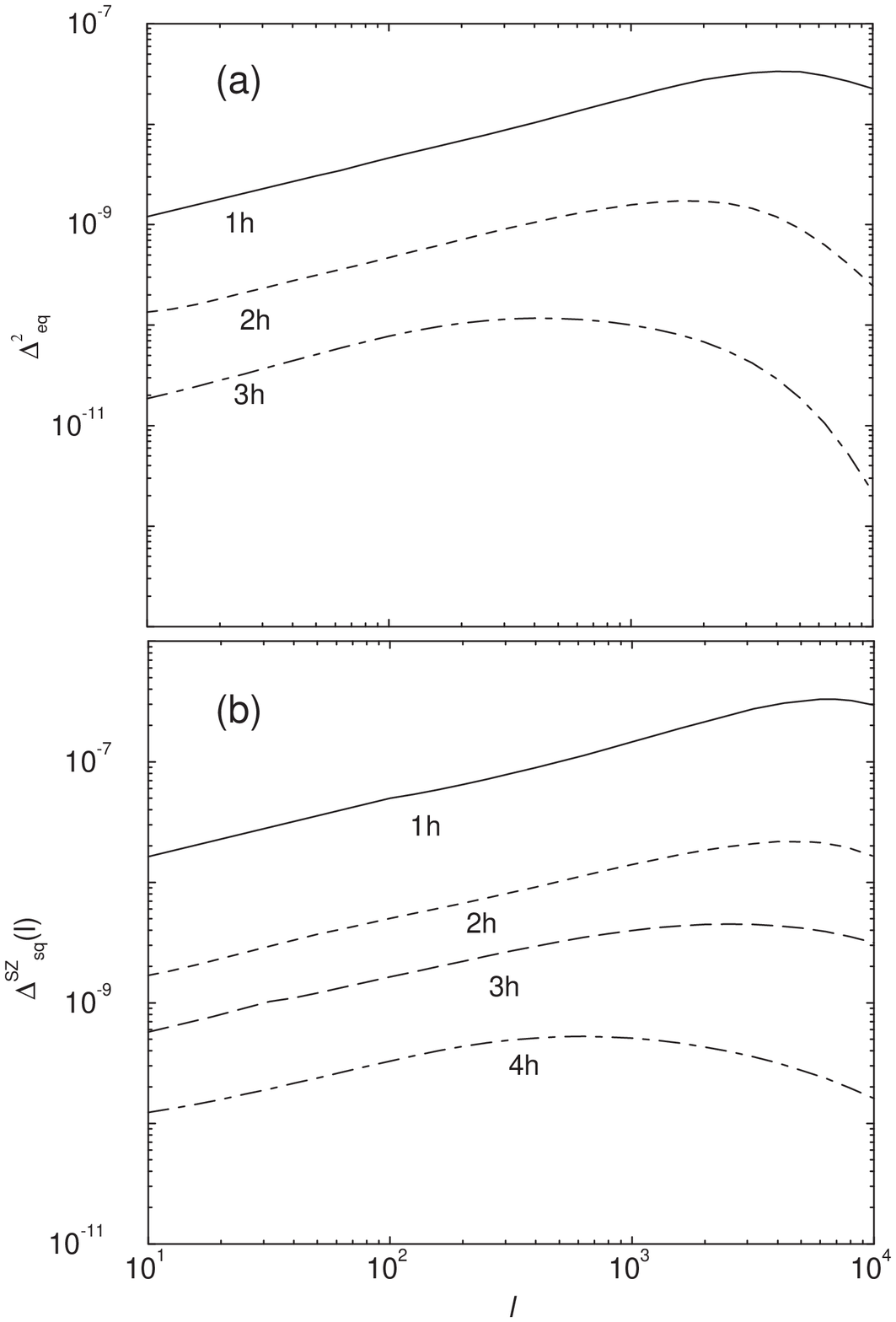}
\end{center}
\caption[SZ bispectrum and the trispectrum]{SZ bispectrum (a) and trispectrum (b) for configurations
that involve equilateral triangles and squares, respectively, under
the halo model. As shown, single halo term dominates the contribution
to bispectrum and trispectrum at all multipoles ranging from large
angular scales to small angular scales. The dependence on the single
halo term is consistent with the general non-Gaussian behavior of the
SZ effect and its significant non-linearity.} 
\label{fig:sztri}
\end{figure}

In addition to the halo model, we can also consider another
semianalytic approach to calculate the SZ power spectrum using a
filtered version of the non-linear dark matter power spectrum.
It is well known that the non-linear effects generally enhances power
at small scales, though, due to pressure cut-off  clustering of
pressure is not expected to occur down to smallest scales. In fact, Pen (1999), based on his
simulations and numerical calculations, argue that large scale
structure gas does not cluster at scales less than $\sim$ 850 kpc.
Similar cutoffs due to pressure has also  been studied in Gnedin \&
Hui (1998) using various filtering mechanisms. To get a qualitative
understanding of any filtering effects of the gas power spectrum on
the SZ effect, we consider a simple filtered version of the gas power
spectrum given by
\begin{equation}
P^{\rm fil}_\Pi(k) = \exp(-k^2/k_F^2)b^2_\Pi P_\delta(k)
\end{equation}
where $k_F$ is the filtering scale and $b_\Pi$ is the large scale bias
of pressure related to the non-linear dark matter power
spectrum. Here, we use the constant bias value at large scales.
The resulting temperature fluctuations for the SZ power spectrum for various numerical
values of $k_F$ are shown in Fig.~\ref{fig:szfilpower}. For
comparison, we also show prediction based on the halo model and
predictions generally from numerical simulations; Note that 
simulations underestimate the SZ effect at large scales due to
numerical issues related to volume of the simulation and resolution.
As shown, a constant bias with a filtering scheme for the non-linear
power spectrum does not fully produce the halo model. Of course, if
one uses the scale dependent bias (as shown in
Fig.~\ref{fig:pressurepower}), the non-linear dark matter power
spectrum produces the SZ effect with no filtering. In
Fig.~\ref{fig:szfilpower}, we also show a recent preliminary detection
of the temperature fluctuations at small angular scales by
\cite{Dawetal00} (2000). With improvements in the experimental side,
it is likely that future measurements of the SZ power spectrum will
greatly constrain the underlying physics that generate the temperature
fluctuations. We will discuss such a possibility with the generation
of the full SZ covariance, including non-Gaussianities.

\begin{figure}[!h]
\begin{center}
\includegraphics[width=3.5in,angle=-90]{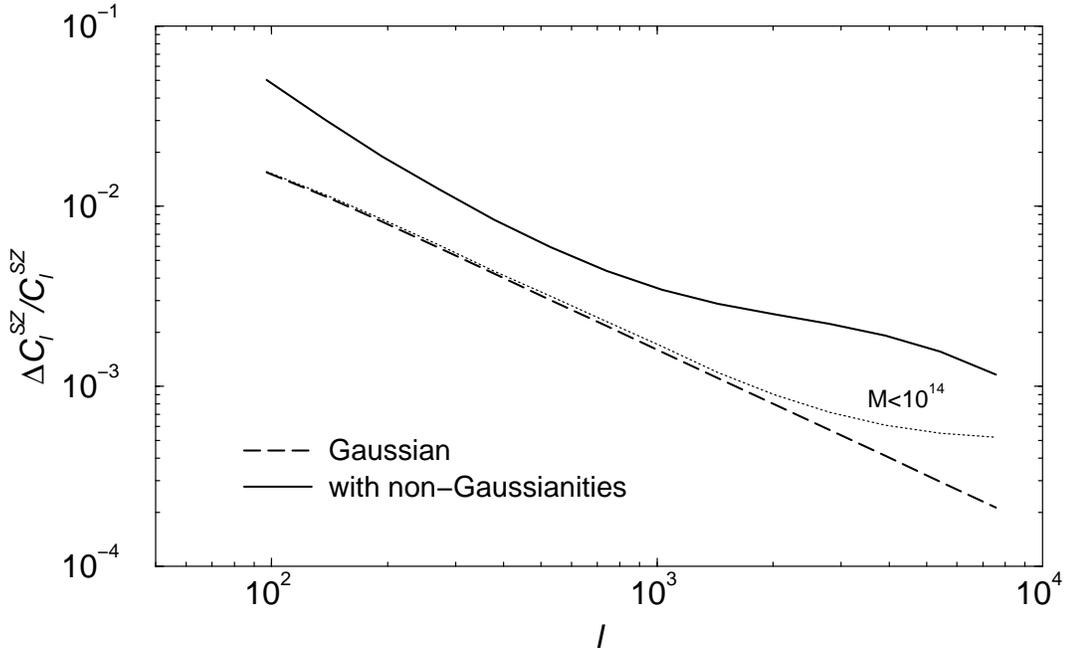}
\end{center}
\caption[Fractional error on the SZ power spectrum]{The fractional errors in the measurements of the SZ
band powers. Here, we show the fractional errors under the Gaussian
approximation, and the total including non-Gaussianities. As shown,
the total
contribution as a function of mass is sensitive to the presence of
most massive halos in the universe. The non-Gaussian term is
essentially dominated by the single halo term.}
\label{fig:trivariance}
\end{figure}

\begin{figure}[!h]
\begin{center}
\includegraphics[width=4.2in,angle=-90]{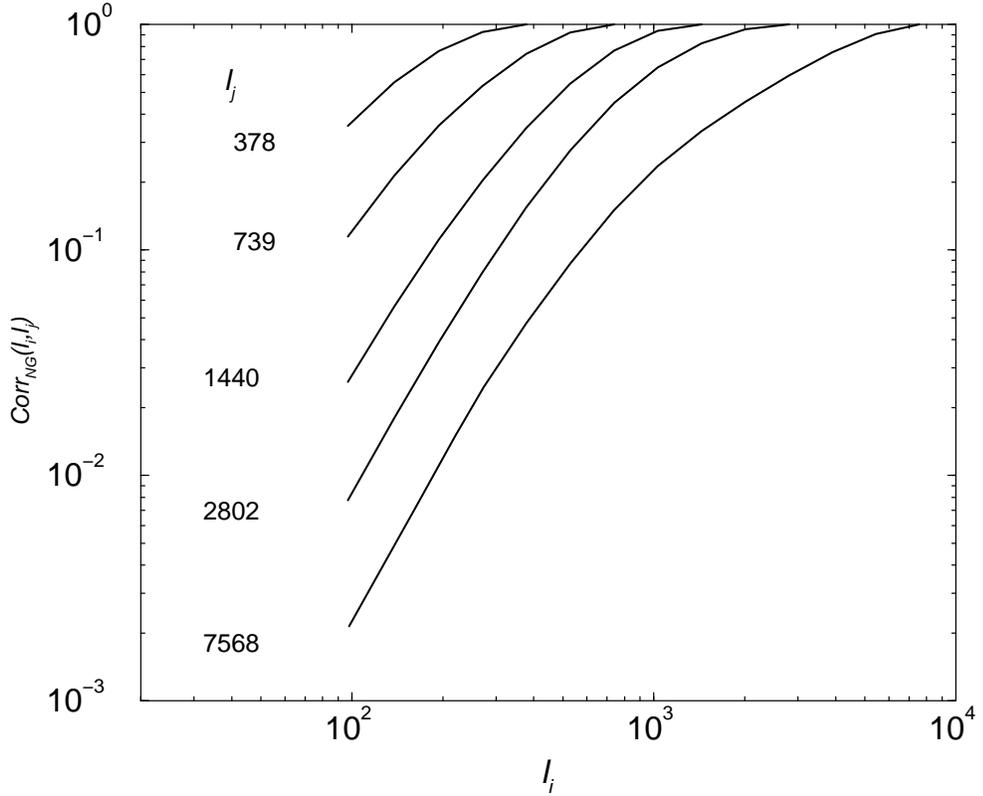}
\end{center}
\caption[SZ correlation coefficients from the trispectrum]{The 
non-Gaussian correlation coefficient $\hat{C}_{ij}^\ngau$,
of the SZ power spectrum, involving only the configuration of the SZ
trispectrum that contribute to the SZ power spectrum covariance (see,
Eq.~\ref{eqn:szng}). The correlations are such that they tend to 1 as
$l_i \rightarrow l_j$ and is fully described by the contribution to
the trispectrum by the single halo term.}
\label{fig:szcorr}
\end{figure}

\section{SZ Power Spectrum Covariance}
\label{sec:szcovariance}

For the purpose of this calculation, we assume that upcoming weak
lensing convergence power spectrum will measure binned logarithmic
band
powers at several $l_i$'s in multipole space with bins of
thickness $\delta l_i$.
\begin{equation}
\bp_i =
\int_{\shell i}
{d^2 l \over{A_{\shell i}}}
\frac{l^2}{2\pi} y(\bf l) y(-\bf l) \, ,
\end{equation}
where $A_\shell(l_i) = \int d^2 l$ is the area of the two-dimensional shell in
multipole and can be written as $A_\shell(l_i) = 2 \pi l_i \delta l_i
+ \pi (\delta l_i)^2$.

We can now write the signal covariance matrix
as
\begin{eqnarray}
C_{ij} &=& {1 \over A} \left[ {(2\pi)^2 \over A_{\shell i}} 2 \bp_i^2
+ T^\sz_{ij}\right]\,,\\
T^\sz_{ij}&=&
\int {d^2 l_i \over A_{\shell i}}
\int {d^2 l_j \over A_{\shell j}} {l_i^2 l_j^2 \over (2\pi)^2}
T^\sz(\bfl_i,-\bfl_i,\bfl_j,-\bfl_j)\,,
\label{eqn:szvariance}
\end{eqnarray}
where
$A$ is the area of the survey in steradians.  Again the first
term is the Gaussian contribution to the sample variance and the
second term is the non-Gaussian contribution.
A realistic survey will also have an additional noise variance due to
the instrumental effects and a covariance resulting from the
uncertainties associated with the separation of the SZ effect from
thermal
CMB and other foregrounds.

\begin{table}[!h]
\begin{flushleft}
\begin{tabular}{cccccccccc}
\hline
$\ell_{\rm bin}$ &
       529  & 739 & 1031& 1440 & 2012
& 2802 & 3905 & 5432 & 7568\\
\hline
   529&1.00 & 0.88 & 0.69 &
0.48  &0.30 & 0.17 & 0.17 & 0.11 & 0.07\\
   739 &(0.00)& 1.00 & 0.88 &
0.68  & 0.49 & 0.38 & 0.28 & 0.20& 0.13\\
  1031 &(0.00)& (0.00)& 1.00
& 0.87&0.68  & 0.56 & 0.43 & 0.31 & 0.21\\
  1440 &(0.00)& (0.00)&
(0.00)& 1.00 &0.88 &0.69 & 0.61 & 0.45 & 0.31\\
  2012 &(0.00)& (0.00)&
(0.01)&(0.05)&1.00& 0.88 & 0.69 & 0.60 & 0.42\\
  2802 &(0.00)&
(0.00)&(0.02)&(0.09)&(0.39)&1.00  & 0.88  & 0.70 & 0.56\\
  3905 &(0.00)&
(0.00)&(0.02)&(0.08)&(0.36)&(0.84)& 1.00  & 0.87 &0.70\\
  5432 &(0.00)&
(0.00)&(0.01)&(0.06)&(0.29)&(0.65)&(0.86)& 1.00 & 0.88\\
  7568 &(0.00)&
(0.00)&(0.01)&(0.04)&(0.20)&(0.53)    &(0.70)  & (0.88)    & 1.00  \\
\hline
\end{tabular}
\caption[SZ Covariance matrix]{
Covariance of the binned power spectrum for  the SZ effect.
Upper triangle displays the covariance found when a perfect frequency
cleaned SZ map is used to determine the SZ power spectrum.
Lower triangle (parenthetical numbers) displays the covariance found
when the variance is dominated by the primary anisotropy contribution,
as in a measurement of the SZ power spectrum in a CMB primary
fluctuations  dominated map.} 
\label{tab:szcov}
\end{flushleft}
\end{table}

In Fig~\ref{fig:sztri}(b), we show the scaled trispectrum
\begin{equation}
\Delta^\sz_{\rm sq}(l) = \frac{l^2}{2\pi}
T^\sz(\vecl,-\vecl,\vecl_\perp,-\vecl_\perp)^{1/3} \, .
\end{equation}
where $l_\perp=l$ and $\vecl \cdot \vecl_\perp=0$.
The projected SZ trispectrum again shows the same behavior as the
pressure  trispectrum with similar conditions on $\veck_i$'s. As
shown, the contributions to the trispectrum essentially comes from the
single halo term at all multipoles. This is consistent with our
observation that SZ power spectrum is essentially dominated by the
correlations of pressure within halos. As discussed in \cite{Coo00}
(2000), and shown in Fig.~\ref{fig:sztri}(a), the SZ bispectrum, shown
here for the equilateral triangular configuration such that
$l_1=l_2=l_3=l$, is also dominated by the single halo term. Given this
dependence on the single halo term, for the
rest of the discussion involving SZ covariance, 
we will only use the single halo
term and ignore the contributions arising from large scale
correlations associated with halos.

We can now use this trispectrum to study the
contributions to the covariance, which is what we are primarily
concerned here. In Fig.~\ref{fig:trivariance}, we show the
fractional error,
\begin{equation}
{\Delta \bp_i  \over \bp_i} \equiv {\sqrt{C_{ii}}  \over \bp_i} \, ,
\end{equation}
for bands $l_i$ given in Table~\ref{tab:szcov} following the
binning scheme used in \cite{CooHu01b} (2001b) for the weak lensing
power spectrum.

In Fig.~\ref{fig:trivariance}, the dashed line shows the 
Gaussian error while the solid line shows the total covariance with
the addition of the SZ trispectrum (Eq.~\ref{eqn:szvariance}). At all
multipoles, the non-Gaussianities from the trispectrum dominates the
variance. As we discussed for the power spectrum, however, a reduction
in the maximum mass of the halos used for the SZ calculation leads to
a sharp decreases in the non-Gaussianities. With a mass cut at
10$^{14}$ $M_{\sun}$, shown by the dotted line, we see that the total
variance is consistent with the Gaussian variance out to $l \sim
1000$.

We can now write the correlation between the bands
as
\begin{equation}
\hat C_{ij} \equiv \frac{C_{ij}}{\sqrt{C_{ii} C_{jj}}} \, .
\end{equation}
In Table \ref{tab:szcov} we tabulate the SZ correlations under the
assumption that the SZ power spectrum is measured independently, say
in a frequency cleaned map, (upper triangle) and is measured in the
CMB primary dominated map (lower triangle).
The correlations along individual columns increase (as one goes to
large $l$'s or small angular scales) and the maximum values are
reached at $l \sim 5000$ consistent with the general
behavior of the trispectrum.

In Fig.~\ref{fig:szcorr}, we show the non-Gaussian trispectrum
correlation coefficient given by
\begin{equation}
\hat C^{\rm NG}_{ij} =
\frac{T_{ij}}{\sqrt{T_{ii} T_{ij}}} \,.
\label{eqn:szng}
\end{equation}
As shown here,
the coefficient increases to higher $l$ to a maximum value of unity.
The gradual increase is consistent with the fact that at all scales
its the single halo term that dominates the non-Gaussian contribution.
Since the power spectrum is dominated by correlations in single halos,
the fixed profile of the halos correlate the power in all the
modes and the correlations between adjacent modes are significant.

The calculation, or experimental measurement,
 of the full SZ covariance is necessary for the interpretation of
observational results on the power spectrum.
The upcoming SZ surveys, where the power spectrum will be measured, is
likely to be  limited to a small area on the sky.
Thus, in the absence of many fields where the covariance can be
estimated directly from the data, the halo model based approach
suggested here provides a useful, albeit model dependent,
quantification of the
covariance.  As suggested for weak lensing observations
in \cite{HuWhi00} (2000) and discussed in \cite{CooHu01b} (2001b),
as a practical approach one could imagine
taking the variances estimated from the survey under
a Gaussian approximation,  after accounting for uneven
sampling and edge effects, and scaling it up by the non-Gaussian
to Gaussian variance ratio of the halo model along with
inclusion of the band power correlations. Additionally, using the
covariance
as the one calculated here, one can use the approach
well known in the fields of CMB and galaxy power spectrum measurements
to decorrelate band powers (e.g., \cite{Ham97} 1997; \cite{HamTeg00}
2000). 

\section{Astrophysical Uses of the SZ Power Spectrum}
\label{sec:szparameters}

The calculation of the full covariance matrix now allows us to study
how well the SZ power spectrum measures certain astrophysical and
cosmological parameters. The upcoming CMB power spectrum
measurements, complemented by the related local universe observations
such as the galaxy power spectrum or supernovae, are expected to
constrain most of the cosmological parameters to a
reasonable accuracy (e.g., \cite{Eisetal00} 2000). Thus,
we ignore the possibility that the SZ effect can be used as
a probe of cosmology and only concentrate on the astrophysical uses of
the SZ effect. This is a reasonable approach to take since
there are many unknown astrophysics associated with the SZ effect
involving the clustering of gas and its temperature. Such an approach
allows us not to complicate the parameter measurements by adding both
astrophysical
and cosmological parameters. Assuming the cosmology will be safely
known, we now ask the question what additional astrophysical 
parameters one can hope to extract from the SZ effect under the
present halo model. 

There are many approaches to parameterize the unknown astrophysics of
the SZ power spectrum. Some possibilities have already been suggested
in the literature, essentially involving the gas evolution (e.g.,
\cite{HolCar99}
1999; \cite{Maj01} 2001). Since the SZ effect involves both gas
and temperature as a product, ie. the pressure, one may be led to
conclude that it is not possible to separate effects associated with
temperature from those associated with gas. Given the dependence of
temperature on the pressure profile, independent of gas, however, 
it is expected that this degeneracy between gas properties and 
temperature effects  can be partly broken.

\begin{figure}[!h]
\begin{center}
\includegraphics[width=4.2in,angle=-90]{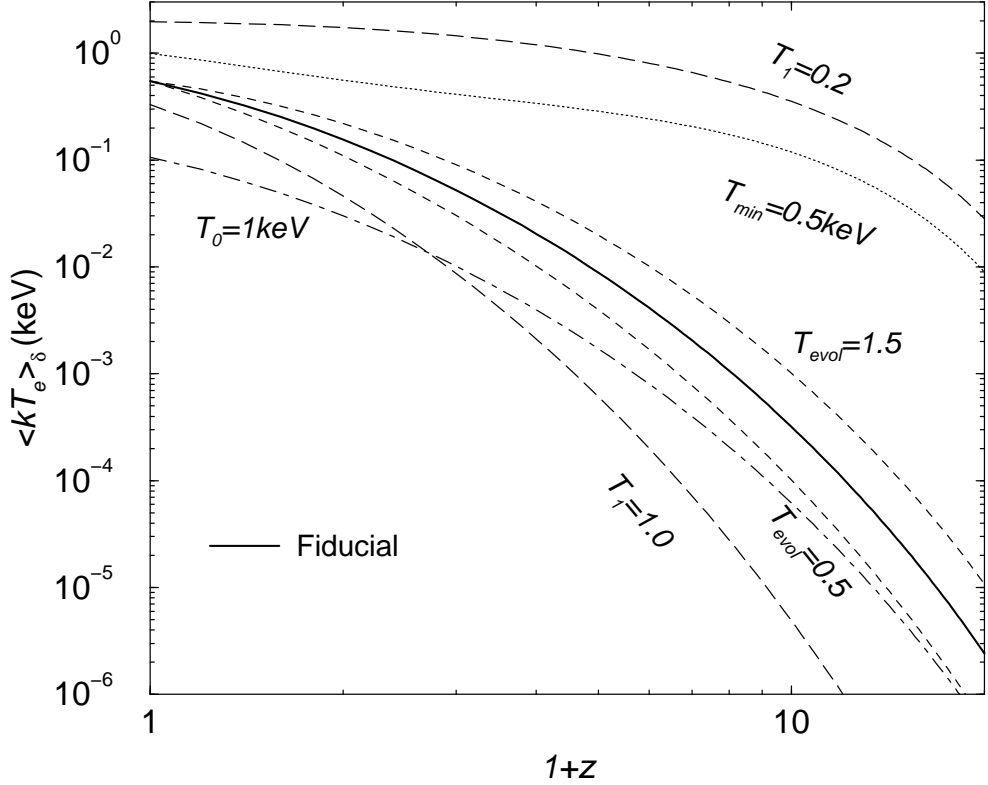}
\end{center}
\caption[Redshift evolution of density weighted temperature]{The variation in the density weighted temperature  of
electron as a function of redshift. The solid line shows the
redshift evolution of the temperature under the fiducial model while
variations about this model are shown as labeled.}
\label{fig:etemp}
\end{figure}

As discussed earlier, the clustering of pressure power spectrum has a
turnover
corresponding to an equivalent scale radius of pressure.
 Through the gas pressure profile, this turn over 
can be characterized by the parameter $b$ and the dark matter scale
radius $r_s$. Note that $b \propto 1/T_e$, so its measurement
is essentially a probe of the electron temperature, though, it is
unlikely that one can obtain all information on temperature and its
evolution from one parameter measurement. Thus, instead of $b$, we
take temperature itself to be one interesting astrophysical parameter
and consider its evolution such that
\begin{equation}
T(M,z) = T_0 \left(\frac{M}{10^{15} h^{-1} M_{\sun}}\right)^{T_1}
(1+z)^{T_{\rm evol}} + T_{\rm min} \, .
\end{equation}
Here, the four parameters represent the temperature-mass
normalization, $T_0$, which in the fiducial case has the value given
by the
virial equation, the mass dependence slope, $T_1$, with a fiducial
value of $2/3$, an
redshift dependent evolutionary parameter, $T_{\rm evol}$, with a 
fiducial value of 1, and a minimum temperature for gas independent of
mass and redshift $T_{\rm min}$, with a value of zero in the fiducial
case.  This latter parameter accounts for any possible preheating of gas
before virializing in halos due to effects
associated with some unknown astrophysics, such as the reionization
process. A measurement of $T_{\rm min}$ would be interesting given
that observational data from clusters to cluster groups suggest
possible preheating of gas before virialization in halos. 
We note here
that a redshift independent value  for $T_{\rm min}$ may be too extreme
since one expects preheating temperature to vary with redshift and
mass such that all three parameters, $T_0$, $T_1$ and $T_{\rm evol}$,
are affected. Still, we interested in the possibility of knowing how
well we can establish a mass independent temperature value such as
$T_{\rm min}$ through the SZ angular power spectrum. 
In Fig.~\ref{fig:etemp}, we show the variation in the redshift evolution
of the density weighted temperature of electrons about the fiducial
model. The density weight temperature  was calculated following
Eq.~\ref{eqn:etemp}.

In addition to the temperature, the
SZ effect also depends on the number density of electrons in clusters.
So far, we have considered this number through the universal baryon
fraction in the universe such that $f_g \equiv
M_g/M_\delta=\Omega_g/\Omega_m$.
This assumption ignores any possible effects associated with the
evolution of the gas fraction in halos, independent of any evolution
that may be associated with temperature. It'll be interesting to study
to what
extent future observations will allow the measurement of the fraction
of baryons that is responsible for the SZ effect, and any evolution
that may be associated with this fraction. Thus, a second set of
parameters one  can hope to extract from SZ observations involves 
gas mass fraction of halos and its evolution.

To study such gas properties, we parameterize the gas mass fraction
such that
\begin{equation}
f_g = f_0 \left(\frac{M}{10^{15} h^{-1} M_{\sun}}\right)^{f_1}
(1+z)^{f_{\rm evol}} \, .
\end{equation}
In a recent paper, \cite{Maj01} (2001) has suggested the strong
possibility a measurement of any mass and redshift dependence of gas
mass fraction in clusters,
given that the SZ power  spectrum was observed to vary 
significantly with changes in these two parameters.
Since the SZ power spectrum essentially is sensitive to 
$\sim f_g^2 T_e^2$, however,
such a suggestion for measurement of gas evolution is not
independent of any variations associated with temperature, which was
ignored in the study of \cite{Maj01} (2001; also,
\cite{HolCar99} 1999). Our general
parameterization above involving both temperature and gas allows us to
quantify how well independent statements can be made on possible
measurement of gas density and temperature evolution, under the assumption
that cosmology is known. Note that 
gas evolution is not present in our fiducial model since we take
the gas fraction to be independent of mass and redshift 
with $f_1=0$ and $f_{\rm evol}=0$, respectively.

We now have a total of seven parameters we wish to extract from a
measurement of the SZ power spectrum. 
In order to perform this calculation  we take a  Fisher matrix based
approach. The Fisher matrix is simply a projection of the covariance
matrix onto the basis of astrophysical parameters $p_i$ (see, Eq.~\ref{eqn:fisher}).
Note that under the approximation of Gaussian shot
noise, the covariance reduces to replacing $C^\sz_l \rightarrow
C^\sz_l + C^{\rm Noise}_l$ in the expressions leading up
to the covariance Eq.~\ref{eqn:szvariance}. 
The noise power spectrum includes the noise associated with detectors,
beam size and variance resulting from the separation of the SZ effect
from other temperature fluctuations in multifrequency data.

Under the approximation that there are a sufficient number
of modes in the band powers that the distribution of power
spectrum estimates is approximately Gaussian, the Fisher matrix
quantifies the best possible errors on cosmological parameters that
can
be achieved by a given survey.  In particular $F^{-1}$ is
the optimal covariance matrix of the parameters and
$(F^{-1})_{ii}^{1/2}$
is the optimal error on the $i$th parameter.
Implicit in this approximation of the Fisher matrix is the neglect of
information from the parameter dependence of the covariance matrix
of the band powers themselves. We neglect this information due to
computational restrictions on the
calculation of covariance for all variations in parameters within
a reasonable amount of time. We do not expect this exclusion to change
our
results significantly. Also, here, we are mostly interested  in an
order
of magnitude estimate on how well the SZ power spectrum can constrain
astrophysics associated with large scale pressure.

The Fisher matrix approach allows us to
address how well degeneracies are broken in the parameter space and
under the assumption of a  fiducial model for the parameters. 
For the purpose of this calculation, we take binned measurements of
the SZ power spectrum following the binning scheme in
Table~2. We consider a perfect SZ experiment with  no noise
contribution to the covariance and observations out to 
$l \sim 10^4$. To consider a real world scenario, we also study the
astrophysical uses of the SZ power spectrum that can be extracted from
the Planck mission. Here, we use the SZ noise
power spectrum calculated for Planck with detector noise and
uncertainties in the separation of SZ from CMB and other foreground
in \cite{Cooetal00a} (2000a). This SZ noise power spectrum is
shown in Fig.~\ref{fig:szpower}(b).

%\begin{table}
%\begin{flushleft}
%\begin{tabular}{ccc} \hline
%field & RA$^{a}$ & Dec$^{a}$ \\
%\hline
%1 & 05:32:34.5 & $-$05:19:37.0 \\
%2 & 05:35:31.5 & $-$04:47:52.0 \\
%3 & 05:37:38.5 & $-$05:27:07.0 \\
%4 & 05:35:12.5 & $-$05:58:43.0 \\
%\hline
%\end{tabular}
%\caption[Title for ToC]{Title for ToC, followed by more words \\
%$^a$ Manually done footnote
%\label{needs_to_be_within_caption_to_work} }
%\end{flushleft}
%\end{table}

\begin{table}[!h]
\begin{center}
\begin{tabular}{lrrrrrrr}
\hline
$p_{i}$
        & $T_0$     & $T_1$      & $T_{\rm evol}$     & $T_{\rm min}$
& $f_0$ & $f_1$ & $f_{\rm evol}$\\
\hline
$T_0$ & 8.80 & 1.32  & 3.67 & -1.93 & -0.21 & -1.69 & -3.36\\
$T_1$&  & 0.51  & 1.08  & -0.18 & -0.04 & -0.41 & -0.07\\
$T_{\rm evol}$ & & & 2.69  & -0.62 & -0.11 & -0.94 & -2.11\\
$T_{\rm min}$ & & & & 0.48 & 0.04 & 0.29 & 0.67\\
$f_0$ & & & & & 0.006 & 0.05 & 0.09\\
$f_1$ & & & & & & 0.39 & 0.73\\
$f_{\rm evol}$ & & & & & & & 1.67\\
\hline
\end{tabular}
\caption[Inverse Fisher matrix for SZ parameter determination]{
Inverse Fisher matrix ($\times 10^2$) 
for the SZ effect with seven parameters and full
non-Gaussian errors. The error on an individual parameter is the
square root of the diagonal element of the Fisher matrix for the
parameter while off-diagonal entries of the inverse Fisher matrix
shows correlations, and, thus, degeneracies, between parameters. We
have assumed a perfect experiment with a 
full sky survey ($f_\sky=1$). The seven parameters are
described in \S~\ref{sec:parameters}.}
\label{tab:szfisher}
\end{center}
\end{table}

\begin{figure}[!h]
\begin{center}
\includegraphics[width=5.6in]{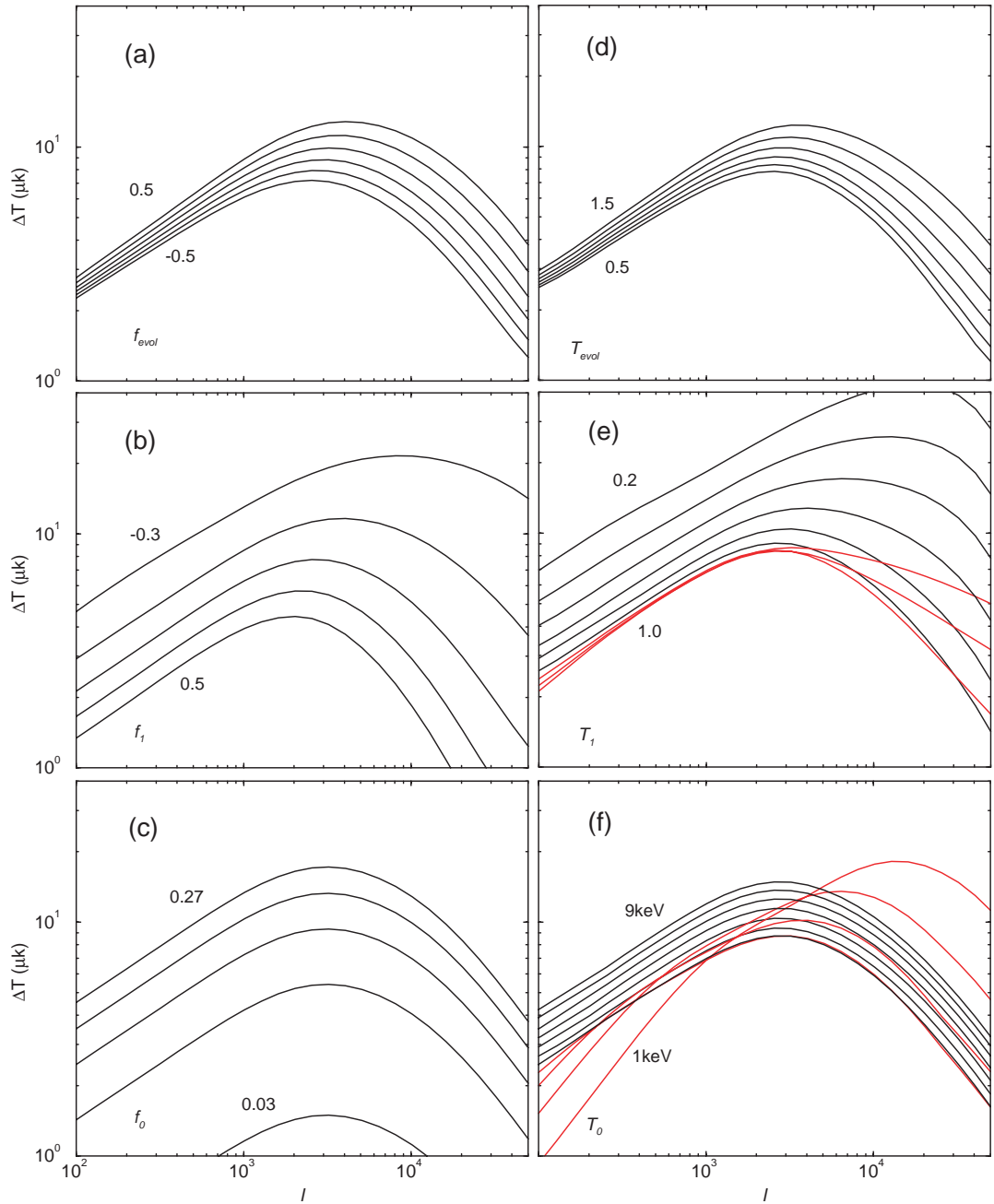}
\end{center}
\caption[Variations in the SZ band power]{The temperature fluctuations of the SZ effect through
variations in the astrophysical parameters under the halo model. From
(a) to (c), we show the variations associated with gas evolution while
from (d) to (f), we show variations involved with temperature. The
parameters are described in \S~\ref{sec:parameters}.}
\label{fig:szparams}
\end{figure}

\subsection{Discussion}

In Fig.~\ref{fig:szparams}, we show the variation associated with SZ
temperature fluctuations written such that
$\Delta T = \sqrt{l(l+1)/(2\pi) C_l} T_{\rm CMB}$  for six of the
seven parameters involved with gas, from (a) to (c), and temperature,
(d) to (f), evolution. These plots allow us to understand some of the
degeneracies associated with the description. For example, as shown,
the gas and temperature redshift evolution essentially predicts
similar behavior for the SZ temperature fluctuation, though there are
minor differences due to the temperature dependence on the pressure
profiles of halos. For the most part, variations due to temperature
evolution is due to the normalization and not due to variations in the
profile shape. In (b) and (e), we show variations due to the mass
slope of the gas evolution and temperature evolution,
respectively. Here again, we see similar behavior. When the slope of
the mass-temperature relation, as a function of mass, is greater than
0.7, we see significant differences, especially involving an increase
in temperature fluctuations at small scales. This is due to the
relatively
increasing weighing of small mass halos.

 In (c) and (f), we show
variations associated with gas evolution normalization $f_0$ and
temperature-mass normalization $T_0$. The variation associated with
$f_0$ is easily understood since the effect is only a change in the
overall normalization of the power spectrum. The variation with
temperature-mass normalization shows both effects due to normalization
and the profile. When the normalization is low, gas clusters to small
radii in low mass halos leading to an increase in power at small
scales. As the temperature normalization is increased, gas profile
varies such that there is a reduction in small scale power and the
angular multipole of the turn-over scale shifts to low values. When
the temperature normalization is sufficiently high, the overall
weighing resulting from the overall temperature  multiplicative factor
becomes important. Now, the power spectrum behaves as a simple
normalization change, similar to the variation in power due to gas
evolution normalization.
As shown in Fig.~\ref{fig:szparams}(a) to (f), there are significant
degeneracies involved with astrophysical parameters that lead to the
SZ effect.

In Table~\ref{tab:szerrors}, we tabulate the errors on these seven
parameters using the inverse Fisher matrix for a possible SZ power
spectrum measurement. Here, we have considered the possibility that
parameter extraction will be limited to 3, 5 and 7 parameters.
The increase in number of parameters to be measured from a SZ power
spectrum increases degeneracies associated with the set of parameters
resulting in their accuracies. In the case of the 3 parameters
involving temperature-mass normalization, $T_0$, a minimum temperature
for
all halos $T_{\rm min}$, and the gas mass fraction $f_0$, in a perfect
experiment, all three parameters can be extracted such that they will
be provide essentially
very strong constraints. For example, the error on $f_0$  is such that
one
can identify the gas fraction of clusters responsible for SZ effect 
from the cosmic mean of $\Omega_g/\Omega_m=0.05/0.3$ with an error of
$4 \times 10^{-4}$. With Planck, one can constrain the preheating temperature at the level
of $\sim$ 0.75 keV, and since current predictions for possible
preheating is also at the level of few tenths keV, Planck SZ power
spectrum can either confirm or put a useful limit on preheating
temperature at current expectations.

As tabulated, however, the accuracy to which parameters can be
determined from SZ power spectrum reduces significantly when the
number of parameters to be determined is increased. For example,
Planck mission will only set a limit at $\sim$ 1.7 keV, if one were to
study both the  mass and redshift dependence of electron temperatures. Such
an upper limit is unlikely to be useful for current studies related to
preheating of gas. Given that we cannot obtain useful errors with
Planck for 5 parameters, we suggest that Planck may not be useful for
the purpose of studying the full parameter space suggested here. This
is understandable since Planck only allows the measurement of the SZ
power spectrum out to $l \sim$ 1500, while most of the variations due
to parameters under discussion here happens at $l \sim 5000$ where the
turnover in the SZ power spectrum is observed.  The Planck mission,
however, allows one to obtain reasonable errors on parameters which
generally define the normalization of the power spectrum, such as the
temperature-mass normalization or the normalization of gas mass
fraction.
The normalization for gas mass fraction from Planck will be useful for
the purpose of understanding what fraction of cosmic baryons reside in
massive halos and contribute to the SZ effect and to look for any
discrepancy of such a value from the total baryon content
predicted by big bang nucleosynthesis arguments.
In order to obtain reliable measurements of evolution of gas and
temperature, a small scale experiment sensitive to multipoles out to
$l \sim 10^4$ will be necessary.

For  a perfect experiment, we show the errors on
seven parameters also in Table~\ref{tab:szerrors}. The inverse Fisher
matrix in this case is tabulated in
Table~\ref{tab:szfisher}. The diagonals of the inverse Fisher matrix
show the variance of individual parameters, while, more importantly,
the off diagonals show the covariance between parameters. These
covariances allow one to understand the degeneracies between
parameters.
In Table~\ref{tab:szerrors}, we show the full extent to which parameters
degrade the accuracies by tabulating degradation factors associated with the
seven parameters. The degradation factor list the increase in
parameter error
from what can be achieved if all other parameters are known to what
can be achieved when all parameters are to be retrieved from data. 
The degradation factors are at the level of one hundred or more for some
parameters, suggesting that there are significant degeneracies
associated with the parameterization of the temperature and gas
fraction as a function of mass and redshift. Our result generally
suggest that significant estimations of gas evolutionary properties,
in the presence of
unknown temperature properties, is not possible. 

In addition to the parameter degeneracies, the non-Gaussianities
associated with the SZ effect also increase the errors on
parameters. For example, for the seven parameters under discussion
here and again for a perfect and full sky experiment, we list the
errors on parameters one can obtain if one were to ignore the
non-Gaussian contributions to the covariance. As tabulated,
non-Gaussianities increase the error on parameters by up to factors of
1.5, suggesting that the ignoring the non-Gaussianities will lead to a
significant underestimate of the errors in parameters. This should be
considered under the context that the SZ effect is significantly
non-Gaussian at all scales of interest and that ability to distinguish
parameters happen only at multipoles of a few thousand where the
non-Gaussianities in fact dominate.

\begin{table}[!h]
\begin{center}
\begin{tabular}{lrrrrrrr}
\hline
      & $T_0$     & $T_1$  & $T_{\rm evol}$  & $T_{\rm min}$ & $f_0$ &
$f_1$ & $f_{\rm evol}$ \\
\hline
Perfect  & 0.04  &        &                 & 0.002 & 0.0004 &       &
 \\
Planck & 0.79     &        &                 & 0.75 & 0.03  &       &
\\
\hline
Perfect  & 0.13  & 0.02   & 0.06 & 0.05 & 0.002 &  & \\
Planck & 1.39 & 0.41 & 1.22 & 1.37 & 0.05 &  & \\
\hline
Perfect & 0.30 & 0.07 & 0.17 & 0.07 & 0.008 & 0.06 & 0.13 \\
Degradation & 47 & 184 & 133 & 34 & 82 &  240 & 130\\
Gaussian & 0.18 & 0.04 & 0.10 & 0.04 & 0.005 & 0.04 & 0.08 \\
Increase (\%) & 64 & 70 & 77 & 81 & 73 & 60 & 68\\
\hline
\end{tabular}
\caption[SZ Parameter errors]{
Parameter errors, $(F^{-1})_{ii}^{1/2}$, using the halo model and the
full covariance for the SZ effect. We tabulate these errors for a
perfect experiment with no instrumental noise and full sky
observations out to $l \sim 10^4$. We also show the expected errors for
Planck mission with a useful sky fraction of 65\% ($f_\sky=0.65$),
and with the noise power spectrum shown in Fig.~\ref{fig:szpower}(b).
The parameters are described in \S~\ref{sec:parameters}.
We break the parameter estimation to consider recovery of 3, 5 and 7
parameters.
Under ``Degradation'' we tabulate the degradation factors,
$(F)_{ii}^{-1/2}/(F^{-1})_{ii}^{1/2}$, due to 
parameter degeneracies. We also list the parameter errors expected if
one were to assume Gaussian sample variance only for the SZ power
spectrum and were to ignore the non-Gaussian covariance. The increase
in error on individual parameters, with the introduction of the full
covariance matrix, ranges from 40\% to nearly 100\%.}
\label{tab:szerrors}
\end{center}
\end{table}

\section{Weak Lensing-SZ Correlation: Non-Gaussianities
in CMB}

Large-scale structure deflects CMB photons in transit from the
last scattering surface.  These structures also give rise
to secondary anisotropies.   The result is a correlation between
the temperature fluctuations and deflection angles.
This effect cannot be seen in the two point function since
gravitational lensing preserves surface brightness: deflections
only alter the temperature field on the sky in the presence of
intrinsic, primary, anisotropies in the unlensed distribution.
The lowest order contribution thus comes from the three-point
function or bispectrum.

In weak gravitational lensing, the deflection angle on the sky is
given by the angular gradient of the lensing
potential which is itself a projection of the gravitational
potential (see e.g. \cite{Kai92} 1992),
\begin{eqnarray}
\Theta(\bm)
&=&
- 2 \int_0^{\rad_0} d\rad
\frac{\da(\rad_0-\rad)}{\da(\rad)\da(\rad_0)}
                \Phi (\rad,\hat{{\bf m}}\rad ) \,.
\end{eqnarray}
This quantity is simply related to the more familiar
convergence
\begin{eqnarray}
\kappa(\bm) & = &{1 \over 2} \nabla^2 \Theta(\bm) \\
            & = &-\int_0^{\rad_0} d\rad
\frac{\da(\rad)\da(\rad_0-\rad)}{\da(\rad_0)}
\nabla_{\perp}^2 \Phi (\rad ,\hat{{\bf m}}\rad) \, , \nonumber\\
\nonumber
\end{eqnarray}
where note that the 2D Laplacian operating on $\Phi$ is
a spatial and not an angular Laplacian.
The two terms $\kappa$ and $\Theta$ contain superficial differences
in their radial and wavenumber weights which we shall see cancel
in the appropriate Limber approximation.  In particular,
their spherical harmonic moments are simply proportional
\begin{eqnarray}
\Theta_{l m} &=&
             -{2 \over l(l+1)} \kappa_{l m} =
                 \int d {\bn} \Ylmn{}^*(\bn) \Theta(\bn) \nonumber\\
             &=& i^l \int {d^3 {\bf k}\over 2\pi^2} \delta({\bf k})
                \Ylmn{}^* (\bk) I_\ell^\ang(k)
\label{eqn:GSSZequiv}
\end{eqnarray}
with
\begin{eqnarray}
I_\ell^\ang(k)& =&
                \int_0^{\rad_0} d\rad W^\ang(k,r)
                 j_l(k\rad)  \,,\nonumber\\
W^\ang(k,r)& =&
                -3 \Omega_m \left({H_0 \over k}\right)^2
                F(r) {\da(\rad_0 - \rad) \over
                \da(\rad)\da(\rad_0)}\,.
\label{eqn:lensint}
\end{eqnarray}
Here, we have used the Rayleigh expansion of a plane wave
(Eq.~\ref{eqn:Rayleigh}),
and the fact that $\nabla^2 \Ylmn = -l(l+1) \Ylmn$.  In an open
universe, one simply replaces the spherical Bessel functions with
ultraspherical Bessel functions in expressions such as
Eq. (\ref{eqn:lensint}).  

As pointed out by \cite{GolSpe99} (1999), it does however have
an effect on the bispectrum which is in principle observable.
The lensed temperature fluctuation in a given direction is the sum of
the primary fluctuation in a different direction plus the secondary
anisotropy
\begin{eqnarray}
T(\bn) &=& T^{\rm P}(\bn + \nabla \Theta) + T^{\rm S}(\bn)  \\
       &\approx&
        \sum_{lm} \Big[ (\almn^{\rm P}+\almn^{\rm S}
                  )\Ylmn(\hat{\bf n})
                  +\almn^{\rm P}  
           \nabla\Theta(\hat{\bf n})\cdot\nabla \Ylmn(\hat{\bf n}) \Big]
\, ,
\nonumber
\end{eqnarray}
or
\begin{eqnarray}
\almn &=& \almn^{\rm P} + \almn^{\rm S}
        + \sum_{l'm'}
        a_{l' m'}^{\rm P}
        \int d \bn \Ylmn{}^* (\bn)
        \nabla\Theta(\hat{\bf n})\cdot
        \nabla Y_{l'}^{m'}(\hat{\bf n})
        \,.
\end{eqnarray}
Utilizing the definition of the bispectrum
in Eq.~\ref{eqn:bispectrum}, we obtain
\begin{eqnarray}
\bi &=& \sum_{m_1 m_2 m_3} \wjm
\nonumber \\
&&\times \int d\hat{\bf m} d\hat{\bf n}
\Ylm{2}{}^*(\bm)
\Ylm{3}{}^*(\bn) C_{l_1}
\nabla \Ylm{1}{}^*(\bm)
\cdot
\langle \nabla\Theta(\bm)
T^{\rm S}(\hat{\bf n}) \rangle  + {\rm Perm.}
\end{eqnarray}
where the five permutations are with respect to the ordering of
$(l_1,l_2,l_3)$.

Integrating by parts and simplifying further following
\cite{GolSpe99} (1999) leads to a
bispectrum of the form:
\begin{eqnarray}
&& \bi = -\wj
\sqrt{ \frac{(2l_1 +1)(2 l_2+1)(2 l_3+1)}{4 \pi}}
\nonumber \\
&\times&
\left[\frac{l_2(l_2+1)-l_1(l_1+1)-l_3(l_3+1)}{2} C_{l_1}b^{\rm
S}_{l_3}+ {\rm Perm.}\right]\,, \nonumber \\
\end{eqnarray}
where we have employed Eq.~\ref{eqn:harmonicsproduct} to perform the
angular integration.

\begin{figure}[!h]
\begin{center}
\includegraphics[width=4.2in,angle=-90]{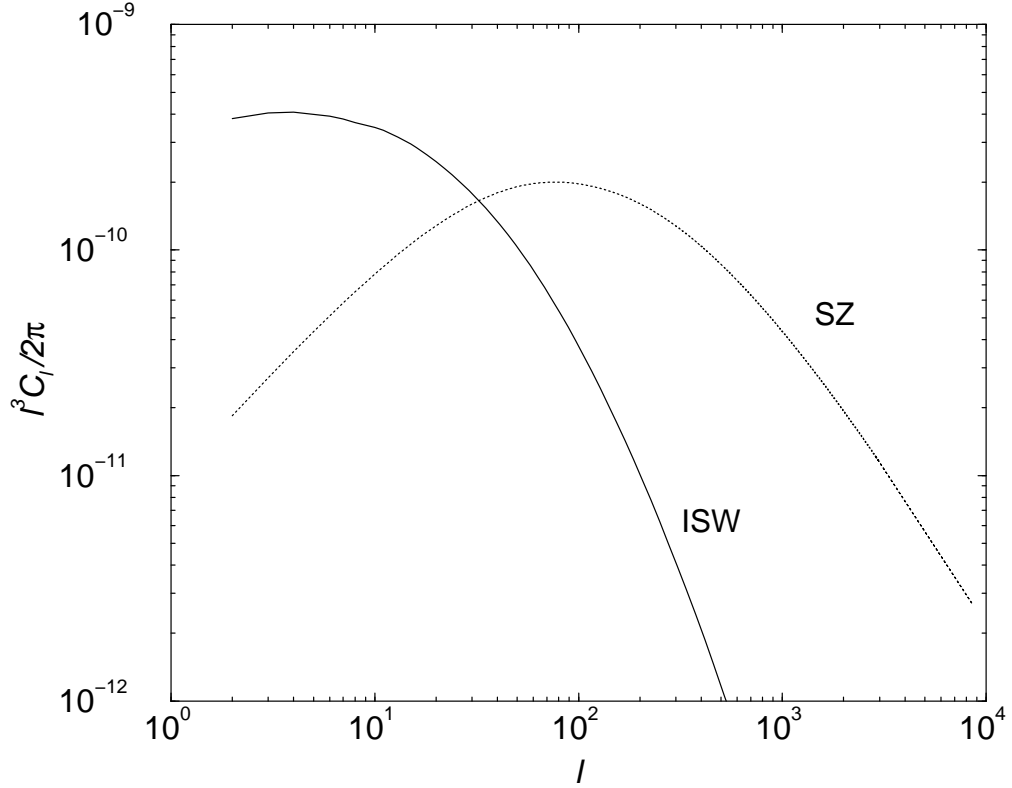}
\end{center}
\caption[Weak Gravitational Lensing-Secondary Correlation]{
The power spectra of correlation between lensing deflections in CMB
and the integrated Sachs-Wolfe effect (\cite{SacWol67} 1967) and the
SZ thermal effect. The correlations are such that potentials
responsible for the ISW effect correlate with lensing at large
angular scales while SZ effect correlates with lensing at medium
angular scales.}
\label{fig:blterms}
\end{figure}

The quantity of interest  here is the correlation between the
deflection potential and the SZ effect, or any other secondary effect,
\begin{equation}
T^\sz(\bn) = \sum \almn^\sz \Ylmn(\bn),
\end{equation}
which becomes
\begin{eqnarray}
\langle \Theta(\hat{\bf n})T^\sz(\hat{\bf m})\rangle&=&
\sum_{l m}
        \left< \Theta_{l m}^* \almn^\sz \right>
           \Ylmn{}^*(\bn) \Ylmn (\bm)\,.
\end{eqnarray}

Statistical isotropy guarantees that we may write the correlation
as
\begin{eqnarray}
        \left< \Theta_{l m}^* \almn^\sz \right> & \equiv &
b_l
         \equiv  {-2\over l(l+1)} C_l^{\sz \kappa} \,, \nonumber\\
\label{eqn:bl}
         &=& \frac{2}{\pi} \int k^2 dk P_{\Pi\delta}(k) I_l^\sz(k)
                I_l^\ang(k) \,, \\
         &\approx&
        \int \frac{d\rad}{d_A^2}
        W^\sz(\rad) W^\ang(\rad)
        P_{\Pi\delta}\left({ l \over d_A}\right) \,, \nonumber
\end{eqnarray}
where we have used Eq.~\ref{eqn:GSSZequiv} to
relate the power spectrum
$b_l$ defined by \cite{GolSpe99} (1999) and the
$\kappa$-secondary
cross power spectrum defined by \cite{SelZal99} (1999).  The last line
represents the Limber approximation.

The full signal-to-noise ratio of the bispectrum is
\begin{equation}
\left( {S \over N} \right)^2 =
f_\sky \sum_{l_1,l_2,l_3}
        {\bi^2 \over
                 6 C_{l_1}^\tot   C_{l_2}^\tot   C_{l_3}^\tot   } \, ,
\label{eqn:bispecnoise}
\end{equation}
where
\begin{eqnarray}
C_l^{\rm tot} = C_l^\sz + N_l\,.
\label{eqn:cltot}
\end{eqnarray}
Recall that the residual noise $N_l$
was defined in equation (\ref{eqn:nl}) and includes contributions
from detector noise.
We plot the bispectrum
cumulative signal-to-noise  as a function of signal $l_3$, summed over
$l_1$ and $l_2$. We refer the reader to \cite{CooHu00} (2000) for a detailed discussion
on the bispectrum, its variance and the calculation of signal-to-noise ratio.

\section{Discussion}

The SZ effect and weak
gravitational lensing of the CMB both trace large-scale structure in
the underlying density field.
By measuring the correlation, one can directly test the
manner in which gas pressure fluctuations trace the dark matter
density fluctuations.
The correlation vanishes in the two-point functions since the
lensing does not affect an isotropic CMB due to conservation
of surface brightness.

The same correlation manifests itself as a
non-vanishing bispectrum in the CMB at RJ
frequencies (\cite{GolSpe99} 1999; \cite{CooHu00} 2000).
Again the cosmic variance from the primary anisotropies presents an
obstacle for detection of the effect above the several arcminute
scale ($l\sim 2000$).
With the multifrequency cleaning of the
SZ map presented here one can enhance the detectability of the
effect.

\begin{figure}[!h]
\begin{center}
\includegraphics[width=4.2in,angle=-90]{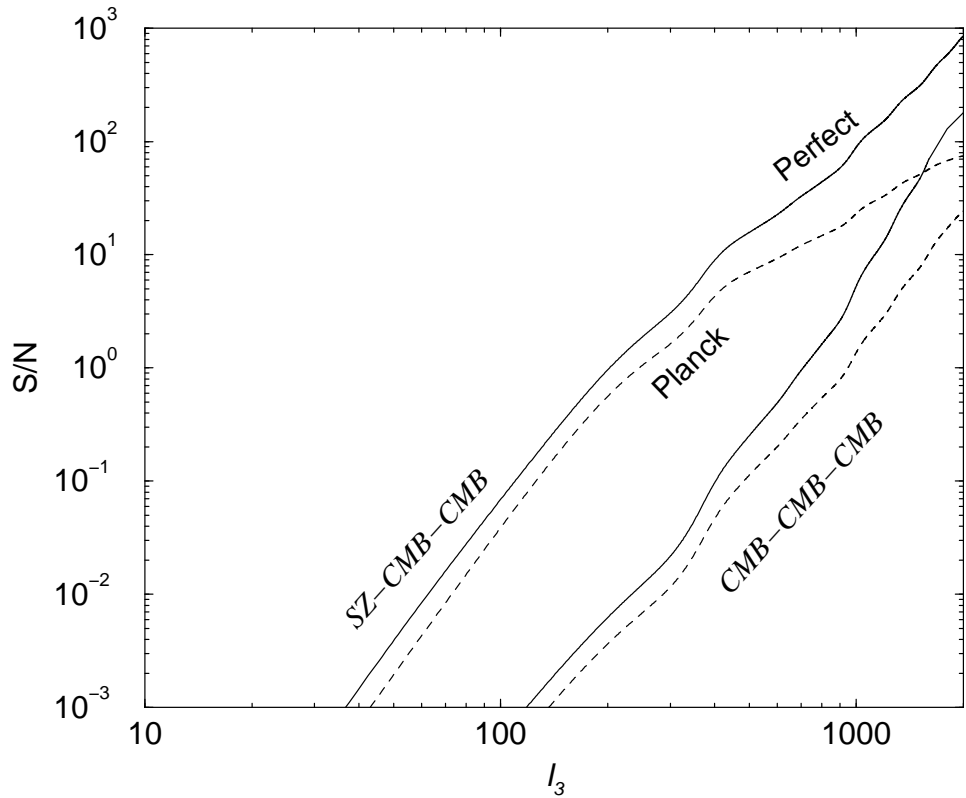}
\end{center}
\caption[Signal-to-noise for the lensing-SZ bispectrum]{
Cumulative signal-to-noise in the measurement of the  SZ-weak
gravitational lensing cross-correlation through the bispectrum
measurement in CMB data.  Compared are the
expected signal-to-noise with (SZ-CMB-CMB) and without (CMB-CMB-CMB)
multifrequency isolation
of the SZ effect for Planck and a perfect/cosmic variance limited
experiment.
Multifrequency isolation provides additional
signal-to-noise and the opportunity to uniquely identify the
bispectrum
contribution with the SZ effect.}
\label{fig:szlens}
\end{figure}

Consider the bispectrum composed of one $a_{l m}$ from the cleaned
SZ map and the other two from the CMB maps.  Call this the SZ-CMB-CMB
bispectrum.   The noise variance of this term will be reduced
by a factor of $C_l^\tot / C_l^\cmb$ compared with the
CMB-CMB-CMB bispectrum.  As one can see from Fig.~\ref{fig:clean}
this can be up to a factor of $10^3$ in the variance.
Details for the calculation of the CMB-CMB-CMB bispectrum
are given in \cite{CooHu00} (2000).  Here we
have updated the normalization for SZ effect, taken
$f_{\sky}=0.65$ for Planck's useful sky coverage, and
compared the $S/N$ of the two bispectra.
As shown in Fig.~\ref{fig:szlens}, 
the measurement using foreground cleaned Planck SZ and CMB
maps has a substantially higher signal-to-noise than  that from
using the Planck CMB map alone for multipoles $l \sim 10^2-10^3$.

Our simple model assumes that the pressure is correlated with lensing
potentials through the halo model. Thus, to the extent that the
lensing and SZ signals can be determined
separately from other measurements, the cross-correlation can
be used to constrain the stochastic nature of the bias.

Beyond the improvement in signal-to-noise, however,
there is an important
advantage in constructing the SZ-lensing bispectrum using SZ and CMB
maps. A mere measurement of the bispectrum in CMB data can lead to
simultaneous detection of non-Gaussianities through processes other
than just
SZ-lensing cross-correlation. As discussed in \cite{GolSpe99} (1999)
and extended in \cite{CooHu00} (2000), gravitational lensing
also correlates with other late time secondary anisotropy contributors
such as integrated Sachs-Wolfe (ISW; \cite{SacWol67} 1967) effect
and the reionized Doppler effect. In addition to lensing correlations,
non-Gaussianities can also  be generated through reionization and
non-linear growth of perturbations (\cite{SpeGol99} 1999;
\cite{GolSpe99} 1999; \cite{CooHu00} 2000).
Bispectrum measurements at a signle frequency can result in a
confusion as to the relative contribution from
each of these scenarios. In \cite{CooHu00} (2000), we
suggested the possibility of using differences in individual bispectra
as
a function of multipoles,
however, such a separation can be problematic
given that these differences are subtle (e.g.,
Fig~6 of \cite{CooHu00} 2000).

The construction
of the SZ-lensing bispectrum using SZ and CMB maps has the advantage
that one eliminates all possibilities, other than SZ, that result in a
bispectrum. For effects related to SZ,
the cross-correlation of lensing and  SZ should produce the dominant
signal; as shown in \cite{CooHu00} (2000), bispectra signal through SZ
and reionization effects, such as Ostriker-Vishniac (OV;
\cite{OstVis86} 1986), are considerably smaller.

Conversely, multifrequency cleaning also eliminates the SZ
contribution from the CMB maps and hence a main contaminant of
the CMB-CMB-CMB bispectrum.
This assists in the detection of smaller signals such as
the
ISW-lensing correlation, Doppler-lensing correlation or the
non-Gaussianity of the initial conditions.
Such an approach is
highly desirable and Planck will allow such detailed studies to be
carried out.

A potential caveat is that as noted above, the full bispectrum
in an all-sky satellite experiment will be difficult to measure.
\cite{ZalSel99} (1999) have developed a reduced set of three-point
statistics
optimized for lensing studies, based on a two point reconstruction of
the lensing-convergence maps from the products of
temperature gradients.
They show that most of the information
is retained in these statistics.  Multifrequency cleaning improves
the signal-to-noise for these statistics by exactly the same factor
as for the full bispectrum.

The cross-correlation coefficient between the SZ effect and CMB weak
lensing is relatively modest ($\sim$ 0.5, see \cite{Seletal00} 2000). This is due to
the fact that the SZ effect is a tracer of the nearby universe while
CMB lensing is maximally sensitive to structure at $z\sim 3$. A higher
correlation is expected if SZ is cross-correlated with an
external probe of low redshift structure.
\cite{PeiSpe00} (2000) suggested the cross-correlation of MAP CMB data and
Sloan galaxy data.
An improved approach would be to use the Planck derived SZ map
rather than a CMB map.
Using a SZ map reduces noise from the primary anisotropies
and guarantees that any detection is due to correlations with the SZ effect.
Extending the calculations in \cite{PeiSpe00} (2000) with the Planck generated
SZ map, we find signal-to-noise ratios which are on average greater by a factor
of $\sim$ 10 when compared to signal-to-noise values using MAP CMB map.
In fact with redshifts for galaxies, Planck SZ map can be
cross-correlated in redshifts bins to study the
redshift evolution of the gas.
Other promising possibilities include cross correlation with
soft X-ray background measurements,
as well as ultraviolet and soft X-ray absorption line studies.

\chapter{Kinetic Sunyaev-Zel'dovich effect}

\section{Introduction}

Extending our calculation on the contribution of large scale
structure gas distribution to CMB anisotropies through SZ effect, 
we also study an 
associated effect involving baryons associated with halos in the large
scale structure.
It is well known that the peculiar velocity of galaxy clusters, along
the line of sight, also lead to a contribution to temperature
anisotropies. This effect is commonly known as the kinetic, or
kinematic,
Sunyaev-Zel'dovich effect and arises from the baryon density
modulation of
the Doppler effect associated with the velocity field (\cite{SunZel80}
1980).
Given that both density and velocity fields are involved,
the kinetic SZ effect is essentially second order, where as the
thermal SZ effect, involving scattering of CMB photons is first order.
Though the kinetic SZ effect was first described in \cite{SunZel80}
(1980)
using massive galaxy clusters,  the same effect has been introduced
under a
different context by Ostriker and Vishniac (OV; \cite{OstVis86} 1986;
also, \cite{Vis87} 1987).
The OV effect has been described as the
contribution to temperature anisotropies due to baryon modulated
Doppler effect in the linear regime of fluctuations.
The kinetic SZ effect can be considered as the OV effect extended to
the non-linear regime of baryon fluctuations, however,
it should be understood that the basic physical
mechanism responsible for the two effects is the same and
that there is no reason to describe them as separate contributions.
For the purpose of this presentation, we will treat both OV effect and
the SZ kinetic effect as one contribution, though it may be easier to
think of OV as the linear contribution while kinetic SZ,
extending to non-linear regime will contain the total contribution.
Such a description has been provided in \cite{Hu00a} (2000a).
 
We calculate the kinetic SZ/OV effect, hereafter simply referred to as
the kinetic SZ effect, using the model we developed to
study the thermal SZ effect. We further extend this calculation to
consider the correlation between SZ thermal and SZ kinetic effects.
Since there is no first order cross-correlation, the lowest order
contribution to the correlation comes from a three-point function, or
a bispectrum. We discussed this bispectrum in \cite{CooHu00} (2000).
Here, we consider an additional possibility to measure the SZ
thermal-kinetic cross-correlation via a two-point correlation function
which involves squares of the temperature, instead of the usual
temperature itself. The power spectrum of squared temperatures probes
one aspect of the trispectrum resulting through the pressure-baryon
cross-correlation probed separately by the thermal SZ and kinetic SZ
effects, respectively. Here, we show that 
there is adequate signal-to-noise for a reliable measurement
of the SZ thermal-SZ kinetic squared power spectrum measurement in
upcoming small angular scale experiments.

The Ostriker-Vishniac effect arises from the second-order
modulation of the Doppler effect by density fluctuations
(Ostriker \& Vishniac 1986; Vishniac 1987).  Its nonlinear
analogue is the kinetic SZ effect from large-scale
structure (Hu 1999). Due to its density weighting,
the kinetic SZ effect peaks at small scales: arcminutes for
$\Lambda$CDM.
For a fully ionized universe, contributions are broadly distributed in
redshift\
 so that the
power spectra are moderately dependent on the optical depth
$\tau$. Here, we assume an optical depth to ionization of 0.1,
consistent with current upper limits on the reionization redshift from
CMB (e.g., \cite{Grietal99} 1999) and
other observational data (see, e.g., \cite{HaiKno99} 1999 and
references therein).

\section{Calculational Method}

The kinetic SZ temperature fluctuations, denoted as $\dsz$,
 can be written as a product of the
line of sight velocity and density 
\begin{equation}
T^\dsz(\hat{\bf n})=  \int d\rad
        g(r) \hat{\bf n} \cdot {\bf v}_g(r,\bn r) n_e(r, \bn r) \, .
\end{equation}
The first order contribution here now comes from the velocity field
with the mean number density of electrons, $n_e  = \bar{n}_e$.
This contribution is discussed in \cite{Kai84} (1984) where it was
shown to be insignificant at small scales due to significant
mode cancellations. It should be understood that, contrary to the
common thought,  this does not mean that the contribution to
temperature fluctuations from the
velocity field is exactly zero. As discussed in
\cite{Kai84} (1984) and 
\cite{CooHu00} (2000), double scattering effects, which are again
due to the the velocity fluctuations, leave a non-zero signal at 
large scales. Also, single scattering effects are sensitive to how one models the transition to reionization; if the transition is
instantaneous mode cancellations are not exact leaving behind a
non-zero signal. The contribution we show as ``Doppler'' in 
Fig.~\ref{fig:szpower} due to velocity fields assume reionization at a redshift of $\sim$ 13 ($\tau=0.1$), and a width $\Delta z$ of 0.1
(see, \cite{CooHu00} 2000).

Including density fluctuations, the full  contribution is
\begin{eqnarray}
T^\dsz(\hat{\bf n})&=&  \int d\rad
        g(r) \hat{\bf n} \cdot {\bf v}_g(r,\bn r) \bar{n}_e (1+\delta_e)(r, \bn r)
\nonumber\\
&=&-i \int d\rad g 
\int \frac{d^3{\bf k}}{(2\pi)^3} \int \frac{d^3{\bf k}'}{(2\pi)^{3}}
 \delta_v({\bf k}-{\bf k}')\delta_g({\bf k'})
e^{i{\bf k}\cdot \hat{\bf n}\rad} \left[ \hat{\bf n} \cdot 
\frac{\veck - \veck'}{|\veck - \veck'|^2}\right] \, , \nonumber \\
\end{eqnarray}
Note that one can use the linear theory to obtain the
large scale velocity field in terms of the linear dark matter density field. The
multiplication between the velocity and density fields in real space
has been converted to a convolution between the two fields in Fourier
space. The second line only includes the contribution from
$v \delta$ term since the $v$ term is the linear Doppler effect.

We can now expand out the temperature perturbation due to kinetic SZ
effect, $T^\dsz$, into
spherical harmonics:
\begin{eqnarray}
a_{lm}^\dsz &=& -i \int d\hat{\bf n}
\int d\rad\; (g)
\int \frac{d^3{\bf k_1}}{(2\pi)^3}\int \frac{d^3{\bf
k_2}}{(2\pi)^3}
\delta_v({\bf k_1})\delta_g({\bf k_2}) \nonumber \\
&&\times e^{i({\bf k_1+k_2})
\cdot \hat{\bf n}\rad} \left[ \frac{\hat{\bf n} \cdot \veck_1}{k_1^2} \right]
Y_l^{m\ast}(\hat{\bf n}) \, ,
\end{eqnarray}
where we have symmetrizised by using $\veck_1$ and $\veck_2$
to represent $\veck-\veck'$ and $\veck'$ respectively.
Using
\begin{equation}
\hat{\bf n} \cdot \veck = \sum_{m'} \frac{4\pi}{3} k
Y_1^{m'}(\hat{\bf n}) Y_1^{m'\ast}(\hat{\veck}) \, ,
\end{equation}
and the Rayleigh expansion (Eq.~\ref{eqn:Rayleigh}),
 we can further simplify and rewrite the multipole moments as
\begin{eqnarray}
&&a_{lm}^\dsz = -i \frac{(4 \pi)^3}{3}
\int d\rad
\int \frac{d^3{\bf k}_1}{(2\pi)^3} \int \frac{d^3{\bf
k}_2}{(2\pi)^3}
\sum_{l_1 m_1}\sum_{l_2 m_2}\sum_{m'} \nonumber\\
&& \times
(i)^{l_1+l_2} g
\frac{j_{l_1}(k_1\rad)}{k_1}
j_{l_2}(k_2\rad)
\delta_v({\bf k_1})\delta_g({\bf k_2})
Y_{l_1}^{m_1}(\hat{\veck}_1) Y_1^{m'}(\hat{\veck}_1)
Y_{l_2}^{m_2}(\hat{\veck}_2) \nonumber \\
&\times& \int d\hat{\bf n}
Y_l^{m\ast}(\hat{\bf n}) Y_{l_1}^{m_1\ast}(\hat{\bf n})
Y_{l_2}^{m_2\ast}(\hat{\bf n})
Y_1^{m'\ast}(\hat{\bf n})\,.
\label{eqn:almdsz}
\end{eqnarray}

We can construct the angular power spectrum by considering
$\langle a_{l_1m_1} a^*_{l_2m_2} \rangle$.
Under the  assumption that the temperature field is statistically
isotropic, the correlation is independent of $m$, and we can write the
angular power spectrum as
\begin{eqnarray}
\langle \alm{1}^{*, \dsz} \alm{2}^\dsz\rangle = \deld_{l_1 l_2} \deld_{m_1 m_2}
        C_{l_1}^\dsz\, .
\end{eqnarray}

The correlation can be written using
\begin{eqnarray}
&& \langle a^{\ast, \dsz}_{l_1m_1} a^\dsz_{l_2m_2} \rangle= \frac{(4 \pi)^6}{9}
\int d\rad_1 g \int d\rad_2 g   \nonumber \\
&\times& \int \frac{d^3{\bf k_1}}{(2\pi)^3}\frac{d^3{\bf k_2}}{(2\pi)^3}
\frac{d^3{\bf k_1'}}{(2\pi)^3}\frac{d^3{\bf k_2'}}{(2\pi)^3}
\langle \delta_v({\bf k_1'})\delta_g({\bf k_2'})
\delta_\delta^{\ast \lin}({\bf k_1})\delta_g^\ast({\bf k_2})  \rangle \nonumber \\
&\times& \sum_{l_1'm_1' l_1'' m_1'' m_1''' l_2'm_2' l_2'' m_2''
m_2'''}  (-i)^{l_1'+l_1''} (i)^{l_2'+l_2''}
j_{l_2'}(k_1'\rad_2) \frac{j_{l_2''}(k_2'\rad_2)}{k_2'}
\frac{j_{l_1'}(k_1\rad_1)}{k_1}
j_{l_1''}(k_2\rad_1) \nonumber \\
&\times& \int d\hat{\bf m}
Y_{l_2m_2}(\hat{\bf m}) Y_{l_2'm_2'}^\ast(\hat{\bf m})
Y_{l_2''m_2''}^\ast(\hat{\bf m}) Y_{1m_2'''}^\ast(\hat{\bf m})
\nonumber \\
&\times& \int d\hat{\bf n}
Y_{l_1m_1}^\ast(\hat{\bf n}) Y_{l_1'm_1'}(\hat{\bf n})
Y_{l_1''m_1''}(\hat{\bf n})
Y_{1m_1'''}(\hat{\bf n})  \nonumber \\
&\times& \int d\hat{\veck_1'} \int d\hat{\veck_2'}
 Y_{l_2'm_2'}(\hat{\veck_1'}) Y_{1m_2'''}(\hat{\veck_2'})
Y_{l_2''m_2''}(\hat{\veck_1'}) \nonumber \\
&\times&\int d\hat{\veck_1} \int d\hat{\veck_2}
Y_{l_1'm_1'}^\ast(\hat{\veck_1})
Y_{1m_1'''}^\ast(\hat{\veck_1}) Y_{l_1''m_1''}^\ast(\hat{\veck_2}) \,
.
\end{eqnarray}
We can separate out the contributions such that the total is made of
correlations
following $\langle v_g v_g\rangle \langle \delta_g \delta_g \rangle$
and $\langle v_g \delta_g \rangle \langle v_g \delta_g \rangle$
depending on
whether we consider cumulants by combining $\veck_1$ with $\veck_1'$
or $\veck_2'$ respectively. After some straightforward but tedious
algebra, and noting that 
\begin{equation} 
\sum_{m_1' m_2'} 
\left(
\begin{array}{ccc}
l_1' & l_2' & l_1 \\
m_1' & m_2'  &  m_1
\end{array}
\right)
\left(
\begin{array}{ccc}
l_1' & l_2' & l_2 \\
m_1' & m_2'  &  m_2
\end{array}
\right) 
= \frac{\delta_{m_1 m_2} \delta_{l_1 l_2}}{2l_1+1}
\end{equation} 
we can write
\begin{eqnarray}
&&C_l^\dsz = \frac{2^2}{\pi^2} \sum_{l_1 l_2}
\left[\frac{(2l_1+1)(2l_2+1)}{4\pi}\right]
\left(
\begin{array}{ccc}
l & l_1 & l_2 \\
0 & 0  &  0
\end{array}
\right)^2 \nonumber \\
&\times& \int d\rad_1 g 
\int d\rad_2 g 
\int k_1^2 dk_1 \int k_2^2 dk_2 \nonumber \\
&\times& \Big( P_{vv}(k_1)
P_{gg}(k_2)j_{l_1}(k_2\rad_2) j_{l_1}(k_2\rad_1) 
\frac{j_{l_2}'(k_1\rad_1)}{k_1} \frac{j_{l_2}'(k_1\rad_2)}{k_1}
\nonumber \\
&+& P_{v g}(k_1) P_{v g}(k_2) j_{l_2}(k_2\rad_1)
\frac{j_{l_1}'(k_1\rad_1)}{k_1}j_{l_1}(k_1\rad_2)
\frac{j_{l_2}'(k_2\rad_2)}{k_2}
  \Big) \, . \nonumber \\
\end{eqnarray}

Here, the first term represents the contribution from $\langle v_g v_g
\rangle
\langle \delta_g  \delta_g \rangle$  while the second term is
the $\langle v_g \delta_g \rangle \langle v_g \delta_g \rangle$
contribution, respectively.
In simplifying the integrals involving spherical harmonics, we have
made use of the properties of Clebsh-Gordon coefficients, in
particular, those involving $l=1$.
The integral involves two distances
and two Fourier modes and is summed over the Wigner-3$j$ symbol to
obtain the power spectrum. The above equation
for the all-sky angular power spectrum of kinetic SZ, or OV effect, is exact, in that we have made no assumptions or simplifications in the
derivation, as have been done in the prior calculations. 
Note that the integrals over spherical Bessel functions and their
derivatives, and sums over Wigner symbols,
 are equivalent to the mode-coupling integrals
one usually encounters in flat-sky coordintes. We will further
explore the relation between the two approaches later.

We are primary interested in the contribution at small
angular scales here. We will model the velocity field using linear
theory. The assumption then is that the on-linear contribution to the
velocity field due to virial motions within halos do not contribute to the temperature fluctuations; this is true since virial motions are
random.  Also, the correlation between large scale bulk flows and
non-linear density field within halos is subdominant. 
At small scales, we only
 consider the contribution that results from baryon-baryon and
linear density-density correlations (related to velocities). 
In fact, under the halo description provided here, there
is no correlation of the baryon field within halos and the velocity
field traced by individual halos. Thus, contribution to the
baryon-velocity correlation only comes from the 2-halo term of the
density field-baryon correlation. This correlation is suppressed at
small scales and is not a significant contributor to the kinetic SZ
power spectrum (see, \cite{Hu00a} 2000a). 

Similar to the Limber approximation (\cite{Lim54} 1954),
in order to simplify the calculation associated with $\langle v_g v_g
\rangle \langle \delta_g \delta_g \rangle$, we use an equation
involving completeness of spherical Bessel functions
(Eq.~\ref{eqn:ovlimber}) and apply 
it to the integral over $k_2$  to obtain
\begin{eqnarray}
&&C_l^\dsz = \frac{2}{\pi} \sum_{l_1 l_2}
\left[\frac{(2l_1+1)(2l_2+1)}{4\pi}\right]
\left(
\begin{array}{ccc}
l & l_1 & l_2 \\
0 & 0  &  0
\end{array}
\right)^2 \nonumber \\
&\times& \int d\rad_1 \frac{(g \dot{G})^2}{d_A^2}
\int k_1^2 dk_1 
P_{\delta\delta}^\lin(k_1) P_{gg}\left[ \frac{l_1}{d_A}; \rad_1 \right]
\left(\frac{j_{l_2}'(k_1\rad_1)}{k_1}\right)^2\, . \nonumber \\
\label{eqn:redallsky}
\end{eqnarray}

The alternative approach, which has been the calculational method in many
of the previous papers (\cite{Vis87} 1987; \cite{Efs88} 1988;
\cite{JafKam98} 1998; \cite{DodJub95} 1995; \cite{Hu00a} 2000a)
is to use a specific coordinate frame with the z-axis along $\vec{\bf
k}$. This allows one to simplify the SZ kinetic power spectrum to:
\begin{eqnarray}
&&C_l^\dsz = \frac{1}{8\pi^2} \int d\rad \frac{(g \dot{G}G)^2}{d_A^2} P_{\delta\delta}(k)^2
I_v\left(k=\frac{l}{d_A}\right) \, ,
\end{eqnarray}
with the mode-coupling integral given by
\begin{eqnarray}
I_v(k) = \int dk_1 \int_{-1}^{+1} d\mu \frac{(1-\mu^2)(1-2\mu
y_1)}{y_2^2} \frac{P_{\delta\delta}(k y_1)}{P_{\delta\delta}(k)}
\frac{P_{\delta\delta}(k y_2)}{P_{\delta\delta}(k)} \, . \nonumber \\
\end{eqnarray}
We refer the reader to \cite{Vis87} (1987) and \cite{DodJub95} (1995)
for details on this derivation. In above, 
$\mu = \hat{\bf k} \cdot \hat{\bf k_1}$, $y_1 = k_1/k$ and
$y_2 = k_2/k = \sqrt{1-2\mu y_1+y_1^2}$. This flat-sky approximation
makes use of the Limber approximation (\cite{Lim54} 1954) to further simplify
the calculation with the replacement of $k = l/d_A$. The power spectra here
represent the baryon field power spectrum 
and the velocity field power spectrum; the former assumed to trace the
dark matter density field while the latter  is generally 
related to the linear dark matter density field through the use of linear
theory arguments. 

The correspondence between the flat-sky and all-sky
formulation can be obtained by noting that in the small scale limit
contributions to the flat-sky effect comes when $k_2 = |\veck -
\veck_1| \sim k$ such that $y_1 \ll 1$. In this limit, the flat sky
Ostriker-Vishniac effect reduces to a simple form given by
(\cite{Hu00a} 2000a)
\begin{equation}
C_l^\dsz = \frac{1}{3}
\int d\rad \frac{(g \dot{G} G)^2}{d_A^2} P_{gg}(k) v_\rms^2 \, .
\label{eqn:redflatsky}
\end{equation}
Here, $v_\rms^2$ is the rms of the uniform bulk velocity
form
large scales
\begin{equation}
v_\rms^2 = \int dk \frac{P_{\delta\delta}(k)}{2\pi^2} \, .
\end{equation}
The $1/3$ arises from the fact that rms in each component is $1/3$rd
of the total velocity.

The above statement, first used in \cite{Hu00a} (2000a),
is equivalent to the fact that the non-linear 
momentum density  field of the large scale structure, relevant to
small angular scales, is equivalent
to
\begin{equation}
P_{pp}(k)  = P_{\delta \delta}(k)\int dk \frac{P_{vv}(k)}{2\pi^2} \, .
\end{equation}
Sheth et al. (in prepartion) has numnerically tested this
statement and has been found to agree with N-body simulations.
There is an additional contribution to the momentum power spectrum
from the trispectrum formed by the density and velocity fields.
Note that in the halo description, the linear velocity fields are
uncorrelated with the non-linear density field within halos.
Thus, contributions from such a higher order term only comes at large
scales from terms in the trispectrum that involve more than one halo.
Also, the momentum field contributes through a non-linear analogue of
the integrated Sachs-Wolfe effect (ISW; \cite{SacWol67} 1967).
We will present a detailed description of the relevant 
contributionto both kinetic SZ and non-linear ISW from momentum
density field in a separate paper.

In the same small scale limit, to be consistent with the flat sky
expression, we can reduce the all-sky expression
such that contributions come from a term that looks like
\begin{eqnarray}
&&C_l^\dsz =\int d\rad\frac{(g \dot{G})^2}{d_A^2}
P_{gg}\left[ \frac{l}{d_A}; \rad_1 \right] \frac{1}{3} v_{\rm rms}^2
\, .
\end{eqnarray}

A comparison of the reduced all-sky (Eq.~\ref{eqn:redallsky}) 
and flat-sky (Eq.~\ref{eqn:redflatsky}) formula in the
small-scale
limit suggests that the correspondence between the two arises when
\begin{equation}
\sum_{l_1 l_2}
(2l_1+1)(2l_2+1)
\left(
\begin{array}{ccc}
l & l_1 & l_2 \\
0 & 0  &  0
\end{array}
\right)^2 
\left[j_{l_2}'(k\rad)\right]^2 =\frac{1}{3} \, .
\label{eqn:simplify}
\end{equation}
Numerically, we determined this to be true as long as $l_2 \ll l$,
however, we have not been able to prove
this relation analytically. We leave this as a challenge to our readers.
It should be understood that the right hand side of the above
expression denotes the all-sky equivalent of the integral that
leads to a 1/3rd of rms of a randomly directed quantity along one
particular line of sight.

\begin{figure}[!h]
\begin{center}
\includegraphics[width=4.2in,angle=-90]{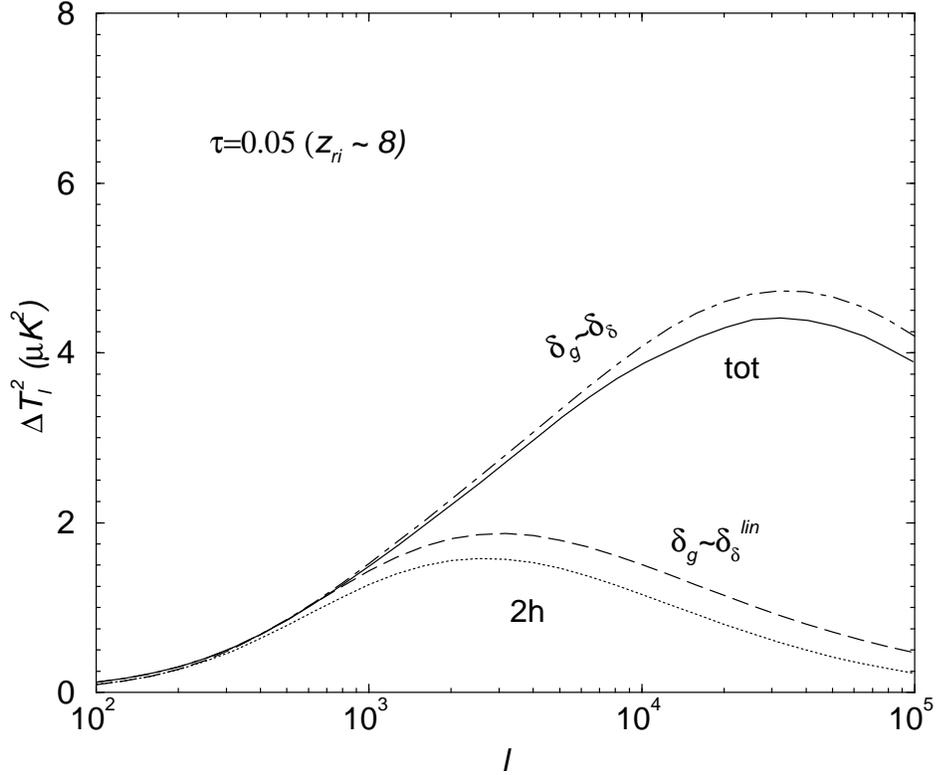}
\end{center}
\caption[kinetic SZ Temperature Fluctuations]{The temperature fluctuation power $(\Delta T^2_l =
l(l+1)/(2\pi) C_l T_{\rm CMB}^2)$ for a variety of methods to
calculate the kinetic SZ effect. Here, we show the contribution for a
reionization redshift of $\sim$ 8 and an optical depth to reionization
of 0.05. The contributions are calculated under the assumption that
the baryon field traces the non-linear dark matter ($P_g(k) =
P_\delta(k)$ with $P_\delta(k)$ predicted by the halo model), 
the linear density field ($P_g(k) = P^\lin(k)$), and the halo model
for gas, with total and the 2-halo contributions  shown separately.
For the most part, the kinetic SZ effect can be described using linear
theory, and the non-linearities only increase the temperature
fluctuation power by a factor of a few at $l \sim 10^5$.}
\label{fig:ovtemp}
\end{figure}

\section{Discussion}

In Fig.~\ref{fig:szpower}, we show our prediction for the SZ kinetic
effect and a comparison with the SZ thermal contribution.
As shown, the SZ kinetic contribution is roughly an order of magnitude
smaller than the kinetic SZ contribution. 
There is also a more fundamental difference between the two:
the SZ thermal effect, due to its dependence on highest temperature
electrons is more dependent on the most massive halos in the universe,
while the SZ kinetic effect arises more clearly due to 
large scale correlations of the halos that make the
large scale structure. The difference arises from that fact that
kinetic SZ effect is mainly due to the baryons and not the temperature
weighted baryons that trace the pressure responsible for the thermal
effect. Contributions to the SZ kinetic effect comes from baryons
tracing all scales and down to small mass halos.
The difference associated with mass dependence 
between the two effects suggests that a wide-field 
SZ thermal effect map and a wide-field 
SZ kinetic effect map will be different from each other in that
massive halos, or clusters, will be clearly visible in a SZ thermal
map while the large scale structure will be more evident in
a SZ kinetic effect map.  Numerical simulations are in fact consistent with
this picture (e.g., \cite{Spretal00} 2000).

As shown in Fig.~\ref{fig:szpower}(b), the
variations in maximum mass used in the calculation does not lead to
orders of magnitude changes in the total kinetic SZ contribution,
which is considerably less than the changes in the total thermal SZ
contribution as a function of maximum mass. This again is consistent with our basic
result that most contributions come from the large scale linear velocity
modulated by baryons in halos. 
Consequently, while the thermal SZ effect is dominated by shot-noise contributions, and is
heavily affected by the sample variance, the same is not true for the
kinetic SZ effect. 

In Fig.~\ref{fig:ovtemp}, we show several additional predictions for the kinetic
SZ effect, following the discussion in \cite{Hu00a} (2000a). Here, 
we have calculated the kinetic SZ power spectrum under several
assumptions, including the case when gas is assumed to trace 
the non-linear density field and the linear density field. 
We compare predictions based on such assumptions
to those calculated using the halo model. As shown, the halo model
calculation shows slightly less power than when using the non-linear dark
matter density field to describe clustering of
baryons. This difference arises from the fact that baryons do not
fully trace the dark matter in halos. Due to small differences,
one can safely use the non-linear dark matter power spectrum to describe
baryons. Using the linear theory only, however, leads to an underestimate of
power at a factor of 3 to 4 at scales corresponding to multipoles of
$l \sim 10^4$ to $10^5$ and may not provide an accurate description of
the total kinetic SZ effect.

Since our model for baryons only include those present in halos and 
given that SZ thermal effect arises essentially  from clustering of
gravitationally heated baryons in single halos, our model
may be more applicable to it. Given that we have ignored the filamentary
structure of the large scale structure and associated smaller
overdensities,  our halo description of
baryons to describe the SZ kinetic effect, which includes
contributions from all mass scales, may likely to be an incomplete description.
Therefore, numerical simulations will certainly be
necessary to improve our calculations on the SZ kinetic effect with the inclusion of diffuse
baryons in smaller overdensities.  Such simulations could also aid in
calibration purposes of the halo model predictions only involving
virialized halos.

The interesting experimental possibility here is whether one can obtain
an wide-field map of the SZ kinetic effect. Since it is now well
known that the unique spectral dependence of the thermal SZ effect can
be used to separate its contribution (\cite{Cooetal00a} 2000a), at smaller
angular scales, it is likely that after the separation, SZ kinetic effect will be the
dominant signal, even after accounting for the lensed CMB
contribution. For such a separation of the SZ thermal effect to be
carried out and such that a detection of the kinetic SZ effect will be possible, 
observations, at multifrequencies, are needed to arcminute scales. 
Upcoming interferometers and similar experiments
will allow such studies to be eventually carried out. A wide-field
kinetic SZ map of the large scale structure will eventually allow an understating of
the large scale velocity field of baryons, as the  density fluctuations
can be identified through cross-correlation of such a map with a
similar thermal SZ map. We now discuss the existence of correlations
between the SZ thermal and SZ kinetic effect.

\chapter{SZ Thermal-SZ kinetic Correlation}

\section{Introduction}

The SZ thermal and SZ kinetic effects both trace the large scale
structure baryons. One can study a correlation between these two
effect to probe  the manner in which baryons are distributed in the
large scale structure. For example, such a correlation study may allow
one to answer
to what extent diffuse baryons contribute to thermal SZ when compared
to their contribution to kinetic SZ. Given that the SZ kinetic effect
is second order in fluctuations, there is no direct two-point
correlation function between the temperature anisotropies produced by
SZ thermal and kinetic effects. As discussed in \cite{CooHu00} (2000),
to the highest order,  the correlation between kinetic SZ and thermal
SZ manifests as a
nonvanishing bispectrum in temperature fluctuations and can be studied
by considering a three-point correlation function or associated
statistics, such as bispectrum, the Fourier space analog of the three
point function, or skewness, a collapsed measurement of the
bispectrum.

\section{SZ thermal-SZ thermal- SZ kinetic bispectrum}

The bispectrum formed by the SZ thermal effect and the SZ kinetic
effect can be derived following \cite{CooHu00} (2000). Note that in
\cite{CooHu00} (2000), we identified this bispectrum as SZ-SZ-OV.
Using the multipole expansion of the kinetic SZ effect given in
Eq.~\ref{eqn:almdsz} and the multipole moments of the SZ effect as
\begin{eqnarray}
\almn^\sz &=&
     i^l \int {d^3{\bf k } \over 2\pi^2} \Pi({\bf k})
     \Ylmn{}^*(\bk)  I_l^\sz(k)  \, , \nonumber\\
I_l^\sz(k) &=& \int d\rad  W^{\rm sz}(\rad) j_{l}(k\rad) \, ,
\label{eqn:szsource}
\end{eqnarray}
we can write the cumulant formed by
$\langle a_{l_1 m_1}^\sz a_{l_2 m_2}^\sz a_{l_3 m_3}^{\ast \dsz} \rangle$
as
\begin{eqnarray}
\langle a_{l_1 m_1}^\sz a_{l_2 m_2}^\sz a_{l_3 m_3}^{\rm \dsz}\rangle &=&
\frac{2^2}{\pi^2}\int k_1^2 dk_1 \int k_2^2 dk_2 P_{\delta \Pi}(k_1)
P_{g \Pi}(k_2) I_{l_1}^\sz(k_1)\nonumber \\
&&\times I_{l_2}^\sz(k_2)
[I_{l_1,l_2}^\dsz(k_1,k_2) +
I_{l_1,l_2}^\dsz(k_2,k_1)]
\nonumber\\
&&\times  \int d\hat{\bf n} Y_{l_3}^{m_3}(\hat{\bf n})
Y_{l_2}^{m_2*}(\hat{\bf n})
Y_{l_1}^{m_1*}(\hat{\bf n})\,,
\label{eqn:ovtriplet}
\end{eqnarray}
where
\begin{eqnarray}
I^\dsz_{l_1,l_2}(k_1,k_2) &=& \int d\rad W^\dsz j_{l_2}(k_2\rad)
j'_{l_1}(k_1\rad) \nonumber\,,\\
W^\dsz(k_1,r) &=& -{1 \over k_1} g\dot{G}G \,.
\end{eqnarray}
In simplifying the integrals involving spherical harmonics,
we have made use of the properties of Clebsch-Gordon coefficients, in
particular, those involving $l=1$. 

\begin{figure}[!h]
\begin{center}
\includegraphics[width=4.0in]{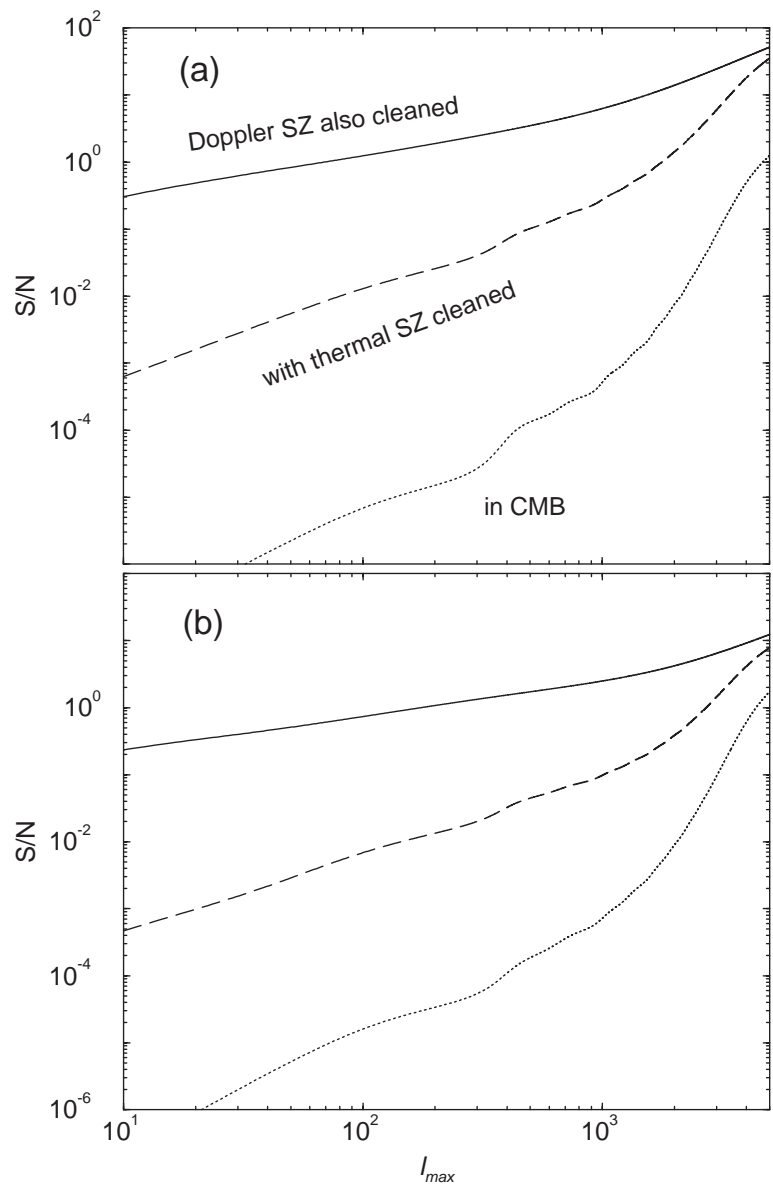}
\end{center}
\caption[Cumulative signal-to-noise for the detection of the SZ
thermal-SZ thermal-SZ kinetic
bispectrum and skewness]{Cumulative signal-to-noise for the detection of SZ thermal-SZ
thermal-SZ kinetic bispectrum (a) and skewness (b). The
dotted line is for the detection of SZ thermal-SZ kinetic
correlation using CMB data alone, while the dashed line is
the same when the SZ thermal effect has been separated from other CMB
contributions and the measurement now involves two points from the SZ
map and one point from the CMB.
Finally, the solid line is the maximum signal-to-noise for achievable
with the separation of the SZ kinetic effect
from all contributors to CMB  anisotropy.}
\label{fig:bisn}
\end{figure}

In order to construct the bispectrum, note that
\begin{equation}
\langle a_{l_1 m_1}^\sz a_{l_2 m_2}^\sz a_{l_3 m_3}^\dsz
\rangle =
(-1)^{l_3}\langle a_{l_1 m_1}^\sz a_{l_2 m_2}^\sz a_{l_3
-m_3}^\dsz \rangle \; .
\end{equation}
The bispectrum then becomes
\begin{eqnarray}
B_{l_1 l_2 l_3} &=& \sum_{m_1 m_2 m_3} \wjm \Big(
\langle a^\sz_{l_1 m_1}a^{\rm SZ}_{l_2 m_2}a^{\rm kSZ}_{l_3 m_3}
\rangle+\nonumber\\
&& \times
\langle a^{\rm SZ}_{l_2 m_2}a^{\rm SZ}_{l_3 m_3}a^{\rm kSZ}_{l_1 m_1}
\rangle+
\langle a^{\rm SZ}_{l_3 m_3}a^{\rm SZ}_{l_1 m_1}a^{\rm kSZ}_{l_2 m_2}
\rangle\Big)\nonumber\\
&=& \sqrt{\frac{(2l_1 +1)(2 l_2+1)(2l_3+1)}{4 \pi}}
\left(
\begin{array}{ccc}
l_1 & l_2 & l_3 \\
0 & 0  &  0
\end{array}
\right)  [ b^{\se-\se}_{l_1,l_2} + {\rm Perm.}] \, ,
\label{eqn:ovbidefn}
\end{eqnarray}
with
\begin{eqnarray}
 b^{\sz-\sz}_{l_1,l_2}
&=&\frac{2^2}{\pi^2} \int k_1^2 dk_1 \int k_2^2 dk_2 P_{\delta \Pi}(k_1)
P_{g \Pi}(k_2) \nonumber \\
&&\times I^\dsz_{l_1,l_2}(k_1,k_2) I_{l_1}^{\rm SZ}(k_1) I_{l_2}^{\rm
SZ}(k_2) \, .
\label{eqn:ovintegral}
\end{eqnarray}
Note that we have rewritten the $k_1 \rightarrow k_2$ term
in Eq.~\ref{eqn:ovtriplet}
as an $l_1 \rightarrow l_2$ interchange so that in
Eq.~\ref{eqn:ovbidefn}
``Perm.'' means a sum over the
remaining 5 permutations of ($l_1$,$l_2$,$l_3$) as usual.

In general, Eq.~\ref{eqn:ovintegral} involves
five integrations,
three over radial distances and two over wavenumbers.
These integrals can be
simplified using the Limber approximation for sufficiently large
$(l_1,l_2)$ and we 
employ 
the completeness relation of spherical Bessel functions in
Eq.~\ref{eqn:ovlimber}. Applying this to the integral over $k_2$ yields
\begin{eqnarray}
&&b^{\sz-\sz}_{l_1,l_2}
=\frac{2}{\pi} \int \frac{d\rad}{d_A^2} W^\sz(\rad) g
P_{g\Pi}\left(\frac{l_2}{d_A};\rad\right)
\nonumber \\
&\times&
\int d\rad_1 \int k_1 dk_1 P_{\delta\Pi}^{2h}(k_1;\rad_1)
W^\sz(\rad_1) j_{l_1}'(k_1\rad_1) j_{l_1}(k_1\rad) \, , \nonumber \\
\label{eqn:finalintegral}
\end{eqnarray}
which we will use to evaluate the SZ thermal-SZ thermal-SZ kinetic bispectrum.

\begin{figure}[!h]
\begin{center}
\includegraphics[width=4.2in]{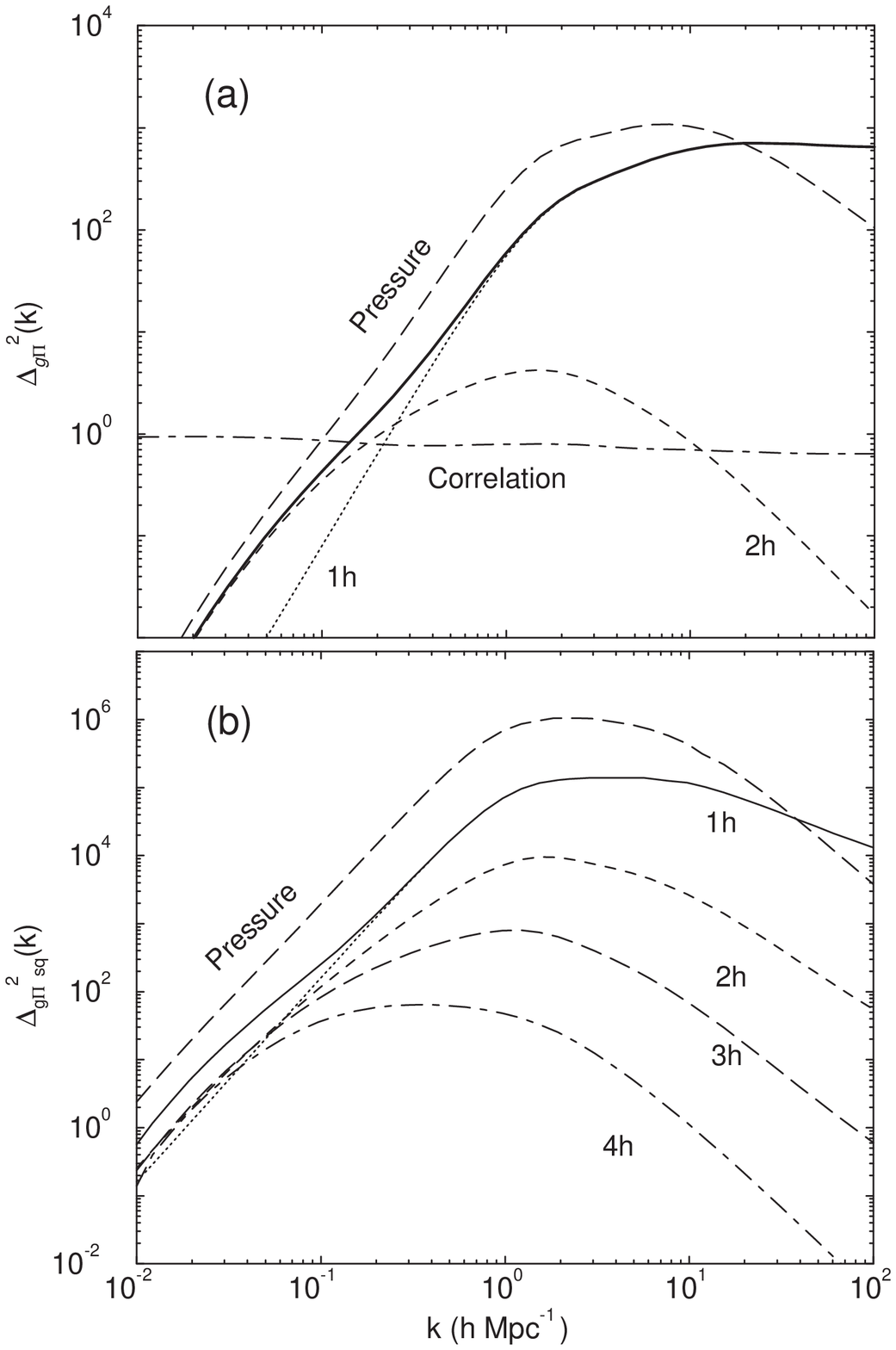}
\end{center}
\caption[The baryon-pressure power spectrum and trispectrum]{The baryon-pressure (a) power spectrum (b) trispectrum
today ($z=0$)
broken into individual contributions under the halo description.
The line labeled 'correlation' shows the correlation coefficient of
gas-pressure correlation with respect to gas-gas and
pressure-pressure. For reference, we also show the pressure power
spectrum and the trispectrum.}
\label{fig:gaspressurepower}
\end{figure}

In \cite{CooHu00} (2000), we assumed that pressure traces dark matter
to calculate the SZ-SZ-OV bispectrum. Using our halo model, we can
now update this calculation to include the bispectrum formed between
SZ thermal-SZ thermal-SZ kinetic effects. Additionally, we can
investigate the improvements in the signal-to-noise for the bispectrum
detections when the SZ thermal effect is separated from CMB. The
maximum signal-to-noise for this bispectrum can only be achieved when
the SZ kinetic effect is also separated from CMB, though, given that the
two effects have the same frequency dependence, it is unlikely that
the kinetic SZ effect can be separated from thermal CMB anisotropies.
With SZ separated from CMB, however, it
is likely that the SZ kinetic effect will dominate the small angular
scale signal in the temperature anisotropies. Thus, one can use small
angular scale thermal CMB temperature fluctuations for the
cross-correlation purposes between thermal SZ and kinetic SZ effects.

In Eq.~\ref{eqn:finalintegral}, 
$P_{g\Pi}$ is the baryon-pressure power spectrum while the
$P_{\delta\Pi}$ is the density-pressure power spectrum, with the
density field tracing the linear regime of fluctuations. Since there
is no contribution coming from the non-linear regime (ie. the 1-halo
term), we model this as the large scale density-pressure correlations
in the
linear regime described by the 2-halo term.

Following \cite{CooHu00} (2000), the signal-to-noise for the detection
of the bispectrum is
\begin{equation}
\left(\frac{{\rm S}}{{\rm N}}\right)^2 \equiv {1 \over \sigma^2(A)} =
\sum_{l_3\ge l_2 \ge l_1}
        \frac{\bi^2}{C_{l_1}^\tot C_{l_2}^\tot C_{l_3}^\tot}\,,
\label{eqn:chisq}
\end{equation}
where
\begin{equation}
C_l^\tot = C_l^{\rm CMB}+C_l^{\rm sec}+C_l^{\rm Noise} \, .
\end{equation}

In \cite{Cooetal00a} (2000a), we suggested that
multifrequency cleaning of SZ effect can be a useful tool for higher
order correlation studies and discussed how the signal-to-noise for
the
detection of SZ-lensing correlation, again through a bispectrum, can
be improved by using CMB primary anisotropy separated SZ map. We
present the same approach here, where we study the possibility for a
detection of the SZ thermal-SZ kinetic correlation by using a
frequency cleaned SZ thermal map, which will provide two measurements,
and the remainder, which will contain CMB primary, SZ kinetic and
other secondary effects and proving a single measurement for the
bispectrum.

In Fig.~\ref{fig:bisn}(a), we update results for the bispectrum
given in \cite{CooHu00} (2000), where
we only studied the possible detection in CMB data alone and with no
consideration for separation of effects, especially the SZ thermal
effect.  The separation allows a
decrease in cosmic variance, as the noise is no longer dominated by
CMB primary anisotropies. This leads to an increase in the cumulative
signal-to-noise. With SZ thermal effect separated, we see that the
signal-to-noise increases by roughly two orders of magnitude. In
\cite{Cooetal00a} (2000a), we showed how one can obtain an order of
magnitude
in signal-to-noise when a CMB separated SZ thermal map is used for a
detection of SZ-lensing correlation. Here, we obtain another order of
magnitude improvement, since the SZ thermal-SZ kinetic correlation is
present with two SZ thermal measurements, instead of one in the case
SZ thermal-lensing correlation (in SZ thermal-CMB-CMB bispectrum).
Note that one cannot use multifrequency data to separate SZ kinetic
from rest of the contributions. Thus, CMB primary anisotropies and
other secondary effects still contribute to the variance.

If one can separate the SZ kinetic such that a perfect SZ kinetic map,
as well as a perfect SZ thermal map, is available, then one can
improve the
signal-to-noise for detection significantly such that a detection is
possible. Since SZ kinetic is expected to dominate anisotropies at
small angular scales, when SZ thermal is removed, an opportunity to
detect the SZ thermal-SZ kinetic correlation will likely come from
small angular scale multifrequency experiments. One can also improve
the possibility of detecting this correlation by noting that the
configuration for the bispectrum is such that it peaks for highly
flattened triangles (see, \cite{CooHu00} 2000). Thus, in addition to
small
angular scale experiments, information from large angular scale
observations may also be necessary for a detection of this
correlation. It is likely that progress in experimental studies will
continue to a level where such studies will eventually be possible.

\subsection{Skewness}
Since the bispectrum may be hard to calculate from observational data,
we also consider a real space statistic that probes the non-Gaussian
information at the three point level.
The simplest aspect of the bispectrum that can be measured in real
space is the
third moment of the map smoothed on some scale with a window
$W(\sigma)$
\begin{eqnarray}
\left< \Theta^3(\bn;\sigma) \right> &=&
                {1 \over 4\pi} \sum_{l_1 l_2 l_3}
                \sqrt{(2l_1+1)(2l_2+1)(2l_3+1) \over 4\pi} \nonumber\\
                &&\times \wj \bi
W_{l_1}(\sigma)W_{l_2}(\sigma)W_{l_3}(\sigma)
                \,, \nonumber \\
\end{eqnarray}
where $W_l$ are the multipole moments (or Fourier transform in a
flat-sky approximation) of the window. Note that the skewness can then
be
calculated as $s_3 = \left< \Theta^3(\bn;\sigma) \right> /
\left< \Theta^2(\bn;\sigma) \right>^2$.

The overall signal-to-noise for the measurement of the third moment is
\begin{equation}
\left( {S \over N} \right)^2 = {f_\sky}
                { \left< \Theta^3(\bn;\sigma) \right>^2 \over {\rm
Var}}
\end{equation}
where the variance, assuming Gaussian statistics, is given by
\begin{eqnarray}
{\rm  Var} &=& { 1 \over (4\pi)^2 } \sum_{l_1 l_2 l_3}
                {(2l_1+1)(2l_2+1)(2l_3+1) \over 4\pi} \wj^2
\nonumber\\
                && \times W_{l_1}^2(\sigma) W_{l_2}^2(\sigma)
W_{l_3}^2(\sigma)
                                6 C_{l_1}^\tot C_{l_2}^\tot
C_{l_3}^\tot\,.
\label{eqn:t3var}
\end{eqnarray}

In Fig.~\ref{fig:bisn}(b), we show the signal to noise for the
detection of the third moment. Here, we use a
top-hat window in multipole space out to $l_{\rm max}$ so that direct
comparison is possible with the signal-to-noise calculation involving
the bispectrum. As shown, we find that there is considerably less
signal-to-noise in the skewness when compared to the full bispectrum
itself. This results from the fact that bispectrum contains all
information at the three point level, while with the third moment
results in a loss of information. This can also be understood by
noting that the signal-to-noise for the bispectrum and skewness is
such that in the case of the bispectrum signal-to-noise is calculated
for each mode and summed up while for the skewness signal-to-noise is
calculated after summing the signal and noise separately over all modes.

\begin{figure}[!h]
\begin{center}
\includegraphics[width=4.2in]{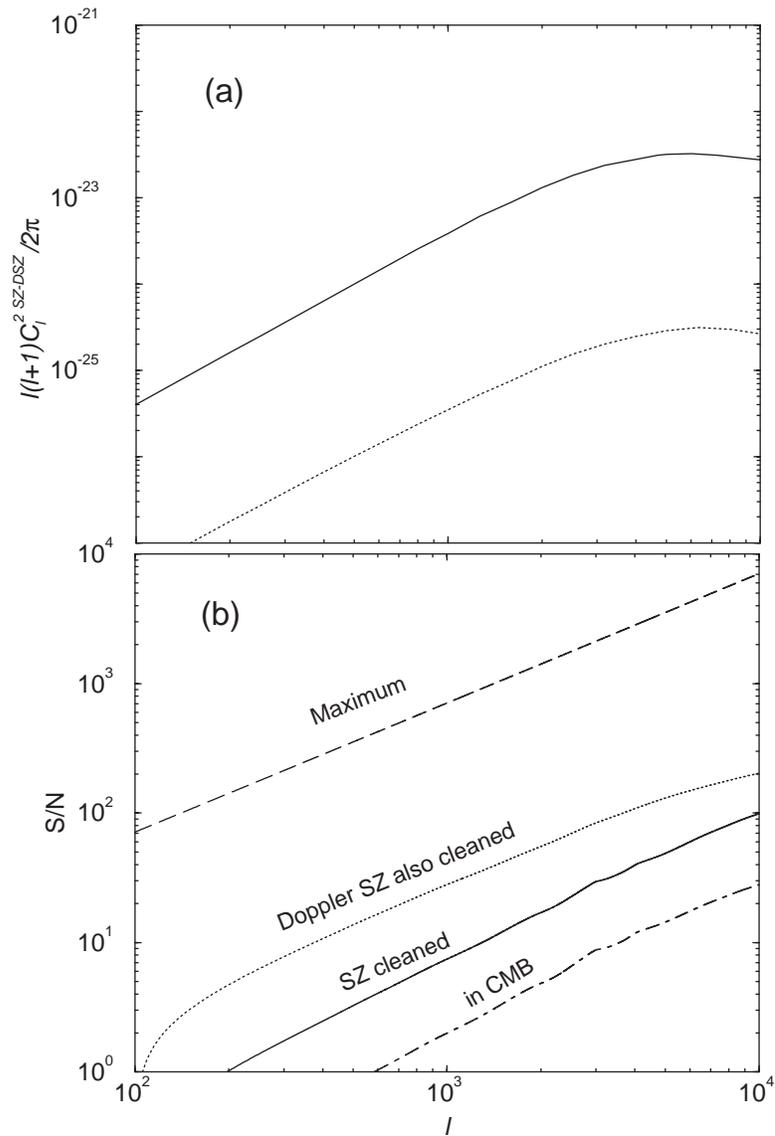}
\end{center}
\caption[The SZ thermal-SZ kinetic squared temperature power spectrum
and signal-to-noise for detection]{ (a) The SZ thermal-SZ kinetic power spectrum of squared
temperatures. Here, we show the contribution to the power spectrum
when only Gaussian terms (dotted line) and when
when non-Gaussianities are introduced (solid line). In (b), we show the
cumulative signal-to-noise for the detection of the SZ thermal-SZ
kinetic squared temperature power spectrum using information in
multipoles from 2000 to 10000 and assuming no instrumental or any
other noise contributions to the covariance. The signal-to-noise is
calculated assuming the power spectrum is measured in CMB data (dot-dashed line),
with a perfect frequency cleaned SZ thermal map (solid line) and with
a perfect SZ thermal and SZ kinetic effect maps (dotted line).}
\label{fig:sz2ov2}
\end{figure}

\section{The SZ Thermal$^2$-SZ kinetic$^2$ Power Spectrum}

In addition to the SZ thermal-SZ kinetic-SZ kinetic bispectrum, we can
introduce higher order correlations involving the SZ thermal and SZ
kinetic effect that probe the correlation between the two.
One such a possibility is the trispectrum formed by the SZ thermal and
SZ kinetic effect.
Given that we do not have a reliable method to measure the bispectrum
even,
the measurement of such a higher order correlations in experimental
data is likely to be challenging.

Here, we focus on a statistic
that captures the correlation information coming from
higher order, essentially from a trispectrum, but is easily measurable
in
experimental data since it only involves only
a power spectrum. Such a possibility
involves the power spectrum of squared
temperatures instead of the usual temperature itself.
Our motivation for such a statistic came when we inspected the
published maps of the large scale SZ thermal and SZ kinetic
effects in simulations by \cite{Spretal00} and realized
that there is a significant correlation between the two
effects. Since the temperature fluctuations produced by the SZ kinetic
effect
oscillates between positive and negative values depending on the
direction of the velocity field along the line of sight, as stated
earlier,
a direct two point correlation involving the temperature results in
no contribution.  A non-zero correlation between the SZ thermal and SZ
kinetic effects still manifests if the absolute
value of the temperature fluctuation due to kinetic SZ effect is
considered.
Since absolute value of temperature is equivalent to squaring the
temperature,
we consider the cross-correlation of SZ thermal and SZ kinetic effects
involving the power spectrum of squared temperatures here.

In order to calculate the SZ thermal$^2$-SZ kinetic$^2$
power spectrum, we first note
that the spherical harmonic  coefficient of the squared can be written
through a convolution of the spherical moments of the fluctuations
\begin{eqnarray}
a_{lm}^{2} &=& \int d\hat{\bf n} Y_{l}^{\ast m} T^2(\hat{\bf n})
\nonumber \\
&=& \sum_{l_1 m_1} \sum_{l_2 m_2} a_{l_1 m_1} a^\ast_{l_2 m_2}
 \int d\hat{\bf n} Y_l^{\ast m}(\hat{\bf n})  Y_{l_1}^{
m_1}(\hat{\bf n}) Y_{l_2}^{\ast m_2}(\hat{\bf n})  \, ,
\end{eqnarray}
where
\begin{equation}
T(\hat{\bf n}) = \sum_{l m} a_{l m} Y_l^{\ast m}
\end{equation}
Note that the integral over three spherical harmonic coefficients can
be written through the use of the Gaunt integral.

Using this identity, we can now construct the power spectrum of
thermal SZ$^2$-kinetic SZ$^2$ as
\begin{eqnarray}
&& \langle a_{lm}^{\dsz^2} a_{l' m'}^{\ast SZ^2} \rangle = C_{l}^{2
\dsz-\sz}
\delta_{l,l'} \delta_{m,m'} \nonumber \\
&=& \sum_{l_1 m_1 l_2 m_2} \sum_{l_3 m_3 l_4 m_4}
\langle a_{l_1 m_1}^{\dsz} a_{l_2
m_2}^\dsz a_{l_3 m_3}^{\ast SZ} a_{l_4 m_4}^{SZ} \rangle^\tot
\nonumber \\
&\times& \int d\hat{\bf n} Y_l^{\ast m}(\hat{\bf n})
Y_{l_1}^{m_1}(\hat{\bf n})
Y_{l_2}^{\ast m_2}(\hat{\bf n}) \int d\hat{\bf m} Y_{l'}^{m'}(\hat{\bf m})  Y_{l_3}^{\ast
m_3}(\hat{\bf m})
Y_{l_4}^{m_4}(\hat{\bf m}) \, , \nonumber \\
\end{eqnarray}
involving two cumulants of the SZ thermal and SZ kinetic effect,
respectively.

After the multipole moments of the SZ effect in Eq.~\ref{eqn:szsource}
and  the kinetic SZ effect from
Eq.~\ref{eqn:almdsz}, we can write the cumulant involving the four
moments as
\begin{eqnarray}
&& \langle a_{l_1 m_1}^\dsz a_{l_2
m_2}^\dsz a_{l_3 m_3}^{\ast SZ} a_{l_4 m_4}^{SZ} \rangle  \nonumber \\
&=& \frac{(4 \pi)^8}{9} \int d\rad_1 ... \int
d\rad_4
\int {d^3{\bf k_1 } \over (2\pi)^3} ...
\int {d^3{\bf k_6 } \over (2\pi)^3} 
 \sum_{l_1' m_1' l_1'' m_1'' m_1'''}
\sum_{l_2' m_2' l_2'' m_2'' m_2'''}
i^{l_1'+l_1''+l_4}(-i)^{l_2'+l_2''+l_3}
\nonumber \\
&\times& (g\dot{G}G)_{\rad_1} (g\dot{G}G)_{\rad_2}
W^{\rm sz}(\rad_3) W^{\rm sz}(\rad_4)  
 \langle \delta_\delta({\bf k_1})\delta_g({\bf k_2})
\delta_\delta({\bf k_3})\delta_g({\bf k_4})
\delta_\Pi({\bf k_5}) \delta_\Pi({\bf k_6}) \rangle \nonumber \\
&\times&
\frac{j_{l_1'}(k_1\rad_1)}{k_1}j_{l_1''}(k_2\rad_1)
\frac{j_{l_2'}(k_3\rad_2)}{k_3}j_{l_2''}(k_4\rad_2)
j_{l_3}(k_5\rad_3) j_{l_4}(k_6\rad_4) \nonumber \\
&\times&
Y_{l_1'}^{m_1'}(\hat{\veck}_1) Y_1^{m_1'''}(\hat{\veck}_1)
Y_{l_1''}^{m_1''}(\hat{\veck}_2) 
 Y_{l_2'}^{m_2'}(\hat{\veck}_3) Y_1^{m_2'''}(\hat{\veck}_3)
Y_{l_2''}^{m_2''}(\hat{\veck}_4)
Y_{l_3}^{m_3}(\hat{\veck}_5)   Y_{l_4}^{m_4}(\hat{\veck}_6) 
 \nonumber \\ 
&\times& \int d\hat{\bf n}
Y_{l_1}^{m_1\ast}(\hat{\bf n}) Y_{l_1'}^{m_1'\ast}(\hat{\bf n})
Y_{l_1''}^{m_1''\ast}(\hat{\bf n})
Y_1^{m_1'''\ast}(\hat{\bf n}) \nonumber \\
&\times& \int d\hat{\bf m}
Y_{l_2}^{m_2\ast}(\hat{\bf m}) Y_{l_2'}^{m_2'\ast}(\hat{\bf m})
Y_{l_2''}^{m_2''\ast}(\hat{\bf m})
Y_1^{m_2'''\ast}(\hat{\bf m}) \, . \nonumber \\
\end{eqnarray}

The cumulant involving six density, baryon and pressure fluctuations
can be broken in to two parts involving a Gaussian term, with
contributions coming from power spectra of velocities and
pressure-density correlations, and a non-Gaussian term, with the
velocity power spectrum and the pressure-pressure-baryon-baryon
trispectrum.
Here, we ignore the correlations between pressure and velocity or
between baryons and velocities as the scales for such correlations
 do not match, especially in the small angular scale of interest here.
Thus, we write
\begin{eqnarray}
&&\langle \delta_\delta({\bf k_1})\delta_g({\bf k_2})
\delta_\delta({\bf k_3})\delta_g({\bf k_4})
\delta_\Pi({\bf k_5}) \delta_\Pi({\bf k_6}) \rangle  \nonumber \\
&=& \langle  \delta_\delta({\bf k_1}) \delta_\delta({\bf k_3}) \rangle
\langle  \delta_g({\bf k_2}) \delta_\Pi({\bf k_5}) \rangle
\langle  \delta_g({\bf k_4}) \delta_\Pi({\bf k_6}) \rangle \nonumber
\\
&+& \langle  \delta_\delta({\bf k_1}) \delta_\delta({\bf k_3}) \rangle
\langle  \delta_g({\bf k_2}) \delta_\Pi({\bf k_5})
\delta_g({\bf k_4}) \delta_\Pi({\bf k_6}) \rangle \, . \nonumber \\
\end{eqnarray}

We first discuss the Gaussian piece given in the first line above.
This term, and a permutation,
contributes to the cross-correlation and involves the linear density
field
power spectrum and the non-linear cross-correlation between baryon
and pressure fields. Keeping
track of the permutation, and after several simplifications, we write
\begin{eqnarray}
&& \langle a_{l_1 m_1}^\dsz a_{l_2
m_2}^\dsz a_{l_3 m_3}^{\ast SZ} a_{l_4 m_4}^{SZ} \rangle^\g  \nonumber
\\
&=& \frac{(4 \pi)^8}{9} \int d\rad_1 ... \int
d\rad_4 \int {d^3{\bf k_1 } \over (2\pi)^3}
\int {d^3{\bf k_2 } \over (2\pi)^3}
\int {d^3{\bf k_4 } \over (2\pi)^3}
 \sum_{l_1' m_1' m_1'''}
\sum_{l_2' m_2' m_2'''}  i^{l_1'+l_3+l_4}(-i)^{l_2'+l_3+l_4}
\nonumber \\
&\times& (g\dot{G}G)_{\rad_1} (g\dot{G}G)_{\rad_2}
W^{\rm sz}(\rad_3) W^{\rm sz}(\rad_4)  
 P_{\delta\delta}^\lin(k_1) P_{g\Pi}(k_2)P_{g\Pi}(k_4)
\nonumber \\
&\times&
\frac{j_{l_1'}(k_1\rad_1)}{k_1}j_{l_3}(k_2\rad_1)
\frac{j_{l_2'}(k_1\rad_2)}{k_1}j_{l_4}(k_4\rad_2)
j_{l_3}(k_2\rad_3) j_{l_4}(k_4\rad_4) \nonumber \\
&\times&
Y_{l_1'}^{m_1'}(\hat{\veck}_1) Y_1^{m_1'''}(\hat{\veck}_1)
Y_{l_2'}^{m_2'}(\hat{\veck}_1) Y_1^{m_2'''}(\hat{\veck}_1) \nonumber
\\
&\times& \int d\hat{\bf n}
Y_{l_1}^{m_1\ast}(\hat{\bf n}) Y_{l_1'}^{m_1'\ast}(\hat{\bf n})
Y_{l_3}^{m_3\ast}(\hat{\bf n})
Y_1^{m_1'''\ast}(\hat{\bf n}) \nonumber \\
&\times& \int d\hat{\bf m}
Y_{l_2}^{m_2\ast}(\hat{\bf m}) Y_{l_2'}^{m_2'\ast}(\hat{\bf m})
Y_{l_4}^{m_4\ast}(\hat{\bf m})
Y_1^{m_2'''\ast}(\hat{\bf m}) \, . \nonumber \\
\end{eqnarray}

Following our derivation of the kinetic SZ power spectrum,
we can simplify to obtain
\begin{eqnarray}
&& \langle a_{l_1 m_1}^\dsz a_{l_2
m_2}^\dsz a_{l_3 m_3}^{\ast SZ} a_{l_4 m_4}^{SZ} \rangle^\g 
= \frac{2^3}{\pi^3} \int d\rad_1 ... \int
d\rad_4  \int k_1^2 dk_1
\int k_2^2 dk_2
\int k_4^2 dk_4 \nonumber \\
&\times& \sum_{l_1'}
(g\dot{G}G)_{\rad_1} (g\dot{G}G)_{\rad_2}
W^{\rm sz}(\rad_3) W^{\rm sz}(\rad_4)  
 P_{\delta\delta}^\lin(k_1) P_{g\Pi}(k_2)P_{g\Pi}(k_4)
\nonumber \\
&\times&
\frac{j_{l_1'}'(k_1\rad_1)}{k_1}j_{l_3}(k_2\rad_1)
\frac{j_{l_1'}'(k_1\rad_2)}{k_1}j_{l_4}(k_4\rad_2)
j_{l_3}(k_2\rad_3) j_{l_4}(k_4\rad_4) \nonumber \\
&\times& \int d\hat{\bf n}
Y_{l_1}^{m_1\ast}(\hat{\bf n}) Y_{l_1'}^{m_1'\ast}(\hat{\bf n})
Y_{l_3}^{m_3\ast}(\hat{\bf n}) \int d\hat{\bf m}
Y_{l_2}^{m_2\ast}(\hat{\bf m}) Y_{l_1'}^{m_1'\ast}(\hat{\bf m})
Y_{l_4}^{m_4\ast}(\hat{\bf m})\, . \nonumber \\
\end{eqnarray}

Finally, collecting all terms, we write the Gaussian piece of the
cross-correlation power
between the squared temperatures between thermal SZ and kinetic SZ as
\begin{eqnarray}
&& \langle a_{lm}^{\dsz^2} a_{l' m'}^{\ast SZ^2} \rangle^\g = C_{l}^\g
\delta_{l,l'} \delta{m,m'} \nonumber \\
&=& \sum_{l_1 m_1 l_2 m_2} \sum_{l_3 m_3 l_4 m_4} \sum_{l_1'}
\frac{2^3}{\pi^3} \int d\rad_1 ... \int
d\rad_4 \int k_1^2 dk_1
\int k_2^2 dk_2
\int k_4^2 dk_4 \nonumber \\
&\times&
(g\dot{G}G)_{\rad_1} (g\dot{G}G)_{\rad_2}
W^{\rm sz}(\rad_3) W^{\rm sz}(\rad_4) 
 P_{\delta\delta}^\lin(k_1) P_{g\Pi}(k_2)P_{g\Pi}(k_4)
\nonumber \\
&\times&
\frac{j_{l_1'}'(k_1\rad_1)}{k_1}j_{l_3}(k_2\rad_1)
\frac{j_{l_1'}'(k_1\rad_2)}{k_1}j_{l_4}(k_4\rad_2)
j_{l_3}(k_2\rad_3) j_{l_4}(k_4\rad_4) \nonumber \\
&\times& \int d\hat{\bf n}
Y_{l_1}^{m_1\ast}(\hat{\bf n}) Y_{l_1'}^{m_1'\ast}(\hat{\bf n})
Y_{l_3}^{m_3\ast}(\hat{\bf n}) \int d\hat{\bf m}
Y_{l_2}^{m_2\ast}(\hat{\bf m}) Y_{l_1'}^{m_1'\ast}(\hat{\bf m})
Y_{l_4}^{m_4\ast}(\hat{\bf m})\nonumber \\
&\times& \int d\hat{\bf n} Y_l^{\ast m}(\hat{\bf n})
Y_{l_1}^{m_1}(\hat{\bf n})
Y_{l_2}^{\ast m_2}(\hat{\bf n}) 
 \int d\hat{\bf m} Y_{l'}^{m'}(\hat{\bf m})  Y_{l_3}^{\ast
m_3}(\hat{\bf m})
Y_{l_4}^{m_4}(\hat{\bf m}) \, . \nonumber \\
\end{eqnarray}

The last four integrals lead to a term that is
\begin{eqnarray}
&&\sum_{m_1 m_2 m_3 m_4}
\left(
\begin{array}{ccc}
l_1 & l_1' & l_3 \\
m_1 & m_1'  &  m_3
\end{array}
\right)
\left(
\begin{array}{ccc}
l_2 & l_1' & l_4 \\
m_2 & m_1'  &  m_4
\end{array}
\right)
\left(
\begin{array}{ccc}
l & l_1 & l_2 \\
m & m_1  &  m_2
\end{array}
\right)\nonumber \\
&&\left(
\begin{array}{ccc}
l' & l_3 & l_4 \\
m' & m_3  &  m_4
\end{array}
\right)  = \frac{\delta_{l l'} \delta_{m m'}}{2l+1}
\left\{
\begin{array}{ccc}
l_4 & l_2 & l \\
l_1 & l_3  &  l_1'
\end{array}
\right\}  \nonumber \\
\end{eqnarray}

Thus, simplifying we obtain
\begin{eqnarray}
&& C_{l}^\g = \sum_{l_1 l_2 l_3 l_4 l_1'}
\frac{\prod_{i=1}^{4}(2l_i+1)(2l_1'+1)}{4\pi} \nonumber \\
&\times& \left(
\begin{array}{ccc}
l_1 & l_1' & l_3 \\
0 & 0  &  0
\end{array}
\right) \left(
\begin{array}{ccc}
l_2 & l_1' & l_4 \\
0 & 0  &  0
\end{array}
\right) \left(
\begin{array}{ccc}
l & l_1 & l_2 \\
0 & 0  &  0
\end{array}
\right)  \left(
\begin{array}{ccc}
l & l_3 & l_4 \\
0 & 0  &  0
\end{array}
\right) \left\{
\begin{array}{ccc}
l_4 & l_2 & l \\
l_1 & l_3  &  l_1'
\end{array}
\right\}  \nonumber \\
&\times& \frac{2^3}{\pi^3} \int d\rad_1 ... \int
d\rad_4 \int k_1^2 dk_1
\int k_2^2 dk_2
\int k_4^2 dk_4 \nonumber \\
&\times&
(g\dot{G}G)_{\rad_1} (g\dot{G}G)_{\rad_2}
W^{\rm sz}(\rad_3) W^{\rm sz}(\rad_4) 
 P_{\delta\delta}^\lin(k_1) P_{g\Pi}(k_2)P_{g\Pi}(k_4)
\nonumber \\
&\times&
\frac{j_{l_1'}'(k_1\rad_1)}{k_1}j_{l_3}(k_2\rad_1)
\frac{j_{l_1'}'(k_1\rad_2)}{k_1}j_{l_4}(k_4\rad_2)
j_{l_3}(k_2\rad_3) j_{l_4}(k_4\rad_4) \, . \nonumber \\
\end{eqnarray}
Note that there is an additional term, due to a permutation, which
involves by interchanging $l_3$ and $l_4$ (with $l_1'$).

Similar to the Limber approximation used with the derivation of the
kinetic SZ power spectrum, we can integrate over Bessel functions and
simplify to obtain
\begin{eqnarray}
&& C_{l}^\g = \sum_{l_1 l_2 l_3 l_4 l_1'}
\frac{\prod_{i=1}^{4}(2l_i+1)(2l_1'+1)}{4\pi} \nonumber \\
&\times& \left(
\begin{array}{ccc}
l_1 & l_1' & l_3 \\
0 & 0  &  0
\end{array}
\right) \left(
\begin{array}{ccc}
l_2 & l_1' & l_4 \\
0 & 0  &  0
\end{array}
\right) \left(
\begin{array}{ccc}
l & l_1 & l_2 \\
0 & 0  &  0
\end{array}
\right)  \left(
\begin{array}{ccc}
l & l_3 & l_4 \\
0 & 0  &  0
\end{array}
\right) \left\{
\begin{array}{ccc}
l_4 & l_2 & l \\
l_1 & l_3  &  l_1'
\end{array}
\right\}  \nonumber \\
&\times& \frac{2}{\pi} \int \frac{d\rad_1}{d_A^2} \int
\frac{d\rad_2}{d_A^2} \int k_1^2 dk_1 
(g\dot{G})_{\rad_1} (g\dot{G})_{\rad_2}
W^{\rm sz}(\rad_1) W^{\rm sz}(\rad_2)  \nonumber \\
&\times& P_{\delta\delta}^\lin(k_1)
P_{g\Pi}\left(\frac{l_3}{d_A};\rad_1\right)
P_{g\Pi}\left(\frac{l_4}{d_A};\rad_2\right)
\frac{j_{l_1'}'(k_1\rad_1)}{k_1}
\frac{j_{l_1'}'(k_1\rad_2)}{k_1}\, . \nonumber \\
\end{eqnarray}

The non-Gaussian piece takes a similar form.  After introducing the
trispectrum of pressure-pressure-baryon-baryon fluctuations and the
power spectrum of velocity correlations, we write
\begin{eqnarray}
&& \langle a_{l_1 m_1}^\dsz a_{l_2 m_2}^\dsz a_{l_3 m_3}^{\ast SZ}
a_{l_4 m_4}^{SZ} \rangle^\ngau
=\frac{(4 \pi)^8}{9} \int d\rad_1 ... \int
d\rad_4 \nonumber \\
&\times& \int {d^3{\bf k_1 } \over (2\pi)^3}
\int {d^3{\bf k_2} \over (2\pi)^3}
\int {d^3{\bf k_4} \over (2\pi)^3}
\int {d^3{\bf k_5} \over (2\pi)^3}
\int {d^3{\bf k_6} \over (2\pi)^3} \nonumber \\
&\times& \sum_{l_1' m_1' l_1'' m_1'' m_1'''}
\sum_{l_2' m_2' l_2'' m_2'' m_2'''}
i^{l_1'+l_3+l_4}(-i)^{l_2'+l_3+l_4} 
(g\dot{G}G)_{\rad_1} (g\dot{G}G)_{\rad_2}
W^{\rm sz}(\rad_3) W^{\rm sz}(\rad_4)  \nonumber \\
&\times& P_{\delta\delta}^\lin(k_1) (2\pi)^3
T_{\Pi g \Pi g}(\veck_2,\veck_4,\veck_5,\veck_6)
\delta_D(\veck_2+\veck_4+\veck_5+\veck_6) \nonumber \\
&\times& \frac{j_{l_1'}(k_1\rad_1)}{k_1}j_{l_1''}(k_2\rad_1)
\frac{j_{l_2'}(k_1\rad_2)}{k_1}j_{l_2''}(k_4\rad_2)
j_{l_3}(k_2\rad_3) j_{l_4}(k_4\rad_4) \nonumber \\
&\times& Y_{l_1'}^{m_1'}(\hat{\veck}_1) Y_1^{m_1'''}(\hat{\veck}_1)
Y_{l_2'}^{m_2'}(\hat{\veck}_1) Y_1^{m_2'''}(\hat{\veck}_1) 
 Y_{l_1''}^{m_1''}(\hat{\veck}_2)
Y_{l_2''}^{m_2''}(\hat{\veck}_4)
Y_{l_3}^{m_3}(\hat{\veck}_5) Y_{l_4}^{m_4}(\hat{\veck}_6) \nonumber \\
&\times& \int d\hat{\bf n}
Y_{l_1}^{m_1\ast}(\hat{\bf n}) Y_{l_1'}^{m_1'\ast}(\hat{\bf n})
Y_{l_1''}^{m_1''\ast}(\hat{\bf n})
Y_1^{m_1'''\ast}(\hat{\bf n}) \nonumber \\
&\times& \int d\hat{\bf m}
Y_{l_2}^{m_2\ast}(\hat{\bf m}) Y_{l_2'}^{m_2'\ast}(\hat{\bf m})
Y_{l_2''}^{m_2''\ast}(\hat{\bf m})
Y_1^{m_2'''\ast}(\hat{\bf m}) \, . \nonumber \\
\end{eqnarray}

To simplify, we expand the delta function associated with the
trispectrum in to two separate triangular parts:
\begin{eqnarray}
&&\delta_D(\veck_2+\veck_4+\veck_5+\veck_6)
\nonumber \\
&=& \int \frac{d^3\veck'}{(2\pi)^3} (2\pi)^3
\delta_D(\veck_2+\veck_4+\veck')  \delta_D(\veck_5+\veck_6-\veck')
\nonumber \\
&=&\int \frac{d^3\vecx_1}{(2\pi)^3}
e^{i \vecx_1 \cdot \left( \veck_2+\veck_4+\veck' \right)}
\int \frac{d^3\vecx_2}{(2\pi)^3}
e^{i \vecx_2 \cdot \left( \veck_5+\veck_6-\veck' \right)} \,
. \nonumber \\
\end{eqnarray}
The assumption here is that through this vector expansion, the
vectorial representation of the quadrilateral formed by  the
trispectrum can be expressed through
a vectorial configuration of two triangles involving two sides and
the diagonal, respectively. With this, the trispectrum is expressed to
be dependent only
on the magnitude of the vectors $\veck_2,\veck_4,\veck_5,\veck_6$ and
not on their directions.
Using the Rayleigh expansion (Eq.~\ref{eqn:Rayleigh}) in above, we
simplify
to obtain
\begin{eqnarray}
&& \langle a_{l_1 m_1}^\dsz a_{l_2
m_2}^\dsz a_{l_3 m_3}^{\ast SZ} a_{l_4 m_4}^{SZ} \rangle^\ngau
= \frac{(4 \pi)^{14}}{9} \int d\rad_1 ... \int
d\rad_4 \sum_{l_1' m_1' l_1'' m_1'' m_1'''}
\sum_{l_2' m_2' l_2'' m_2'' m_2'''} \sum_{L M}
\nonumber \\
&\times& \int {d^3{\bf k_1 } \over (2\pi)^3}
\int {d^3{\bf k_2} \over (2\pi)^3}
\int {d^3{\bf k_4} \over (2\pi)^3}
\int {d^3{\bf k_5} \over (2\pi)^3}
\int {d^3{\bf k_6} \over (2\pi)^3}
\int {d^3{\bf k'} \over (2\pi)^3} \nonumber \\
&\times&
i^{l_1'+l_3+l_4+L}(-i)^{l_2'+l_3+l_4-L} i^{l_1''+l_2''+l_3+l_4}
\nonumber \\
&\times& (g\dot{G}G)_{\rad_1} (g\dot{G}G)_{\rad_2}
W^{\rm sz}(\rad_3) W^{\rm sz}(\rad_4)  
 P_{\delta\delta}^\lin(k_1)
T_{\Pi g \Pi g}(k_2,k_4,k_5,k_6)
\nonumber \\
&\times&
\frac{j_{l_1'}(k_1\rad_1)}{k_1}j_{l_1''}(k_2\rad_1)
\frac{j_{l_2'}(k_1\rad_2)}{k_1}j_{l_2''}(k_4\rad_2)
j_{l_3}(k_2\rad_3) j_{l_4}(k_4\rad_4) \nonumber \\
&\times& \int x_1^2 dx_1 j_{l_1''}(k_2 x_1) j_{l_2''}(k_4 x_1)
j_{L}(k' x_1)  \int x_2^2 dx_2 j_{l_3}(k_5 x_2) j_{l_4}(k_6 x_2)
j_{L}(k' x_2) \nonumber \\
&\times&
Y_{l_1'}^{m_1'}(\hat{\veck}_1) Y_1^{m_1'''}(\hat{\veck}_1)
Y_{l_2'}^{m_2'}(\hat{\veck}_1) Y_1^{m_2'''}(\hat{\veck}_1) \nonumber
\\
&\times&  \int d\hat{\bf x_1}
Y_{l_1''}^{m_1''}(\hat{\bf x_1}) Y_{l_2''}^{m_2''}(\hat{\bf x_1})
Y_{L}^{M}(\hat{\bf x_1}) 
 \int d\hat{\bf x_2}
Y_{l_1''}^{m_1''}(\hat{\bf x_2}) Y_{l_2''}^{m_2''}(\hat{\bf x_2})
Y_{L}^{M}(\hat{\bf x_2})  \nonumber \\
&\times& \int d\hat{\bf n}
Y_{l_1}^{m_1\ast}(\hat{\bf n}) Y_{l_1'}^{m_1'\ast}(\hat{\bf n})
Y_{l_1''}^{m_1''\ast}(\hat{\bf n})
Y_1^{m_1'''\ast}(\hat{\bf n}) \nonumber \\
&\times& \int d\hat{\bf m}
Y_{l_2}^{m_2\ast}(\hat{\bf m}) Y_{l_2'}^{m_2'\ast}(\hat{\bf m})
Y_{l_2''}^{m_2''\ast}(\hat{\bf m})
Y_1^{m_2'''\ast}(\hat{\bf m}) \, . \nonumber \\
\end{eqnarray}

Following our derivation of the kinetic SZ power spectrum, and employing the
Limber approximation on the Bessel functions,
we can simplify further and write the non-Gaussian piece of the
correlation
between the squared temperatures of SZ thermal and SZ kinetic effects
as
\begin{eqnarray}
&&\langle a_{lm}^{\dsz^2} a_{l' m'}^{\ast SZ^2} \rangle^\ngau =
C_{l}^\ngau
\delta_{l,l'} \delta{m,m'} \nonumber \\
&=& \frac{2}{\pi}\int \frac{d\rad}{d_A^6} \int k_1^2 dk_1 \sum_{l_1'
m_1' l_1''\
 m_1''}
\sum_{l_2'' m_2'' L M} \left[(g\dot{G})W^{\rm sz}(\rad)
\right]^2\nonumber \\
&\times& P_{\delta\delta}^\lin(k_1) T_{\Pi g \Pi
g}\left(\frac{l_1''}{d_A},
\frac{l_2''}{d_A},\frac{l_3}{d_A},\frac{l_4}{d_A};\rad\right)
\left(\frac{j'_{l_1'}(k_1\rad)}{k_1}\right)^2 \nonumber \\
&\times& \int d\hat{\bf x_1}
Y_{l_1''}^{m_1''}(\hat{\bf x_1}) Y_{l_2''}^{m_2''}(\hat{\bf x_1})
Y_{L}^{M}(\hat{\bf x_1})  \int d\hat{\bf x_2}
Y_{l_1''}^{m_1''}(\hat{\bf x_2}) Y_{l_2''}^{m_2''}(\hat{\bf x_2})
Y_{L}^{M}(\hat{\bf x_2})  \nonumber \\
&\times& \int d\hat{\bf n}
Y_{l_1}^{m_1\ast}(\hat{\bf n}) Y_{l_1'}^{m_1'\ast}(\hat{\bf n})
Y_{l_1''}^{m_1''\ast}(\hat{\bf n})
 \int d\hat{\bf m}
Y_{l_2}^{m_2\ast}(\hat{\bf m}) Y_{l_1'}^{m_2'\ast}(\hat{\bf m})
Y_{l_2''}^{m_2''\ast}(\hat{\bf m}) \nonumber \\
&\times& \int d\hat{\bf n} Y_l^{\ast m}(\hat{\bf n})
Y_{l_1}^{m_1}(\hat{\bf n})
Y_{l_2}^{\ast m_2}(\hat{\bf n}) \int d\hat{\bf m} Y_{l'}^{m'}(\hat{\bf m})  Y_{l_3}^{\ast
m_3}(\hat{\bf m})
Y_{l_4}^{m_4}(\hat{\bf m}) \, .\nonumber \\
\end{eqnarray}

Following simplifications used in the case of the Gaussian part, we
find
\begin{eqnarray}
&& C_{l}^\ngau = \sum_{l_1 l_2 l_3 l_4 l_1' l_1'' l_2''}
\frac{\prod_{i=1}^4(2l_i+1)(2l_1'+1)(2l_1''+1)(2l_2''+1)}{(4\pi)^3}
\nonumber \\
\
&\times& \left(
\begin{array}{ccc}
l_1 & l_1' & l_1'' \\
0 & 0  &  0
\end{array}
\right) \left(
\begin{array}{ccc}
l_2 & l_1' & l_2'' \\
0 & 0  &  0
\end{array}
\right) \left(
\begin{array}{ccc}
l & l_1 & l_2 \\
0 & 0  &  0
\end{array}
\right) \left(
\begin{array}{ccc}
l & l_1'' & l_2'' \\
0 & 0  &  0
\end{array}
\right) \left(
\begin{array}{ccc}
l & l_3 & l_4 \\
0 & 0  &  0
\end{array}
\right)^2 \nonumber \\
&\times& \left\{
\begin{array}{ccc}
l_1 & l_1' & l \\
l_2'' & l_1''  &  l_2
\end{array}
\right\}  \frac{2}{\pi}\int \frac{d\rad}{d_A^6} \left[(g\dot{G})W^{\rm
sz}(\rad) \right]^2 \nonumber \\
&\times& T_{\Pi g \Pi g}\left(\frac{l_1''}{d_A},
\frac{l_2''}{d_A},\frac{l_3}{d_A},\frac{l_4}{d_A};\rad\right)
\int k_1^2 dk_1  P_{\delta\delta}^\lin(k_1)
\left(\frac{j'_{l_1'}(k_1\rad)}{k_1}\right)^2 \, . \nonumber \\
\end{eqnarray}

Note that the total contribution to the thermal SZ$^2$-kinetic SZ$^2$
power
spectrum is
\begin{equation}
C_l^{2 \dsz-\sz} = C_l^\g + C_l^\ngau \, .
\end{equation}
Since the full calculation of the squared power spectra is
computationally time consuming, we make several simplifications as
outline in the next subsection. These simplifications make use of the
fact that at small angular scales, we can utilize flat-sky
approximations and that at the same scales, the velocity field is
completely independent of the baryon field itself.

\subsection{Flat-sky approach}
Similarly, we can derived the temperature squared power spectrum of
kinetic and thermal SZ effects in the flat-sky limit. In the same
limit, we also take the velocity field of the kinetic SZ effect to be
independent of the baryon fluctuations. Following our previous
definitions, we define the flat sky temperature squared power spectrum
as
\begin{equation}
\langle \cmb^{2\dsz}(\vecl) \cmb^{2\sz}(\vecl') \rangle = (2\pi)^2
\delta_D(\vecl+\vecl') C_l^{2\dsz-\sz} \, ,
\end{equation}
where the Fourier transform of the squared temperature can be
written as a convolution of the temperature transforms
\begin{equation}
\cmb^2(\vecl) = \int \frac{d\vecl_1}{(2\pi)^2}
\cmb(\vecl_1)\cmb(\vecl-\vecl_1) \, .
\end{equation}
Here, it should be understood that $\cmb^2(\vecl)$ refers to the
Fourier transform of the square of the temperature rather than square
of the
Fourier transform of temperature. We will denote the latter as
$[\cmb(\vecl)]^2$. To compute the square of the SZ thermal and
SZ kinetic temperature power spectrum, we take
\begin{eqnarray}
&&\langle \cmb^{2 \dsz}(\vecl) \cmb^{2 \sz}(\vecl') \rangle = (2\pi)^2
\delta_D(\vecl+\vecl') C_l \nonumber \\
&=& \int \frac{d\vecl_1}{(2\pi)^2}\int \frac{d\vecl_2}{(2\pi)^2}
\langle \cmb^\dsz(\vecl_1)\cmb^\dsz(\vecl-\vecl_1)
\cmb^\sz(\vecl_2)\cmb^\sz(\vecl'-\vecl_2) \rangle \, .\nonumber \\
\label{eqn:flat}
\end{eqnarray}
Note that the Fourier transform of
the temperature fluctuations in the flat sky is
\begin{equation}
\cmb(\vecl) = \int d^2\theta e^{-i\vecl \cdot \theta} T(\theta) \, .
\end{equation}

In the small scale limit where the density field is separated from the
velocity field, we can write the above cumulant again in two parts
with one involving power spectrum and another with the trispectrum.
We first write down the Gaussian-like piece as
\begin{eqnarray}
&& \langle \cmb^\dsz(\vecl_1)\cmb^\dsz(\vecl-\vecl_1)
\cmb^\sz(\vecl_2)\cmb^\sz(\vecl'-\vecl_2) \rangle^\g =  \nonumber \\
&& \int \frac{d\rad_1}{d_A^4} \int \frac{d\rad_2}{d_A^4}
\left[g(\rad_1) W^\sz(\rad_2) \right]^2
\frac{1}{3} v_{\rm rms}^2 \nonumber \\
&\times& \int \frac{dk_1}{(2\pi)} \int \frac{dk_2}{(2\pi)}
e^{ik_1 \rad_1} e^{ik_2 \rad_2}
\left\langle \delta_g\left[\frac{\vecl_1}{d_A},k_1\right]
\Pi\left[\frac{\vecl_2}{d_A},k_2\right] \right\rangle \nonumber \\
&\times& \int \frac{dk_3}{(2\pi)} \int \frac{dk_4}{(2\pi)}
e^{ik_3 \rad_1} e^{ik_4 \rad_2}
\left\langle \delta_g\left[\frac{\vecl-\vecl_1}{d_A},k_3\right]
\Pi\left[\frac{\vecl'-\vecl_2}{d_A},k_4\right] \right\rangle \,
,\nonumber \\
\end{eqnarray}
where we have taken $\langle (\hat{\bf{\theta}} \cdot {\bf v})
(\hat{\bf{\theta}}' \cdot {\bf v'}) \rangle \sim 1/3 v_{\rm rms}^2$
with $1/3$ coming from the fact that only a third of the velocity
component contribute to the line of sight rms.
We can now introduce the power spectra in above
correlators such that
\begin{eqnarray}
&& \langle \cmb^\dsz(\vecl_1)\cmb^\dsz(\vecl-\vecl_1)
\cmb^\sz(\vecl_2)\cmb^\sz(\vecl'-\vecl_2) \rangle^\g = \nonumber \\
&& \int \frac{d\rad_1}{d_A^4} \int \frac{d\rad_2}{d_A^4}
\left[g(\rad_1) W^\sz(\rad_2) \right]^2
\frac{1}{3}v_{\rm rms}^2 \nonumber \\
&\times& \int \frac{dk_1}{(2\pi)} e^{ik_1 \left(\rad_1 -
\rad_2\right)}
(2\pi)^2\delta_D\left(\frac{\vecl_1}{d_A}+\frac{\vecl_2}{d_A}\right)
P_{g\Pi}\left[\sqrt{\frac{l_1^2}{\rad_1^2}+k_1^2}\right] \nonumber \\
&\times&\int \frac{dk_3}{(2\pi)} e^{ik_3 \left(\rad_1 - \rad_2\right)}
(2\pi)^2\delta_D\left(\frac{\vecl-\vecl_1}{d_A}+\frac{\vecl'-\vecl_2}{d_A}\right)
P_{g\Pi}\left[\sqrt{\frac{|l-l_1|^2}{\rad_1^2}+k_3^2}\right]\,
,\nonumber \\
\end{eqnarray}
and the integrals
over the line-of-sight wavevectors behave such that only perpendicular
Fourier modes contribute to the projected field, such that
$l^2/d_A^2 \gg k^2$.  This is the so-called
Limber approximation (\cite{Lim54} 1954). Doing the integral over the
wavevector, then, results in a delta function in $(\rad_1-\rad_2)$
such
that only contributions come from the same redshift.
Putting the correlator back in the power spectrum equation
(Eq.~\ref{eqn:flat}), we now get
\begin{eqnarray}
&& C_{l}^\g = \int \frac{d^2\vecl_1}{(2\pi)^2} \int
\frac{d\rad}{d_A^4}
(g\dot{G})^2 W^{\rm sz}(\rad)^2 2 P_{g\Pi}\left(\frac{l_1}{d_A};\rad\right)
P_{g\Pi}\left(\frac{|\vecl-\vecl_1|}{d_A};\rad\right)\frac{1}{3}v_\rms^2
\, ,\nonumber \\
\end{eqnarray}
where we have introduced a factor of 2 account for the additional
permutation
involved in the baryon density-pressure correlation.

Similarly,  the non-Gaussian piece follows as
\begin{eqnarray}
&& C_{l}^\ngau = \int \frac{d^2\vecl_1}{(2\pi)^2} \int
\frac{d^2\vecl_2}{(2\pi)^2} \int \frac{d\rad}{d_A^6}
(g\dot{G})^2 W^{\rm sz}(\rad)^2 \frac{1}{3}v_\rms^2 \nonumber \\
&\times& T_{g \Pi g \Pi}\left[\left(\frac{l_1}{d_A}\right),
\left(\frac{|\vecl-\vecl_1|}{d_A}\right),
\left(\frac{l_2}{d_A}\right),\left(\frac{|\vecl+\vecl_2|}{d_A}\right);\rad\right]\,.\nonumber \\
\end{eqnarray}
Since we only use the single halo term to calculate $T_{g \Pi g\Pi}$,
the arguments are simply scalars and does not dependent on the
orientation of the quadrilateral. In general, however, the trispectrum
depends on the length of the four sides plus the orientation of at
least one of the diagonals.

Comparing the flat sky power spectra written above and the ones
derived under the all-sky assumption, we note that the two are related
under similar approximations as the ones suggested for comparison
between the all-sky and flat-sky kinetic SZ power spectra. For
example, for the Gaussian piece,
we can obtain the correspondence
by setting $l_1 \sim l_4$ and $l_2 \sim l_3$,
and further simplifying to separate
\begin{eqnarray}
&& C_{l}^\g = \sum_{l_3 l_4} \frac{(2l_3+1)(2l_4+1)}{4\pi}
\left(
\begin{array}{ccc}
l & l_3 & l_4 \\
0 & 0  &  0
\end{array}
\right)^2 \nonumber \\
&\times& \int \frac{d\rad}{d_A^4} (g\dot{G}G)^2
W^{\rm sz}(\rad)^2 P_{g\Pi}\left(\frac{l_3}{d_A}\right)
P_{g\Pi}\left(\frac{l_4}{d_A}\right) \frac{1}{3}v_\rms^2 \,
. \nonumber \\
\end{eqnarray}
This requires the sum
\begin{eqnarray}
&&\sum_{l_1 l_2 l_1'} (2l_1+1)(2l_2+1)(2l_1'+1)
\left(
\begin{array}{ccc}
l_1 & l_1' & l_2 \\
0 & 0  &  0
\end{array}
\right)^2  \left\{
\begin{array}{ccc}
l_1 & l_2 & l \\
l_1 & l_2  &  l_1'
\end{array}
\right\}  [j'_{l_1'}(k\rad)]^2 = \frac{1}{3} \, .\nonumber \\
\end{eqnarray}
We note here that the above relation and the one suggested in
Eq.~\ref{eqn:simplify} agree if we set the index associated with the
Bessel function to either $l_1$ or $l_2$ such that
the $l_1'$ sum reduces to
\begin{equation}
\sum_{l_1'} (2l_1'+1) \left\{\begin{array}{ccc}
l_1 & l_2 & l \\
l_1 & l_2  &  l_1'
\end{array}
\right\}  =1
\end{equation}
when $l_1,l_2,l$ satisfy the triangular condition. This condition is,
in fact,
imposed by the Wigner-3$j$ symbol in above.

Finally, the correspondence between the flat sky
$\vecl$, $\vecl_1$ and all sky $l,l_3,l_4$ can be noted by
introducing $\vecl_3 = \vecl-\vecl_1$ and expanding the delta function
of $\delta(\vecl-\vecl_1-\vecl_3)$ to obtain a Wigner-3$j$ squared
symbol in $l,l_1,l_3$ which correspond to $l,l_3,l_4$ in the flat sky
formulation. We refer the reader to
\cite{Hu00b} (2000b) for further details in such a reduction.
Similarly, we can obtain the correspondence between the flat-sky
non-Gaussian expression and the all-sky non-Gaussian expression
through a simplification as above for the Wigner-6$j$ symbol.
To obtain the correspondence between $\vecl,\vecl_1,\vecl_2$ in
flat-sky and $l,l_1,l_2,l_3,l_4$ we can introduce $\vecl_3 =
\vecl-\vecl_1$ and $\vecl_4=\vecl+\vecl_2$ to the flat sky expression
and break the delta function
formed by the flat sky trispectrum involving
$\vecl_1+\vecl_2+\vecl_3+\vecl_4$
to be two parts involving two sides and the diagonal formed by the
$\vecl$ and expand that as suggested above from \cite{Hu00b} (2000b).
Following simplifications used in the case of the Gaussian part, we
find the non-Gaussian piece to be
\begin{eqnarray}
&& C_{l}^\ngau = \sum_{l_1 l_2 l_3 l_4}
\frac{\prod_{i=1}^4(2l_i+1)}{(4\pi)^2}
\left(
\begin{array}{ccc}
l & l_1 & l_2 \\
0 & 0  &  0
\end{array}
\right)^2 \left(
\begin{array}{ccc}
l & l_3 & l_4 \\
0 & 0  &  0
\end{array}
\right)^2 \nonumber \\
&\times& \int \frac{d\rad}{d_A^6} \left[(g\dot{G})W^{\rm
sz}(\rad) \right]^2 T_{\Pi g \Pi g}\left(\frac{l_1''}{d_A},
\frac{l_2''}{d_A},\frac{l_3}{d_A},\frac{l_4}{d_A};\rad\right)
\frac{1}{3}v_{\rm rms}^2\, .\nonumber \\
\end{eqnarray}

\begin{figure}[!h]
\begin{center}
\includegraphics[width=3.8in]{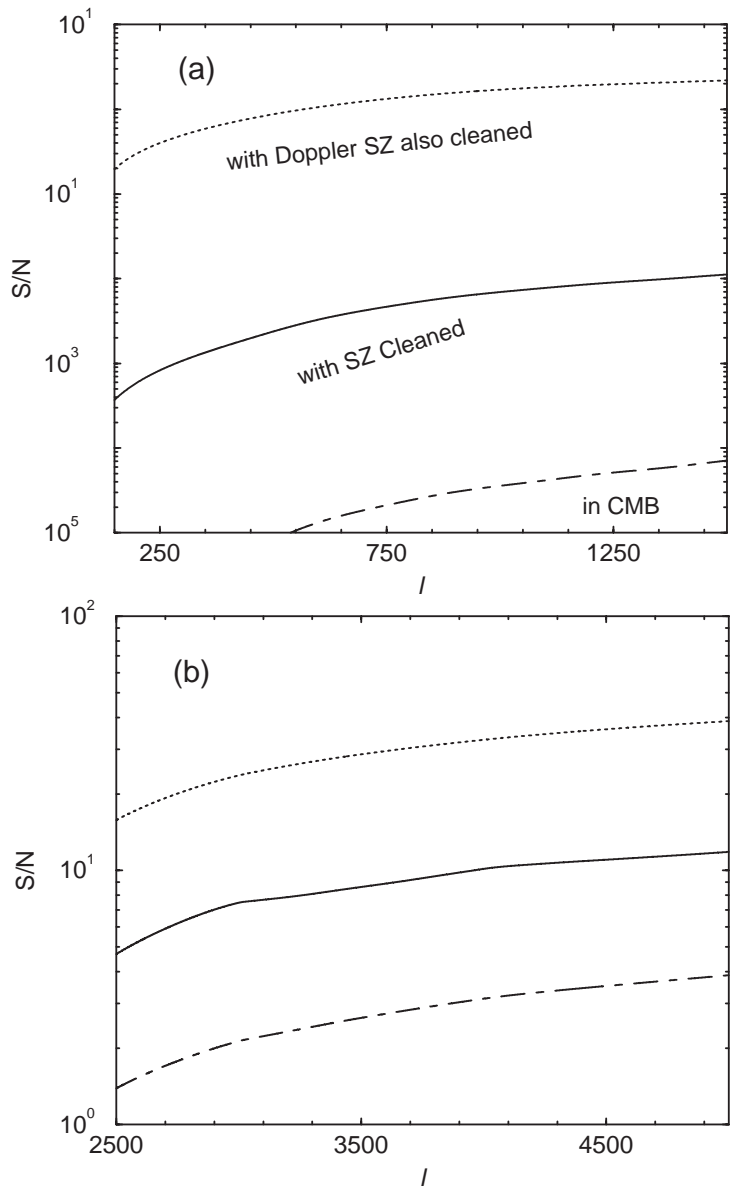}
\end{center}
\caption[Signal-to-noise for the squared power spectrum for a large
and small angular scale experiment]{
The cumulative signal-to-noise for the detection of the
thermal SZ-kinetic SZ squared temperature power spectrum. In (a), we consider
a large angular scale experiment. The cumulative
signal-to-noise, even with a perfectly cleaned SZ map,
is significantly less than 1.  In (b), we show the cumulative
signal-to-noise for a small angular scale experiment;
There is now adequate signal-to-noise.}
\label{fig:sz2ov2planck}
\end{figure}

\subsection{Signal-to-Noise}

In order to calculate the possibility for a detection of the thermal
SZ$^2$-kinetic SZ$^2$ power spectrum, we calculate the signal-to-noise
for the detection in couple of experimental possibilities. First, we
need to covariance of the estimator involved with the measurement of
the squared power spectrum:
\begin{equation}
\hat{C}_l^{2\dsz-\sz} = \frac{A_f}{(2\pi)^2} \int
\frac{d^2\vecl}{A_\shell}\cmb^{2\dsz}(\vecl)
\cmb^{2\sz}(-\vecl) \, .
\end{equation}
Here, $A_\shell = \int d^2\vecl$  is the area in the two-dimensional
 shell in
Fourier space over which the integral is done and $A_f$ is the total
area of
the survey in Fourier space and can be written as $A_f =
(2\pi)^2/\Omega$ with a total survey area on the sky of $\Omega$.
Following \cite{Zal00} (2000), we can write down the covariance of our
estimator as
\begin{eqnarray}
&&\Cov\left[\left(\hat{C}_l^{2\dsz-\sz}\right)^2\right] 
= \frac{A_f}{A_\shell}
\left[\left(C_l^{2\dsz-\sz}\right)^2 + C_l^{2\dsz-\dsz}
C_l^{2\sz-\sz}\right] \, , \nonumber \\
\label{eqn:sz2ov2cov}
\end{eqnarray}
where $C_l^{2\dsz-\dsz}$ is the squared power spectrum of kinetic SZ
and thermal SZ while $C_l^{2\dsz-\dsz}$ and $C_l^{2\sz-\sz}$ are the
squared power spectra of kinetic SZ and thermal SZ respectively.
Here, we assume that squared fields are Gaussian.
To calculate $C_l^{2\dsz-\dsz}$ and $C_l^{2\sz-\sz}$ we make several
assumptions: We assume that the temperature squared power spectrum
will be measured using two maps involving frequency separated SZ
contribution (which will have $\cmb^\sz+\cmb^\noise$) and a map with
kinetic SZ contribution with the CMB primary component, such that the
will be composed of $\cmb^{\rm primary}+\cmb^\dsz+\cmb^{\noise '}$.
Following such a separation we can write the $C_l^{2\sz-\sz}$ and
$C_l^{2\dsz-\dsz}$ as
\begin{eqnarray}
&&\langle \cmb^{2\sz}(\vecl) \cmb^{2\sz}(\vecl') \rangle = (2\pi)^2
\delta_D(\vecl+\vecl') C_l^{2\sz-\sz} \nonumber \\
&=& \int \frac{d\vecl_1}{(2\pi)^2}\int \frac{d\vecl_2}{(2\pi)^2}
\langle \cmb^\sz(\vecl_1)\cmb^\sz(\vecl-\vecl_1)
\cmb^\sz(\vecl_2)\cmb^\sz(\vecl'-\vecl_2) \rangle \, .\nonumber \\
&=& \int \frac{d\vecl_1}{(2\pi)^2} \left [ 2 C_{l_1}^\sz
C_{|\vecl-\vecl_1|}^\sz + \int \frac{d\vecl_2}{(2\pi)^2}
T^\sz(\vecl_1,\vecl-\vecl_1,\vecl_2,-\vecl-\vecl_2)\right] \, ,
\nonumber \\
\label{eqn:conv}
\end{eqnarray}
where the contributions now come from a Gaussian piece involving the
SZ power spectra and a non-Gaussian piece through the SZ
trispectrum. Since the primary component fluctuations dominate the
kinetic  SZ temperature, and that there is no measurable trispectrum
for this
component under current adiabatic CDM predictions we ignore any
non-Gaussian contribution to $C_l^{2\dsz-\dsz}$, and write it as the
one with the Gaussian piece in above. This assumption is also safe at
small angular scales when $\cmb^\dsz > \cmb^{\rm primary}$ since the
kinetic SZ effect, under our halo description, is dominated by the
large scale correlations and not the single halo term out to $l
\sim 10^4$.

In the Eq.~\ref{eqn:sz2ov2cov}, the ratio of $A_\shell/A_f$ is the
total number of modes that measures the squared power
spectrum independently and can be approximated such that $A_\shell/A_f
=
f_\sky(2l+1)$. To calculate the signal-to-noise involved in the
detection of the squared temperature power spectrum, we consider an
optimized estimator with weighing factor $W_l$ such that
\begin{equation}
\hat{Y} = \sum_l W_l \hat{C}_l^{2\dsz-\sz} \, ,
\end{equation}
and write the signal-to-noise as
\begin{equation}
\sn =
\left[\frac{\langle\hat{Y}\rangle^2}{\Cov(\hat{Y}^2)}\right]^{1/2} \,
.
\end{equation}
The weight $W_l$ that maximizes the signal-to-noise is $W_l =
C_l^{2\dsz-\sz}/\Cov[(C_l^{2\dsz-\sz})^2]$ (\cite{Zal00} 2000) and we
can write the required signal-to-noise as
\begin{eqnarray}
&&\sn = \left[ f_{\rm sky} \sum_l (2l+1)
\frac{\left(C_l^{2\dsz-\sz}\right)^2}{
\left(C_l^{2\dsz-\sz}\right)^2 + C_l^{2\dsz-\dsz}
C_l^{2\sz-\sz}}\right]^{1/2} \, .\nonumber \\
\end{eqnarray}

\section{Discussion}

In Fig.~\ref{fig:sz2ov2}(a), we show  the power spectrum of squared
temperatures for the SZ thermal and SZ kinetic effects using the halo
term. Here, we have separated the Gaussian and non-Gaussian
contribution  to the squared power spectrum. As shown, the
non-Gaussian contribution to the power spectrum is significantly
higher than the Gaussian contributions. 

The Gaussian contribution to the squared power spectrum traces the
pressure-baryon density field power spectrum, which is shown in
Fig.~\ref{fig:gaspressurepower}(a) using the halo model. In the same
figure, for comparison, we also show the pressure-pressure power
spectrum and the correlation coefficient for pressure-baryon with
respect to the pressure-pressure and baryon-baryon power spectra. The
correlation behaves such that pressure and baryons trace each other
at very large scales while the correlation is decreased at small
scales due to the turn over in the pressure power spectrum. This is
equivalent to the statement that there is no low mass halo
contribution to the pressure power spectrum; these halos continue to
contribute to the baryon density field power spectrum. 

The non-Gaussian contribution to the thermal SZ-kinetic SZ squared
temperature power spectrum traces the trispectrum formed by pressure
and density field. We show this in Fig.~\ref{fig:gaspressurepower}(b)
following the halo model. For comparison, we also show the trispectrum
formed by pressure alone in the same figure. The pressure-baryon
trispectrum  is such that at large scales, corresponding to linear
scales, significant contributions come from the correlations between
halos instead of the single halo term. If there are significant
contributions coming to the squared temperature power spectrum from
such linear scales, the Gaussian part of the power spectrum should
dominate. Since all contributions to the
squared temperature power spectrum comes from small angular scales
corresponding to non-linear scales in the pressure-baryon
trispectrum, we only use the single halo contribution in calculating
the non-Gaussian part of  the squared temperature power spectrum.
In both Gaussian and non-Gaussian parts of the power spectrum,
the velocity field of the halos are taken to be the large scale bulk flows through
the linear theory.

In order to assess the maximum possibility for a measurement of the
temperature squared power spectrum involving kinetic SZ and thermal
SZ, here, we ignore the detector and beam noise contributions to the
covariance. Also, we assume full-sky experiments with $f_\sky=1$.

As  written in Eq.~\ref{eqn:conv}, the contribution to the
covariance comes as a convolution in Fourier space. Thus, even at
small angular scales corresponding to high multipoles,
noise contributions can come from large angular scales or low
wavelength modes. Such modes do not have any signal and by only
contributing to the variance, they can reduce the
effective signal-to-noise in the measurement. Since the squared power
spectrum effectively peaks at multipoles of $\sim$ $10^4$, we can
essentially ignore any contribution to the signal, as well as noise,
from multipoles less than few thousand. These are the same multipoles
in which the CMB primary anisotropies dominate, thereby, increasing
the effective noise for the measurement. In order to remove the low
multipoles,  we introduce a filtering scheme to the spherical or
Fourier transform of the temperature anisotropy measurements and
suppress the low multipole data such that the filter essentially acts
as a high pass filter above some $l > l_{\rm min}$.

In Fig.~\ref{fig:sz2ov2}(b), we show the cumulative signal to noise
for the measurement of the thermal SZ-kinetic SZ squared temperature
power spectrum. Here, we have assumed an experiment, with no
instrumental noise, such that information is only used in multipoles
of 2000 to 10000. We find significant signal-to-noise for the
detection of the squared temperature power spectrum, especially when
using a frequency cleaned SZ map with a CMB map, which has no SZ
contribution. One can in fact use the CMB map itself especially if
multifrequency information is not available for SZ separation. Since an
experiment with multipolar information out to $l \sim 10^4$ will not
readily be available, we consider two separated realistic cases,
involving a large angular scale experiment, similar to Planck, and a
small angular scale experiment similar to the ones proposed for the
study of SZ effect. We summarize our results in
Fig.~\ref{fig:sz2ov2planck}. As shown in (a), an experiment only
sensitive to multipolar information ranging from 100 to 1500 does not
have any signal-to-noise for a detection of the squared power
spectrum. With a perfect SZ separated map in this multipolar range,
the cumulative signal-to-noise for the squared power spectrum is in the
order of $\sim$ 0.01. Thus, it is unlikely that Planck data will be useful for
this study. Since we have not included any instrumental noise in
calculating signal-to-noise, the realistic signal-to-noise for Planck
would be even lower. 

Going to smaller angular scales, we find that the
signal increases significantly such that an experiment only sensitive
to the range of $l \sim 2000$ to 5000 has adequate signal-to-noise for
a detection of the squared power spectrum. A multifrequency experiment
in the arcminute scales can use its frequency  cleaned SZ map to
cross-correlate with the CMB map and obtain the squared power spectrum
with a cumulative signal-to-noise of order few tens. A comparison to
Fig.~\ref{fig:sz2ov2}(b), suggests that going to lower scales beyond
5000 increases the signal-to-noise, and this is due to the fact that
SZ thermal-SZ kinetic squared power spectrum peaks at multipoles of
$\sim$ 7000 to 8000, suggesting that for an optimal detection of the
squared power spectrum, one should also include observations out to
such high multipoles. The increasing activity in the experimental
front suggest that such possibilities will soon be available.

\chapter{Angular Power Spectrum of Dark Matter Halos as a Cosmological Test}

\section{Introduction}

A significant number of observational attempts are now underway to
image the large scale structure of the universe out to a redshift of a
few. These wide-field surveys involving many tens to thousands of
square degrees of angular area on the sky include  some of the unique
experimental attempts in astronomy: the ongoing Sloan Digital 
Sky Survey\footnote{http://www.sdss.org}, 
which measures the distribution of galaxies, the planned weak
gravitational lensing shear (see, review by \cite{BarSch00} 2000)
observations with dedicated instruments such as
the Large Synoptic Survey Telescope (LSST; \cite{TysAng00} 2000) and
the Sunyaev-Zel'dovich (SZ; \cite{SunZel80} 1980) 
effect on the temperature fluctuations of the
cosmic microwave background (CMB) at the combined BIMA/OVRO  array
(CARMA; John Carlstrom, private communication) and with other instruments
(e.g., BOLOCAM at CSO; Andrew Lange, private communication).  
In addition to measurement of properties involved with each of these
effects, such as galaxy power spectrum in the case of Sloan or 
the shear correlations in the case of lensing, these surveys
also probe the underlying dark matter distribution of the local
universe that define the large scale structure.

Under the context of cold dark matter (CDM) models for structure
formation, the dark matter distribution within and between halos has
properties that have been intensely studied through analytical
techniques and numerical simulations. In particular, analytical
descriptions now exist for the abundance (e.g., Press-Schechter mass
function; \cite{PreSch74} 1974), profile (e.g., Navarro-Frenk-White
profile; \cite{Navetal96} 1996) and correlations of
halos of a given mass. Under this halo approach to non-linear
clustering, the underlying dark
matter distribution of the large scale structure can be described
through halos and these halos trace the linear density field power
spectrum with a mass and redshift dependent bias given by the halo
bias model of \cite{MoWhi96} (1996). Using such a halo description, 
several authors have now shown how to construct the
clustering properties of the large scale structure  dark matter and
other physical properties (e.g., \cite{Sel00} 2000; \cite{Cooetal00b}
2000b; \cite{MaFry00a} 2000a; \cite{Scoetal00} 2000; \cite{Coo00} 2000).

In \cite{Cooetal00b} (2000b), \cite{CooHu01a} (2001a) and
\cite{CooHu01b} (2001b), we studied the statistical properties,
mainly the power spectrum, bispectrum and the trispectrum, of weak
gravitational lensing convergence of large scale structure through
halos.  In these papers, we have shown how to construct the
statistical  measurements of lensing through properties of the dark
matter
distribution within halos and large scale correlations between
halos. The
upcoming wide-field weak lensing observations, through galaxy shear
data, will essentially measure properties studied in these
papers. In addition to statistical properties of weak lensing,
the wide-field maps created in the data reduction process
is expected to contain additional information such as
the spatial distribution of dark matter halos (see, e.g.,
\cite{KruSch99} 1999). Similarly, with galaxy imaging data in surveys
such as Sloan, one can use the spatial distribution and surface
density of galaxies to construct halo catalogs (e.g., \cite{Kepetal98}
1998).

Thus, it is expected that in wide-field images from current and 
upcoming surveys,
one can identify and produce catalogs of  halos responsible
for large scale clustering. In fact, with some effects such as the SZ,
one should be able to readily identify the halos; 
in the case of the SZ effect, 
this is due to fact that large scale pressure power
spectrum is mainly due to the massive halos in the universe with
highest temperature electrons (see, \cite{Coo00} 2000). Thus, 
instead of measuring correlation function of the effect
itself, one can use the number counts of detected SZ halos
as a probe of cosmology (e.g., \cite{Haietal00} 2000).
In addition to the number counts,
important cosmological and astrophysical information can also
be obtained through the spatial distribution of halos. 

Making an additional use of the spatial distribution of halos in
wide-field
surveys, we propose the measurement of the comoving angular diameter
distance
through the angular power spectrum associated with clustering.
As in many other classical tests of cosmology, the standard ruler of
the proposed test is the shape of the linear power spectrum, which
is what will be measured most accurately
with upcoming CMB anisotropy observations.
With adequate redshift information for halos, such that halo redshift
distribution can be binned reliably, we show how one can essentially
measure the angular diameter distance to each of the bins through
information related to projected  linear density power spectrum.
Under a cosmological description of the distance, and the evolution of
growth, one can determine
parameters such as an equation of state for an additional energy
density component.

Our proposed cosmological test can be performed easily with upcoming
wide-field
surveys and will provide an additional use  of the data.
To a large extent, the method is independent of
assumptions such as the mass function which is necessary in many
cosmological
tests related to number counts. To fully exploit the method, however,
a proper understanding  of halo bias is needed.
Since dark matter halos are involved, necessary information on bias
comes directly from N-body simulations. The method can be performed
with sources such as galaxies, though, fully understanding galaxy bias
can be a
problem due to uncertainties associated with galaxy formation
etc. Also, since more than one galaxy is present in halos,
redshifts for halos can be accurately obtained through photometric
techniques than for an individual galaxy.
Here, we make
the assumption that halo bias is scale independent, though, we will
investigate the effect of a scale dependence  and
suggest that a sharp
variation in the halo bias from constant linear value at large scales
to a scale-dependent value, say due to the onset of on-linearities,
produce
a sharp feature in the halo angular power spectrum; such features
increase  the ability of the halo angular
power spectrum as a probe of cosmology. Also, features such as the
oscillations due to baryons provide a direct method to calibrate the
distance scale independent of other measures of distance. Thus, the
technique has the added advantage that it may provide absolute
measures of distance.

In the next section, we make  predictions for the halo angular power
and show that it can be detected in upcoming wide-field surveys.
For the illustration of
our results, we take a $\Lambda$CDM cosmology with parameters
$\Omega_m=0.35$ for the matter density, $\Omega_b=0.05$ for the baryon
density, $\Omega_\Lambda=0.65$ for the cosmological constant, $h=0.65$
for the dimensionless Hubble constant, and a scale-invariant spectrum
of primordial fluctuations, normalized to galaxy cluster abundances
($\sigma_8=0.9$; see, \cite{ViaLid99}).
For the linear power spectrum, we take the
fitting formula for the transfer function given in \cite{EisHu99}
(1999). We use Press-Schechter
(PS; \cite{PreSch74} 1974) based theory to define halo redshift
distribution. For the illustration of results, we
assume a survey of 4000 degrees$^2$ with a
sensitivity  to a minimum halo mass of $10^{14}$ M$_{\sun}$;
The survey area and threshold mass are consistent with planned
upcoming lensing and SZ effect surveys (for discussions, see
\cite{KruSch99} 1999; \cite{Holetal00} 2000; Joffre et al. in
preparation).

\begin{figure}[!h]
\begin{center}
\includegraphics[width=4in,angle=-90]{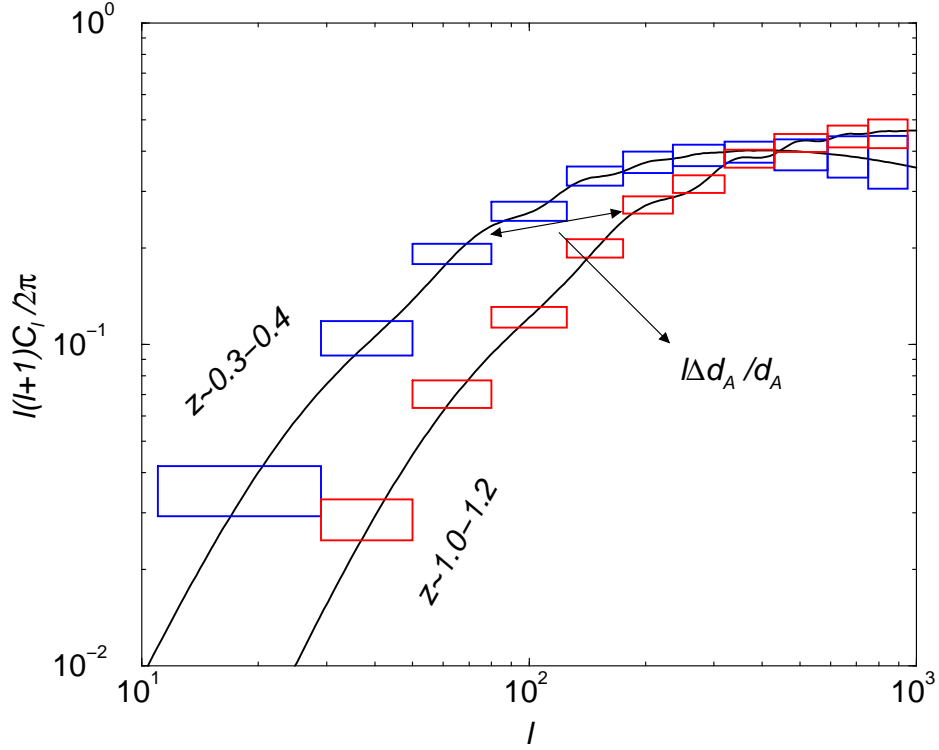}
\end{center}
\caption[Angular power spectrum of halos]{
Angular power spectrum of halos in a wide-field survey
in redshift bins of 0.3 to 0.4 and 1.0 and 1.2.
The binned errors assume a survey of 4000 deg$^2$, consistent with
upcoming weak lensing and SZ surveys. The angular power spectrum
at high redshifts is shifted towards the right essentially
by the relative change in the comoving
angular diameter distance. The oscillations in the angular power
spectra are
due to baryons in our fiducial cosmological model as calculated
numerically with CMBFAST (\cite{SelZal96} 1996). }
\label{fig:halo}
\end{figure}

\section{Angular Power Spectrum}

The halo angular power spectrum probes the statistical properties of
the halo density field on the sky which is a weighted projection of
the matter distribution along the line of sight.
We define the angular power spectrum of halos in the flat sky
approximation in the usual way
\begin{equation}
\left< n_h(\bfl_1)n_h(\bfl_2)\right> =
        (2\pi)^2 \delta_\dirac(\bfl_1+\bfl_2) C_l \, .
\end{equation}
Through radial distance projection, we can write (e.g., \cite{Kai92}
1992)
\begin{equation}
C_l = \int d\rad {W^2(\rad) \over d_A^2} P_{hh}\left(\frac{l}{d_A};
\rad\right)\
 \,.
\label{eqn:cl}
\end{equation}
Here, $\rad$ is the comoving radial distance, $d_A$ is the comoving
angular
diameter distance\footnote{Note
that we use the comoving coordinates throughout. There is an
additional factor of $(1+z)$ involved between our definition of $d_A$
and the one commonly called angular diameter distance.
The comoving angular diameter distance has
also been called the angular size distance by \cite{Pee93} (1993) and
the transverse comoving distance  by \cite{Hog99} (1999).}, and
$W(\rad)$ is the normalized radial distribution of the halos.
In deriving Eq.~\ref{eqn:cl}, we have utilized the Limber
approximation (\cite{Lim54} 1954) by
setting $k=l/d_A$.

Assuming that halos trace the linear density field,
we can write the large scale halo power spectrum as
\begin{equation}
P_{hh}(k;\rad) = \left< b_M \right>^2(\rad) P^\lin(k;\rad)
\end{equation}
where the mass-averaged bias is
\begin{equation}
\left< b_M \right>(z) = \frac{1}{\bar{n}_h(z)}
      \int_{M=M_{\rm min}}^{M_{\rm max}} dM \frac{dn(M,z)}{dM}
b_h(M,z)
\end{equation}
with the mean number density of halos, as a function of
redshift, given  by $\bar{n}_h(z) = \int dM
dn(M,z)/dM$.  The halo bias parameters follow from \cite{MoWhi96}
(1996) and \cite{Moetal97} (1997):
\begin{equation}
b_h(M,z) = 1+ \frac{\left[\nu^2(M,z) - 1\right]}{\delta_c}\, ,
\end{equation}
where $\nu(M,z) = \delta_c/\sigma(M,z)$ is the peak-height threshold.
$\sigma(M,z)$ is the rms fluctuation within a top-hat filter at the
virial radius corresponding to mass $M$, and $\delta_c$ is the
threshold overdensity of spherical collapse. Useful fitting functions
and additional information on these quantities could be found in
\cite{Hen00} (2000). The halo mass distribution as a function of
redshift $[dn(M,z)/dM]$ is determined through the Press-Schechter (PS;
\cite{PreSch74} 1974) formalism. 

 Using bias and the fact that in linear
theory  the  density
field is simply scales to higher redshifts through  the growth
function $G(z)$ where $\delta^\lin(k;\rad) = G(\rad)\delta^\lin(k;0)$
(\cite{Pee80} 1980), we can write the angular power spectrum of
halos between redshift $z_i$ and $z_i+\Delta z_i$ as
\begin{equation}
C_l^i = \int dz W_i(z)^2 F_i(z) P^\lin\left(\frac{l}{d_A^i}\right) \,,
\label{eqn:cz}
\end{equation}
where the function $F_i(z)$, associated with redshift bin $i$,
now contains information related to
 halo bias, growth related to linear evolution,
the power spectrum normalization, and terms
related to 3-d to 2-d projection (such as a $1/d_A^2$ term; see,
Eq.~\ref{eqn:cl}).

As written in Eq.~\ref{eqn:cz},
we can now understand how the halo angular power spectrum
allows a determination of the angular diameter distance. Here,
$W_i(z)$
involves the redshift distribution of halos and this information comes
straight from observations while $F_i(z)$ contains all other unknowns
associated
with bias, and thus, the halo mass function, and power spectrum
normalization.
Our main interest involves the effect of 3-d density field
power spectrum and its projection to 2-d angular power spectrum
through the Limber form involving
$k=l/d_A^i$. For a given $k$, since $l$ increases as $d_A$ is
increased, we propose the measurement of angular diameter distance
$d_A^i$
through the relative shift in the halo
angular power spectrum in multipole, or Fourier, space.

\subsection{Theoretical Expectations}

In order to discuss the extent to which halos can be used to measure
$d_A$, we use simple predictions based on the  PS
description of halo mass function $[dn(M,z)/dM]$.
We use the linear mass-averaged bias, $\left<b_M\right>(z)$,
following the description from \cite{MoWhi96}
(1996). Since information related to bias is contained within
$F_i(z)$, its
exact value is not important since we construct both the angular
diameter distance $d_A^i$ and $F_i(z)$. The exact
properties of bias will be necessary, say, if one were to separately
study bias from linear evolution of growth.  We will later use
reasonable priors on bias to estimate errors on
$w$, the equation of state of the dark-energy component, using both
comoving angular diameter distance and linear evolution of growth.

Following previous studies on the abilities of large structure
observations as a probe of cosmology
(e.g., \cite{HuTeg99} 1999), we can write the  error on the
determination of the  angular power spectrum of halos as
\begin{equation}
\Delta C_l = \sqrt{2 \over (2l +1)\fsky}\left(C_l + C_l^\sn\right) \,,
\label{eqn:delta}
\end{equation}
where $\fsky = \Theta_{\rm deg}^2 \pi/129600 $ is the fraction of the
sky covered by a survey of dimension $\Theta_{\rm deg}$ in degrees and
$C^\sn$ is the shot-noise power spectrum of halos.
Here, we have made the assumption that halo angular power spectrum
covariance is described by Gaussian statistics. This is a reasonable
assumption to take since we are considering the large scale
clustering of
halos and that we do not expect non-Gaussianities due to a
trispectrum
to dominate covariance at such large scales (see,
\cite{CooHu01b} (2001b) for a discussion of non-Gaussian contribution
to
the covariance in the case of weak gravitational lensing).

The shot-noise  power spectrum is given by the surface-density of
halos $C_l^\sn \equiv 1/\bar{N}$, which will be available from
observations. For calculational purposes here, we again use PS theory
to calculate $\bar{N}$
\begin{equation}
\bar{N} = \int dz \frac{d^2V}{d\Omega dz} \left[ \int dM
\frac{dn(M,z)}{dM} \right] \, .
\end{equation}
Though we calculate the shot-noise term using PS mass function, it is
simply the surface density of halos observed in the survey.
The shot noise power varies with the mass thresholds detectable in a
survey, but in all interesting cases, sets the $l$ at which shot noise
from the finite number of halos becomes important.
The first term is simply the sampling error assuming Gaussian
statistics
for the underlying field and makes the fractional errors of order
unity
at the scale of the survey $l \sim 2\pi/\Theta_{\rm deg}$.

In Fig.~\ref{fig:halo}, we illustrate the measurement of the
angular diameter distance. The two curves show the halo angular power
spectra in two arbitrarily chosen redshift bins of 0.3 to 0.4 and 1.0
to 1.2. The curve corresponding to the higher redshift bin is clearly
shifted relative to the one for the lower redshift bin.
Note that the oscillatory features corresponding to baryons in
our fiducial model is horizontally shifted and this shift simply
corresponds
to the relative change in the
angular diameter distance. In addition to specific features
such as oscillations, which may not be easily
detectable, one can use the broad feature involving the turnover of
the angular halo power spectrum for the purpose of distance
determination.  Since the scale of baryon oscillations will
be known in Mpc$^{-1}$ from CMB and will be observed through
projection in
$h$ Mpc${^-1}$ with halos, one can use clustering to absolutely
calibrate the distance scale through a precise measurement of $h$.
Since there may be complications associated with bias
and the shot-noise removal, we ignore cosmological information
contained in features and only concentrate on the use of overall
shape.

\begin{figure}[!h]
\begin{center}
\includegraphics[width=4.2in,angle=0]{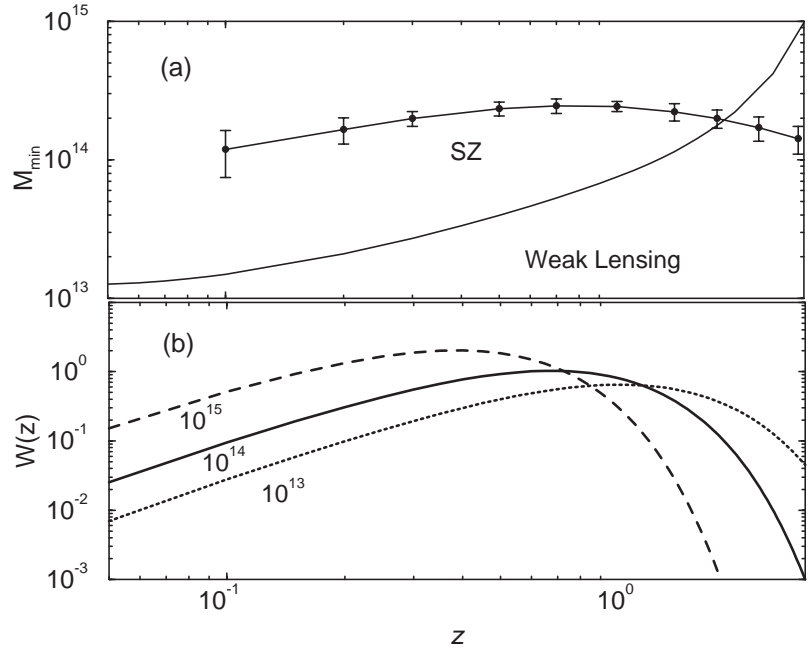}
\end{center}
\caption[The halo redshift distribution]{
(a) The limiting mass for a weak lensing survey (Joffre et
al. in preparation) and for a SZ survey (\cite{Holetal00} 2000). For
the
purpose of this calculation, we take a constant minimum mass of
$10^{14}$ M$_{\sun}$ out to a redshift of 2.
In (b), for reference we show the redshift distribution of
halos, as calculated using PS mass function, with minimum masses as
noted.}
\label{fig:dcl}
\end{figure}

\begin{figure}[!h]
\begin{center}
\includegraphics[width=4in,angle=0]{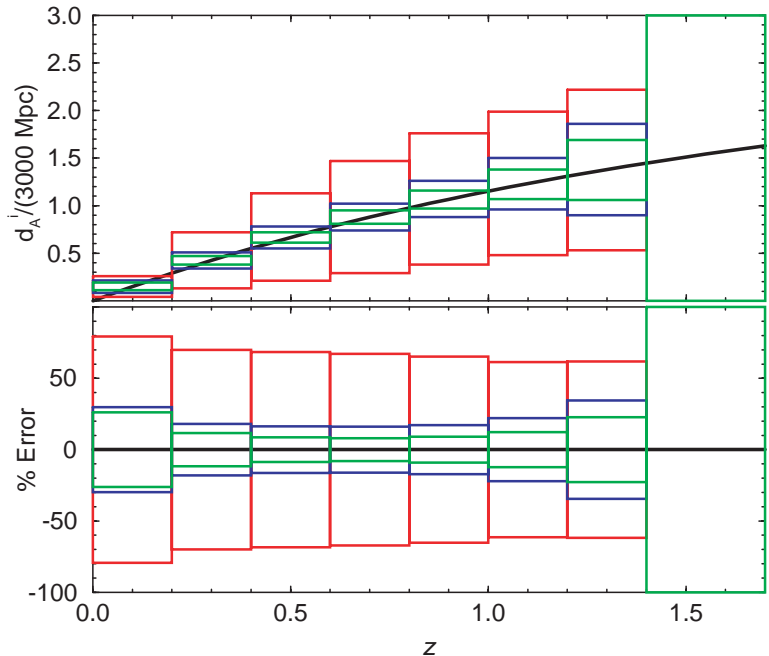}
\end{center}
\caption[Errors on Comoving Angular Diameter Distance]{
(a) The error on angular diameter distance as a function of redshift.
We have binned the halos to 9 redshift bins between 0 and
2. The larger
errors are with no prior assumption on the cosmological parameters
that define the transfer function while the smaller errors are with
weak and strong priors (see text). We have taken $l_{\rm max}$ of
1000.
In (b), we show relative errors on distance.}
\label{fig:da}
\end{figure}

\section{Parameter Estimation}

To estimate how well halo clustering can recover certain cosmological
and astrophysical parameters, we construct the Fisher matrix:
\begin{equation}
{\bf F}_{\alpha \beta} = -\left< \partial^2 \ln L \over \partial 
p_\alpha \partial p_\beta \right>_{\bf x} \,,
\end{equation}
where $L$ is the likelihood of observing a data set ${\bf x}$
given the true parameters
$p_1 \ldots p_n$.
With equation~(\ref{eqn:delta}), the Fisher
matrix for halo power becomes
\begin{equation}
{\bf F}_{\alpha \beta} = \sum_{i=1}^{{\rm N_{\rm bins}}}
\sum_{l=l_{\rm min}}^{l_{\rm max}}
        {\fsky (l + 1/2) \over (C_l^i + C_\sn^i)}
        {\partial C_l^i \over \partial p_\alpha}
        {\partial C_l^i \over \partial p_\beta}\,,
\label{eqn:Fisher}
\end{equation}
with the $i$ sum representing the sum over the total number of redshit
bins.
Since the variance of an unbiased estimator of a
parameter $p_\alpha$ cannot be less than $({\bf F}^{-1})_{\alpha
\alpha}$,
the Fisher matrix
quantifies the best statistical errors on parameters possible with a
given
data set.

We choose $l_{\rm min}=2\pi/\Theta_{\rm deg}$ when evaluating
equation~(\ref{eqn:Fisher}) as it corresponds roughly to the
survey size.
The precise value does not matter for parameter estimation due to the
increase in sample variance on the survey scale.  We choose a
value for $l_{\rm max}$ where $C_l^\sn = C_l$, which generally ranges
from 200 at low redshift bins to 400 at high redshift bins.
At low redshifts, this cut off is slightly in the non-linear regime
though at redshifts greater than 0.8 or so, one is well within the
linear regime. We also use a fix $l_{\rm max}$ of 1000 in all bins, in
order to
understand the gain in information going to such a high $l$.
Using information out to
such a high multipole value for cosmological purposes will require a
detailed understanding of shot-noise substraction.

Due to the dependence on the linear density field power spectrum,
all cosmological parameters that change the shape of the power
spectrum across the scales probed by halos also affect the
measurement of distance. The shape of the transfer
function is determined by $\Omega_mh^2$ and $\Omega_bh^2$ while the
overall tilt is determined by a scalar tilt $n_s$ which we defined to
be around a fixed scale.  Note that these are  the cosmological
parameters that will be easily determined from CMB anisotropy
observations.
When estimating expected errors on distance, we will consider two sets
of
priors on these cosmological parameters: a prior set
consistent with expected errors from MAP missions with $\sigma(\ln
\Omega_m h^2)=0.2$ and $\sigma(n_s)$= 0.11,
and a prior set consistent with Planck mission
with $\sigma(\ln \Omega_m h^2)=0.064$ and $\sigma(n_s)$=0.041
(see, \cite{Eisetal00} 2000).
In constructing the equation of state for an dark energy, $w$, we
make use of the information present in both $d_A$ and $G$ by
separating $F(z)$ to bias, growth, amplitude of the normalization and
information related to radial projection.
Since we do not consider information present in baryon oscillations,
 our predictions for angular diameter distance or associated cosmology
is not strongly sensitive to $\Omega_b h^2$.

\begin{figure}[t]
\begin{center}
\includegraphics[width=3.8in,angle=0]{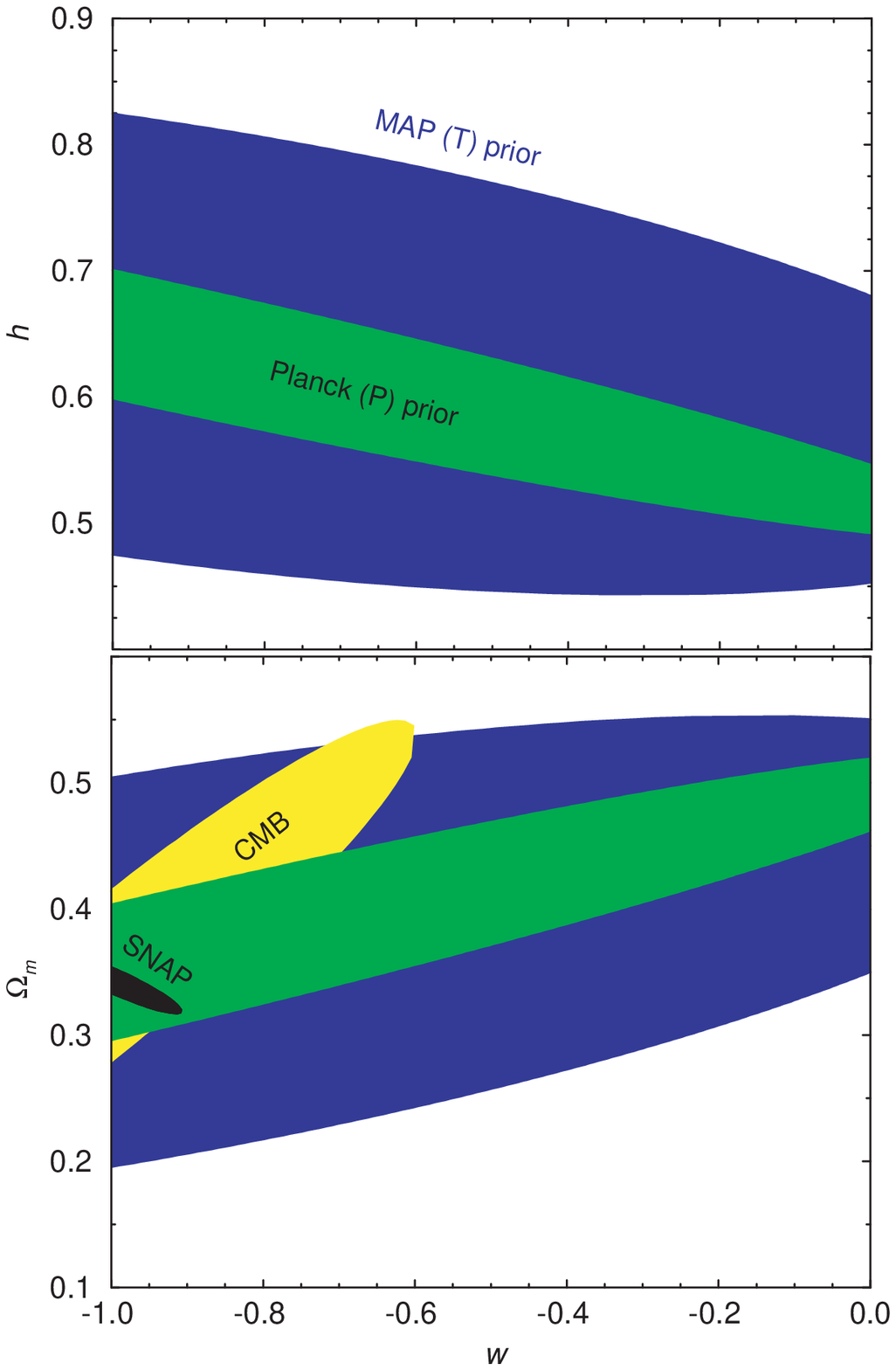}
\end{center}
\caption[Error on $w-h$ and $w-\Omega_m$]{
(a) The error on parameter $w$, the equation of state for the
additional energy density, and $h$ and (b) $w$ and $\Omega_m$. 
We use distance information only.
Here, we e show errors for halos with priors following MAP Temp and
Planck Temp+Pol.In (b), for reference,
we also show errors on $\Omega_m$ and $w$ from CMB (involving
Planck temperature and polarization) and Type Ia SNe with SNAP
mission.}
\label{fig:w1}
\end{figure}

\begin{figure}[t]
\begin{center}
\includegraphics[width=3.8in,angle=-90]{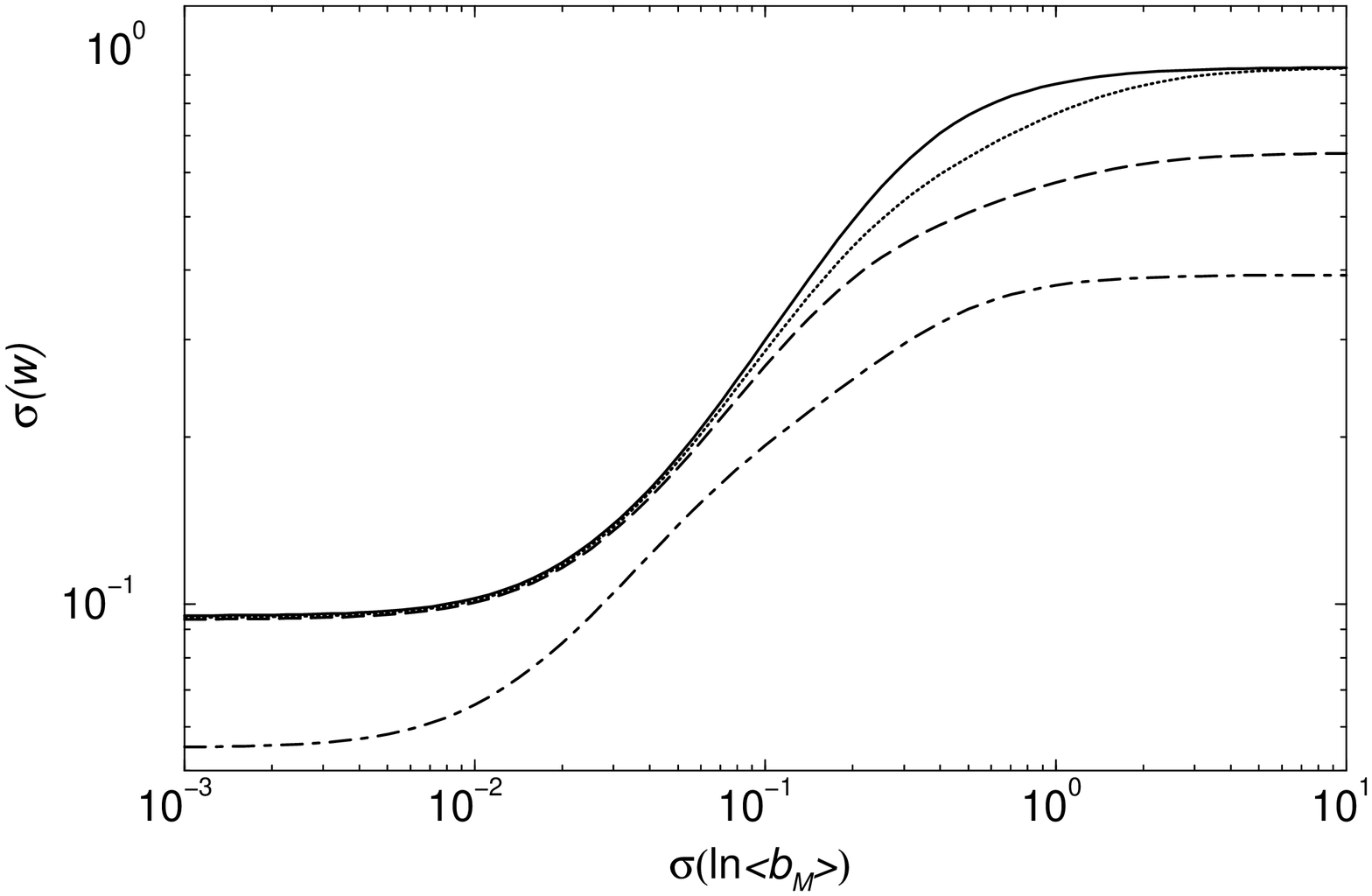}
\end{center}
\caption[Error on $w$ as a function of prior on bias]
{The 1-$\sigma$ error on $w$ as a function of the prior
on the produce of normalization and bias. The four curves assume
Planck (Pol) priors on $\Omega_mh^2$, $n_s$ and $\Omega_bh^2$ to
define the linear power spectrum. The soild line is with no prior on
$\ln A$, and $h$. The dotted line includes a prior of 0.2 in $\ln A$,
while the long dashed line is with an addition prior of 0.1 in $h$.
The dot-dashed line is an highly optimistic scenario with  exact
$A$, a prior of 0.1 in $h$ and  using halo angular power spectrum
information out to $l_{\rm max}$ of 1000.}
\label{fig:wg}
\end{figure}

\section{Results \& Discussion}

As the first step in distance construction through halo clustering,
we consider a non-parametric approach and study the possibility for
a measurement of comoving angular diameter
distance $d_A^i$ to each redshift bin $i$. In addition to the distance
measurement, we also estimate errors
on $F_i(z)$, the normalization, which includes details on bias, growth
etc. The non-parametric approach to construct $d_A^i$ is such that
it is independent of a particular cosmological model
(e.g., open or flat)  or the form of the additional
energy density at low redshifts, such as a cosmological constant.
At each redshift bin, we consider a
mean comoving angular diameter distance value valid for that bin
$\left< d_A^i \right>$ and replace this mean value in
Eq.~\ref{eqn:cz}.
This mean distance parameter can take any value in each of
the bins, independent of its value in adjacent redshift bins.
Note that the normalization of the power
spectrum, in addition to bias and growth, is included in $F_i(z)$ and
we do not need to specify them separately here.

In Fig.~\ref{fig:da}(a), we show three sets of errors: the largest
errors assume no prior knowledge on the transfer function, while
the smaller errors correspond to priors from MAP (Temp) and Planck
(Pol) respectively. In Fig.~\ref{fig:da}(b), we show the
fractional percentage errors on the distance. The errors in the lowest
bins
tells us how well one can estimate the Hubble constant, while
the slope of $d_A(z)$ around $z \sim 1$ provides information on
cosmology. Though we have parametrized the distance in each of the
redshift bins as an independent quantity, through gloabl parameters
such as $\Omega_m h^2$, the errors on each of these distance estimates
are correlated with each at the 5\% level.

In a realistic cosmology $d_A(z)$ is expected to be sufficiently
smooth so
as to be adequately parameterized by a small number of parameters
at the redshifts in question.  Since $\Omega_m h^2$ is already taken
as
a parameter, we choose the remaining parameters to be the Hubble
constant
$H_0$ and the equation of state of the dark energy $w$ assuming a
flat
Universe.

To compare with other cosmological probes we also show the
constraints
in the $h$-$w$ and $\Omega_m$-$w$ plane in Fig.~\ref{fig:w1}.
The constraints from halos comes directly on the $h$-$w$ plane.
As shown in Fig.~\ref{fig:w1}(b), note that different linear
combinations of $\Omega_m$ and $w$ will be determined by halos
(which
probe $d_A$ at $z\sim 1$) and the CMB (which probes $d_A$ at $z\sim
1000$), which in the case of CMB is shown for the Planck (with
information through polarization also).
Although each constraint alone may not be able to pin down
$w$, the combined region on $\Omega_m$ and $w$ probed by
halos and CMB allow very interesting
constraints even under our conservative assumptions. We bin halos
following the binning scheme in Fig.~\ref{fig:da}, which is somewhar
arbitrarily chosen in equal redshift bin sizes at low redshifts.
We expect that more
information, however, can be gained by optimizing the binning
strategies; we 
relegate these issues to future work.

We now investigate how the relative linear evolution of growth improve
constraints on $\Omega_m$ and $w$. To obtain interesting
constraints, we need to impose the prior on bias,
$\left< b_M \right>$, and the power spectrum normalization, $A$,
separately.
In Fig.~\ref{fig:wg}, we
plot marginalized error on $w$ as a function of the
assumed fractional prior on the quantity of $\ln \left<b_M\right>$
and for various assumptions with regards to how well we will know
$\ln A$ and $h$.
Note that the power spectrum normalization, $A$, results in an overall
scaling, while the bias, which we parametrize by independent values in
each redshift bin, allows for relative variations;  Since $A$ affects
only an overall scaling and cosmological information from growth comes
from
relative variations, the effect of an unknown $A$ is minimal. The
difference between an unknown $A$ and a prior of 0.2 in $\ln A$ only
results in a maximum change of $\sim$ 25\% in the error in $w$.

Finally let us address the issue of the scale-dependence of the
bias.  A scale-dependent bias that can be {\it predicted} in fact
aids in the determination of angular diameter distances: the
scale-dependence
acts as another standardizable ruler for the test.  Indeed, 
the scale-dependence of the bias as a function of halo mass
is something that can be precisely
determined from $N$-body simulations \cite{KraKly99} (1999). 
A more subtle problem is introduced by uncertainties in the mass
threshold
or selection function.  Since the bias is also mass dependent,
this translates into uncertainties
in the predictions for scale-dependence.
To investigate the effects of this sort of uncertainty we can add in
an additional parameter and marginalize its effects.  We take
\begin{equation}
b_M(k,z) = \left< b_M \right>(z) \left[1 + f \left(
\sqrt{\frac{P^{\rm NL}(k;z)}{P^\lin(k;z)}} - 1\right)\right] \, .
\label{eqn:scalebias}
\end{equation}
where $f$ is dimensionless parameter meant to interpolate between
the linear and non-linear mass power spectra $P^{\rm NL}$
\cite{PeaDod96} (1996)
which has an
inflection at the non-linear scale.  Under the halo 
model, a halo power spectrum equal to the non-linear mass power
spectrum
is the extreme limit since there would then be no room for intrahalo
power
associated with their profiles.
Taking a fiducial model with $f=0$ to be conservative, we find
that, in the case with MAP (Temp) priors, error on $w$ increases by
less than ten percent from
0.86 to 0.92.

In addition to the scale-dependent bias, the proposed test can be
affected by any process that changes the shape of the power spectrum
as a function of redshift, e.g. an eV mass neutrino and a running
tilt.  
For neutrinos, while a precise measurement of the linear
power spectrum at low redshifts will help resolve any ambiguities,
we do not expect a running tilt to be a problem since the decade of
scales in the power spectrum
probed by CMB is also what is used for the present test.

In spite of these caveats, it is clear that future surveys which can
identify 
dark matter halos as a function of redshift contain valuable
information
beyond the evolution in their number abundance.  As the theoretical 
modelling of the halo distribution and empirical modelling of the
selection process improves, the correlation function of the halos
can provide not only the angular diameter distance measures
emphasized
here but also direct measures of the growth of large-scale structure
itself.

\appendix
\chapter{Trispectrum Under the Halo Model}
% contents of appendix A

In this appendix, we discuss the derivation of the trispectrum under
the halo model. Even though we limit this discussion to the
trispectrum, this derivation can easily be extended to any n-point
correlation function in Fourier space.
For the purpose of this discussion, we take
the approach presented in \cite{SchBer91} (1991) and write the
connected four-point correlation function in real space and take the
Fourier transform to construct the trispectrum.

First, we write the dark matter is distribution in 
halos such that the density at location $\vecx$ is
$\rho_h(\vecx;M)$ and the overdensity as $y(\vecx;M)$. 
We also take a mass function  for halos given by
$\nm$. The Fourier transform of the dark matter distribution within a
halo of mass $M$ is
\begin{equation}
\hat{\rho}_h(\veck;M) = \int d^3\vecx \rho_h(\vecx; M) e^{-i\veck\cdot
\vecx}
\, ,
\label{eqn:fourier}
\end{equation}
and we take this to be related to the fourier transform of the
overdensity: $y(k,M) = \hat{\rho}_h(\veck;M)/\rho_b$, with background
mean density of $\rho_b$. In addition to dark matter density field,
this description applies to any property
associated with halos, such as pressure. One has
to simply substitute the Fourier transform of the relevant
distribution function for $y(k,M)$ for the property of interest. 
In the rest of the discussion, we will generalize the derivation with
an index $i$ in $y_i$.

In our simplifications, we will make use of  the fact that
\begin{eqnarray}
\delta_D(\veck_{1234}) = \int
\frac{d^3\vecx}{(2\pi)^3} e^{- i \vecx \cdot (\veck_1 + ... +
\veck_4)} \, ,
\end{eqnarray}
where $\veck_{1234} = \veck_1+\veck_2+\veck_3+\veck_4$.

A fundamental assumption in the halo approach to describe non-linear
clustering is that halos themselves are clustered following
fluctuations in the linear density field.  
Using the halo power spectrum, $P_{hh}(k)$, bispectrum,
$B_{hhh}(\veck_1,\veck_2,\veck_3)$ and trispectrum,
$T_{hhhh}(\veck_1,\veck_2,\veck_3,\veck_4)$, 
 we can write the two-point 
\begin{eqnarray}
&&\zeta^2_{hh}(M_1,M_2;\vecx_1,\vecx_2) = \int \frac{d^3\veck^i}{(2\pi)^3} (2\pi)^3 
P_{hh}(k^i;M_1,M_2) e^{i \veck^i \cdot (\vecx_1 - \vecx_2)} \,
\end{eqnarray}
three point
\begin{eqnarray}
&&\zeta^3_{hhh}(M_1,M_2,M_3;\vecx_1,\vecx_2,\vecx_3) = 
\int \frac{d^3\veck^i}{(2\pi)^3} ... \int \frac{d^3\veck^k}{(2\pi)^3}
\nonumber \\
&&B_{hhh}(\veck^i,\veck^j,\veck^k;M_1,M_2,M_3) 
(2\pi)^3 \delta_D(\veck_{ijk}) e^{i (\veck^i \cdot
\vecx_1 + ... + \veck^k \cdot \vecx_3)} \, , \nonumber \\
\end{eqnarray}
and four-point 
\begin{eqnarray}
&&\zeta^4_{hhhh}(M_1,M_2,M_3,M_4;\vecx_1,\vecx_2,\vecx_3,\vecx_4) =
\int \frac{d^3\veck^i}{(2\pi)^3} ... \int
\frac{d^3\veck^l}{(2\pi)^3} 
\nonumber \\
&&
T_{hhhh}(\veck^i,\veck^j,\veck^k,\veck^l;M_1,M_2,M_3,M_4) 
 (2\pi)^3 \delta_D(\veck_{ijkl})
e^{i(\veck^i \cdot \vecx_1 + ... + \veck^l \cdot \vecx_4)} \, , \nonumber \\
\end{eqnarray}
correlation functions of halos in real space. 

Following the arguments presented in  \S~\ref{sec:halomodel}, we
take halos to be clustered following the linear theory, such that
the halo power spectrum 
\begin{eqnarray}
P_{hh}(k;M_1,M_2) = \prod_{i=1}^{2}b_i(M_i)P^\lin(k)  \, ,
\end{eqnarray}
bispectrum
\begin{eqnarray}
&&B_{hhh}(\veck_1,\veck_2,\veck_3;M_1,M_2,M_3) = \nonumber \\
&&\prod_{i=1}^{3}b_i(M_i)
 \left[B^\lin(\veck_1,\veck_2,\veck_3)  
+  \frac{b_2(M_3)}{b_1(M_3)}P^\lin(k_1)P^\lin(k_2)\right]\, , \nonumber \\
\end{eqnarray}
and, trispectrum
\begin{eqnarray}
&&T_{hhhh}(\veck_1,\veck_2,\veck_3,\veck_4;M_1,M_2,M_3,M_4) =
\prod_{i=1}^{4}b_i(M_i) \nonumber \\
&\times& \left[T^\lin(\veck_1,\veck_2,\veck_3,\veck_4)  
+
\frac{b_2(M_4)}{b_1(M_4)}P^\lin(k_1)P^\lin(k_2)P^\lin(k_3)\right]\, ,
\end{eqnarray}
trace the linear theory power spectra and higher order correlations
with an $i$th-order bias $b_i(M)$ for a halo of 
mass $M$ relative to the linear dark matter density field. These
biases could be found in \cite{Moetal97} (1997). The 
linear density field power spectrum comes directly from linear
perturbation theory, while the bispectrum and trispectrum
requires second-order perturbation theory 
(see, e.g., \cite{Fry84} 1984; \cite{Goretal86} 1986).

Given a profile, a mass function, and a description of halo
clustering, we can now write the four-point correlation functions in
real
space. We will then consider the Fourier transform of these
correlations functions to obtain the trispectrum.

\subsection{Real Space Four Point Correlation}

Following \cite{SchBer91} (1991), we can separate contributions to the
four point correlation
function as those arising from one to four halos. Thus,
\begin{eqnarray}
&& \eta(\vecr_1,\vecr_2,\vecr_3,\vecr_4) =
\eta^{1h}(\vecr_1,\vecr_2,\vecr_3,\vecr_4)
+\eta^{2h}(\vecr_1,\vecr_2,\vecr_3,\vecr_4) \nonumber \\
&+&\eta^{3h}(\vecr_1,\vecr_2,\vecr_3,\vecr_4) +
\eta^{4h}(\vecr_1,\vecr_2,\vecr_3,\vecr_4) \,.
\end{eqnarray}
Here, $\vecr_i$ are position vectors from an arbitrary origin.
It should be understood that this is the non-Gaussian part of
the four point function and that there are no contribution from any
Gaussian terms.

The correlation functions involving one to four halos can now be
written as integrals over the spatial distribution of the number of
halos involved and their mass functions. We first write the four point
correlations within a single halo
\begin{eqnarray}
&& \eta^{1h}(\vecr_1,\vecr_2,\vecr_3,\vecr_4)  =  
\int d^3\vecx_1 \int dM_1 \nma y_i(\vecr_1-\vecx_1; M_1) ...
y_i(\vecr_4-\vecx_1; M_1)  \nonumber \\
\end{eqnarray}

The two-halo term contains two parts with one involving three point in
one halo and the fourth in the second halo and another part with two
points in each of the two halos. We write the 3-1 combination first
and then the 2-2 combination:
\begin{eqnarray}
&& \eta^{2h}(\vecr_1,\vecr_2,\vecr_3,\vecr_4)  = 
\Big[\int d^3\vecx_1 \int dM_1 \nma y_i(\vecr_1-\vecx_1;M_1)
... y_i(\vecr_3-\vecx_1;M_1)
\nonumber \\
&\times&  \int d^3\vecx_2 \int dM_2 \nma
y_i(\vecr_4-\vecx_2;M_2) \nonumber \\ 
&+& \int d^3\vecx_1 \int dM_1 \nma y_i(\vecr_1-\vecx_1;M_1)
y_i(\vecr_2-\vecx_1;M_1)  \int d^3\vecx_2 \nonumber \\
&\times& \int dM_2 \nmb y_i(\vecr_3-\vecx_2;M_2)
y_i(\vecr_4-\vecx_2;M_2)\Big] 
\zeta^2_{hh}(M_1,M_2;\vecx_1,\vecx_2)  \, ,
\end{eqnarray}
with $\zeta^2_{hh}$ as defined above. Through permutations, the 3
points in one and one point in 2nd term contains a total of four terms
while the second part with two point each contains a total of three
terms (see, \cite{SchBer91} 1991).
 
The three halo term can be written with two points in one halo and a
point in each of the other two halos
\begin{eqnarray}
&& \eta^{3h}(\vecr_1,\vecr_2,\vecr_3,\vecr_4)  = 
\int d^3\vecx_1 \int dM_1 \nma y_i(\vecr_1-\vecx_1;M_1)
y_i(\vecr_2-\vecx_1;M_1)
\nonumber \\
&\times&  \int d^3\vecx_2 \int dM_2 \nmb y_i(\vecr_3-\vecx_2;M_2)
 \int d^3\vecx_3 \int dM_3 \nmc y_i(\vecr_4-\vecx_3;M_3) \nonumber \\
&\times& \zeta^3_{hhh}(M_1,M_2,M_3;\vecx_1,\vecx_2,\vecx_3) \, ,
\end{eqnarray}
with the halo three-point correlation function
$\zeta^3_{hhh}$. Through permutations, with respect to the ordering of
$\vecr_1$ to $\vecr_4$, there are total of 6 terms involving three
halos that contribute to the four point correlations function.
 
The four halo term involves a point in each of the four halos
\begin{eqnarray}
&& \eta^{4h}(\vecr_1,\vecr_2,\vecr_3,\vecr_4)  = 
\int d^3\vecx_1 \int dM_1 \nma y_i(\vecr_1-\vecx_1;M_1)
\nonumber \\
&\times&  \int d^3\vecx_2 \int dM_2 \nmb y_i(\vecr_3-\vecx_2;M_2)
  \int d^3\vecx_3 \int dM_3 \nmc y_i(\vecr_4-\vecx_3;M_3)
\nonumber \\
&\times&  \int d^3\vecx_4 \int dM_4 \nmd y_i(\vecr_4-\vecx_4;M_4)
\zeta^4_{hhhh}(M_1,M_2,M_3,M_4;\vecx_1,\vecx_2,\vecx_3,\vecx_4) \, , 
\nonumber \\
\end{eqnarray}
and is proportional to the halo four point correlation function
$\zeta^4_{hhhh}$.
 
\subsection{Trispectrum}
 
In order to derive the dark matter trispectrum using halos, we again
break the contributions to four parts involving one to four halos
\begin{eqnarray}
&&T(\veck_1,\veck_2,\veck_3,\veck_4) =
T^{1h}(\veck_1,\veck_2,\veck_3,\veck_4)
+T^{2h}(\veck_1,\veck_2,\veck_3,\veck_4) \nonumber \\
&+&T^{3h}(\veck_1,\veck_2,\veck_3,\veck_4) +
T^{4h}(\veck_1,\veck_2,\veck_3,\veck_4)
\end{eqnarray}
and define the trispectrum such that
\begin{eqnarray}
&&\eta(\vecr_1,\vecr_2,\vecr_3,\vecr_4) = \int
\frac{d^3\veck_1}{(2\pi)^3} ...
\int \frac{d^3\veck_4}{(2\pi)^3} 
 T(\veck_1,\veck_2,\veck_3,\veck_4) \delta_D(\veck_{1234})
e^{i(\veck_1 \cdot \vecr_1 + ... + \veck_4 \cdot \vecr_4)} \, . \nonumber \\
\end{eqnarray}

We now  take the Fourier transform of each of the four terms
associated with the real space four-point correlation function.

First, the term involving one halo can be written as 
\begin{eqnarray}
&& \eta^{1h}(\vecr_1,\vecr_2,\vecr_3,\vecr_4)  =  
\int \frac{d^3\veck_1}{(2\pi)^3} ... \int
\frac{d^3\veck_4}{(2\pi)^3}
\int dM_1 \nma \hat{y}_i(\veck_1;M_1)
... \hat{y}_i(\veck_4;M_1) \nonumber \\
&\times&e^{i (\veck_1 \cdot \vecr_1 + ... + \veck_4 \cdot \vecr_4)} 
\int d^3\vecx_1 e^{-i \vecx_1 \cdot (\veck_1+...+\veck_4)} \, 
\end{eqnarray}
which can be used to write the one-halo contribution to the
trispectrum as
\begin{eqnarray}
&& T^{1h}(\veck_1,\veck_2,\veck_3,\veck_4)  = \int dM_1 \nma
\hat{y}_i(\veck_1;M_1) ...
\hat{y}_i(\veck_4;M_1) \nonumber \\
&&\equiv I^0_{4,iiii}(k_1,k_2,k_3,k_4)
\end{eqnarray}
  
Similarly, the term involving two halos is
\begin{eqnarray}
&& \eta^{2h}(\vecr_1,\vecr_2,\vecr_3,\vecr_4)  = 
\int \frac{d^3\veck_1}{(2\pi)^3} ... \int \frac{d^3\veck_4}{(2\pi)^3}
\nonumber \\
&&\int dM_1 \nma \hat{y}_i(\veck_1;M_1) ... \hat{y}_i(\veck_3;M_1) 
\int dM_2 \nmb \hat{y}_i(\veck_4;M_2) \nonumber \\
&\times& \int d^3\vecx_1 e^{-i \vecx_1 \cdot (\veck_1+...+\veck_3)}
\int d^3\vecx_2 e^{-i \vecx_2 \cdot \veck_4} 
\nonumber \\
&\times& e^{i (\veck_1 \cdot \vecr_1 + ... + \veck_4 \cdot \vecr_4)} 
\int \frac{d^3\veck}{(2\pi)^3}  P_{hh}(k,M_1,M_2)
e^{-i\veck \cdot (\vecx_1 - \vecx_2)} \nonumber \\
&+&  \int \frac{d^3\veck_1}{(2\pi)^3} ...
\int \frac{d^3\veck_4}{(2\pi)^3} \int dM_1 \nma
\hat{y}_i(\veck_1;M_1)
\hat{y}_i(\veck_2;M_1) \nonumber \\
&\times&\int dM_2 \nmb \hat{y}_i(\veck_3;M_2)  \hat{y}_i(\veck_4;M_2)
\nonumber \\
&\times&\int d^3\vecx_1 e^{-i \vecx_1 \cdot (\veck_1+\veck_2)}
\int d^3\vecx_2 e^{-i \vecx_2 \cdot (\veck_3+\veck_4)} \nonumber \\
&\times& e^{i (\veck_1 \cdot \vecr_1 + ... + \veck_4 \cdot \vecr_4)} 
\int \frac{d^3\veck}{(2\pi)^3}  P_{hh}(k,M_1,M_2)
e^{-i\veck \cdot (\vecx_1 - \vecx_2)} \,
\end{eqnarray}
where we have also expanded the two-point correlation function in real
space.
Integrating over $\vecx_2$ first and then $\veck$, we write
\begin{eqnarray}
&& \eta^{2h}(\vecr_1,\vecr_2,\vecr_3,\vecr_4)  = 
\int\frac{d^3\veck_1}{(2\pi)^3} ... \int \frac{d^3\veck_4}{(2\pi)^3}
\nonumber \\
&\times&\int dM_1 \nma \hat{y}_i(\veck_1;M_1)
... \hat{y}_i(\veck_3;M_1) 
\int dM_2 \nmb \hat{y}_i(\veck_4;M_2) \nonumber \\
&\times& \int d^3\vecx_1 e^{-i \vecx_1 \cdot (\veck_1+ ... +\veck_4)} 
e^{i (\veck_1 \cdot \vecr_1 + ... + \veck_4 \cdot \vecr_4)} 
P_{hh}(k_4,M_1,M_2) \nonumber \\
&+& \int \frac{d^3\veck_1}{(2\pi)^3} ... \int
\frac{d^3\veck_4}{(2\pi)^3} 
\int dM_1 \nma \hat{y}_i(\veck_1;M_1) \hat{y}_i(\veck_2;M_1)
\nonumber \\
&\times& \int dM_2 \nmb \hat{y}_i(\veck_3;M_2)  \hat{y}_i(\veck_4;M_2)
\int d^3\vecx_1 e^{-i \vecx_1 \cdot (\veck_1+ ... +\veck_4)}
\nonumber \\
&\times& e^{i (\veck_1 \cdot \vecr_1 + ... + \veck_4 \cdot \vecr_4)} 
P_{hh}(|\veck_3+\veck_4|,M_1,M_2) \, . 
\end{eqnarray}
 
We can now write the contribution to the two halo part of the
trispectrum as
\begin{eqnarray}
&&T^{2h}(\veck_1,\veck_2,\veck_3,\veck_4)  = 
\int dM_1 \nma \hat{y}_i(\veck_1;M_1) ... \hat{y}_i(\veck_3;M_1)
\nonumber \\
&\times&\int dM_2 \nmb \hat{y}_i(\veck_4;M_2) P_{hh}(k_4,M_1,M_2)
\nonumber \\
&+& \int dM_1 \nma \hat{y}_i(\veck_1;M_1) \hat{y}_i(\veck_2;M_1)
\nonumber \\
&\times&\int dM_2 \nmb \hat{y}_i(\veck_3;M_2)  \hat{y}_i(\veck_4;M_2)
P_{hh}(|\veck_3+\veck_4|,M_1,M_2) \nonumber \\
&\equiv& I^1_3(k_1,k_2,k_3)I_1^1(k_4)P^\lin(k_4) +
I_2^1(k_1,k_2)I_2^1(k_3,k_4)P^\lin(|\veck_{34}|) \nonumber \\
\end{eqnarray}

The term involving three halos is
\begin{eqnarray}
&& \eta^{3h}(\vecr_1,\vecr_2,\vecr_3,\vecr_4)  = \int
\frac{d^3\veck_1}{(2\pi)^3} ...
\int \frac{d^3\veck_4}{(2\pi)^3} \nonumber \\
&\times&\int dM_1 \nma \hat{y}_i(\veck_1;M_1)
\hat{y}_i(\veck_2;M_1)  
\int dM_2 \nmb \hat{y}_i(\veck_3;M_2) \nonumber \\
&\times&\int dM_3 \nm(M_3) \hat{y}_i(\veck_4;M_3) \nonumber \\
&\times&\int d^3\vecx_1 e^{-i \vecx_1 \cdot (\veck_1+\veck_2)} 
\int d^3\vecx_2 e^{-i \vecx_2 \cdot \veck_3} 
\int d^3\vecx_3 e^{-i \vecx_3 \cdot \veck_4} \nonumber \\
&\times&e^{i (\veck_1 \cdot \vecr_1 + ... +
\veck_4 \cdot \vecr_4)} \int \frac{d^3\veck^i}{(2\pi)^3} ...
\int \frac{d^3\veck^k}{(2\pi)^3} \delta_D(\veck_{ijk})  \nonumber \\
&\times&B_{hhh}(\veck^i,\veck^j,\veck^k;M_1,M_2,M_3) 
e^{i (\veck^i \cdot \vecx_1 + ... + \veck^k \cdot \vecx_3)} \, ,
\end{eqnarray}
where we have also expanded the three-point correlation function.
Integrating over $\vecx_1$, $\vecx_2$ and $\vecx_3$ simultaneously and
then $\veck^i$, $\veck^j$ and $\veck^k$, we write
\begin{eqnarray}
&& \eta^{3h}(\vecr_1,\vecr_2,\vecr_3,\vecr_4)  = 
\int\frac{d^3\veck_1}{(2\pi)^3} ... \int \frac{d^3\veck_4}{(2\pi)^3}
\nonumber \\
&\times&\int dM_1 \nma \hat{y}_i(\veck_1;M_1) \hat{y}_i(\veck_2;M_1)
\nonumber \\
&\times&\int dM_2 \nmb \hat{y}_i(\veck_3;M_2) 
\int dM_3 \nm(M_3) \hat{y}_i(\veck_4;M_3) \nonumber \\
&\times& e^{i (\veck_1 \cdot \vecr_1 + ... + \veck_4 \cdot \vecr_4)} 
\delta_D(\veck_{1234})
B_{hhh}(\veck_1+\veck_2,\veck_3,\veck_4;M_1,M_2,M_3) \nonumber \\
\end{eqnarray}
We can now write the contribution to the three halo part of the
trispectrum as
\begin{eqnarray}
&& T^{3h}(\veck_1,\veck_2,\veck_3,\veck_4)  = \int dM_1 \nma
\hat{y}_i(\veck_1;M_1)
\hat{y}_i(\veck_2;M_1) \nonumber \\
&\times&\int dM_2 \nmb \hat{y}_i(\veck_3;M_2) 
\int dM_3 \nmc \hat{y}_i(\veck_4;M_3) \nonumber \\
&\times&B_{hhh}(\veck_1+\veck_2,\veck_3,\veck_4;M_1,M_2,M_3)
\nonumber \\
&\equiv&
I_2^1(k_1,k_2)I_1^1(k_3)I_1^1(k_4)\left[B^\lin(|\veck_{12}|,k_3,k_4)
+ \frac{I_2^2}{I_2^1}P^\lin(k_3)P^\lin(k_4)\right] \nonumber \\
\end{eqnarray}
  
Finally, the term involving four halos is
\begin{eqnarray}
&& \eta^{4h}(\vecr_1,\vecr_2,\vecr_3,\vecr_4)  = \int
\frac{d^3\veck_1}{(2\pi)^3}
... \int \frac{d^3\veck_4}{(2\pi)^3} \nonumber \\
&\times&\int dM_1 \nma \hat{y}_i(\veck_1;M_1) 
\int dM_4 \nmd \hat{y}_i(\veck_4;M_4) \nonumber \\
&\times& \int d^3\vecx_1 e^{-i \vecx_1 \cdot \veck_1}
... \int d^3\vecx_4 e^{-i \vecx_4 \cdot \veck_4} \nonumber \\
&\times& e^{i (\veck_1 \cdot \vecr_1 + ... +
\veck_4 \cdot \vecr_4)} \int \frac{d^3\veck^i}{(2\pi)^3}
... \int \frac{d^3\veck^l}{(2\pi)^3} 
e^{i (\veck^i \cdot \vecx_1 + ... + \veck^l \cdot \veck_4)} \nonumber
\\
&\times& \delta_D(\veck_{ijkl})
T_{hhhh}(\veck^i,\veck^j,\veck^k,\veck^l;M_1,M_2,M_3,M_4) \, ,
\end{eqnarray}
where we have also expanded the four point correlation function of
halos.
Integrating over $\vecx_1$, $\vecx_2$, $\vecx_3$, $\vecx_4$
simultaneously and
then $\veck^i$, $\veck^j$, $\veck^k$ and $\veck^l$, we write
\begin{eqnarray}
&& \eta^{4h}(\vecr_1,\vecr_2,\vecr_3,\vecr_4)  = 
\int \frac{d^3\veck_1}{(2\pi)^3} ... \int \frac{d^3\veck_4}{(2\pi)^3}
\nonumber \\
&\times&\int dM_1 \nma \hat{y}_i(\veck_1;M_1)
... \int dM_4 \nmd \hat{y}_i(\veck_4;M_4) \nonumber \\
&\times& e^{i (\veck_1 \cdot \vecr_1 + ... +
\veck_4 \cdot \vecr_4)} \delta_D(\veck_{1234})
T_{hhhh}(\veck_1,\veck_2,\veck_3,\veck_4;M_1,M_2,M_3,M_4) \nonumber \\
\end{eqnarray}
 
We can now write the contribution to the four halo part of the
trispectrum as
\begin{eqnarray}
&& T^{4h}(\veck_1,\veck_2,\veck_3,\veck_4)  =  \nonumber \\
&&\int dM_1 \nma \hat{y}_i(\veck_1;M_1) 
... \int dM_4 \nmd \hat{y}_i(\veck_4;M_4) \nonumber \\
&\times& T_{hhhh}(\veck_1,\veck_2,\veck_3,\veck_4;M_1,M_2,M_3,M_4)
\nonumber \\
&\equiv&
I^1_1(k_1)I_1^1(k_2)I_1^1(k_3)I_1^1(k_4)\Big[T^\lin(\veck_1,\veck_2,\veck_3,\veck_4)
\nonumber \\
&+& \frac{I_1^2(k_1)}{I_1^1(k_1)}P(k_2)P(k_3)P(k_4)\Big] \, .
\end{eqnarray}

%\chapter{title of appendix B}
% contents of appendix B
% they may decide to give you crap about how this appears in the table
% of contents, because it just says "A" and "B".
% use the \nofiles trick described above, and add this line to the
% thesis.toc file in the right place:
% \contentsline {chapter}{Appendix } {}

%
% References (the thesis office prefers that to Bibliography...)
%

\newpage % added by bph, otherwise bibliography was one page too early
	 % in the Table of Contents.  

\addcontentsline{toc}{chapter}{References}

\begin{singlespace}  % added by bph, the references have to be single-spaced
% BiBTeX stuff, uncomment to use
%\bibliography{listofbibs}  % filename of bibtex file, do not include
			    % .bib

%\bibliographystyle{aas}  % the aas.bst file should be included with
			  % this template file.  
			  % This style file will occasionally cause 
			  % an over-full hbox error.  The text will
			  % then not fit within the margins.  To fix
			  % this I used \nocite{} and wrote the 
			  % citation in by hand.

%\bibliographystyle{apj}  % this is the other option.  It uses the
			  % AASTEX 5.0 style citiations (\citep,
			  % \citet, etc.) which makes it easier to go
			  % from your thesis to paper.  To use this,
			  % you need the stuff at the beginning of the
			  % template. 

% otherwise, standard citations in LaTeX

\end{singlespace}

\end{document}